\documentclass[12pt]{ruthesis}

\usepackage{algorithm}
\usepackage{algorithmic}
\usepackage{amsmath}
\usepackage{amsfonts}
\usepackage{amssymb}
\usepackage{bbm}
\usepackage{cite}
\usepackage{comment}
\usepackage{enumitem}
\usepackage{epsf}
\usepackage{epsfig}
\usepackage{latexsym}
\usepackage{graphics}
\usepackage{graphicx}
\usepackage{listings}
\usepackage{rotating}
\usepackage{subcaption}
\usepackage{theorem}
\usepackage{todonotes}
\usepackage{wrapfig}
\usepackage{xcolor}
\usepackage{hyperref}
\hypersetup{
	colorlinks=true,
	citecolor=red,
	linkcolor=blue,
	breaklinks
}

\title{Automata-Based Quantitative Reasoning}
\ctitle{Title here}
\author{Suguman Bansal}
\department{Computer Science}
\school{Rice University}
\degree{Doctor of Philosophy}

\committee {
         Rajeev Alur \\
         Zisman Family Professor of Computer and Information Science, University of Pennsylvania\and
         Konstantinos Mamouras\\
         Assistant Professor of Computer Science,  Rice University   \and 
         Moshe Y. Vardi, Chair \\
         University Professor, Karen Ostrum George Distinguished Service Professor in Computational Engineering, Rice University      
         \and 
         Peter J. Varman\\
         Professor of Electrical and Computer Engineering, Rice University 
         }

\address{Houston, Texas}
\donemonth{June} \doneyear{2020} \makeindex

\theoremheaderfont{\bfseries}

\newtheorem{lemma}{Lemma}[chapter] 
\newtheorem{corollary}{Corollary}[chapter] 
\newtheorem{conjecture}{Conjecture}[chapter] 
\newtheorem{theorem}{Theorem}[chapter] {\theorembodyfont{\rmfamily}  \newtheorem{proof}{Proof}}

{\theorembodyfont{\rmfamily}\newtheorem{Def}{Definition}[chapter]}

 \newcommand{\qed}{\hfill $\blacksquare$}

\usepackage[graphicx]{realboxes}
\usepackage{tikz}
\usetikzlibrary{automata,positioning}
\usepackage{tikz-qtree}
\usepackage{float}
\usepackage{floatflt}

\newcommand*{\cc}[1]{\mathsf{#1}}
\newcommand*{\cct}[1]{\textsf{#1}}
\renewcommand{\Re}{\mathbb{R}}

\newcommand*{\N}{\mathbb{N}}
\newcommand*{\Q}{\mathbb{Q}}
\renewcommand*{\L}{\mathcal{L}}
\newcommand*{\I}{\mathcal{I}}
\newcommand*{\Z}{\mathbb{Z}}
\renewcommand*{\O}{\mathcal{O}}
\newcommand{\GA}{\mathsf{GA}}
\newcommand{\GT}{\mathcal{G}_{\mathsf{T}}}
\newcommand{\ap}{\mathit{AP}}

\newcommand{\Java}{\mathsf{JAVA}}
\newcommand{\QuantIP}{\mathsf{QuIP}}

\newcommand{\cplus}{\textsf{C++}}
\newcommand{\QuIP}{\textsf{QuIP}}
\newcommand{\RABITt}{\textsf{RABIT}}
\newcommand{\glpsol}{\textsf{GLPSOL }}
\newcommand{\optimize}{\mathsf{VIOptimal}}
\newcommand{\compSatisfice}{\mathsf{CompSatisfice}}
\newcommand{\visatisfice}{\mathsf{VISatisfice}}

\newcommand*{\A}{\mathcal{A}}
\newcommand*{\G}{\mathcal{G}}
\newcommand*{\Obs}{\mathit{O}}

\newcommand*{\post}[2]{\mathsf{post}_{#1}(#2)}
\newcommand*{\Statess}{\mathit{S}}
\newcommand*{\Final}{\mathcal{F}}
\newcommand{\F}{\mathcal{F}}
\newcommand*{\StartState}{{\mathit{Init}}}

\newcommand*{\State}{S}
\newcommand*{\Start}{s_I}
\newcommand*{\wt}{\mathit{wt}}
\newcommand*{\sink}{\mathsf{sink}}
\newcommand*{\init}{v_\mathit{init}}
\newcommand{\compL}{\mathsf{compLow}}
\newcommand{\compU}{\mathsf{compUpper}}

\newcommand*{\LS}[1]{\mathsf{LimSup(#1)}}

\newcommand*{\DSum}[2]{\mathit{DS}({#1}, {#2})}

\newcommand*{\DSsucceqFord}[1]{#1}
\newcommand*{\DSsucceqFordf}[1]{\succ_{DS(d)}}
\newcommand*{\MaxC}{\mathit{maxC}}
\newcommand*{\MaxX}{\mathit{maxX}}
\newcommand*{\transwt}{\mathsf{cost}}
\newcommand*{\opt}{W}

\newcommand{\gapL}[1]{\mathsf{gapLower}(#1,k,p)}
\newcommand{\gapU}[1]{\mathsf{gapUpper}(#1,k,p)}
\newcommand{\DSumL}[1]{\mathsf{DSLow}(#1,k,p)}
\newcommand{\DSumU}[1]{\mathsf{DSUpper}(#1,k,p)}
\newcommand{\roundL}[1]{\mathsf{roundLow}(#1,k,p)}
\newcommand{\roundU}[1]{\mathsf{roundUpper}(#1,k,p)}

\newcommand*{\DSumDiff}[1]{\mathit{DS}^{-}(B, A, d, #1)}

\newcommand*{\Res}{\mathit{Res}}
\newcommand*{\floor}[1]{\lfloor #1 \rfloor}
\newcommand*{\ceil}[1]{\lceil #1 \rceil}
\newcommand*{\CSum}{\mathit{CSum}}

\newcommand{\univ}{baseline }
\newcommand{\U}{\mathcal{B}}

\renewcommand*{\lim}[3]{\text{lim}_{#1 \rightarrow #2} #3}
\renewcommand*{\liminf}[3]{\text{lim inf}_{#1 \rightarrow #2} #3}
\renewcommand*{\limsup}[3]{\text{lim sup}_{#1 \rightarrow #2} #3}
\newcommand*{\LA}[1]{\mathsf{LimAvg}(#1)}
\newcommand*{\PLA}[1]{\mathsf{PrefixAvg}(#1)}
\newcommand*{\LASup}[1]{\mathsf{LimAvgSup}(#1)}
\newcommand*{\LAInf}[1]{\mathsf{LimAvgInf}(#1)}
\newcommand*{\Av}[1]{\mathsf{Avg}(#1)}
\newcommand*{\PASuc}{\mathsf{PA}}

\newcommand*{\Sum}[1]{\mathsf{Sum}(#1)}
\newcommand*{\ExistFin}[3]{\exists^{f} #1,  \Sum{#2[0,#1-1]} \geq \Sum{#3[0,#1-1]}}
\newcommand*{\ExistInf}[3]{\exists^{\infty} #1,  \Sum{#2[0,#1-1]} > \Sum{#3[0,#1-1]}}

\newcommand*{\less}{\mathit{DomProof}}
\newcommand*{\first}{\mathit{Dom}}

\newcommand*{\DSInclusion}{\mathsf{InclusionRegular}}
\newcommand*{\AugmentWtAndLabel}{\mathsf{AugmentWtAndLabel}}
\newcommand*{\AugmentWt}{\mathsf{AugmentWt}}
\newcommand*{\MakeProduct}{\mathsf{MakeProduct}}
\newcommand*{\Intersect}{\mathsf{Intersect}}
\newcommand*{\Project}{\mathsf{FirstProject}}
\newcommand*{\ProjectOutWt}{\mathsf{ProjectOutWt}}

\newcommand*{\SepRuns}{\mathsf{UniqueId}}
\newcommand*{\CompareRuns}{\mathsf{Compare}}
\newcommand*{\EnsureRuns}{\mathsf{Ensure}}

\newcommand*{\lessI}{\mathit{DomWithWitness}}
\newcommand*{\firstI}{\mathit{Dom}}
\newcommand*{\MakeProductI}{\mathsf{MakeProductSameAlpha}}

\newcommand*{\BCVt}{\textsf{BCV }}
\newcommand*{\DetLPt}{\textsf{DetLP }}
\newcommand*{\BCV}{\mathsf{BCV}}

\newcommand*{\LI}{\textsf{LI}}
\newcommand*{\LP}{\textsf{LP}}
\newcommand*{\DS}{\textsf{DS}}

\newcommand*{\maximal}[1]{\mathsf{Maximal}(#1)}

\newcommand*{\counter}[1]{\mathsf{Counterexample}(#1)}
\newcommand{\strict}{\mathsf{strct}}
\newcommand{\nstrict}{\mathsf{nstrct}}
\newcommand{\R}{\mathsf{inc}}

\newcommand*{\bad}[1]{\mathit{bad}(#1)}
\newcommand*{\pre}[2]{#1[#2]}
\newcommand*{\suf}[2]{#1[#2\dots]}
\newcommand{\gap}[1]{\mathsf{gap}(#1,d)}
\newcommand{\thresh}{\mathsf{T}}
\newcommand{\baut}[3]{\mathsf{bad}(#1, #2,#3)}

\newcommand*{\approxInc}{\mathsf{anytimeInclusion}}
\newcommand*{\rationalInc}{\mathsf{DSInclusion}}

\newcommand{\lowerInc}{\mathsf{lowApproxDSInc}}
\newcommand{\upperInc}{\mathsf{upperApproxDSInc}}
\newcommand{\MakeDifProduct}{\mathsf{productDif}}
\newcommand{\dominatedWitness}{\mathsf{dominatedWitness}}
\newcommand{\dominated}{\mathsf{dominated}}

\newcommand*{\FSum}[1]{\mathsf{Sum}({#1})}

\begin{document}

  \begin{frontmatter}
   \pagenumbering{roman}
   \maketitle
   \newpage
\thispagestyle{plain}
\pagenumbering{gobble}

 \par\vspace*{.3\textheight}{\centering Dedicated to my family,\vspace*{0.3in}\\My mother, Anita, who now resides among the stars\\My father, Rajeev\\ My brother, Sanchit\par}
  
   \thispagestyle{empty}
\begin{abstract}

The analysis of  {\em quantitative properties} of computing systems, or {\em quantitative analysis} in short,  is an emerging area in automated formal analysis. Such properties address aspects such as costs and rewards, quality measures, resource consumption, distance metrics, and the like. So far, several applications of quantitative analysis have been identified, including formal guarantees for reinforcement learning, planning under resource constraints, and verification of (multi-agent) on-line economic protocols.

Existing solution approaches for problems in quantitative analysis suffer from  two challenges that adversely impact the theoretical understanding of quantitative analysis, and large-scale applicability due to limitations on scalability. These are the {\em lack of generalizability}, and {\em separation-of-techniques}. Lack of generalizability refers to the issue that solution approaches are often specialized to the underlying {\em cost model} that evaluates the quantitative property. Different cost models deploy such disparate algorithms that there is no transfer of knowledge from one cost model to another. Separation-of-techniques refers to the inherent dichotomy in solving problems in quantitative analysis. Most algorithms comprise of two phases: A {\em structural phase}, which reasons about the structure of the quantitative system(s) using techniques from automata or graphs; and a {\em numerical phase}, which reasons about the quantitative dimension/cost model using numerical methods. The techniques used in both phases are so unlike each other that they are difficult to combine, forcing the phases to be performed sequentially, thereby impacting scalability.

This thesis contributes towards a novel framework that addresses these challenges. The introduced framework, called {\em comparator automata} or {\em comparators} in short, builds on automata-theoretic foundations to generalize across a variety of cost models. The crux of comparators is that they enable automata-based methods in the numerical phase, hence eradicating the dependence on numerical methods. In doing so, comparators are able to integrate the structural and numerical phases. On the theoretical front, we demonstrate that comparator-based solutions have the advantage of generalizable results, and yield complexity-theoretic improvements over a range of problems in quantitative analysis. On the practical front, we demonstrate through empirical analysis that comparator-based solutions render more efficient, scalable, and robust performance, and hold the ability to integrate quantitative with qualitative objectives. 

\end{abstract}

   \chapter*{List of Publications}

This thesis is based on the following publications.  Publications~\ref{one}-\ref{two} are yet to appear  in an official proceedings at the time of thesis submission. 

\begin{enumerate}

	\item\label{one} \href{}{\textbf{\textcolor{blue}{On the analysis of quantitative games}}}\\
		Suguman Bansal, Krishnendu Chatterjee, and Moshe Y. Vardi 
		
	\item\label{two} \href{}{\textbf{\textcolor{blue}{Anytime discounted-sum inclusion}}}\\
	Suguman Bansal and Moshe Y. Vardi

	\item \href{https://www.cs.rice.edu/~sb55/Papers/CAV19.pdf}{\textbf{\textcolor{blue}{Safety and co-safety comparator automata for discounted-sum inclusion}}}\\
	Suguman Bansal and Moshe Y. Vardi \\
	In Proceedings of International Conference on Computer-Aided Verification (CAV) 2019
	
	\item \href{https://www.cs.rice.edu/~sb55/Papers/CAV18b.pdf}{\textbf{\textcolor{blue}{Automata vs linear-programming discounted-sum inclusion}}}\\
	Suguman Bansal, Swarat Chaudhuri, and Moshe Y. Vardi \\
	In Proceedings of International Conference on Computer-Aided Verification (CAV) 2018 

	\item \href{https://www.cs.rice.edu/~sb55/Papers/FoSSaCS18.pdf}{\textbf{\textcolor{blue}{Comparator automata in quantitative verification}}} \\
	Suguman Bansal, Swarat Chaudhuri, and Moshe Y. Vardi\\
	In Proceedings of International Conference on Foundations of Software Science and Computation Structures (FoSSaCS) 2018\\
	(\href{https://arxiv.org/abs/1812.06569}{\textbf{\textcolor{blue}{Extended version}}} with additional results on Arxiv)
	
\end{enumerate}

   \chapter*{Acknowledgements}

Foremost, I would like to thank my advisor and mentor Moshe Vardi for his unwavering support, constant guidance, and encouragement to pursue my ideas. He took me under his wings during the most difficult times I have faced, both academic and personal. 
Needless to say, there has been no looking back since. Thank you for giving me the freedom for ample exploration while also nudging me towards the right path. Most importantly, thank you for believing in me. I hope to be the advisor you have been to me to another dreamer someday.

I am fortunate to have worked with an excellent thesis committee: Rajeev Alur, Konstantinos  Mamouras, and Peter Varman. 
Their suggestions, feedback, and thorough evaluations of this thesis have led to numerous improvements and pursuits for future work. 
Thanks go to Krishnendu Chatterjee and Swarat Chaudhuri, collaborators on parts of this thesis.
Their feedback on my doctoral research from the very beginning has been crucial in shaping the course. 

I am grateful to have found mentors in Swarat and Kedar Namjoshi. 
Right from building my foundations  in Computer Science, Swarat has helped me navigate through the bigger career decisions.
Incidentally, it was upon his advice that I took up the internship offer from Bell Labs to work with Kedar, which has turned out to be one of my best collaborative experiences. The daily, intense brainstorming sessions with Kedar were tiring yet immensely fulfilling.
That said, a respite from the sweltering Houston weather takes the cherry on the cake for both of my summers at Bell Labs.

During the Ph.D., I had the opportunity to collaborate with several brilliant researchers from around the globe: Rajeev Alur, Shaull Almagor, Swarat Chaudhuri, Krishnendu Chatterjee, Dror Fried, Yong Li, Kuldeep Meel, Kedar Namjoshi, Yaniv Sa'ar, Lucas Tabajara, Moshe Vardi, Andrew Wells.
I am, perhaps, most partial to collaborations with my peers at Rice - Yong Li and Lucas Tabajara - where discussions would begin in our offices but end in Valhalla (\textit{Viva Valhalla!}).

The fact that there is a large overlap between my friends and (extended) research group LAPIS is a testament to how instrumental they have been in this journey. Dror Fried has been so much more than a friend to me. I am eternally grateful to his wife Sagit, him, and their three little kids (not so little anymore) for giving me an abode in their hearts. 
In a discipline that suffers from a dearth of women, I have been very lucky to have Afsaneh Rahbar and Shufang Zhu by my side. 
These women have truly made my successes and failures their own, as we continue to inspire each other with our \textit{``We can do it"} chant. 
Because Aditya and I share a history of borrowing each other's sentences, 
I'll paraphrase him here - Thank you to Yong Li and Aditya Shrotri for keeping me company during the late nights and weekends in the office, the car rides home, coffee breaks, lunches, and dinners.
I am humbled to share a special siblings-like bond with Vu Phan. I have relied on Jeffery Dudek, Antonio Di Stasio, my officemate Lucas Tabajara, Kevin Smith, Abhinav Verma, and Zhiwei Zhang for cerebral discussions and simple reasons for laughter.

My life outside the colorful walls of Duncan Hall would have been so lack-luster had it not been for the friendships of Priyadarsini (Priya) Dasari, Vaideesh Logan, Rakesh Malladi, Sushma Sri Pamulapati, and the {\em folks@IMT}.
In addition to keeping our apartment in pristine condition, my flatmate Priya has made me a better cook, and counseled me through personal and work-related turmoils;  all of this over a steaming cup of {\em chai}.
The {\em folks@IMT} have been a source of strength in the times of social-distancing/isolation during the ongoing Covid-19 pandemic. In these uncertain times, their small gestures have gone a long way to ease the anxiety around the global upheaval. Thank you all for making Houston my home. 

My friends from undergraduate and school have been available in moments of despair and self-doubt. I cannot thank Saheli, Siddharth, Siddhesh, and Visu enough for having been just a phone call away for almost a decade.

Of all the things I had imagined graduate school would entail, meeting the love of my life was not on the agenda. Yet, here we are. Whether we are ten centimeters or ten thousand miles apart, Kuldeep can always make me smile. Thank you, {\em ghano saaro}, for brightening my day, every day.

Finally but most importantly, 
I owe my deepest gratitude to my parents, Anita Bansal and Rajeev Kumar, and brother Sanchit Bansal. 
Right since I can remember, they filled me with curiosity, creativity, and imagination, and
have shunned societal norms so that I could run, stumble, and chase after my dreams. Their love and support have been unconditional. It is to them that I dedicate this thesis. 

Yet there is a pit in my stomach. We lost my mother, Anita, four years ago.     
She was my loudest champion and strictest critic. 
Never had I imagined crossing this milestone without her cheering from the stands. 
But I am not upset, because I know that wherever she is, she is very proud of her \textit{guriya}. {\em We miss you every day, Mummy!}

   \tableofcontents
   \listoffigures
   \listofalgorithms
   
  \end{frontmatter}
\pagenumbering{arabic}

\linespacing{1.7}

\chapter{Introduction}
\label{ch:intro}

\sloppy
{\em Formal methods} offers rigorous mathematical guarantees about properties of systems. Two cornerstones of formal methods are (a) {\em Verification}: Given a system and a property, does the system satisfy the property, and (b) {\em Synthesis}:  Given a property, does there exist a system that satisfies it? These strong guarantees come at the price of high computational complexity and practical intractability. 
Yet, diligent efforts of the last few decades have resulted in the emergence of formal methods as an integral asset in the development of modern computing systems. These systems range from hardware and software, to  security and cryptographic protocols, to safe autonomy.
In most of these use-cases, systems are evaluated on their {\em functional properties} that describe temporal system behaviors, safety or liveness conditions and so on. 

This thesis looks into the extension of formal methods with {\em quantitative properties} of systems, leading towards scalable and efficient verification and synthesis {\em with quantitative properties}. 


\section{Quantitative analysis}
The notion of correctness with functional properties is Boolean.
The richness of modern systems justifies properties that are {\em quantitative}. These can be thought to extend functional properties, since in this case executions are assigned to a real-valued cost using a {\em cost model} that captures the quantitative dimension of interest.
Quantitative properties can reason about aspects such as quality measures, cost- and resource- consumption, distance metrics and the like~\cite{almagor2013formalizing, alur2017introduction,baierprobabilistic,Kwi07}, which functional properties cannot easily express. 
For example, whether an arbiter grants every request is a functional property; But the promptness with which an arbiter grants requests is  a quantitative property.

The analysis of  quantitative properties of computing systems, or {\em quantitative analysis} in short,  is an emerging area in automated formal analysis.
It enables the formal reasoning about features such as timeliness, quality, reliability and so on. 
Quantitative properties have been used in several forms. These include probabilistic guarantees about the correctness of hardware or software~\cite{baierprobabilistic,Kwi07,kwiatkowska2010advances}; use of distance-metrics to produce low-cost program repair~\cite{d2016qlose,hu2018syntax}, generate feedback with few corrections in computer-aided education~\cite{singh2013automated}, and automated generation of questions of similar difficulty for MOOCs~\cite{ahmed2013automatically}; 
in planning for robots with resource-constraints~\cite{he2017reactive}, with hard and soft constraints where the soft constraints are expressed quantitatively~\cite{lahijanian2015time,wakankar2019dcsynth}, and generate plans of higher quality~\cite{almagor2013formalizing,bloem2009better}.

The cornerstones of formal analysis of quantitative properties of systems are (a). Verification of quantitative properties, and (b) Synthesis from quantiative properties. This thesis focuses on the two problems that form the fundamentals of these two cornerstones. These problems are {\em Quantiative inclusion}
and  {\em Solving quantitative games}, respectively, as detailed below: 

\paragraph{From ``Verification of quantitative properties" to ``Quantiative inclusion".}

In the verification of functional properties, the task is to verify if a system $S$ satisfies a functional property  $P$, possibly represented by a temporal logic~\cite{pnueli1977temporal}. 
Traditionally, $S$ and $P$ are interpreted as sets of (infinite-length) executions, and $S$ is determined to satisfy $P$ if $S \subseteq P$. 
The task, however, can also be framed in terms of a comparison between executions in $S$ and $P$. Suppose
an execution $w$ is assigned a cost of 1 if it belongs to the language of the system
or property, and 0 otherwise. Then determining if $S\subseteq P$ amounts to checking whether the cost of every execution in $S$ is less than or equal to its cost in $P$~\cite{baier2008principles,vardi1986automata}.

The verification of quantitative properties is an extension of the same from functional properties. In the quantitative scenario, executions in the system $S$ and property $P$ are assigned a real-valued cost, as opposed to a 0 or 1 cost. Then, as earlier, verification of quantitative properties amounts to checking whether the cost of every execution in $S$ is less than or equal to its cost in $P$. 
When the system and property are modeled/represented by a finite-state (quantitative) abstraction, this problem is referred to as {\em quantitative inclusion}~\cite{chatterjee2010quantitative,filiot2012quantitative}. Formally speaking, {\em given two finite-state (quantitative) abstractions $S$ and $P$, quantitative inclusion determines whether the cost of every execution in $S$ is less than or equal to its cost in $P$.}

 \begin{figure}[t]
 	\begin{subfigure}{0.45\textwidth}
 		
 		\centering
 		\begin{tikzpicture}[shorten >=1pt,node distance=2.5cm,on grid,auto] 
 		\node[state,initial, accepting] (q_0)   {\textsf{OFF}}; 
 		\node[state, accepting] (q_1) [right=of q_0] {\textsf{ON}}; 
 		\path[->] 
 		(q_0)   edge [loop above ] node { \textsf{off, 0}} ()
 		edge [bend left = 20]node { \textsf{on, slow, 10}}  (q_1)
 		(q_1) edge  [loop above] node  {\textsf{on,2}} ()
 		edge [bend left = 20]node {\textsf{off, slow, 10}} (q_0);	
 		\end{tikzpicture}
 		\caption{Motor prototype 1}
 		\label{Fig:Motor1}

 	\end{subfigure}
 	\hfill
 	\begin{subfigure}{0.54\textwidth}
 		
 		\centering
 		\begin{tikzpicture}[shorten >=1pt,node distance=3cm,on grid,auto] 
 		
 		\node[state,initial, accepting] (q_0)   { \textsf{OFF}}; 
 		\node[state, accepting] (q_2) [above left =of q_0] {\textsf{slow}}; 
 		\node[state, accepting] (q_1) [above right =of q_0] {\textsf{ON}}; 
 		
 		\path[->] 
 		(q_0)   edge [loop below] node { \textsf{off, 0}} ()
 		edge [bend left = 5] node { \textsf{on, 10}}  (q_1)
 		edge [bend left = 20] node {\textsf{slow, 5}} (q_2)
 		(q_1) edge  [loop above] node  {\textsf{on,2}} ()
 		edge [bend left = 20] node { \textsf{off, 10}} (q_0)
 		edge [bend right = 35] node {\textsf{slow, 5}} (q_2)
 		(q_2) edge  [bend left = 3]  node [above]{ \textsf{off, 5}} (q_0)
 		edge [bend right = 5] node  {\textsf{on, 5}} (q_1)
 		edge [loop above ] node [above]{\textsf{slow, 1}} ();
 		
 		\end{tikzpicture}
 		\caption{Motor prototype 2}
 		\label{Fig:Motor2}	
 	\end{subfigure}
 	\caption[Quantitative inclusion: Which motor prototype is more efficient?]{Quantitative inclusion: Which motor prototype is more efficient? Weights on transitions indicate amount of energy consumed in that single step. Cost of an execution is the limit-average of costs along all transition on the execution, i.e., cost model is limit-average}
 	\label{Fig:MotorPrototype}
 	
 \end{figure}
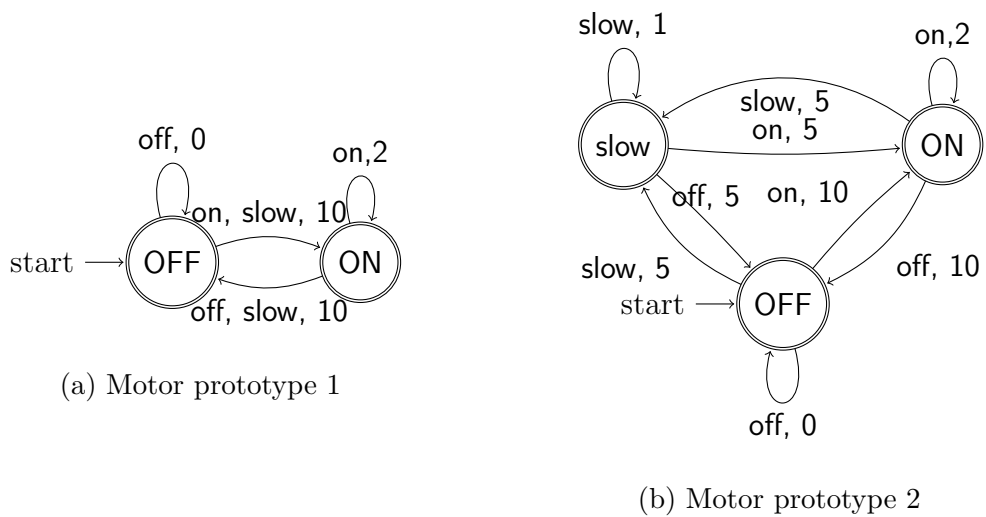
 
 For verification purposes, the systems are often modeled by finite-state (quantitative) abstractions. Even though properties are typically represented by quantitative logics, several of them can be converted to the aforementioned abstractions~\cite{de2004model,degorre2010energy}. Therefore, verification from quantitative properties reduces to quantitative inclusion. More generally, 
quantitative inclusion is applied to compare systems on their quantitative dimensions (Illustration in Fig~\ref{Fig:MotorPrototype}), and in the analysis of rational behaviors in multi-agent systems with reward-maximizing agents~\cite{abreu1988theory,MastersThesis}.
\qed

\paragraph{From ``Synthesis from quantitative properties" to ``Solving quantitative games".} 
Synthesis from a functional properties in an uncertain environment is perceived as a game between an uncontrollable environment and the system that we wish to designed. In this game, the environment and the system have the opposing objectives of guaranteeing that the resulting play flouts and satisfies the functional property, respectively.
The system can be constructed if it is able to win the game against the environment. As earlier, synthesis from quantitative properties extends the later by assigning real-valued costs to the plays~\cite{he2017reactive,bloem2009better,chakrabarti2005verifying}.  

As a concrete instantiation, suppose we are to design a plan for a robot to patrol a grid without colliding into another (movable) robot present on the grid. 
Now, as our robot moves around the grid, it expends the charge on its battery, and will have to enter the recharge stations every now and then.  Therefore, while planning for our robot, we need to guarantee that its battery charge never goes below level 0. Note that this criteria is a quantitative constraint.

Such problems of planning in an uncertain environment with quantitative constraints can be reduced to solving a min-max optimization on a two-player {\em quantitative graph game}. The two players refer to the system for which the plan is being designed, and an uncontrollable environment which is assumed to be adversarial to the objective of the system player.
In these games, players take turns to pass a token along the {\em transition relation} between the states. As the token is pushed around, the play accumulates weights along the transitions using the underlying cost model $f:\mathbb{Z}^\omega\rightarrow \Re$. 
One player maximizes the cost while the other minimizes it. 
\qed

\paragraph{}
Both of the problems extend their functional counterparts. As a result, quantitative analysis inherits their high complexity and practical intractability~\cite{almagor2011whats,baier2018model}. However, algorithms that solve over functional properties cannot be directly applied to solve with quantitative properties. Therefore, the developments in functional properties have limited impact on solving with qualitative properties. To this end, this thesis focuses on development of efficient and scalable solutions for quantitative analysis.

\section{Challenges in quantitative analysis}

This thesis begins with identifying two broad challenges that obstruct the theoretical understanding and algorithmic development in quantitative analysis.

\paragraph{Challenge 1. {Lack of generalizability}} 
 refers to the issue that solution approaches of problems in quantitative analysis are  specialized to the underlying cost model. Different cost models deploy disparate techniques to solve. This disparity restricts the transfer of knowledge of solving problems over one cost model to another cost model. 
 Furthermore, owning to the large variety of cost models, soon it will become too cumbersome to digest progress made across cost models. 
 
A concrete instantiation of this challenges appears in problems over {\em weighted $\omega$-automata}~\cite{droste2009handbook}.  Weighted $\omega$-automaton~\cite{droste2009handbook} are a well-established finite-state (quantiative) abstraction used to model quantitative system, as illustrated in Fig~\ref{Fig:MotorPrototype}. 
An execution is an infinite sequence of labels that arise from an infinite sequence of subsequent transitions. 
Weighted $\omega$-automata assign a real-valued cost to all executions over the machine using a cost model $f:\mathbb{Z}^\omega\rightarrow\Re$. The cost of an execution is assigned by applying the cost model $f $ to the weight-sequence arising from transitions along the execution. In case the transition relation is non-deterministic, each execution may have multiple runs. In these cases, the cost of the execution is resolved by taking the infimum/supremum of cost along all runs.

In case of quantitative inclusion over weighted $\omega$-automata, the lack of generalizable is apparent from the diversity in complexity-theoretic results. While quantitative inclusion is \cct{PSPACE}-hard, there is ample variance in upper bounds. Quantitative inclusion is \cct{PSPACE}-complete under limsup/liminf~\cite{chatterjee2010quantitative}, undecidable for limit-average~\cite{degorre2010energy}, and decidability is unknown for discounted-sum in general but its decidable fragments are \cct{EXPTIME}~\cite{boker2014exact,chatterjee2010quantitative}.

Yet another well-studied problem over weighted $\omega$-automata is to determine if there exists an execution with cost that exceeds a constant threshold value~\cite{droste2009handbook,de2004linear}. The problem finds applications in verification from quantitative logics~\cite{de2004model,degorre2010energy}, planning with quantitative constraints in closed environments~\cite{lahijanian2015time}, low-cost program repair.	
The landscape is even less uniform on this problem. Known solutions involve diverse techniques ranging from linear-programming for discounted-sum~\cite{andersson2006improved} to negative-weight cycle detection for limit-average~\cite{karp1978characterization} to  computation of maximum weight of cycles for limsup/liminf~\cite{chatterjee2009expressiveness}.

{\em The question that this thesis asks is whether one can develop a unifying theory for quantitative analysis that generalizes across a variety of cost models}. It would be too navie to expect that all cost models can be brought under a single theoretical framework. So, the natural question to ask is about boundary conditions over cost models that can and cannot be generalized. 
\qed

\paragraph{Challenge 2. {Separation-of-techniques}} refers to the inherent dichotomy in existing algorithms for problems in quantitative analysis.
Most algorithms comprise of two phases. In the first phase, called the {\em structural phase}, the algorithm reasons about the structure of the quantitative system(s) using techniques from automata or graphs. In the second phase, called the {\em numerical phase}, the algorithm reasons about the quantitative dimension/cost model using numerical methods. The techniques used in both these phases are so unlike each other that it is difficult to solve them {\em in tandem}, and hence the phases have to be performed sequentially.
The issue with this is it isn't uncommon to observe an exponential blow-up in the first phase. 
This poses a computational barrier to solvers of numerical methods in the second phase, that despite being of industrial strength are still unable to operate on exponentially larger inputs. Hence, this separation-of-techniques affects the scalability of algorithms in quantitative analysis. 

As a concrete instantiation, consider the problem of planning in an uncertain environment under quantitative constraints. 
Here, the {structural phase} consists of the reduction from planning to the {\em quantitative graph game}, while the numerical phase consists of solving the min-max optimization on the game. 
In case of our robot planning example, the quantitative game has as many states as there are configurations of the grid. So, if the grid has a size of $n\times n$, then the quantitative graph game will have $\O(n^4)$ states. Therefore, the size of the min-max optimization problems grows rapidly, and will limit scalability of planning.

{\em The question this thesis asks is whether one can design an algorithmic approach for quantitative analysis that {\em integrates} the two phases as opposed to the existing separation-of-techniques , and whether an integrated approach will lead to efficient and scalable solutions}.
 
Existing work on {\em probabilistic verification} presents initial evidence of the existence of integrated approaches~\cite{KNP04b,Par02}.
Probabilistic verification incorporates state-of-the-art symbolic data structures and algorithms based on BDDs (Binary Decision Diagrams)~\cite{bryant1986graph} or MTBDD (Multi-Terminal Binary Decision Diagrams). These Decision Diagrams can encode and manipulate both the the structural and numerical aspects of underlying problem at once, and hence are an integrated method. The disadvantage of Decision Diagrams is that their performance is unpredictable, as they are known to exhibit large variance in runtime and memory consumption. Therefore, existing integrated approaches fall short of wholesome improvement in  the theory and practice of problems in quantitative analysis. 
\qed

\section{Thesis contributions}

Existing solution approaches suffer from the lack of generalizability and separation-of-techniques, adversely impacting a clear theoretical understanding of quantitative analysis, and large-scale applicability due to limitations on scalability. 
This thesis contributes towards a novel theoretical framework
that addresses both of the challenges, and demonstrates  utility on problems of quantitative inclusion and solving quantitative games.
 The introduced framework, called {\em comparator automata} or {\em comparators} in short, builds on automata-theoretic foundations to generalize across a variety of cost models. The crux of comparators is that they substitute the numerical analysis phase with automata-based methods, and hence naturally offer an integrated method for quantitative analysis. 
 In all, we show that comparator-based algorithms have the advantages of generalizable results, 
yield complexity-theoretic and algorithmic development, and  render practically scalable solutions. 

The detailed contributions are as follows:

\begin{enumerate}
	\item \textbf{Comparator automata}:  The approach takes the view that the comparison of costs of system executions with the costs on other executions is the fundamental operation in quantitative reasoning, and hence that should be brought to the forefront.   
	To this end, {\em comparator automata} as an automata-theoretic formulation of this form of comparison between weighted sequences. Specifically, comparator
	automata (comparators, in short), is a class of automata that read pairs of infinite weight sequences synchronously, and compare their costs in an online manner. We show that cost models such as limsup/liminf and discounted-sum with integer discount factors permit comparators that require af finite amount of memory, while comparators for limit-average require an infinite-amount of memory.
	We show that results and algorithms over weighted $\omega$-automata and  quantitative games generalize over cost models that permit comparators with finite memory.
	Lastly, since comparators are automata-theoretic, these algorithms are integrated in approach. 
\end{enumerate} 
We illustrate further benefits of comparator-based approaches by investigating the discounted-sum cost model, which is a fundamental cost model in decision-making  and planning domains including reinforcement learning, game theory, economics
\begin{enumerate}[resume]
	
	\item \textbf{Quantitative inclusion over 
	discounted-sum}: The decidability of quantitative inclusion over discounted-sum, DS inclusion, is unknown when the discount factor is an integer. This has been an open problem for almost 15 years now. When the discount factor  is an integer, DS inclusion is known to have an \cct{EXPTIME} upper bound and \cct{PSPACE} lower bound. Hence, its exact complexity is unknown. 

	Comparator-based arguments make resolutions towards both of these questions. We are able to prove that when the discount factor is an integer, then DS inclusion is indeed \cct{PSPACE}-complete. When the discount factor is not an integer, then we are not able to resolve the decidability debate. However we are able to provide an anytime algorithm for the same, hence rendering a pragmatic solution. 

	Last but not the least, we demonstrate that by leveraging its automata-theoretic foundations, comparator-based algorithms for DS inclusion are more scalable and efficient in practice that existing methods. 
	
	\item \textbf{Quantitative games over discounted-sum}:
	The benefits of $\omega$-regular and safety/co-safety properties of comparators for discounted-sum continue in quantitative games as well. We show that comparator-based solutions to quantitative games are more efficient in theory and more efficient in practice. In addition, they broaden the scope of quantitative games by extending with temporal goals.

\end{enumerate}

\section{Outline}

Chapter~\ref{ch:background} introduces the necessary notation and background, and should be treated as an index.
The technical sections of this thesis have been split into three parts.

\begin{itemize}[label=Part ]
	\item \ref{part:framework} 	
	lays down the theoretical foundations of {\em comparator automata}, our automata-based technique to obtain integrated methods to solve problems in quantitative reasoning. 
	The formal introduction of comparator automata, and utility of $\omega$-regular comparators in designing generalizable solutions is in Chapter~\ref{ch:Comparator}.  Chapter~\ref{ch:applicationcomparator} furthers the study by investigating aggregate functions that permit $\omega$-regular functions. 
	
	\item~\ref{part:DSinclusion} undertakes an investigation of discounted-sum automata, following the result that DS comparators are $\omega$-regular iff their discount factor is an integer. The Part is split into three chapters. Chapter~\ref{ch:empirical} and Chapter~\ref{ch:safetycosafety} are concerned with DS inclusion under integer discount factors. Safety and co-safety comparators for DS are introduced in Chapter~\ref{ch:safetycosafety}. Chapter~\ref{ch:anytime} is concerned with designing anytime algorithm for DS inclusion with non-integer discount factors.

	\item \ref{part:games} consists of Chapter~\ref{ch:integergames}. It studies the advantages of safety/co-safety automata comparator automata in discounted-sum games. 
	
\end{itemize}
Last but not the least, the thesis concludes with a discussion of future directions in Chapter~\ref{ch:conclude}.
\chapter{Background}
\label{ch:background}

\section{Automata and formal languages}

\subsection*{B\"uchi automaton}

B\"uchi automaton correspond to acceptors of words that require a finite-amount of memory.
Formally,  a (finite-state) {\em B\"uchi automaton}~\cite{thomas2002automata} is a tuple
  $\A = (\Statess$, $\Sigma$, $\delta$, $\StartState$, $\Final)$, where
  $ \Statess $ is a finite set of {\em states}, $ \Sigma $ is a finite {\em input alphabet},   $ \delta \subseteq (\Statess \times \Sigma \times \Statess) $ is the   {\em transition relation}, $ \StartState \subseteq \Statess $ is the set of {\em initial states}, and $ \Final \subseteq \Statess $ is the set of {\em accepting states}~\cite{thomas2002automata}.

A B\"uchi automaton is {\em deterministic} if for all states $ s $ and
inputs $a$, $ |\{s'|(s, a, s') \in \delta \textrm{ for some $s'$} \}|
\leq 1 $ and $|\StartState|=1$. Otherwise, it is {\em nondeterministic}.
A B\"uchi automaton is {\em complete} if for all states $ s $ and
inputs $a$, $ |\{s'|(s, a, s') \in \delta \textrm{ for some $s'$} \}| \geq 1$. 
For a word $ w = w_0w_1\dots \in \Sigma^{\omega} $, a {\em run} $ \rho$ of $ w $ is a sequence of states $s_0s_1\dots$ s.t.
$ s_0 \in \StartState$, and $ \tau_i =(s_i, w_i, s_{i+1}) \in \delta $ for all $i$.
Let $ \mathit{inf}(\rho) $ denote the set of states that occur infinitely
often in run ${\rho}$. 
A run $\rho$ is an {\em accepting run} if $ \mathit{inf}(\rho)\cap \Final \neq \emptyset $. A word $w$ is an accepting word if it has an accepting run. 
The language $\L(\A)$ of B\"uchi automaton $\A$ is the set of all words accepted by it. 
Languages accepted by these automata are called {\em $ \omega $-regular   languages}.
 B\"uchi automata are known to be closed under set-theoretic union,
intersection, and complementation~\cite{thomas2002automata}. 
For B\"uchi automata $A$ and $B $, the {\em language-equivalence} and {\em language-inclusion} are whether $ \L(A) \equiv \L(B)$ and $\L(A)\subseteq \L(B)$, resp.

\subsection*{B\"uchi pushdown automaton}

B\"uchi pushdown automaton correspond to acceptors of words that require an infinite-amount of memory.
     A \emph{B\"uchi pushdown automaton}~\cite{cohen1977theory} is a tuple
    $\A = (\Statess, \Sigma, \Gamma, \delta, \StartState, Z_0, \Final)$, where $\Statess$, $\Sigma$, $\Gamma$, and $\Final$ are finite sets of {\em states}, {\em input alphabet},  {\em pushdown alphabet} and {\em accepting states}, respectively.
    $ \delta \subseteq (\Statess \times \Gamma\times (\Sigma\cup \{\epsilon\}) \times \Statess \times \Gamma) $ is the
    {\em transition relation}, $ \StartState \subseteq \Statess $ is a
    set of {\em initial states}, $Z_0 \in \Gamma$ is the {\em start symbol}.
  
 A \emph{run} $\rho$ on a word $ w = w_0 w_1\dots\in\Sigma^{\omega} $ of a B\"uchi PDA $\A$ is a sequence of configurations $(s_0,\gamma_0),(s_1,\gamma_1)\dots$ satisfying (1) $ s_0 \in \StartState$, $\gamma_0 = Z_0$, and (2) ($s_i, \gamma_i, w_i, s_{i+1}, \gamma_{i+1}) \in\delta$ for   all $i$.
 B\"uchi PDA consists of a {\em stack}, elements of which are the tokens $\Gamma$, and initial element $Z_0$.  
Transitions {\em push}  or {\em pop} token(s) to/from the top of the stack. 
  Let $ \mathit{inf}(\rho) $ be the set of states that occur infinitely
  often in state sequence  $s_0s_1\dots$ of run $\rho$. 
 A run $\rho$ is  an {\em accepting run} in B\"uchi PDA if $ \mathit{inf}(\rho)\cap \Final \neq \emptyset $. 
  A word $w$ is an {\em accepting word} if it has an accepting run.
Languages accepted by B\"uchi PDA are called {\em $ \omega $-context-free languages} ({$\omega$-CFL}).

\subsection*{Weighted $ \omega$-automaton }
A {\em weighted automaton}~\cite{chatterjee2010quantitative,mohri2009weighted} over infinite words is a tuple $\A= (\mathcal{M},  \gamma, f)$, where $\mathcal{M} = (\Statess, \Sigma, \delta, \StartState, \Statess) $ is a B\"uchi automaton with all states as accepting, $\gamma: \delta \rightarrow \Q$ is a {\em weight function}, and $f: \Q \rightarrow \Re$ is the {\em aggregate function}~\cite{chatterjee2010quantitative,mohri2009weighted}.

{\em Words} and {\em runs} in weighted automata are defined as they are in B\"uchi automata.    
The {\em weight-sequence} of run $\rho = s_0 s_1 \dots$ of word $w = w_0w_1\dots$ is given by $wt_{\rho} = n_0 n_1 n_2\dots$ where $n_i = \gamma(s_i, w_i, s_{i+1})$ for all $i$. 
The {\em weight of a run}  $\rho$, denoted by $f(\rho)$, is given by $f(wt_{\rho})$. 
Here the {\em weight of a word} $w \in \Sigma^{\omega}$ in weighted automata is defined as $wt_{\A}(w) = sup \{f(\rho) | \rho$ is a run of $w$ in $\A \}$. 
In general, weight of a word can also be defined as the infimum of the weight of all its runs. 
By convention, if a word $w \notin \L(\mathcal{M})$ its weight $wt_\A(w) = -\infty$.

\section{Games over graphs}

Games over graphs are popular in formal methods to study interactions between two players. More specifically, these are extensively used in the context of planning and synthesis. A graph game consists two players. It is played over a directed graph in which the states are partitioned between the two players. A play consists of the players moving a token along edges in the graph: The play begins in the initial state, and the token is pushed to the next state by the player that own the current state. This play goes on till infinitum. The outcome of which player wins a play is determined by an {\em acceptance condition} over the play itself. Reachability, safety, and parity are few examples of such conditions, and are described in detail below. In other cases, the plays could be associated with a quantitative value. In these cases, the winning criteria will be a quantitative property, as described later in the section.

\subsection*{Reachability and safety games}
Both, {\em reachability} and  {\em safety games}, are defined over the structure  $G = (V  = V_0 \uplus V_1, \init, E, \F)$~\cite{thomas2002automata}, where $V$, $V_0$, $V_1$, $\init$, and $E$ are defined as above.
We assume that every state has at least one outgoing edge, i.e, $vE  \neq \emptyset$ for all $v \in V$.
The non-empty set of states $ \F \subseteq V$ is called the {\em accepting} and {\em rejecting} states in reachability and safety games, resp.

A play $\rho = v_0v_1v_2\dots$ is defined as earlier. 
A play is  {\em winning for player $P_0$} in a reachability game if no state in the play is an accepting state, and {\em winning for player $P_1$} otherwise. 
The opposite holds in safety games. A play is winning for player $P_0$ if it visits the rejecting states, and winning for $P_1$ otherwise. 

Strategies for players are defined as earlier as well.
A strategy $\Pi_i$ is winning for player $P_i$ if strategies of the opponent player $P_{1-i}$, all resulting plays are winning for $P_i$. To {\em solve} a reachability/safety game refers to  determining whether there exists a winning strategy for player $P_1$ in the game. Both reachability and safety games are solved in linear time to their size, i.e., $\O(|V|+|E|)$.

\subsubsection{Parity games}
A {\em parity game} is defined over the structure  $G = (V  = V_0 \uplus V_1, \init, E, c)$~\cite{thomas2002automata}, where $V$, $V_0$, $V_1$, $\init$, and $E$ are defined as above.
We assume that every state has at least one outgoing edge, i.e, $vE  \neq \emptyset$ for all $v \in V$.
The {\em coloring function} $c:V\rightarrow \N$ assigns a natural-valued {\em color} to states in the game. 

Plays and strategies are defined as earlier. A color sequence of a play $c_\rho = c_0c_1c_2\dots$ where $c_i = c(v_i)$ for  all $i\geq 0$. is Wlog, a play is said to be {\em winning} if the maximum color appearing infinitely often in its color sequence is even.  
As earlier, a strategy $\Pi_i$ is winning for player $P_i$ if strategies of the opponent player $P_{1-i}$, all resulting plays are winning for $P_i$. To {\em solve} a parity game refers to  determining whether there exists a winning strategy for player $P_1$.

\subsection*{Quantitative games with complete information}
A {\em quantitative graph with complete information game}, referred to as quantitative game in short,  is defined over a structure $G = (V  = V_0 \uplus V_1, \init, E, \gamma, f)$.
It consists of a directed graph $(V,E)$,
and a partition $(V_0, V_1)$ of its set of states $V$. 
State $\init$ is the {\em initial state} of the game. 
$vE $ designates the set $\{w \in V | (v, w) \in E\}$  to indicate the successor states of state $v \in V$. For convenience, we assume that every state has at least one outgoing edge, i.e, $vE  \neq \emptyset$ for all $v \in V$.
Each transition of the game is associated with a {\em cost} determined by the {\em cost function} $\gamma: E \rightarrow \Z$. Finally, $f:\Z^\omega \rightarrow \Re$
is the {\em aggregate function}.

A {\em play} of a game involves two players, denoted by $P_0$ and $P_1$, that form an infinite path by moving a token along the transitions as follows:
At the beginning, the token is at the initial state. If the current
position $v$ belongs to $V_0$, then $P_0$ chooses the successor state $w$ from 
$vE$; otherwise, if $v \in V_1$, then $P_1$ chooses the next state from $vE$. 
Formally,
a play $\rho = v_0v_1v_2\dots$ is an infinite sequence of states such that the first state $v_0 = v_{\mathsf{init}}$, and each pair of successive states is a transition, i.e., $(v_k, v_{k+1}) \in E$ for all $k\geq 0$. 
The {\em cost sequence} of a play $\rho$ is the sequence of costs $w_0w_1w_2\dots$ such that  $w_k = \gamma((v_k, v_{k+1}))$ for all $i\geq 0$.
Given the aggregate function $f:\Z^\omega\rightarrow \Re$, the {\em cost of play} $\rho$, denoted $\wt(\rho)$, is the aggregate function applied to the cost sequence, i.e., $\wt(\rho) = f{\rho}$.
W.l.o.g, we assume that the players have opposing objectives, one player attempts to maximize the cost from plays while the other attempts to minimize the cost.

A {\em strategy} for player $P_i$ is a
is a partial function $\Pi_i : V^*V_i\rightarrow V$ such that $v = \Pi_i(v_0v_1\dots v_k)$ belongs to $v_kE$. Intuitively, strategy $\Pi_i$ directs the player $P_i$ which state to go to next based on the history of the play such that it is consistent with the transitions. A player $P_i$ is said to follow a strategy 
$\Pi$ on a play $\rho$ if for all $k$-length prefixes $v_0v_1\dots v_{k-1}$ of $\rho$ if
$v_{k-1} \in V_i$ then $v_k = \Pi(v_0v_1\dots v_{k-1})$.

\subsection*{Quantitative game with incomplete information}
An {\em incomplete-information quantitative game} is a tuple $\G = (S,s_{\mathcal{I}} ,\Obs, \Sigma, \delta, \gamma, f)$, where $S$, $\Obs$, $\Sigma$ are sets of {\em states}, {\em observations}, and {\em actions}, respectively, $s_{\mathcal{I}} \in S$ is the {\em initial state}, $\delta \subseteq S \times \Sigma \times S$ is the {\em transition relation},  $\gamma : S \rightarrow \Z\times\Z$ is the {\em weight function}, and $f: \Z^{\omega}\rightarrow \mathbb{R} $ is the {\em aggregate function}. 

The transition relation $\delta$ is {\em complete}, i.e., for all states $p$ and actions $a $, there exists a state $q $ s.t. $(p, a, q)\in \delta$. 
A {\em play} $\rho$ is a sequence $s_0a_0s_1a_1\dots$, where $\tau_i=(s_i, a_i, s_{i+1}) \in \delta$. 
The {\em observation of state} $s $ is denoted by $\Obs(s) \in \Obs$.
The {\em observed play} $o_{\rho}$ of $\rho$ is the sequence
$o_0a_0o_1aa_1\dots$, where $o_i = \Obs(s_i)$. 
Player $P_0$ has incomplete information about the game $\G$; it only perceives the observation play $o_{\rho}$. Player $P_1$ receives full information and witnesses play $\rho$.   
Plays begin in the initial state $s_0 = s_\mathcal{I}$. For $i\geq 0$, Player $P_0$ selects action $a_i$. Next, player $P_1$ selects the state $s_{i+1}$, such that $(s_i, a_i, s_{i+1}) \in \delta$.
The {\em weight of state} $s$ is the pair of payoffs $\gamma(s) =  (\gamma(s)_0, \gamma(s)_1)$. 
The {\em weight sequence} $wt_i$ of player $P_i$ along $\rho$ is given by $\gamma(s_0)_i \gamma(s_1)_i  \dots$, 
and its payoff from $\rho$ is given by $f(wt_i)$ for aggregate function $f$, denoted by $f(\rho_i)$, for simplicity. 
A play on which a player receives a greater payoff is said to be a {\em winning play} for the player.
 A strategy for player $P_0$ is given by a function $\alpha : \Obs^* \rightarrow \Sigma$ since it only sees observations.  Player $P_0$ follows strategy $\alpha$ if for all $i$, $a_i = \alpha(o_0\dots o_i)$. A strategy $\alpha$ is said to be a  {\em winning strategy} for player $P_0$ if all plays following $\alpha$ are winning plays for $P_0$. 
      
\section{Aggregate functions}
\label{Sec:AggregateFunctionDef}

Let $A = A[0]A[1]\dots$ be an infinite weight sequence. Let $A[i]$ denote the $i$-th element of the sequence, for all $i\geq 0$. Let $A[0,i-1]$ denote the $i$-length prefix $A[0]\dots A[i-1]$ of a sequence $A$, for $i\geq 0$.

\subsubsection*{Discounted-sum aggregate function}

Discounted-sum (DS) is a commonly appearing mode of aggregation which captures the intuition that weights incurred in the near future are more significant than those incurred later on~\cite{AHR2003}. The {\em discount-factor} is a rational-valued paramater that governs the rate at which weights loose their significance in the DS.

 Let $d >1$ be the rational-valued discount-factor. Then the discounted-sum of an infinite weight sequence $A$ w.r.t. discount-factor $d$, denoted by $\DSum{A}{d}$, is defined as 
\begin{align}
	\DSum{A}{d} &= \sum_{i=0}^{|A|-1} \frac{A[i]}{d^{i}}
\end{align}

DS is a preferred mode of aggregation across several domains including reinforcement learning~\cite{sutton1998introduction}, planning under uncertainty~\cite{puterman1990markov}, and game-theory~\cite{osborne1994course}. One reason behind its popularity is that DS is known to exist for all bounded infinite-length weight sequences, since its value is guaranteed to converge.

\subsubsection*{Limit-average aggregate function}


The limit-average of an infinite is intuitively defined as is the point of convergence of the average of prefixes of the sequence.

Let $\Sum{A[0,n-1]} = \sum_{i=0}^{n-1} A[i]$ denote the sum of the $n$-length prefix of sequence $A$. Then intuitively, the limit-average of a sequence $A$ should be defined as $\lim{n}{\infty} \frac{1}{n}\cdot \Sum{A[0,n-1]}$. However, this is not well-defined since $\frac{1}{n}\cdot \Sum{A[0,n-1]}$ may not converge as $n\rightarrow \infty$. Therefore, limit-average is defined in terms of auxilary functions {\em limit-average infimum }and {\em limit-average supremum}.

The {\em limit-average infimum} of an infinite weight sequence $A$, denoted by $\LAInf{A}$, is defined as 
\begin{align}
	\LAInf{A} &= \liminf{n}{\infty}{\frac{1}{n}\cdot \Sum{A[0,n-1]}}
\end{align}

The {\em limit-average supremum} of an infinite weight sequence $A$, denoted by $\LASup{A}$, is defined as 
\begin{align}
	\LASup{A} &= \limsup{n}{\infty}{\frac{1}{n}\cdot \Sum{A[0,n-1]}}
\end{align}

Then, the limit-average of sequence $A$, denoted by $\LA{A}$ is defined either as the limit-average infinum or supremum. Note, limit-average is well defined {\em only if} the limit-average infimum and limit-average supremum coincide, in which case limit-average is the same as limit-average infinum. In the other case, it is simply defined as either the limit-average supremum or infimum.

\subsubsection*{$\omega$-regular aggregate function}
 
 The  class of $\omega$-regular aggregate function are those functions on which arithmetical operations can  be performed on on automata~\cite{chaudhuri2013regular}.
 
 Let $\beta\geq 2$ be an integer base, Let $\mathsf{Digit}(\beta) =\{0,\dots, \beta-1\}$ be its {\em digit set}. 
 Let $x \in \Re$, then there exist unique words $\mathsf{Int}(x, \beta) = z_0z_1\dots \in \mathsf{Digit}(\beta)^*\cdot 0^\omega $ and $\mathsf{Frac}(x, \beta)=f_0f_1\dots \notin \mathsf{Digit}(\beta)^*\cdot (\beta-1)^\omega$ such that $|x| = \sum_{i=0}^\infty \beta^i \cdot z_i + \sum_{i=0}^\infty \frac{f_i}{\beta^i}$.  
 Thus, $z_i$ and $f_i$ are respectively the $i$-th
 least significant digit in the base $\beta$ representation of the
 integer part of $x$, and the $i$-th most significant digit in
 the base $\beta$ representation of the fractional part of $x$.
 Then, the real-number $x \in \Re$ in base $\beta$ is represented by $\mathsf{rep}(x, \beta) = \mathsf{sign} \cdot  (\mathsf{Int}(x, \beta),\mathsf{Frac}(x, \beta) )$, where $\mathsf{sign} = +$ if $x \geq 0$, $\mathsf{sign} = -$ if $x<0$, and $(\mathsf{Int}(x, \beta),\mathsf{Frac}(x, \beta) )$ is the interleaved word of $\mathsf{Int}(x, \beta)$ and $\mathsf{Frac}(x, \beta)$.
 Clearly, $x = \mathsf{sign}\cdot |x| = \mathsf{sign}\cdot (\sum_{i=0}^\infty \beta^i \cdot z_i + \sum_{i=0}^\infty \frac{f_i}{\beta^i})$.
 For all integer $\beta\geq 2$, we denote the alphabet of representation of real-numbers in base $\beta$ by $\mathsf{AlphaRep}(\beta)$.
 
 \begin{Def}[Aggregate function automaton, $\omega$-Regular aggregate function]
 	\label{def:aggregateregular}
 	Let $\Sigma$ be a finite set, and $\beta\geq 2$ be an integer-valued base. 
 	A B\"uchi automaton $\A$ over alphabet $\Sigma\times\mathsf{AlphaRep}(\beta)$ is an {\em aggregate function automata of type $\Sigma^\omega\rightarrow\Re$} if the following conditions hold:
 	\begin{itemize}
 		\item For all $A \in \Sigma^\omega$, there exists at most one $x \in \Re$ such that $(A,\mathsf{rep}(x,\beta)) \in \L(\A)$, and
 		
 		\item For all $A \in \Sigma^\omega$, there exists an  $x \in \Re$ such that $(A,\mathsf{rep}(x,\beta)) \in \L(\A)$
 	\end{itemize}
 \end{Def}

 $\Sigma$ and $\mathsf{AlphaRep}(\beta)$ are the {\em input} and {\em output} alphabets, respectively. An aggregate function $f:\Sigma^\omega\rightarrow \Re$ is {\em $\omega$-regular} under integer base $\beta \geq 2$ if there exists an aggregate function automaton $\A$ over alphabet $\Sigma\times\mathsf{AlphaRep}(\beta)$ such that for all sequences $A \in \Sigma^\omega$ and  $x\in \Re$, $f(A) = x$ iff $(A,\mathsf{rep}(x,\beta)) \in L(\A)$.

\section{Quantitative inclusion}

\begin{Def}[Quantitative inclusion]
	Let $P$ and $Q$ be  weighted $\omega$-automata with the {\em same} aggregate function $f$. 
	The {\em strict quantitative inclusion problem}, denoted by  $P \subset_f Q$,  asks whether  for all words $w \in \Sigma^{\omega}$, $wt_{P}(w) < wt_{Q}(w)$. 
	The {\em non-strict quantitative inclusion problem},  denoted by  $P \subseteq_f Q$,  asks whether  for all words $w \in \Sigma^{\omega}$, $wt_{P}(w) \leq wt_{Q}(w)$. 
	
\end{Def}

Quantitative inclusion, strict and non-strict, is \cct{PSPACE}-complete for limsup and liminf~\cite{chatterjee2010quantitative}, and undecidable for limit-average~\cite{degorre2010energy}. 
For discounted-sum with integer discount-factor it is in \cct{EXPTIME}~\cite{boker2014exact,chatterjee2010quantitative}, and decidability is unknown for rational discount-factors

Applications of quantitative inclusion are verification from quantitative logics, verification of reward-maximizing agents in multi-agent systems such as online economic protocols, auctions systems and so on, and in comparing systems based on a quantitative dimension.

\section{Solving quantitative games}

\subsubsection{Games with complete information}

Several problems in planning and synthesis with quantitative properties are formulated into an {\em analysis problem} over a quantiative game~\cite{chakrabarti2005verifying}. 
{\em Optimization} over such games is a popular choice of analysis in literature. 

\begin{Def}[Optimization problem]
	Given a quantitative graph game $G$, the optimization problem is to compute the optimal cost from all possible plays from the game, under the assumption that the objectives of $P_0$ and $P_1$ are to maximize and minimize the cost of plays, respectively. 
\end{Def}

Zwick and Patterson have  shown that the optimization problem is pseudo-polynomial, and that the optimal cost can be obtained  via memoryless strategies for both players~\cite{zwick1996complexity}. They also show that a VI algorithm will converge to the optimal cost.  
However, a thorough worst-case analysis of VI is missing.

We argue that the optimal solution may not be required in several tasks. For instance, a solution with {\em minimal} battery consumption may be replaced by one that operates {\em within} the battery life. 
Hence, many tasks can be reformulated into an alternative form of analysis in which the problem is to search for a solution the adheres to a given threshold constraint~\cite{satisficing}.
We call this the {\em satisficing problem}. 

\begin{Def}[Satisficing problem]
	Given a quantitative graph game $G$ and a threshold value $v \in \Q$, the {\em satisficing problem} is to determine whether player $P_1$ has a strategy such that for all possible resulting plays of the game, the cost of all plays is less than (or $\leq$) to the threshold $v$, assuming that the objectives of $P_0$ and $P_1$ are to maximize and minimize the cost of plays, respectively. 
\end{Def}

In several tasks one is interested in combining quantitative games with a temporal goal. For quantitative games, it is known that an optimal solution may not exist when combined with temporal goals~\cite{chatterjee2017quantitative}

\subsubsection{Games with incomplete information}

The main problem in games with incomplete information is to determine whether a player has a winning strategy. If so, one is interested in generating it. 

\part{Theoretical framework}
\label{part:framework}
This part lays down the theoretical foundations of {\em comparator automata}, or {\em comparators } in short.

\begin{itemize}[label=Chapter ]
	\item \ref{ch:Comparator} formally introduces comparator automata, our automata-based technique to obtain integrated methods to solve problems in quantitative reasoning, as opposed to separation-of-techniques used so far. We show that generalizable solutions for quantitative inclusion and solving quantitative games can be designed for all aggregate functions for which  the comparator is represented by a B\"uchi automata. 
	
	\item \ref{ch:applicationcomparator} studies  and constructs comparator automata commonly occurring aggregate functions, namely discounted-sum and limit-average. 
\end{itemize}
\chapter[Comparator automata]{Comparator automata:\\ Foundations of a generalizable and integrated theory for quantitative reasoning}
\label{ch:Comparator}

The landscape of algorithms, complexity, tools and techniques for the quantitative analysis of systems is wide and non-uniform. A part of the reason is that there are several aggregate functions, and for each one of those a different form of reasoning is applied. 
For instance,  consider the problem of quantitative inclusion for aggregate function $f$, called {\em f-inclusion} in short. 
In case of quantitative inclusion, even the complexity-theoretic results are diverse. While quantiative inclusion is \cct{PSPACE}-hard, there is ample variance in upper bounds. Quantitative inclusion is \cct{PSPACE}-complete under limsup/liminf~\cite{chatterjee2010quantitative}, undecidable for limit-average~\cite{degorre2010energy}, and decidability is unknown for discounted-sum in general but its decidable fragments are \cct{EXPTIME}~\cite{boker2014exact,chatterjee2010quantitative}. 
In view of these vast differences, the aforementioned landscape could benefit from a  unified theory that {\em generalizes} across aggregate functions. 
This chapter contributes towards just that: A unified theory for quantitative reasoning that generalizes across a wide class of aggregate functions. 

To this end, we take the view that the notion of {\em comparisons} between systems runs or inputs is central to formal methods especially to quantitative analysis,
 and hence comparison should be brought to the forefront. 
To see why our view holds, first consider the classical {\em model checking problem} of verifying if a system $S$ satisfies a linear-time temporal specification $P$~\cite{clarke1986automatic}. Traditionally, this problem is phrased language-theoretically: $S$ and $P$ are interpreted as sets of (infinite) words, and $S$ is determined to satisfy $P$ if $S \subseteq P$. The problem, however, can also be framed in terms of a {\em comparison} between words in $S$ and $P$.
Suppose a word $w$ is assigned a weight of 1 if it belongs to the language of the system or property, and 0 otherwise. Then determining if $S \subseteq P$ amounts to checking whether the weight of every word in $S$ is less than or equal to its weight in $P$~\cite{baier2008principles}.
The ubiquity of comparisons becomes more pronounced in quantitative analysis: Firstly, because every system execution is assigned a real-valued cost.  W.l.o.g, we can assume that the cost model is an {\em aggregate function} $f: \mathbb{Z}^\omega \rightarrow \Re$. The cost of an execution is related to the aggregate function $f$ applied to the weight-sequence corresponding to the execution; 
Secondly,
because problems in quantitative analysis reduce to comparing the cost of executions to a constant value (such as in quantitative games), or more generally to the cost of another execution (as in quantitative inclusion).

Keeping comparisons at the center, we introduce a language-theoretic/automata-theoretic formulation of the comparison between weighted sequences. Specifically, in Section~\ref{Sec:ComparatorIntro} we introduce {\em comparator automata} (\emph{comparators}, in short), a class of automata that read pairs of infinite weight sequences synchronously, and compare their aggregate values in an online manner. 
Formally, a {\em comparator automata for aggregate function $f$, relation $\mathsf{R}$, and upper bound $\mu>0$} is an automaton that accepts a pair $(A,B) \in (\Sigma\times\Sigma)^\omega$ of sequences  of bounded integers, where $\Sigma =\{-\mu, \mu-1,\dots, \mu\}$, iff $f(A)$ $\mathsf{R}$ $f(B)$, where $\mathsf{R} \in \{ >, <, \geq, \leq, \neq\,=\}$ is an inequality or equality relation. A comparator could be finite-state or  (pushdown) infinite-state. We say a comparator is {\em $\omega$-regular} if it is finite-state and accepts by the B\"uchi condition. 
Similarly, we  say a comparator is {\em $\omega$-pushdown} if it is an $\omega$-context free automaton.

The central result of this chapter is that one can design generalizable solutions for all aggregate functions that permit an $\omega$-regular comparator (Section~\ref{Sec:Generalize}). 
We illustrate this point by designing algorithms for quantitative inclusion (Section~\ref{Sec:InclusionGen}), and solving quantitative games under perfect  and imperfect information (Section~\ref{Sec:PerfectInfo} and Section~\ref{Sec:GraphGames}, respectively) for an aggregate function $f:\mathbb{Z} \rightarrow \Re$ such that the comparator for $f$ is $\omega$-regular. 
What enables these generalizable results is the fact that since B\"uchi automata are closed under all set-theoretic operations,  $\omega$-regular comparators can easily be operated on in the generic algorithms that we design. Since $\omega$-pushdown automata are not closed under all set-theoretic operations, one cannot design such generic algorithms with $\omega$-pushdown comparators. In the next chapter (Chapter~\ref{ch:applicationcomparator}), we give concrete instances of aggregate functions which permit $\omega$-regular or $\omega$-pushdown comparators. 

Another benefit of comparators has to do with the separation-of-techniques challenge of quantitative analysis. Recall from Chapter~\ref{ch:intro} that separation-of-techniques refers to the inherent dichotomy in existing algorithms for problems in quantitative analysis.
Most algorithms comprise of two phases. In the first phase, called the {\em structural analysis phase}, the algorithm reasons about the structure of the quantitative system(s) using techniques from automata or graphs. In the second phase, called the {\em numerical analysis phase}, the algorithm reasons about the quantitative dimension/cost model using numerical methods. The techniques used in both these phases are so unlike each other that they cannot be solved {\em in tandem}.  Hence the phases have to be performed sequentially, hence affecting their scalability.

The advantage our automata-based formulation of comparators in this aspect is that comparators reduce the numerical problem of comparison of aggregation of weight sequences into one of membership in an automaton. This way, enabling both phases, the structural phase and numerical phase, of quantitative analysis to be performed using automata-based techniques. Subsequently, creating an opportunity to design {\em integrated methods }as opposed to separation-of-techniques methods for problems in quantitative analysis. Infact, the  generalizable algorithms that we design when the aggregate function permits $\omega$-regular comparators are based on reducing the problems in quantitative analysis to those in qualitative analysis. In particular, quantitative inclusion is reduced to {\em language inclusion}, and quantitative games under perfect and imperfect information are both reduced to solving {\em parity games}. 

As a result, our novel comparator automata framework not only results in generalizable algorithms for problems in quantitative analysis, but the algorithms designed using comparators are inherently integrated by approach. Thereby, comparators resolve both challenges in quantitative analysis.


\section{Comparison language and comparator automata}
\label{Sec:ComparatorIntro}

This section introduces comparison languages and comparator automata as a a class of langauges/automata that can read pairs of weight sequences synchronously and  establish an equality or inequality relationship between these sequences. 
Formally, we define: 

\begin{Def}[Comparison language]
\label{def:comparisonlanguage}
Let $\Sigma $ be a finite set of rational numbers, and $f:\Q^{\omega}\rightarrow \Re$ denote an aggregate function. 	A {\em comparison language for aggregate function $f$  with inequality or equality relation $\mathsf{R} \in \{<,>,\leq, \geq, =, \neq\}$} is the language over the alphabet $\Sigma\times\Sigma$ that accepts a pair of infinite length sequences $(A,B)$ iff $f(A)$ $\mathsf{R}$ $f(B)$ holds. 

\end{Def}
\begin{Def}[Comparator automata]
\label{def:comparator}
Let $\Sigma $ be a finite set of rational numbers, and $f:\Q^{\omega}\rightarrow \Re$ denote an aggregate function. 	A {\em comparator automaton for aggregate function $f$  with inequality or equality relation $\mathsf{R} \in \{<,>,\leq, \geq, =, \neq\}$} is the automaton that accepts the comparison language for $f$ over alphabet $\Sigma\times\Sigma$ with relation $\mathsf{R}$.
\end{Def}


From now on, unless mentioned otherwise, we assume that all weight sequences are bounded integer sequences. 
First, the boundedness assumption is justified since  the set of weights forming the alphabet of a comparator is bounded. In particular, $\Sigma$ is bounded as it is a finite set. 
Second, for all aggregate functions considered in this thesis, the result of comparison of weight sequences is preserved by a uniform linear transformation that converts  rational-valued weights into integers; justifying the integer assumption. 
Hence, now onward, we will define a comparison language w.r.t an {\em upper bound} $\mu>0$ so that $\Sigma = \{-\mu, -\mu+1,\dots, \mu\}$.

\subsection{$\omega$-regular comparator}
When the comparison language for an aggregate function and a relation can be represented by a B\"uchi automaton, we refer to the comparison language and the comparator automata as {\em $\omega$-regular}. 
Note that we \textbf{do not} refer to the corresponding aggregate function as $\omega$-regular. This is because existing literature already defines $\omega$-regular aggregate functions~\cite{chaudhuri2013regular}. Section~\ref{Sec:RegularAggFunctions} will explore the relationship between $\omega$-regular comparison languages/comparators and aggregation functions. 
$\omega$-regular comparison languages benefit from closure-properties of B\"uchi automata. As a result, $\omega$-regular comparison languages exhibit the following closure properties: 

\begin{theorem}[Closure]
\label{lem:closure}
Let $\mu>$ be an upper bound such that $\Sigma = \{-\mu, -\mu+1,\dots, \mu\}$, and $f:\Z^{\omega}\rightarrow \Re$ be an aggregate function. Let $\mathsf{R} \in \{\leq, \geq, <, >\} $ be an inequality relation. 
\begin{enumerate}
\item\label{item:closedregular} If the comparison language for $f$ over $\Sigma\times\Sigma $ with inequality relation $\mathsf{R}$ is $\omega$-regular, then the comparison language for $f$ over $\Sigma\times\Sigma $ with all inequality and equality relations $\mathsf{R}' \in \{\leq, \geq, <, >, =, \neq\}$ is $\omega$-regular

aggregate function $f$ is $\omega$-regular. 
\item \label{item:closedcomplement} The comparison language for $f$ over $\Sigma\times\Sigma$ with relation $=$ is $\omega$-regular iff the comparison language for $f$ over $\Sigma\times\Sigma$ with relation $\neq$ is $\omega$-regular. 

\item\label{item:closednotregular} If the comparison language for $f$ over $\Sigma\times\Sigma $ with relation $\mathsf{R}$ is not $\omega$-regular, then the comparison language for $f$ over $\Sigma\times\Sigma $ with all inequality  relations $\mathsf{R}' \in \{\leq, \geq, <, >\}$ is not $\omega$-regular.
\end{enumerate}
\end{theorem}
\begin{proof}
Item~\ref{item:closedregular}. 
W.l.o.g, let us assume that the comparison language for $f$ with $\leq$ is $\omega$-regular. This means that automaton representing the mentioned language is a B\"uchi automaton, say it is $\A$. 
Now, since B\"uchi automaton are closed under complementation, the automaton for the complementation of $\A$ is also a B\"uchi automaton. But note that the language of the complementation of $\A$ is the comparison language for $>$. Therefore, comparison language for $>$ is also $\omega$-regular. Now, it is easy to show that $(A,B)$ is a word in the comparison language for $\leq$ iff $(B,A)$ is a word in the comparison language of $\geq$. Now the B\"uchi automaton for comparison language for $\geq$ can be constructed from $\A$ by swapping the alphabet. Therefore, the comparison language for $\geq$ is also $\omega$-regular. By the same reason as above, if the comparison language for $\geq$ is $\omega$-regular, then the comparison language for $ <$ is also $\omega$-regular.
Finally, since the intersection of B\"uchi automaton is also a B\"uchi automaton, the B\"uchi automaton for comparison language of $=$ can be obtained by intersecting those for $\leq$ and $\geq$. Lastly, $\leq$ can be obtained by complementing that for $\neq$.

Hence, one can prove that if the comparison language for any one inequality relation is $\omega$-regular is sufficient to show that the comparison language for all relations is $\omega$-regular.
Item~\ref{item:closedcomplement} and Item~\ref{item:closednotregular} can similarly be shown using closure properties of B\"uchi automaton. 
\qed
\end{proof}

It is worth mentioning that Item(~\ref{item:closedregular}) is a means to demonstrate whether an aggregate function is $\omega$-regular. Chapter~\ref{ch:applicationcomparator} will give concrete examples of aggregate function that are $\omega$-regular. We issue a concrete example below to illustrate these concepts.

\subsection*{Illustrative example: Limsup aggregate function}
\begin{figure}[t]

\centering

\begin{tikzpicture}[shorten >=1pt,node distance=3.5cm,on grid,auto] 
\node[state,initial] (q_0)   {\footnotesize{$s$}};
\node[state, accepting] (q_1)  [right of = q_0] {\footnotesize{$f_k$}};
\node[state] (q_2) [right of=q_1] {\footnotesize{$s_k$}};
\path[->] 
(q_0)   edge [loop above] node {\footnotesize{$(*,*)$}} ()
edge node  {\footnotesize{$ (k, \leq k) $}} (q_1)
(q_1)  edge [loop above] node {\footnotesize{$ (k,\leq k) $}} ()
edge [bend left = 10]  node {\footnotesize{$(\leq k-1, \leq k)$}} (q_2)
(q_2)  edge [loop above] node {\footnotesize{$ (\leq k-1,\leq k) $}} ()
edge [bend left = 10]  node {\footnotesize{$(k, \leq k)$}} (q_1);
\end{tikzpicture}

\caption[Snippet of the Limsup comparator]{Snippet of the Limsup comparator. State $f_k$ is an accepting state.  Automaton $\A_k$ accepts $(A,B)$ iff $\LS{A}=k$, $\LS{B}\leq k$. $*$ denotes $\{-\mu,-\mu+1,\dots, \mu\}$, $\leq m$ denotes $\{-\mu,-\mu+1,\dots, m\}$}
\label{Fig:LSk}

\end{figure}
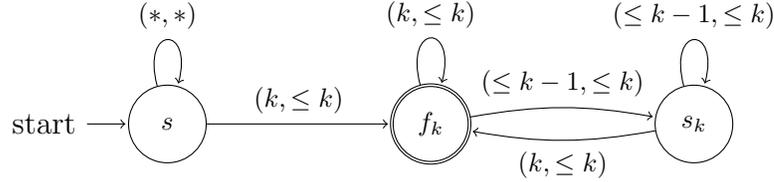 

\renewcommand{\max}{\mu}
We explain comparison languages and comparator automata through an example.  
The aggregate function we consider is the {\em limit supremum} function. Loosely speaking, the limit supremum function determines the maximally appearing value appearing in an infinite sequence. We will construct a B\"uchi automaton corresponding to its comparator automaton for an inequality relation to show that its comparison language/comparator is $\omega$-regular (Lemma~\ref{lem:closure}). 

The {\em limit supremum} (limsup, in short) of a bounded, integer sequence $A$, denoted by $ \LS{A}$, is the largest integer that appears infinitely often in $A$. 
For a given upper bound $\mu>0 $, The {\em limsup comparison language for relation $\geq$} is a language over $\Sigma\times\Sigma$, where $\Sigma = \{-\mu, -\mu+1, \dots, \mu\} $, that accepts the pair $(A,B)$ of sequences  iff  $\LS{A}\geq \LS{B} $. We will show that the limsup comparison languages are $\omega$-regular for all relations. To this end, we prove that the {\em limsup comparator for $\geq$} is a B\"uchi automaton. 

The working of the limsup comparator for relation $\geq$ is based on {\em guessing} the limsup of sequences $A$ and $B$, and then {\em verifying} that $\LS{A}\geq \LS{B}$.
This can be encoded using a non-deterministic B\"uchi automaton, as partially illustrated by $\A_k$ in Fig.~\ref{Fig:LSk}. More specifically, $\A_k$ is the basic building block of the limsup comparator for relation $\geq$.  Automaton $\A_k$ accepts pair $ (A,B)$ of number sequences  iff  $\LS{A} = k$, and $\LS{B}\leq k$, for integer $k$.
 (Lemma~\ref{Lemma:LSFinalState}). 

\begin{lemma}
	\label{Lemma:LSFinalState}
	Let $A$ and $B$ be integer sequences bounded by $\max$. 	
	B\"uchi automaton $\A_k$ (Fig.~\ref{Fig:LSk}) accepts $(A,B)$ iff $\LS{A} = k$, and $\LS{A}\geq\LS{B}$.
\end{lemma}

\begin{proof}
Let $(A,B)$ have an accepting run in $\A_k$. We show that $\LS{A} =k \geq \LS{B}$. 
The accepting run visits state $f_k$ infinitely often. 
Note that all incoming transitions to accepting state $f_k$ occur on alphabet $(k,\leq k)$ while  all transitions between states $f_k$ and $s_k$ occur on alphabet $(\leq k, \leq k)$, where $\leq k$ denotes the set $\{-\mu,-\mu+1,\dots, k\}$. 
So, the integer $k$ must appear infinitely often in $A$ and all elements occurring infinitely often in $A$ and $B$ are less than or equal to $k$. Therefore, if $(A,B)$ is accepted by $\A_k$ then $\LS{A} =k$, and $\LS{B} \leq k$, and $\LS{A}\geq \LS{B}$.

Conversely, let $\LS{A} = k > \LS{B}$. We prove that $(A,B)$ is accepted by $\A_k$.
	For an integer sequence $A$ when $ \LS{A} = k $ integers greater than $k$ can occur only a finite number of times in $A$. Let $l_A$ denote the index of the last occurrence of an integer greater than $k$ in $A$.    
	Similarly, since $\LS{B} \leq k$, let $l_B$ be index of the last occurrence of an integer greater than $k$.
	Therefore, for sequences $A$ and $B$ integers greater than $k$ will not occur  beyond index $l = \mathit{max}(l_A, l_B)$. 
	B\"uchi automaton $\A_k$ (Fig.~\ref{Fig:LSk}) non-deterministically determines $l$. 
	On reading the $l$-th element of input word $(A,B)$, the run of $(A,B)$ exits the start state $s$ and  shifts to accepting state $f_k$. Note that all runs beginning at state $f_k$ occur on alphabet $(a,b)$ where $a,b\leq k$. Therefore, $(A,B)$ can continue its infinite run even after transitioning to $f_k$. To ensure that this is an accepting run, the run must visit accepting state $f_k$ infinitely often. But this must be the case, since $k$ occurs infinitely often in $A$, and all transitions on $(k,b)$, for all $b\leq k$, transition into state $f_k$. Hence, for all integer sequences $A$,$B$ bounded by $\max$, if $\LS{A}=k$, and $\LS{A}\geq \LS{B}$, the automaton accepts $(A,B)$.	
	\qed
\end{proof}
\begin{theorem}
	\label{thrm:limsupcomparator}
	The comparison language for the limsup aggregate function is $\omega$-regular for all relations $\mathsf{R} \in \{\leq, \geq, <, >, =, \neq\}$. 
\end{theorem}
\begin{proof}
Let $\mu>0$ be the upper bound, then $\Sigma = \{-\mu, -\mu+1, \dots, \mu\}$.
To prove the above, we show that the limsup comparator with $\geq$ over alphabet $\Sigma\times\Sigma$ is $\omega$-regular. This is sufficient due to Theorem~\ref{lem:closure}-Item~\ref{item:closedregular}.

Observe that the union of B\"uchi automata $\A_k$ (Fig~\ref{Fig:LSk}, Theorem~\ref{Lemma:LSFinalState}) for $k \in \{-\mu, -\mu+1, \dots,\max\}$ corresponds to the coveted limsup comparator for relation $\geq$.
Since the union of B\"uchi automata is also a B\"uchi automata, this implies that limsup comparator for $\geq$ is $\omega$-regular. 
\qed
\end{proof}

The {\em limit infimum} (liminf, in short) of an integer sequence is the smallest integer that appears infinitely often in it; its comparators will have a similar construction to their limsup counterparts. 
Hence, the comparison language for the liminf aggregate function is also $\omega$-regular for all relations.

\subsection{$\omega$-pushdown comparator}
When the comparison language for an aggregate function and a relation can be represented by a B\"uchi pushdown automaton (and not a B\"uchi automaton), then the comparison language and comparator automata are referred to as {\em $\omega$-pushdown}.  

Unlike $\omega$-regular comparison languages, $\omega$-pushdown comparison languages may not exhibit closure properties since the underlying B\"uchi pushdown automaton are not closed under several operations such as complementation and intersection. 

The next chapter (Chapter~\ref{ch:applicationcomparator}) will illustrate an example of an aggregate function for which the comparison language is $\omega$-pushdown.

\section{Generalizability with $\omega$-regular comparators}
\label{Sec:Generalize}

This section illustrates the generalizability with $\omega$-regular comparators. We demonstrate this over the problems of quantitative inclusion, and solving quantitative games under perfect and imperfect information. By virtue of the automata-based nature of comparator automata, all algorithms we design are integrated in approach.

\subsection{Quantitative inclusion}
\label{Sec:InclusionGen}

The analysis of  quantitative dimensions of computing systems such as cost,  resource consumption, and distance metrics~\cite{alur2017introduction,baierprobabilistic,Kwi07} has been studied thoroughly to design efficient computing systems. 
Cost-aware program-synthesis~\cite{bloem2009better,chakrabarti2005verifying} and low-cost program-repair~\cite{hu2018syntax} have found compelling applications in robotics~\cite{he2017reactive,lahijanian2015time}, 
education~\cite{d2016qlose}, 
and the like.
{\em Quantitative verification} facilitates efficient system design by automatically determining if a system implementation is more efficient than a specification model.

At the core of quantitative verification lies the problem of {\em quantitative inclusion}
which formalizes the goal of determining which of two given systems is more efficient~\cite{chatterjee2010quantitative,filiot2012quantitative,mohri2009weighted}. 
In quantitative inclusion,  quantitative systems are  abstracted as weighted $\omega$-automata~\cite{aminof2010reasoning,droste2009handbook,mohri2002weighted}.
Recall, a run in a weighted $\omega$-automaton is associated with a sequence of weights. The quantitative dimension of these runs is determined by the weight of runs, which is computed by taking an aggregate of the run's weight sequence. The problem of quantitative inclusion between two weighted $\omega$-automata determines whether the weight of all words in one automaton is less that (or equal to) that in the other automaton.
It can be thought of as the quantitative generalization of (qualitative) language inclusion.

In the chapter's preface, we saw how the solution approaches and complexity for $f$-inclusion vary with differences in the aggregate function $f$.
This section presents a generic algorithm (Algorithm~\ref{Alg:DSInclusion})  to solve quantitative inclusion for function which permits $\omega$-regular comparators.
This section focuses on the {\em non-strict} variant of quantitative inclusion. 
Strict quantitative inclusion is similar, hence has been skipped. 

Given weighted $\omega$-automata $P$ and $Q$ with an aggregate function $f$, let whether $P$ is non-strictly $f$-included in $Q$ be denoted by $P \subseteq_f Q$.
For sake of succinctness, we simply say {\em inclusion with an $\omega$-regular comparator} to mean {\em $f$-inclusion where the function $f$ permits an $\omega$-regular comparator}.

\subsubsection*{Running example}
\label{Sec:Example}
\begin{figure*}[t]
	\centering
	\begin{subfigure}{0.48\textwidth}
		\centering
		\begin{tikzpicture}[shorten >=1pt,node distance=2.5cm,on grid,auto] 
		\node[state,accepting, initial] (q_1)   {$p_1$}; 
		\node[state, accepting] (q_2) [right of = q_1] {$p_2$};

		\path[->] 
		(q_1)	edge  node  {$a,1$} (q_2)
		(q_2)   edge [loop right] node  {$a,1$} (q_2);

		\end{tikzpicture}
		\caption{Weighted automaton $P$}
		\label{Fig:WA-P}
	\end{subfigure}
	\hfill
	\centering
		\begin{subfigure}{0.48\textwidth}
		\centering
		\begin{tikzpicture}[shorten >=1pt,node distance=2.5cm,on grid,auto] 
		
		\node[state,accepting, initial] (q_1)   {$q_1$}; 
		\node[state, accepting] (q_2) [right of = q_1] {$q_2$};
		
		\path[->] 
		(q_1)	edge  [bend left] node  {$a,0$} (q_2)
		edge  node  [below] {$a,2$} (q_2)
		(q_2)   edge [loop right] node  {$a,1$} (q_2);
		
		\end{tikzpicture}
		\caption{Weighted automaton $Q$}
		\label{Fig:WA-Q}
	\end{subfigure}
	\hfill
	\caption{Algorithm $\DSInclusion$: Running example}
\end{figure*}

The following running example is used to walk us through the steps of $\DSInclusion$, the generic algorithm  for inclusion with $\omega$-regular comparators presented in  Algorithm~\ref{Alg:DSInclusion}. 

Let weighted $\omega$-automata $P$ and $Q$ be as illustrated in Fig.~\ref{Fig:WA-P}-\ref{Fig:WA-Q} with the limsup aggregate function. From Theorem~\ref{thrm:limsupcomparator} we know that the limsup comparator $\A_{\mathsf{LS}}^\leq$ for $\leq$ is $\omega$-regular. Therefore, the generic algorithm we describe will apply to the example. 

The word $w=a^{\omega}$ has one run $\rho^P_1 = p_1p_2^{\omega}$ with weight sequence $wt^P_1=1^{\omega}$ in $P$ and two runs $\rho^Q_1 = q_1q_2^{\omega}$ with weight sequence $wt^Q_1=0,1^{\omega}$ and run $\rho^Q_2 = q_1q_2^{\omega}$ with weight sequence $wt^Q_2=2,1^{\omega}$. Clearly, $wt_P(w) \leq wt_Q(w$). Therefore $P\subseteq_f Q$. 

This section describes Algorithm~\ref{Alg:DSInclusion} for inclusion for $\omega$-regular comparators with the motivating example as a runnign example. Intuitively, the algorithm must be able to identify that for run $\rho^P_1$ of $w$ in $P$, there exists a run $\rho^Q_2$ in $Q$ s.t. $(wt^P_1, wt^Q_2)$ is accepted by the limsup comparator for $\leq$.

\subsubsection*{Algorithm description and analysis}

\paragraph{Key ideas}

A run $\rho_P$ in $P$ on word $w\in \Sigma^\omega$ is said to be {\em dominated} w.r.t $P\subseteq_f Q$ if there exists a run $\rho_Q$ in $Q$ on the same word $w$ such that $wt_P(\rho_P)\leq wt_Q(\rho_Q)$.
$P \subseteq_f Q$ holds if for every run $\rho_P$ in $P $ is dominated w.r.t. $P\subseteq_f Q$. 

The central construction of 
$\DSInclusion$ is a B\"uchi automaton $\first$ that consists of exactly the domianted runs of $P$ w.r.t $P\subseteq_f Q$. Then, $\DSInclusion$ returns $\mathsf{True}$ iff $\first$ contains all runs of $P$.
In order to construct $\first$, the algorithm first constructs another B\"uchi automaton $\less$ that accepts  word $(\rho_P, \rho_Q)$ iff $\rho_P$ and $\rho_Q$ are runs of the same word in  $P$ and $Q$ respectively, and $wt_P(\rho_P)\leq wt_Q(\rho_Q)$ i.e. if $w_P$ and $w_Q$ are weight sequence of $\rho_P$ and $\rho_Q$, respectively, then $(w_P, w_Q)$ is present in the $\omega$-regular comparator $\A_f^\leq$ for aggregate function $f$ with relation $\leq$. 
The projection of $\less$ on runs of $P$ results in $\first$.

\begin{algorithm}[t]
\caption{ $\DSInclusion(P,Q, \A_f)$, Is $P \subseteq_f Q$?}
\label{Alg:DSInclusion}
\begin{algorithmic}[1]
\STATE \textbf{Input: } Weighted $\omega$-automata $P$ and $Q$ over function $f$, and $\omega$-regular comparator $\A_f$ for function $f$ for the inequality $\leq$
\STATE \textbf{Output: } $\mathsf{True}$ if $P \subseteq_f Q$, $\mathsf{False}$ otherwise

\STATE $\hat{P} \leftarrow  \AugmentWtAndLabel(P)$ \label{alg-line:AugmentP}
\STATE $\hat{Q} \leftarrow  \AugmentWtAndLabel(Q)$  \label{alg-line:AugmentQ}

 \STATE $\hat{P}\times\hat{Q} \leftarrow \MakeProduct(\hat{P}, \hat{Q})$ \label{alg-line:Prod}
 
\STATE $\less \leftarrow  \Intersect(\hat{P}\times\hat{Q}, \A_{\succeq} )$  \label{alg-line:Intersect}

\STATE $\first \leftarrow \Project(\less)$ \label{alg-line:Project}
\RETURN {$\hat{P} \equiv \first$} \label{alg-line:ensure}

\end{algorithmic}
\end{algorithm}
\paragraph{Algorithm details} 
For sake a simplicity, we assume that every word present in $P$ is also present in $Q$ i.e. $P\subseteq Q$ (qualitative inclusion).
$\DSInclusion$ has three steps: 
(a). $\SepRuns$ (Lines~\ref{alg-line:AugmentP}-\ref{alg-line:AugmentQ}):  Enables unique identification of runs in $P$ and $Q$ through {\em labels}. 
(b). $\CompareRuns$ (Lines~\ref{alg-line:Prod}-\ref{alg-line:Project}): Compares weight of runs in $P$ with weight of runs in $Q$, and constructs $\first$. (c). $\EnsureRuns$ (Line~\ref{alg-line:ensure}): Ensures if all runs of $P$ are diminished. 

\begin{enumerate}
\item $\SepRuns$: $\AugmentWtAndLabel$ transforms weighted $\omega$-automaton $\A$ into B\"uchi automaton $\hat{\A}$ by converting transition $\tau = (s, a, t )$ with weight $\gamma(\tau)$ in $ \A$ to transition $\hat{\tau} =  (s, (a, \gamma(\tau), l), t)$ in $\hat{\A}$, where $l$ is  a unique label assigned to transition $\tau$. The word $\hat{\rho} = (a_0,n_0,l_0)(a_1,n_1,l_1)\dots \in \hat{A}$  iff  run $\rho \in \A$ on word $a_0a_1\dots$ with weight sequence $n_0n_1\dots$. 
Labels ensure bijection between runs in $\A$ and words in $\hat{\A}$.
Words of $\hat{A}$ have a single run in $\hat{A}$.
Hence, transformation of weighted $\omega$-automata  $P$ and $Q$ to B\"uchi automata $\hat{P}$ and $\hat{Q}$ enables disambiguation between runs of $P$ and $Q$ (Line~\ref{alg-line:AugmentP}-\ref{alg-line:AugmentQ}). 

The corresponding $\hat{A}$ for weighted $\omega$-automata $P$ and $Q$ from Figure~\ref{Fig:WA-P}-~\ref{Fig:WA-Q} are given in Figure~\ref{Fig:WA-Phat}-~\ref{Fig:WA-Qhat} respectively. 

\begin{figure*}[t]
	\centering
	\begin{subfigure}{0.48\textwidth}
		\centering
		\begin{tikzpicture}[shorten >=1pt,node distance=3cm,on grid,auto] 
		\node[state,accepting, initial] (q_1)   {$p_1$}; 
		\node[state, accepting] (q_2) [right of = q_1] {$p_2$};
		
		\path[->] 
		(q_1)	edge  node  {$(a,1,1)$} (q_2)
		(q_2)   edge [loop above] node  {$(a,1,2)$} (q_2);
		
		\end{tikzpicture}
		\caption{$\hat{P}$}
		\label{Fig:WA-Phat}
	\end{subfigure}
	\hfill
	\centering
	\begin{subfigure}{0.48\textwidth}
		\centering
		\begin{tikzpicture}[shorten >=1pt,node distance=3cm,on grid,auto] 
		
		\node[state,accepting, initial] (q_1)   {$q_1$}; 
		\node[state, accepting] (q_2) [right of = q_1] {$q_2$};
		
		\path[->] 
		(q_1)	edge  [bend left] node  {$(a,0,1)$} (q_2)
		edge   node  [below] {$(a,2,2)$} (q_2)
		(q_2)   edge [loop above] node  {$(a,1,3)$} (q_2);
		
		\end{tikzpicture}
		\caption{$\hat{Q}$}
		\label{Fig:WA-Qhat}
	\end{subfigure}
	\hfill
		\centering
	\begin{subfigure}{0.55\textwidth}
		\centering
		\begin{tikzpicture}[shorten >=1pt,node distance=4cm,on grid,auto] 
		\node[state,accepting, initial] (s_1)   {$p_1,q_1$}; 
		\node[state, accepting] (s_2) [right of = s_1] {$p_2,q_2$};

		\path[->] 
		(s_1)	edge  [bend left] node  {$(a,1,1,0,1)$} (s_2)
		(s_1)	edge  node  [below] {$(a,1,1,2,2)$} (s_2)
		(s_2)   edge [loop, distance=1.5cm] node  [above] {$(a,1,2,1,3)$} (s_2);
		
		\end{tikzpicture}
		\caption{$\hat{P}\times\hat{Q}$}
		\label{Fig:PhatQhat}
	\end{subfigure}
	\hfill 
	\centering
	\begin{subfigure}{0.4\textwidth}
		\centering
		\begin{tikzpicture}[shorten >=1pt,node distance=2.5cm,on grid,auto] 
		\node[state,initial] (s_1)   {$s_1$}; 
		\node[state, accepting] (s_2) [right of = s_1] {$s_2$};
		\node[state] (s_3) [below right of = s_1] {$s_3$};

		\path[->] 
		(s_1)	edge   node  {$(1,2)$} (s_2)
		(s_1)	edge  node  [below,sloped] {$(1,0)$} (s_3)
		(s_2)   edge [loop right] node  [above] {$(1,1)$} (s_2)
		(s_3)   edge [loop right] node  [above]  {$(1,1)$} (s_3);
			
		\end{tikzpicture}
		\caption{Snippet of limsup comparator $\A_{\mathsf{LS}}^\leq$ for relation $\leq$}
		\label{Fig:SupComparator}
	\end{subfigure}
	\hfill
	\centering
	\begin{subfigure}{0.45\textwidth}
		\centering
		\begin{tikzpicture}[shorten >=1pt,node distance=3cm,on grid,auto] 
		\node[state,accepting, initial] (s_1)   {$s_1$}; 
		\node[state, accepting] (s_2) [right of = s_1] {$s_2$};

		\path[->] 
		(s_1)	edge  node  [below] {$(a,1,1,2,2)$} (s_2)
		(s_2)   edge [loop, distance=1.5cm] node  [above] {$(a,1,2,1,3)$} (s_2);
		
		\end{tikzpicture}
		\caption{$\Intersect$}
		\label{Fig:Intersect}
	\end{subfigure}
	\hfill
	\centering
	\begin{subfigure}{0.45\textwidth}
		\centering
		\begin{tikzpicture}[shorten >=1pt,node distance=3cm,on grid,auto] 
		\node[state,accepting, initial] (q_1)   {$p_1$}; 
		\node[state, accepting] (q_2) [right of = q_1] {$p_2$};
		
		\path[->] 
		(q_1)	edge  node  {$(a,1,1)$} (q_2)
		(q_2)   edge [loop above] node  {$(a,1,2)$} (q_2);
			
		\end{tikzpicture}
		\caption{$\first$}
		\label{Fig:Dim}
	\end{subfigure}
	\caption{Algorithm $\DSInclusion$: Steps on the running example}
\end{figure*}

\item $\CompareRuns$:  The output of this step is the B\"uchi automaton $\first$, that contains the word ${\hat{\rho}}\in\hat{P}$ iff $\rho$ is a dominated run in $P$ w.r.t $P\subseteq_f Q$ (Lines~\ref{alg-line:Prod}-\ref{alg-line:Project}).

$\MakeProduct(\hat{P}, \hat{Q})$ constructs $\hat{P}\times \hat{Q}$ s.t. word $(\hat{\rho_P}, \hat{\rho_Q}) \in \hat{P}\times \hat{Q}$ iff $\rho_P$ and $\rho_Q$ are runs of the same word in $P$ and $Q$ respectively (Line~\ref{alg-line:Prod}). 
Concretely, for transition $\hat{\tau_\A}=(s_\A,(a, n_\A, l_\A),t_\A)$ 
in automaton $\A$, where $\A \in \{\hat{P}, \hat{Q}\}$,
 transition $\hat{\tau_P}\times\hat{\tau_Q}=((s_P, s_Q),(a, n_P, l_P, n_Q, l_Q),(t_P, t_Q))$ is in $ \hat{P}\times\hat{Q}$, as shown in Figure~\ref{Fig:PhatQhat}.


$\Intersect$ intersects the weight components of $\hat{P}\times\hat{Q}$ with comparator $\A_f^\leq$ (Line~\ref{alg-line:Intersect}). The resulting  automaton $\less$ accepts word $(\hat{\rho_P}, \hat{\rho_Q})$ iff $f(\rho_P)\leq f(\rho_Q)$, and $\rho_P$ and $\rho_Q$ are runs on the same word in $P$ and $Q$ respectively. The result of $\Intersect$ between $\hat{P}\times\hat{Q}$ with the limsup comparator $\A_{\mathsf{LS}}^\leq$ for relation $\leq$ (Figure~\ref{Fig:SupComparator}) is given in Figure~\ref{Fig:Intersect}.

The projection of $\less$ on the words of $\hat{P}$ returns $\first$ which contains the word $\hat{\rho_P}$ iff $\rho_P$ is a dominated run in $P$ w.r.t $P\subseteq_f Q$ (Line~\ref{alg-line:Project}), as shown in Figure~\ref{Fig:Dim}.

\begin{figure*}[t]

\end{figure*}

\item $\EnsureRuns$: $P\subseteq_f Q$ iff $\hat{P} \equiv \first$ (qualitative equivalence) since $\hat{P}$ consists of all runs of $P$ and $\first$ consists of all domianted runs w.r.t $P\subseteq_f Q$ (Line~\ref{alg-line:ensure}). 
\end{enumerate}

\paragraph{Algorithm details}
We now prove the correctness of the algorithm and determine its complexity. We begin with a proof of correctness. 

\begin{lemma}
	\label{lem:dim}
	B\"uchi automaton $\first$ consists of all domianted runs in $P$ w.r.t $P\subseteq_f Q$.	
\end{lemma}

\begin{proof}
	Let $\A_f^\leq$ be the comparator for $\omega$-regular aggregate function $f$ and relation $\leq$ s.t. $\A_f$ accepts $(A,B)$ iff $f(A)\leq f(B)$. 
	A run $\rho$ over word $w$ with weight sequence $wt$ in $P$ (or $Q$) is represented by the unique word $\hat{\rho} = (w, wt, l)$ in $\hat{P}$ (or $\hat{Q}$) where $l$ is the unique label sequence associated with each run in $P$ (or $Q$). Since every label on each transition is separate, $\hat{P}$ and $\hat{Q}$ are deterministic automata. Now, $\hat{P}\times\hat{Q}$ is constructed by ensuring that two transitions are combined in the product only if their alphabet is the same. 
	Therefore if $(w, wt_1, l_1, wt_2, l_2) \in \hat{P}\times\hat{Q}$, then $\hat{\rho} = (w, wt_1, l_1)\in \hat{P}$, $\hat{\sigma} = (w, wt_2, l_2)\in \hat{Q}$. Hence, there exist runs $\rho$ and $\sigma$ with weight sequences $wt_1$ and $wt_2$ in $P$ and $Q$, respectively.
	Next, $\hat{P}\times\hat{Q}$ is intersected over the weight sequences with $\omega$-regular comparator $\A_f^\leq$ for aggregate function $f$ and relation $\leq$. 
	Therefore $(w, wt_1, l_1, wt_2, l_2) \in \less$
	iff  $f(wt_1) \leq f(wt_2)$. Therefore runs $\rho$ in $P$ and $\sigma$ in $Q$ are runs on the same word s.t. aggregate weight in $P$ is less than or equal to that of $\sigma$ in $Q$. Therefore $\first$ constitutes of these $\hat{\rho}$. 
	Therefore $\first$ consists of $\hat{\rho}$ only if $\rho$ is a dominated run in $P$ w.r.t $P\subseteq_f Q$.
	
	Every step of the algorithm has a two-way implication, hence it is also true that every dominated run in $P$ w.r.t $P\subseteq_f Q$ is present in $\first$.
	\qed
\end{proof}

\begin{lemma}
\label{Lemma:DSInclusionAlg}
Given weighted $\omega$-automata $P$ and $Q$ and their  $\omega$-regular  comparator $\A_f^\leq$ for aggregate function $f$ and relation $\leq$.
$\DSInclusion(P,Q,\A_f)$ returns $\mathsf{True}$ iff $P \subseteq_f Q$.
\end{lemma}
\begin{proof}
	$\hat{P}$ consists of all runs of $P$. $\first$ consists of all dominated run in $P$ w.r.t $P\subseteq_f Q$. $P\subseteq_f Q$ iff every run of $P$ is dominated w.r.t $P\subseteq_f Q$. Therefore $P\subseteq_f Q$ is given by whether $\hat{P} \equiv \first$, where $\equiv$ denotes qualitative equivalence.	
	\qed
\end{proof}

It is worth noting that Algorithm $\DSInclusion$ can be easily adapted to solve {\em strict} quantitative inclusion, denoted $P\subset_f Q$,  by repeating the same procedure with the $\omega$-regular comparator with the inequality relation $<$. In this case,  a run $\rho_P$ in $P$ on word $w\in \Sigma^\omega$ is said to be {\em dominated} w.r.t $P\subset_f Q$ if there exists a run $\rho_Q$ in $Q$ on the same word $w$ such that $wt_P(\rho_P)< wt_Q(\rho_Q)$. A similar adaption will work for quantitative equivalence, denoted  $P \equiv_f Q$.

Lastly, we present the complexity analysis $\DSInclusion$:

\begin{theorem}
	\label{thrm:RegularComplexity}
	
	Let $P$ and $Q$ be weighted $\omega$-automata and $\A_f$ be an $\omega$-regular comparator. 
	Quantitative inclusion problem, quantitative strict-inclusion problem, and quantitative equivalence problem for $\omega$-regular aggregate function $f$ is $\cc{PSPACE}$-complete.

\end{theorem}
\begin{proof}
All operations in $\DSInclusion$ until Line~\ref{alg-line:Project} are polytime operations in the  size of weighted $\omega$-automata $P$, $Q$ and comparator $\A_f$. Hence, $\first$ is polynomial in size of $P$, $Q$ and $\A_f$. Line~\ref{alg-line:ensure} solves a $\cc{PSPACE}$-complete problem. 
Therefore, the quantitative inclusion for $\omega$-regular aggregate function $f$ is in $\cc{PSPACE}$ in size of the inputs $P$, $Q$, and  $\A_f$.

The $\cc{PSPACE}$-hardness of the quantitative inclusion is established via reduction from the {\em qualitative} inclusion problem, which is   $\cc{PSPACE}$-complete. 
The formal reduction is as follows: Let $P$ and $Q$ be B\"uchi automata (with all states as accepting states). Reduce $P$, $ Q$ to weighted automata $\overline{P}$, $\overline{Q}$ by assigning a weight of 1 to each transition. 
Since all runs in $\overline{P}$, $\overline{Q}$ have the same weight sequence, weight of all words in $\overline{P}$ and $\overline{Q}$ is the same for any function $f$.  
It is easy to see $P \subseteq Q$ (qualitative inclusion) iff $\overline{P} \subseteq_f \overline{Q}$ (quantitative inclusion). 
\qed
\end{proof}

Theorem~\ref{thrm:RegularComplexity} extends to weighted $\omega$-automata when weight of words is the {\em infimum} of weight of runs. 
The key idea for $P\subseteq_f Q$ here is to ensure that for every run $\rho_Q$ in $Q$ there exists a run  on the same word in $\rho_P$ in $P$  s.t. $f(\rho_P)\leq f(\rho_Q)$.

\subsubsection*{Representation of counterexamples}
\label{Sec:Represent}

When $P \nsubseteq_f Q$,
there exists word(s) $w \in \Sigma^*$ s.t $wt_P(w)>wt_Q(w)$.
Such a word $w$ is said to be a {\em counterexample word}. 
The ability to extract counterexamples and analyze them has been of immense advantage in verification and synthesis in qualitative systems~\cite{baier2008principles}.

Here, we show how $\DSInclusion$ can be adapted to yields B\"uchi automaton-representations for all counterexamples of an instance of quantitative inclusion for $\omega$-regular comparators. Hopefully, this could be a starting point for counter-example guided frameworks in the quantitative verification of systems:

\begin{theorem}
	\label{Thm:RegularInclusion-Counterexample}
	All counterexamples of the quantitative inclusion problem for an $\omega$-regular aggregate function can be expressed by a B\"uchi automaton. 
	
\end{theorem}
\begin{proof}
	For word $w$ to be a counterexample, it must contain a run in $P$ that is not dominated. Clearly, all non-dominated runs of $P$ w.r.t to the quantitative inclusion are members of $\hat{P}\setminus \first$. The counterexamples words can be obtained from $\hat{P}\setminus \first$ by modifying its alphabet to the alphabet of $P$ by dropping transition weights and their unique labels.
	\qed
\end{proof}

\subsection{Quantitative games with perfect information}
\label{Sec:PerfectInfo}

Recall, a {\em quantitative graph with complete information game}, referred to as graph game in short,  is defined over a structure $G = (V  = V_0 \uplus V_1, \init, E, \gamma, f)$.
It consists of a directed graph $(V,E)$,
and a partition $(V_0, V_1)$ of its set of states $V$. 
State $\init$ is the {\em initial state} of the game. 
$vE $ designates the set $\{w \in V | (v, w) \in E\}$  to indicate the successor states of state $v \in V$. For convenience, we assume that every state has at least one outgoing edge, i.e, $vE  \neq \emptyset$ for all $v \in V$.
Each transition of the game is associated with a {\em cost} determined by the {\em cost function} $\gamma: E \rightarrow \Z$. Finally, $f:\Z^\omega \rightarrow \Re$
is the {\em aggregate function}.

We illustrate how to solve the satisficing problem over quantitative games when the comparator automata for the aggregate function is $\omega$-regular. 

Wlog, let the winning condition for the maximizing player be to ensure that the cost of resulting play is greater than or equal to a given threshold value $v$. Then the winning condition for the minimizing player is to ensure that the cost of plays is less than threshold $v$.

For sake of simplicity, let the threshold value be 0. Further, suppose that $f(B) = 0$ where $B$ is the sequence of all 0s. The we claim that the satisficing problem will reduce to solving a parity game.

Let $\A$ be the comparator automata for the inequality $\geq$. Currently $\A$ accepts a sequence $(A,B)$ iff $f(A)\geq f(B)$. We can convert $\A$ to an NBA that accepts sequence $A$ iff $f(A) \geq 0$ if we fix sequence $B$ to $0^\omega$. Let us denote this by $\A_0$. Note that the size of $\A_0$ is the same as $\A$. Finally, determinize $\A_0$ into a parity automata $P$. Now, the winning condition of the maximizing agent in this game is that the weight sequence of a play must be accepted by the automaton $P$. Therefore, we take synchronized product of the quantitative game with the parity automata to obtain a parity game. Then the maximizing player will win the quantitative game iff it wins in the aforementioned parity game.

\begin{theorem}
	\label{Thrm:GameIncomplete}
	Given a quantitative game with complete information $\G$ and $\omega$-regular comparator  $\A_f$ for  the aggregate function $f$. Let the threshold value be 0 and $f(0^\omega) = 0$. Then determining whether the maximizing agent has a winning strategy in the satisficing game with threshold 0 reduces to solving a parity game with $\O(|\G|\times |D_f|)$ states with $|\A_f|$ colors, where $|D| = |\A_f|^{|\A_f}$.
	\end{theorem}
\begin{proof}
	Since, $D_f$ is the deterministic parity automaton equivalent to $A_f, |D_f|= |\A_f|^{O(|\A_f|)}$. The parity automaton will have $|\A_f|$ colors. Therefore, the product game will be a parity automaton with states equal to the product of the size of the game and the parity automaton, and colors equal to the size of the initial NBA.
	\qed
\end{proof}

\subsection{Quantitative games with imperfect information}
\label{Sec:GraphGames}

Given an incomplete-information quantitative game $\G = (S,s_ {\mathcal{I}},\Obs, \Sigma, \delta, \gamma, f)$, our objective is to determine if player $P_0$ has a winning strategy $ \alpha: \Obs^* \rightarrow \Sigma$ for $\omega$-regular aggregate function $f$. We assume we are given the $\omega$-regular comparator $\A_f$ for function $f$. 
Note that a function $A^*\rightarrow B$ can be treated like a  $B$-labeled $A$-tree, and vice-versa. 
Hence, we proceed by finding a $\Sigma$-labeled $\Obs$-tree -- the {\em winning strategy tree}. Every branch of a winning strategy-tree is an {observed play} $o_{\rho}$ of $\G$ for which every actual play $\rho$ is a winning play for $P_0$.

We first consider all {\em game trees} of $\G$ by interpreting $\G$ as a tree-automaton over $\Sigma$-labeled $S$-trees. Nodes $n \in S^*$ of the game-tree correspond to states in $S$ and labeled by actions in $\Sigma$ taken by player $P_0$. Thus, the {\em root node} $\varepsilon$ corresponds to $s_\I$, and a node $s_{i_0},\ldots,s_{i_k}$ corresponds to the state $s_{i_k}$ reached via $s_{\I},s_{i_0},\ldots,s_{i_{k-1}}$.
Consider now a node $x$ corresponding to state $s$ and labeled by an action $\sigma$. Then $x$ has children $x s_1,\ldots x s_n$, for every $s_i \in S$. If $s_i\in \delta(s,\sigma)$, then we call $x s_i$ a \emph{valid} child, otherwise we call it an \emph{invalid} child. Branches that contain invalid children correspond to invalid plays.


A game-tree $\tau$ is a {\em winning tree} for player $P_0$ if every branch of $\tau$ is either a winning play for $P_0$ or an invalid play of $\G$.  
One can check, using an automata, if a play is invalid by the presence of invalid children.
Furthermore, the winning condition for $P_0$ can be expressed by the $\omega$-regular comparator $\A_f$ that accepts $(A,B)$ iff $f(A) > f(B)$.
To use the comparator $ \A_f$, it is determinized to  parity automaton $D_f$.
Thus, a product of  game $\G$ with $D_f$ is a deterministic parity tree-automaton accepting precisely winning-trees for player $P_0$. 


Winning trees for player $P_0$ are $\Sigma$-labeled $S$-trees. We need to convert them to $\Sigma$-labeled $\Obs$-trees. Recall that every state has a unique observation.
We can simulate these $\Sigma$-labeled $S$-trees on strategy trees using the technique of {\em thinning} states $S$ to observations $\Obs$~\cite{kupfermant2000synthesis}. The resulting alternating parity tree automaton $\mathcal{M}$ will accept a $\Sigma$-labeled $\Obs$-tree $\tau_o$ iff for all actual game-tree $\tau$ of $\tau_o$, $\tau$ is a winning-tree for $P_0$ with respect to the strategy $\tau_o$. 
The problem of existence of winning-strategy for $P_0$ is then reduced to non-emptiness checking of $\mathcal{M}$.


\begin{theorem}
\label{Thrm:GameIncomplete}
Given an incomplete-information quantitative game $\G$ and $\omega$-regular comparator  $\A_f$ for  the aggregate function $f$, the complexity of determining whether $P_0$ has a winning strategy is exponential in ${|\G| \cdot |D_f|} $, where $|D_f| = |\A_f|^{O(|\A_f|)}$.
\end{theorem}
\begin{proof}
Since, $D_f$ is the deterministic parity automaton equivalent to $A_f, |D_f|= |\A_f|^{O(|\A_f|)}$. The thinning operation is linear in size of $|\G\times D_f|$, therefore $|\mathcal{M}| = |\G|\cdot| D_f|$. Non-emptiness checking of alternating parity tree automata is exponential. Therefore, our procedure is doubly exponential in size of the comparator and exponential in size of the game. 
\qed
\end{proof}

 The question of tighter bounds is open.

\section{Chapter summary}
\label{Sec:Conclusion}

This chapter identified a novel mode for comparison in quantitative systems: the online comparison of aggregate values of sequences of quantitative weights. This notion is embodied by comparators automata that read two infinite sequences of weights synchronously and relate their aggregate values.
We showed that $\omega$-regular comparators not only yield generic algorithms for problems including quantitative inclusion and winning strategies in incomplete-information quantitative games, they also result in algorithmic advances. 

We believe comparators, especially $\omega$-regular comparators, can be of significant utility in verification and synthesis of quantitative systems, as demonstrated by the existence of finite-representation of counterexamples of the quantitative inclusion problem. Another potential application is computing equilibria in quantitative games. Applications of the prefix-average comparator, in general $\omega$-context-free comparators, is open to further investigation. Another direction to pursue is to study aggregate functions in more detail, and attempt to solve the conjecture relating $\omega$-regular aggregate functions and $\omega$-regular comparators.

\chapter{On comparators of popular aggregate functions}
\label{ch:applicationcomparator}

Chapter~\ref{ch:Comparator} laid down the theoretical foundations of comparator automata, and demonstrated that for the class of aggregate functions that permit $\omega$-regular comparators, comparator automata result in algorithms that are generalizable as well as integrated in approach. 
This chapter asks a concrete question: {\em Which aggregate functions allow $\omega$-regular comparators?}

We know from Theorem~\ref{thrm:limsupcomparator} that limsup and liminf posses $\omega$-regular comparators. This chapter investigates the comparators for commonly appearing aggregate functions, namely discounted-sum and limit-average. We observe that discounted-sum permits $\omega$-regular comparators iff the discount factor is an integer (Section~\ref{Sec:RationalDF}-~\ref{Sec:IntegerDF}), while limit-average does not permit $\omega$-regular comparators (Section~\ref{Sec:LA}).
In an attempt to establish a broader characterization of which aggregate function permit $\omega$-regular comparators, we investigate the class of $\omega$-regular functions. We deliver positive news: All $\omega$-regular functions will have an $\omega$-regular comparator (Section~\ref{Sec:RegularAggFunctions}).

Last but not the least, despite being a generalizable framework, comparator-based algorithms may be better in complexity that existing approaches. In particular, for the case of DS with integer discount factors, we observe that the comparator-based algorithm is \cct{PSPACE} as opposed to the previously known \cct{EXPTIME} and \cct{EXPSPACE} algorithm. This also establishes that DS inclusion for integer discount factor is \cct{PSPACE}-complete, resolve the open question about its complexity class (Section~\ref{Sec:DSApplication}).

\section{Discounted-sum with non-integer discount factors}
\label{Sec:RationalDF}

Recall, discounted-sum aggregation accumulates diminishing returns. 
This section establishes is that a DS comparison language and DS comparator are not $\omega$-regular when the associated discount factor is not an integer. 

We begin by formally defining DS comparison languages and DS comparator automata.
Given an upper bound $\mu>0$, discount-factor $d>1$, and an inequality or equality relation $\mathsf{R} \in \{<,>,\leq, \geq, =,\neq\}$, the {\em DS comparison language with $\mu$, $d$, and $\mathsf{R}$} is a language over the alphabet $\Sigma\times\Sigma$, where $\Sigma = \{-\mu, -\mu+1, \dots, \mu\}$, that accepts the pair of infinite-length  weight sequences $(A,B)$ iff $\DSum{A}{d}$ $\mathsf{R}$ $\DSum{B}{d}$ holds.
Given an upper bound $\mu>0$, discount-factor $d>1$ and inequality or equality relation $\mathsf{R} \in \{<,>,\leq, \geq, =,\neq\}$, the {\em DS comparator automata (DS comparator, in short) for $\mu$, $d$, and $\mathsf{R}$} is the automaton that accepts the DS comparison language for $\mu$, $d$, and $\mathsf{R}$.

The proof begins with defining a {\em cut-point language}~\cite{chatterjee2009expressiveness}: For a weighted $\omega$-automaton $\A$ and a real number $r \in \Re$, the {\em cut-point language} of $\A$ w.r.t.  $r $ is defined as  $L^{\geq r} = \{w \in L(\A) | wt_\A(w) \geq r \}$. 
When the discount factor is a rational value $1<d<2$, it is known that not all deterministic weighted $\omega$-automaton with discounted-sum aggregate function (DS-automaton, in short) have an $\omega$-regular cut-point language for an $r \in \Re$~\cite{chatterjee2009expressiveness}.
In this section, this prior result is first extended to all non-integer, rational discount factors $d>1$. 
Finally, this extension is utilized to establish that the DS comparison language for a non-integer, rational discount-factors $d>1$ is not $\omega$-regular. This proves that the DS aggregation function is not $\omega$-regular when the discount factor is a non-integer, rational number. 


\begin{theorem}
\label{lem:cutpoint}
Let $d>1$ be a non-integer, rational discount factor.  There exists a deterministic discounted-sum automata $\A$ and a rational value $r \in \Q$ such that its cut-point language w.r.t. $r$ is not $\omega$-regular.
\end{theorem}
\begin{proof}

Since the proof for $1<d<2$ has been presented in~\cite{chatterjee2009expressiveness}, we skip that case. 

The proof presented here extends the earlier result on $1<d<2$ from~\cite{chatterjee2009expressiveness} to all non-integer, rational discount factors $d>1$.

Let $d>2$ be a non-integer, rational discount-factor. Define deterministic discounted-sum automata $\A$ over the alphabet $\{0,1,\dots, \ceil{d}-1\}$ such that the weight of transitions on alphabet $n \in \{0,1,\dots, \ceil{d}-1\}$ is $n$. Therefore, weight of word $w \in \A$ is $\DSum{w}{d}$. Consider its cut-point language $L^{\geq (\ceil{d}-1)}$. We say a finite-length word $w$ is {\em ambiguous} iff $\ceil{d}-1 - \frac{1}{d^{|w|-1}} \cdot \frac{\ceil{d}-1}{d-1} \leq DS(w,d) < \ceil{d}-1$. Intuitively, a finite word $w$ is ambiguous if it can be extended into infinite words in $\A$ such that some extensions lie in $L^{\geq (\ceil{d}-1)}$ and some do not. Clearly, the finite word $w = \ceil{d}-2$ is ambiguous. Note that when $d>2$ is non-integer, rational valued, then $\frac{\ceil{d}-1}{d-1} > 1$, so word $\ceil{d}-2$ falls within the range for ambiguity. We claim that if finite word $w$ is ambiguous, then either $w\cdot (\ceil{d}-2)$ or $w\cdot (\ceil{d}-1)$ is ambiguous. 
To prove ambiguity, we need to show that $\ceil{d}-1 - \frac{1}{d^{|w|}} \cdot \frac{\ceil{d}-1}{d-1}\leq \DSum{w\cdot (\ceil{d}-k)}{d} < \ceil{d}-1$, for $k \in \{1,2\}$. Note, $\DSum{w\cdot (\ceil{d}-2)}{d} = \DSum{w}{d} + \frac{1}{d^{|w|}}\cdot (\ceil{d}-2)$, and $\DSum{w\cdot (\ceil{d}-1)}{d} = \DSum{w}{d} + \frac{1}{d^{|w|}}\cdot (\ceil{d}-1)$.
 Simplifying the expressions for $w\cdot (\ceil{d}-1)$ and $w\cdot (\ceil{d}-2)$, we need to prove 
either $\ceil{d}-1 - \frac{1}{d^{|w|+1}} \cdot \frac{\ceil{d}-1}{d-1}\leq \DSum{w}{d} < \ceil{d}-1 - \frac{1}{d^{|w|}}\cdot (\ceil{d}-1)$ or 
$\ceil{d}-1 - \frac{1}{d^{|w|}} \cdot \frac{\ceil{d}-1}{d-1} - \frac{1}{d^{|w|}}\cdot(\ceil{d}-2)\leq \DSum{w}{d} < \ceil{d}-1 - \frac{1}{d^{|w|}}\cdot (\ceil{d}-2)$. 
Now, this is true if $\ceil{d}-1 - \frac{1}{d^{|w|}}\cdot\frac{\ceil{d}-1}{d-1} - \frac{1}{d^{|w|}}\cdot(\ceil{d}-2) < \ceil{d}-1 - \frac{1}{d^{|w|}}\cdot (\ceil{d}-1)$ which is equivalent  $d>2$ is  a is non-integer, rational discount-factor. Therefore, every ambiguous finite word can be extended to another ambiguous word.  This means there exists an infinite word $w^\geq$ such that $\DSum{w^\geq}{d} = \ceil{d}-1$ and all finite prefixes of $w^\geq$ are ambiguous. 

Let us assume that the language $L^{\geq (\ceil{d}-1)}$ is $\omega$-regular and represented by B\"uchi automaton $\mathcal{B}$. For $n<m$, let the $n$- and $m$-length prefixes of $w^{\geq}$, denoted $w^{\geq}[0,n-1]$ and $w^{\geq}[0,m-1]$, respectively, be such that they reach the same states in $\mathcal{B}$. Then there exists an infinite length word $w_s$ such that $\DSum{w^{\geq}[0,n-1]\cdot w_s}{d} = \DSum{w^{\geq}[0,m-1]\cdot w_s}{d} = \ceil{d}-1$. Now, $\DSum{w^{\geq}[0,n-1]\cdot w_s}{d} = \DSum{w^{\geq}[0,n-1]}{d} + \frac{1}{d^n}\cdot\DSum{w_s}{d}$ and $\DSum{w^{\geq}[0,m-1]\cdot w_s}{d} = \DSum{w^{\geq}[0,m-1]}{d} + \frac{1}{d^m}\cdot\DSum{w_s}{d}$. Eliminating $\DSum{w_s}{d}$ from the equations and simplification, we get:
\[
d^{m-1}\cdot(\DSum{w^{\geq}[0,m-1]}{d} - (\ceil{d}-1)) + d^{n-1}\cdot(\DSum{w^{\geq}[0,n-1]}{d} - (\ceil{d}-1))  = 0
\]
The above is a polynomial over $d$ with degree $m-1$ and integer coefficients. Specifically, $d = \frac{p}{q} > 2$ such that integers $p,q>1$, and $p$ and $q$ are mutually prime. Since $d =\frac{p}{q}$ is a root of the above equation, $q$ must divide co-efficient of the highest degree term, in this case it is $m-1$. The co-efficient of the highest degree term in the polynomial above is $(w^{\geq}[0] - (\ceil{d}-1))$.  Recall from construction of $w^{\geq}$ above, $w^{\geq}[0] = \ceil{d}-2$. So the co-efficient of the highest degree term is $-1$, which is not divisible by integer $q>1$.  Hence, resulting in a contradiction. 
\qed
  
\end{proof}

Finally, we use Theorem~\ref{lem:cutpoint} to prove the DS comparison language is not $\omega$-regular when the discount-factor $d>1$ is not an integer. 

\begin{theorem}
\label{Thrm:DSNotRegular}
	Comparison language for the DS aggregation function is not $\omega$-regular when the discount-factor $d>1$ is a non-integer, rational number.
\end{theorem}
\begin{proof}
	To prove that the DS aggregation function is not $\omega$-regular when the discount factor is not an integer, it is sufficient to prove that DS comparison language for $\geq$ is not $\omega$-regular.
	
	Let $d>1$ be a non-integer, rational discount factor. Let $ \A$ be the weighted $\omega$-automaton as described in proof of Lemma~\ref{lem:cutpoint}. Consider its cut-point language $L^{\geq (\ceil{d}-1)}$. From Lemma~\ref{lem:cutpoint} and~\cite{chatterjee2009expressiveness}, we know that $L^{\geq (\ceil{d}-1)}$ is not an $\omega$-regular language. 
		
	Suppose there exists an $\omega$-regular DS comparator $\A_d^\leq$ for non-integer rational discount factor $d>1$ for relation $\geq$.
	We define the B\"uchi automaton $\mathcal{P}$ s.t. $\L(\mathcal{P}) = \{(w,v) | w \in \L(\A), v = \ceil{d}-1 \cdot 0^{\omega} \}$. Note that $\DSum{\ceil{d}-1\cdot 0^{\omega}}{d} = \ceil{d}-1$. . Then the cut-point language $L^{\geq (\ceil{d}-1)}$ of deterministic discounted-sum automata $\A$ can be constructed by taking the intersection of $\mathcal{P}$ with $\A_d^\geq$. Since all actions are closed under $\omega$-regular operations, $L^{\geq 1}$ can be represented by a B\"uchi automaton. But this contradicts Theorem~\ref{lem:cutpoint}. Hence, our assumption cannot hold. 
	\qed
\end{proof}

Since the DS comparison language with a non-integer discount-factor for any one inequality relation is not $\omega$-regular, due to closure properties of $\omega$-regular (Theorem~\ref{lem:closure}-Item~\ref{item:closednotregular}) this implies that the DS comparison language for all other inequities is also not $\omega$-regular all inequalities is also not $\omega$-regular. Finally,  the cut-point argument can be extended to $=$ and $\neq$ relation to show that their DS comparison languages with non-integer discount-factors is also not $\omega$-regular.

\section{Discounted-sum with integer discount factors}
\label{Sec:IntegerDF}

This section proves that a DS comparison languages is $\omega$-regular when the discount-factor $d>1$ is an integer. Hence, DS aggregation function is $\omega$-regular for integer discount-factors. 
The result is proven by explicitly constructing a B\"uchi automaton that accepts the DS comparison language for an arbitrary upper bound $\mu$, an inequality relation $\mathsf{R}$ and integer discount factor $d >1$. 

An immediate side-effect of this result is that DS inclusion is $\cct{PSPACE}$-complete. Not only does this improve upon the previously best known algorithm that was $\cct{EXPTIME}$ and $\cct{EXPSPACE}$, it also resolves 15-year long open question of the complexity class of DS inclusion with integer discount factor (Section~\ref{Sec:DSApplication}).

\subsubsection{Core intuition} 
Recall the definitions of DS comparison language and DS comparator automata from Section~\ref{Sec:RationalDF}.

Let integer $\mu>0$ be the upper-bound on sequences. 
The core intuition is that bounded sequences can be converted to their value in an integer base $d$ via a finite-state transducer. Lexicographic comparison of the converted sequences renders the desired DS-comparator. Conversion of sequences to base $d$ requires a certain amount of {\em look-ahead} by the transducer.
Here we describe a method that directly incorporates the look-ahead with lexicographic comparison to obtain the B\"uchi automaton corresponding to the DS comparator for integer discount factor $d>1$. 

\subsubsection{Construction details}
We explain the construction in detail now. 
We complete the construction for the relation $<$.
For sake of simplicity, we assume that sequences the weight sequences are {\em positive} integer sequences. Under  this assumption, for a weight  sequence $A$ and integer discount-factor $d>1$, $\DSum{A}{d}$ can be interpreted as a value in base $d$ i.e. $\DSum{A}{d} = A[0] + \frac{A[1]}{d} + \frac{A[2]}{d^2} + \dots =  (A[0].A[1]A[2]\dots)_d$~\cite{chaudhuri2013regular}. 
The proofs can be extended to  integer sequences easily. 

Unlike comparison of numbers in base $d$, the lexicographically larger sequence may not be larger in value since (i) The elements of  weight sequences may be larger in value than base $d$, and (ii) Every value has multiple infinite-sequence representations. 
To overcome the two challenges mentioned above, we resort to arithmetic techniques in base $d$. 
Note that $ \DSum{B}{d} > \DSum{A}{d} $ iff there exists a sequence $C$ such that $\DSum{B}{d} =  \DSum{A}{d} + \DSum{C}{d} $, and $\DSum{C}{d} > 0$.
Therefore, to compare the discounted-sum of $A$ and $B$, the objective is to  
 obtain the sequence $ C $.
 Arithmetic in base $d$ also results in sequence $X$ of carry elements. Then:
 
\begin{lemma}
 \label{Lemma:discountinvariant1}
Let $A, B, C, X$ be weight sequences, $ d > 1 $ be a positive integer such that following equations holds true:
 \begin{enumerate}
 \item  \label{eq:initial} When $ i = 0 $, $ A[0] + C[0] + X[0] = B[0]$
 \item \label{eq:invariant} When $ i\geq 1 $, $A[i] + C[i] + X[i] = B[i] + d \cdot X[i-1]$
 \end{enumerate}
 Then $ \DSum{B}{d} = \DSum{A}{d} + \DSum{C}{d}$.
 \end{lemma}
 \begin{proof}
  $ \DSum{A}{d} + \DSum{C}{d} = \Sigma_{i=0}^{\infty} A[i] \frac1{d^i} + \Sigma_{i=0}^{\infty} C[i] \frac1{d^i} = \Sigma_{i=0}^{\infty} (A[i] + C[i]) \frac1{d^i} = (B[0] - X[0]) + \Sigma_{i=1}^{\infty} (B[i] + d\cdot X[i-1] - X[i]) \frac1{d^i} =  (B[0] - X[0]) + \Sigma_{i=1}^{\infty} (B[i] + d\cdot X[i-1] - X[i]) \frac1{d^i} = \Sigma_{i=0}^{\infty} B[i] \cdot \frac1{d^i} - \Sigma_{i=0}^{\infty} X[i] +  \Sigma_{i=0}^{\infty} X[i]  =  \Sigma_{i=0}^{\infty} B[i] \cdot \frac1{d^i} = \DSum{B}{d}$
  \qed
  \end{proof}
  
Hence to determine $\DSum{B}{d}-\DSum{A}{d}$, systematically guess sequences $ C $ and $ X $ using the equations, element-by-element beginning with the 0-th index and moving rightwards. 
There are two crucial observations here: (i)
Computation of $i$-th element of $C$ and $X$ only depends on $i$-th and $(i-1)$-th elements of $A$ and $B$. Therefore guessing $C[i]$ and $X[i]$ requires {\em finite memory} only. 
(ii) Intuitively, $C$ 
refers to a representation of value $\DSum{B}{d} - \DSum{A}{d}$ in base $d$ and $X$ is the carry-sequence. If we can prove that $X$ and $C$ are also bounded-sequences and can be constructed from a finite-set of integers, we would be able to further proceed to construct a B\"uchi automaton for the desired comparator. 

We proceed by providing an inductive construction of sequences $C$ and $ X$ that satisfy properties in Lemma~\ref{Lemma:discountinvariant1}, and show that these sequences are bounded when $A$ and $B$ are bounded. In particular, when $A $ and $B$ are bounded integer-sequences, then sequences $C$ and  $X$ constructed here are also bounded-integer sequences. Therefore, they are be constructed from a finite-set of integers. Proofs for sequence $C$ are in Lemma~\ref{Lemma:SimplifyResidual}-Lemma~\ref{Lemma:BoundOnC}, and proof for sequence $X$ is in Lemma~\ref{Lemma:BoundOnX}.

We begin with  introducing some notation.
Let $\DSumDiff{i} = \Sigma^i_{j=0} (B[j] - A[j])\cdot \frac{1}{d^j} $ for all index $i\geq 0$. Also, let $ \DSumDiff{\cdot} = \Sigma^{\infty}_{j=0} (B[j] - A[j])\cdot \frac{1}{d^j} = \DSum{B}{d} - \DSum{A}{d} $. Define $ \MaxC = \max\cdot\frac{d}{d-1} $.
 We define the residual function $ \Res: \mathbb{N} \cup \{0\} \mapsto \mathbb{R} $ as follows:
 \[
 \Res(i) = 
 \begin{cases}
 \DSumDiff{\cdot} - \floor{\DSumDiff{\cdot}} & \text{if } i = 0 \\
 \Res(i-1) - \floor{\Res(i-1)\cdot d^i}\cdot \frac{1}{d^i } & \text{otherwise} 
 \end{cases}\]
 Then we define $C[i]$ as follows:
 \[
 C[i] = 
 \begin{cases}
 \floor{\DSumDiff{\cdot}} &\text{if } i = 0 \\
 \floor{\Res(i-1)\cdot d^i } &\text{otherwise} 
 \end{cases}\]
 Intuitively, $ C[i] $ is computed by \emph {stripping off} the value of the $ i $-th digit in a representation of $ \DSumDiff{\cdot} $ in base $ d $.
 $ C[i] $ denotes the numerical value of the $ i $-th position of the difference between $ B $ and $ A $. The residual function denotes the numerical value of the difference remaining after assigning the value of $ C[i] $ until that $ i $. 
 
 We define function $ \CSum(i): \N\cup\{0\} \rightarrow \mathbb{Z}$ s.t. $\CSum(i) = \Sigma_{j=0}^{i} C[j]\cdot \frac{1}{d^j} $.
 Then, we define $  X[i] $ as follows:
 \[
 X[i] = (\DSumDiff{i} - \CSum(i)) \cdot d^i 
 \]
Therefore, we have defined sequences $C$ and $X$ as above. We now prove the desired properties  one-by-one. 

 First, we establish  sequences $C$, $X$ as defined here satisfy Equations~\ref{eq:initial}-\ref{eq:invariant} from Lemma~\ref{Lemma:discountinvariant1}. Therefore, ensuring that $C$ is indeed the difference between sequences $B$ and $A$, and $X$ is their carry-sequence.
  \begin{lemma}
  \label{Lemma:XInvariant}
  Let $A$ and $B$ be bounded integer sequences and $C$ and $X$ be defined as above. Then, 
  \begin{enumerate}
  \item $ B[0] = A[0] + C[0] + X[0] $
  \item For $ i \geq 1 $, $ B[i] + d\cdot X[i-1] = A[i] + C[i] + X[i] $
  \end{enumerate}
  \end{lemma} 
  
  \begin{proof}
  We prove this by induction on $ i $ using definition of function $ X $.
  
  When $ i = 0 $, then $ X[0] = \DSumDiff{0} - \CSum(0) \implies X[0] = B[0] - A[0] - C[0] \implies B[0] = A[0] + C[0] + X[0] $. 
  
  
  When $ i = 1 $, then $ X[1] =  (\DSumDiff{1} - \CSum(1))\cdot d = (B[0] + B[1]\cdot \frac1d ) - (A[0] + A[1]\cdot\frac1d) - (C[0] + C[1]\cdot\frac1d))\cdot d \implies X[1] = B[0]\cdot d + B[1] - (A[0]\cdot d + A[1]) - (C[0]\cdot d + C[1])$. From the above we obtain $X[1] = d \cdot X[0] + B[1] - A[1] - C[1] \implies B[1] + d \cdot X[0] = A[1] + C[1]+ X[1] $. 
  
  
  Suppose the invariant holds true for all $ i\leq n $, we show that it is true for $ n+1 $. $X[n+1] = (\DSumDiff{n+1} - \CSum(n+1))\cdot d^{n+1} \implies X[n+1] = (\DSumDiff{n} - \CSum(n))\cdot d^{n+1} + (B[n+1] - A[n+1] - C[n+1]) \implies X[n+1] = X[n]\cdot d + B[n+1] - A[n+1] - C[n+1] \implies B[n+1] + X[n] \cdot d = A[n+1] + C[n+1] + X[n+1]$. 
  \qed
  \end{proof}
  
  Next, we establish the sequence $C$ is a bounded integer-sequences, therefore it can be represented by a finite-set of integers. First of all, by definition of $C[i]$ it is clear that $C[i]$ is an integer for all $i\geq 0$.  We are left with proving boundedness of $C$. Lemma~\ref{Lemma:SimplifyResidual}-Lemma~\ref{Lemma:BoundOnC} establish boundedness of $C[i]$.
 
  \begin{lemma}
  \label{Lemma:SimplifyResidual}
  For all $ i \geq 0 $, $ \Res(i) = \DSumDiff{\cdot} - \CSum(i) $.
  \end{lemma}
  \begin{proof}
  Proof by simple induction on the definitions of functions $ \Res $ and $ C $.
  
  \begin{enumerate}
  	\item When  $i = 0$,  $\Res(0) = \DSumDiff{\cdot} - \floor{\DSumDiff{\cdot}}$. By definition of $C[0]$, $\Res(0) = \DSumDiff{\cdot} - C[0] \iff \Res(0) = \DSumDiff{\cdot} - \CSum(0)  $.
  	
  	\item Suppose the induction hypothesis is true for all $i<n$. We prove it is true when $i = n$. When $i = n$, $\Res(n) = \Res(n-1) - \floor{\Res(n-1)\cdot d^{n} } \cdot \frac{1}{d^n}$.  By definition of $C[n]$ and I.H, we get 
  	$\Res(n) = (\DSumDiff{\cdot} - \CSum(n-1)) - C[n] \cdot \frac{1}{d^n}$. Therefore
  		$\Res(n) = \DSumDiff{\cdot} - \CSum(n)$.
  \end{enumerate}
  \qed
  \end{proof}
  
  \begin{lemma}
  \label{Lemma:BoundOnRes}
  When $ \DSumDiff{\cdot}\geq 0 $, for all $ i\geq  0 $, $ 0 \leq \Res(i) < \frac{1}{d^i} $.
  \end{lemma}
  \begin{proof}
   Since, $ \DSumDiff{\cdot}\geq 0 $, $ \Res(0) = \DSumDiff{\cdot} - \floor{\DSumDiff{\cdot}} \geq 0 $ and $ \Res(0) = \DSumDiff{\cdot} - \floor{\DSumDiff{\cdot}} < 1 $ . Specifically, $ 0 \leq \Res(0) < 1 $. 
   
   Suppose for all $ i \leq k $, $ 0\leq  \Res(i) < \frac{1}{d^i} $. We show this is true even for $ k+1 $.
   
   Since $ \Res(k)\geq 0 $,  $ \Res(k)\cdot d^{k+1}  \geq 0 $. Let $ \Res(k)\cdot d^{k+1}  = x+f $, for integral $ x\geq 0 $, and fractional $ 0 \leq f < 1 $. Then, from definition of $\Res$, we get $\Res(k+1) = \frac{x+f}{d^{k+1}} - \frac{x}{d^{k+1}} \implies \Res(k+1) < \frac{1}{d^{k+1}} $. 
   
   Also, $ \Res(k+1) \geq 0 $ since $ a - \floor{a } \geq 0 $ for all positive values of $ a $ (Lemma~\ref{Lemma:SimplifyResidual}).
   \qed
  \end{proof}
  
  \begin{lemma}
  \label{Lemma:BoundOnC}
Let $\MaxC = \mu\cdot \frac{d}{d-1}$.  When $ \DSumDiff{\cdot} \geq 0 $, for $ i = 0 $, $0 \leq C(0) \leq \MaxC $, and for $ i \geq 1 $, $0 \leq  C(i) < d $.
  \end{lemma}
  \begin{proof}
  Since both $ A $ and $ B $ are non-negative bounded weight sequences, maximum value of $ \DSumDiff{\cdot} $ is when $ B = \{\max\}_{i} $ and $ A = \{0\}_i $. In this case $ \DSumDiff{\cdot} = \MaxC $. Therefore, $ 0 \leq C[0] \leq \MaxC $.
  
  From Lemma~\ref{Lemma:BoundOnRes}, we know that for all $ i$, $ 0 \leq \Res(i) < \frac{1}{d^i} $. Alternately, when $ i \geq 1 $, $ 0 \leq \Res(i-1) < \frac{1}{d^{i-1}} \implies 0 \leq \Res(i-1) \cdot d^i < \frac{1}{d^{i-1}} \cdot d^i \implies 0 \leq \Res(i-1)\cdot d^i < d \implies 0 \leq \floor{\Res(i-1)\cdot d^i} < d \implies 0 \leq C[i] < d$.
  \qed
  \end{proof}
  
Therefore, we have established  that sequence $C$ is non-negative integer-valued and is bounded by $\MaxC = \mu\cdot \frac{d}{d-1}$.

 Finally,  we prove that sequence $X$ is also a bounded-integer sequence, thereby proving that it is bounded, and can be represented with a finite-set of integers. Note that for all $i\geq 0$, by expanding out the definition of $X[i]$ we get that $X[i]$ is an integer for all $i\geq 0$.
 We are left with proving boundedness of $X$:

 \begin{lemma}
 \label{Lemma:BoundOnX}
 Let $ \MaxX = 1 + \frac{\max}{d-1} $. When $ \DSumDiff{\cdot} \geq 0 $, then for all $ i\geq 0 $, $ |X(i)| \leq \MaxX $.
 \end{lemma}
 \begin{proof}
 From definition of $ X $, we know that $  X(i) = (\DSumDiff{i} - \CSum(i)) \cdot d^i  \implies  X(i) \cdot \frac{1}{d^i} =  \DSumDiff{i} - \CSum(i) $. From Lemma~\ref{Lemma:SimplifyResidual} we get $ X(i) \cdot \frac{1}{d^i} =  \DSumDiff{i} - (\DSumDiff{\cdot} - \Res(i)) \implies X(i) \cdot \frac{1}{d^i} =  \Res(i) - (\DSumDiff{\cdot} - \DSumDiff{i}) \implies  X(i) \cdot \frac{1}{d^i} =  \Res(i) - 
  (\Sigma_{j=i+1}^{\infty}(B[j]-A[j])\cdot \frac{1}{d^j}) \implies  |X(i) \cdot \frac{1}{d^i}| \leq |\Res(i)| + |(\Sigma_{j=i+1}^{\infty}(B[j]-A[j])\cdot \frac{1}{d^j}) | \implies  |X(i) \cdot \frac{1}{d^i}| \leq |\Res(i)| + \frac{1}{d^{i+1}}\cdot|(\Sigma_{j=0}^{\infty}(B[j+i+1]-A[j+i+1])\cdot \frac{1}{d^j}) | \implies  |X(i) \cdot \frac{1}{d^i}| \leq |\Res(i)| + \frac{1}{d^{i+1}}\cdot|\MaxC|$. From Lemma~\ref{Lemma:BoundOnRes}, this implies  $|X(i) \cdot \frac{1}{d^i}| \leq \frac{1}{d^i} + \frac{1}{d^{i+1}}\cdot|\MaxC| \implies  |X(i)| \leq 1 + \frac{1}{d}\cdot|\MaxC| \implies  |X(i)| \leq 1 + \frac{\max}{d-1} \implies |X(i)| \leq \MaxX$.
  \qed
 \end{proof}
 
 We summarize our results from~\ref{Lemma:XInvariant}-Lemma~\ref{Lemma:BoundOnX} as follows:
 \begin{corollary}
 	\label{lem:FiniteAlphabetXandC}
 	Let $d>1$ be an integer discount-factor. Let $A$ and $B$ be non-negative
 	integer sequences bounded by $\mu$, and $\DSum{A}{d} <
 	\DSum{B}{d}$. Then there exists bounded integer-valued sequences $X$ and $C$ that satisfy the conditions in Lemma~\ref{Lemma:discountinvariant1}. Furthermore, $C$ and $X$ are bounded as follows:
 	\begin{enumerate}
 		\item \label{Item1:BoundOnC}  $0 \leq C[0] \leq \mu\cdot \frac{d}{d-1}$ and for all $i\geq 1$, $0\leq C[i] < d$, 
 		\item For all $i \geq 0$, $0 \leq |X[i]| \leq 1 + \frac{\mu}{d-1}$
 	\end{enumerate} 
 \end{corollary}

Intuitively, we construct a B\"uchi automaton $\A_{\DSsucceqFord{d}}^< $ with states of the form $(x,c)$ where $x$ and $c$ range over all possible values of $X$ and $C$, respectively, and a special initial state $s$. Transitions over alphabet $(a,b)$ replicate the equations in Lemma~\ref{Lemma:discountinvariant1}.
i.e.  transitions from the start state $(s,(a,b), (x,c))$ satisfy  $a+c+x = b$ to replicate Equation~\ref{eq:initial} (Lemma~\ref{Lemma:discountinvariant1}) at the 0-th index, and all other transitions $((x_1, c_1), (a,b), (x_2, c_2))$ satisfy $a+ c_2 + x_2 = b + d\cdot x_1$ to replicate Equation~\ref{eq:invariant} (Lemma~\ref{Lemma:discountinvariant1}) at indexes $i>0$. 
The complete construction is as follows:

\renewcommand{\max}{\mu}
\renewcommand{\MaxC}{\mu_C}
\renewcommand{\MaxX}{\mu_X}

\paragraph{Full and final construction} 

Let $d > 1$ be an integer discount factor, $\mu>0$ be the upper bound, and $<$ be the strict inequality relation. 
The DS comparator automata with discount factor $d>1$, upper bound $\mu$, and relation $<$ is constructed as follows:

Let $ \MaxC = \max \cdot \frac{d}{d-1}  $ and $ \MaxX = 1 + \frac{\max}{d-1}$.
Then, construct B\"uchi automaton 
 $ \A_{\DSsucceqFord{d}}^< = (\Statess, \Sigma, \delta_d, \StartState, \Final) $ such that,
\begin{itemize}

\item $ \Statess =  \{s\} \cup \Final \cup S_{\bot} $ where \\ 
$\Final = \{(x,c) ||x| \leq \MaxX, 0 \leq c \leq \MaxC \} $, and \\
$S_{\bot} = \{(x, \bot) | | x| \leq \MaxX\}$ where $\bot$ is a special character, and $ c \in \N$, $x \in \mathbb{Z}$. 


\item State $s$ is the initial state, and  $\Final$ are accepting states

\item $ \Sigma = \{(a,b) : 0 \leq a, b \leq \max \} $ where $ a $ and $ b $ are integers.

\item $\delta_d \subseteq \Statess \times \Sigma \times \Statess$ is defined as follows:
	\begin{enumerate}
	\item Transitions from start state $ s $:
	\begin{enumerate}[label = \roman*]
	\item $ (s ,(a,b), (x,c)) $ for all $ (x,c) \in \Final$ s.t. $ a + x + c = b $ and $ c \neq 0 $
			
	\item $ (s ,(a,b), (x, \bot)) $ for all $ (x, \bot) \in S_{\bot} $ s.t. $ a + x  = b $

	\end{enumerate}
	
	\item Transitions within $ S_{\bot} $: $ ((x, \bot) ,(a,b), (x', \bot) )$ for all $(x, \bot)$, $(x', \bot) \in S_{\bot} $, if  $ a + x' = b + d\cdot x $

	\item Transitions within $ \Final $: $ ((x,c) ,(a,b), (x',c') )$ for all $ (x,c)$, $(x',c') \in \Final $ where $ c' < d $, if $ a + x' + c' = b + d\cdot x $

	\item Transition between $ S_{\bot} $ and $ \Final $: $ ((x, \bot),(a,b), (x',c')) $ for all $ (x,\bot) \in S_{\bot} $, $ (x',c') \in \Final $ where $ 0 < c' < d $, if  $ a + x' + c' = b + d\cdot x $

	\end{enumerate}

\end{itemize}

\begin{theorem}
	\label{thm:Construction}
	Let $d>1$ be an integer discount-factor, and $\mu>1$ be an integer upper-bound. 
	B\"uchi automaton $ \A_{\DSsucceqFord{d}}^<$ represents the DS comparator automata for $\mu$, $d$, and $<$. B\"uchi automaton $ \A_{\DSsucceqFord{d}}^<$ has $\mathcal{O}(\frac{\mu^2}{d})$-many states. 
\end{theorem}  
\begin{proof}
Corollary~\ref{lem:FiniteAlphabetXandC} proves that if $\DSum{A}{d}< \DSum{B}{d}$ then sequence $X$ and $C$ satisfying the integer sequence criteria and bounded-criteria will exist. Let these sequences be $X = X[0]X[1]\dots$ and $C = [0]C[1]\dots$.  
Since $\DSum{C}{d} > 0$, there exists an index $i\geq 0$ where $C[i]>0$. Let the first position where $C[i]>0$ be index $j$.
By construction of $\A_{\DSsucceqFord{d}}^<$, the state sequence given by $s,(X[0],\bot)\dots, (X[j-1],\bot),(X[j], C[j]),(X[j+1], C[j+1])\dots $, where for all $i\geq j$, $C[i]\neq \bot$, forms a run of word $(A,B)$ in the B\"uchi automaton. Furthermore, this run is accepting since state $(x,c)$ where $c \neq \bot$ are accepting states. Therefore, ($A,B) $ is an accepting word in $\A_{\DSsucceqFord{d}}^<$.

 To prove the other direction, suppose the pair of sequence $(A, B)$ has an accepting run with state sequence $s$, $(x_0,\bot),\dots (x_{j-1}, \bot), (x_j,c_j), (x_{j+1}, c_{j+1})\dots $, where for all $i\geq j$, $c_j \neq \bot$. Construct sequence X and C as follows:
 For all $i\geq 0$, $X[i] = x_i$. For all $i<j$, $C[i] = 0$ and for all $i\geq j$ $C[i]=c_i$. 
 Then the transitions of  $\A_{\DSsucceqFord{d}}^<$ guarantees equations Equation~\ref{eq:initial}-~\ref{eq:invariant}
 from   Lemma~\ref{Lemma:discount-invariant1} to hold for sequences $A$,$B$ and $C$,$X$. Therefore, it must be the case that  $\DSum{B}{d} = \DSum{A}{d} + \DSum{C}{d}$.
 Furthermore, since the first transition to accepting states $(x,c)$ where $c \neq \bot$ is possible only if $c>0$, $\DSum{C}{d}>0$. Therefore, $ \DSum{A}{d}<\DSum{B}{d}$.
 Therefore,  $\A_{\DSsucceqFord{d}}^< $ accepts $(A,B)$ if $ \DSum{A}{d}<\DSum{B}{d}$.
 \qed
\end{proof}

\begin{corollary}
	\label{thm:discountedSumComparator}
	Comparison languages for DS aggregation function with an integer discount-factor $d>1$ is $\omega$-regular.
\end{corollary}  
\begin{proof}
	Immediate from Theorem~\ref{thm:Construction}, and closure properties of $\omega$-regular comparison languages (Theorem~\ref{lem:closure}-Item-\ref{item:closedregular}).
	\qed
\end{proof}

Note that the proof of Theorem~\ref{lem:closure}-Item-\ref{item:closedregular} also gives a way to construct the DS comparators for all other equality or inequality relations. However, if those were to be followed, then the resulting DS comparators may have a very large number of states. For instance, the proof of of Theorem~\ref{lem:closure}-Item-\ref{item:closedregular} suggests to construct the DS comparator for $\geq$ by taking the complement of $ \A_{\DSsucceqFord{d}}^<$ constructed above. B\"uchi complementation~\cite{safra1988complexity} will result in an exponential blow-up in the state space. Hence, this method of constructing DS comparators for the remaining relations is not practical. The silver lining here is that the construction of $ \A_{\DSsucceqFord{d}}^<$ can be modified in trivial ways to obtain the DS comparators for all other relations. Hence, 
\begin{theorem}
	\label{thm:ConstructionSize}
	Let $d>1$ be an integer discount-factor, and $\mu>1$ be an integer upper bound. 
	The B\"uchi automaton for the DS comparator for $\mu$, $d$, and $\mathsf{R}$ for $\mathsf{R} \in \{\leq, \geq, <, >, =, \neq\}$ consists of $\mathcal{O}(\frac{\mu^2}{d})$-many states. 
\end{theorem}  
\begin{proof}
One could use the constructions from Theorem~\ref{lem:closure}-Item-\ref{item:closedregular} to obtain the DS comparators for all other inequalities and equalities. However, those constructions would lead to comparators with larger state space than $\mathcal{O}(\frac{\mu^2}{d})$. In this proof, we will illustrate how to construct DS comparators for all other inequality and equalities by modifying the DS comparator for $<$ from Theorem~\ref{thm:discountedSumComparator} such that the resulting comparator has the size size as $\mathcal{O}(\frac{\mu^2}{d})$.

For $>$: The DS comparator for $>$ can be obtained by flipping the order of alphabet on the transitions. Specifically, for every transition in DS comparator for $<$, if the alphabet is $(a,b)$, then switch the alphabet to $(b,a)$ in the DS comparator for $>$.

For $\leq$: The DS comparator for $\leq$  can be constructed from the DS comparator for $<$ by changing the accepting states only. For the DS comparator for $\leq$, let the accepting states be $\Final \cup S_{\bot}$. Intuitively, the states $S_{\bot}$ are those where the discounted-sum of the difference sequence $C$ is 0, since sequence $C = 0^\omega$. Therefore, by adding $S_{\bot}$ to the accepting states of DS comparator for $<$, we have included all those sequence for which $\DSum{C}{d} = 0$. Thereby, converting it to the DS comparator for $\leq$.

For $=$: In this case, set the accepting states in the DS comparator for $< $ to $S_{\bot}$ only. This way, the automaton will accept a sequence $(A,B)$ iff the corresponding $C$ is such that  $\DSum{C}{d} = 0$ since $C = 0\omega$.

For $\geq$: As done for constructing the DS comapartor for $>$ from DS comparator for $<$, simply swap the alphabet in the DS comparator for $\leq$ to get the comparator for $\geq$.

For $\neq$: Intuitively, one would want to take the $S_{\bot}$ states, but make sure they are not accepting. Next, one would appropriately add the $(x,c)$ states from the DS comparators for both $<$ and $>$ in order to accept sequences $(A,B)$ iff either $\DSum{A}{d} > \DSum{B}{d}$ or $\DSum{A}{d} < \DSum{B}{d}$.
\qed
\end{proof}

\subsection{Complexity of DS inclusion with integer discount factors}
\label{Sec:DSApplication}

Finally, we utilize the $\omega$-regularity of DS comparators with integer discount factors to establish that DS inclusion is \cct{PSPACE}-complete when the discount factor is an integer. 
The prior best known algorithm is
\textsf{EXPTIME} and \cct{EXPSPACE}~\cite{boker2014exact,chatterjee2010quantitative}, which does not match with its known \cct{PSAPCE} lower bound.

\begin{theorem}
	\label{Thrm:DSInSEq}
	Let integer $\mu>1$ be the maximum weight on transitions in DS-automata $P$ and $Q$, and $d>1$ be an integer discount-factor. Let $\mu$ and $d$ be represented in unary form. Then DS-inclusion, DS-strict-inclusion, and DS-equivalence between  are $\cc{PSPACE}$-complete.
	
\end{theorem}
\begin{proof}
	Since size of DS-comparator is polynomial w.r.t. to upper bound $\mu$, when represented in unary, (Theorem~\ref{thm:Construction}),  DS-inclusion is $\cc{PSPACE}$ in size of input weighted $\omega$-automata and $\mu$ (Theorem~\ref{thrm:RegularComplexity}). 
	\qed
\end{proof}

By closing the gap between upper and lower bounds for DS inclusion, Theorem~\ref{Thrm:DSInSEq} resolves a 15-year old open problem. 
The earlier known \cct{EXPTIME} upper bound in complexity is based on an exponential determinization construction (subset construction) combined with arithmetical reasoning~\cite{boker2014exact,chatterjee2010quantitative}. We observe that the determinization construction can be performed on-the-fly in $\cc{PSPACE}$. To perform, however, the arithmetical reasoning on-the-fly in \cct{PSPACE} would require essentially using the same bit-level ($(x,c)$-state) techniques that we have used to construct DS-comparator.

\section{Limit-average aggregation function}
\label{Sec:LA}

The limit-average of a sequence refers to the point of convergence of the average of prefixes of the sequence.
Unlike discounted-sum, limit average is known to not converge for all sequences. 
To work around this limitation, most applications simply use limit-average infimum or limit-average supremum of sequences~\cite{brim2011faster,chatterjee2010quantitative,chatterjee2005mean,zwick1996complexity}. 
We argue that the usage of limit-average infimum or limit-average supremum in liu of limit-average for purpose of comparison can be misleading. 
For example, consider sequence $A$ s.t. $\LASup{A} = 2$ and $\LAInf{A} = 0$, and sequence $B$ s.t. $\LA{B} = 1$. Clearly,  limit-average of $A$ does not exist. So while it is true that$ \LAInf{A} < \LAInf{B}$, indicating that at infinitely many indices the average of prefixes of $A$ is lower, this renders an incomplete picture since at infinitely many indices, the average of prefixes of $B$ is greater as $\LASup{A} = 2 $.

Such inaccuracies in limit-average comparison may occur when the limit-average of at least one sequence does not exist. However, it is not easy to distinguish sequences for which limit-average exists from those for which it doesn't. 

We define {\em prefix-average comparison} as a relaxation of limit-average comparison. 
Prefix-average comparison coincides with limit-average comparison when limit-average exists for both sequences. 
Otherwise, it determines whether eventually the average of prefixes of one sequence are greater than those of the other.
This comparison does not require the limit-average to exist to return intuitive results.
Further, we show that the {\em prefix-average comparator} is $\omega$-context-free.

\subsection{Limit-average language and comparison}
\label{Sec:LALanguage}

Let $\Sigma = \{0,1,\dots, \mu\}$ be a finite alphabet with $\mu>0$. The {\em limit-average language} $\L_{LA}$ contains the sequence (word) $A \in \Sigma^{\omega}$ iff its limit-average exists. 
Suppose $\L_{LA}$ were $\omega$-regular, then $\L_{LA} = \bigcup_{i=0}^n U_i\cdot V_i^{\omega}$, where $U_i, V_i\subseteq \Sigma^*$ are regular languages over {\em finite} words. 
The limit-average of sequences is determined by its behavior in the limit, so limit-average of sequences in $V_i^{\omega}$  exists. 
Additionally, the average of all (finite) words in $V_i$ must be the same.
If this were not the case, then two words in $V_i$ with unequal averages $l_1$ and $l_2$, can generate a word $w \in V_i^{\omega}$ s.t~the average of its prefixes oscillates between $l_1$ and $l_2$. This cannot occur, since limit-average of $w$ exists.
Let the average of sequences in $V_i$ be $a_i$, then limit-average of  sequences in $V_i^{\omega}$ and  $U_i \cdot V_i^{\omega}$ is also $a_i$. 
This is contradictory since 
there are sequences with limit-average different from  the $a_i$ (see appendix).
Similarly, since every $\omega$-CFL is  represented by $\bigcup_{i=1}^n U_i \cdot V_i^{\omega}$ for CFLs $U_i, V_i$ over finite words~\cite{cohen1977theory}, a similar argument proves that $\L_{LA}$ is not $\omega$-context-free.

\begin{theorem}
\label{Lemma:LARegularNotExist}
$\L_{LA}$ is neither an $\omega$-regular nor an $\omega$-context-free language.
\end{theorem}
\begin{proof}
We first prove that {$\L_{LA}$ is not $\omega$-regular}.

	Let us assume that the language $\L_{LA}$ is $\omega$-regular. Then there exists a finite number $n$ s.t. $\L_{LA} = \bigcup_{i=0}^n U_i \cdot V_i^{\omega}$, where $U_i$ and $V_i \in \Sigma^{*}$ are regular languages over finite words. 
	 
	 For all $i \in \{0,1,\dots n\}$, the limit-average of any word in $U_i\cdot V_i^{\omega}$ is given by the suffix of the word in $V_i^{\omega}$. Since $U_i\cdot V_i^{\omega} \subseteq \L_{LA}$, limit-average exists for all words in $U_i \cdot V_i^{\omega}$. Therefore, limit-average of all words in $V_i^{\omega}$ must exist. 
	 Now as discussed above, the average of all words in $V_i$ must be the same. Furthermore, the limit-average of all words in $V_i^{\omega}$ must be the same, say $\LA{w} = a_i$ for all $w \in V_i^{\omega}$.

	 
	 Then the limit-average of all words in $\L_{LA}$ is one of $a_0, a_1\dots a_n$. Let $a = \frac{p}{q}$ s.t $p<q$, snd $a \neq a_i$ for $i \in \{0,1,\dots,\mu\}$. Consider the word $w = (1^{p}0^{q-p})^{\omega}$. It is easy to see the $\LA{w} = a$. However, this word is not present in $\L_{LA}$ since the limit-average of all words in $\L_{LA}$ is equal to $a_0$ or $a_1$ \dots or $a_n$.
	 
	 Therefore, our assumption that $\L_{LA}$ is an $\omega$-regular language has been contradicted.

Next we prove that $\L_{LA}$ is not an $\omega$-CFL.

Every $\omega$-context-free language can be written in the form of $\bigcup_{i=0}^n U_i \cdot V_i^{\omega}$ where $U_i$ and $V_i $ are context-free languages over finite words. 
	The rest of this proof is similar to the proof for non-$\omega$-regularity of $\L_{LA}$. 
	\qed
\end{proof}

In the next section, we will define {\em prefix-average comparison} as a relaxation of limit-average comparison. To show how prefix-average comparison relates to limit-average comparison, we will require the following two lemmas:
Quantifiers  $\exists^{\infty}i$ and $\exists^{f}i$ denote the existence of {\em infinitely} many and {\em only finitely} many indices $i$, respectively.
\begin{lemma}
	\label{Lemma:LAExistFull}
Let $A$ and $B$ be sequences s.t.~their limit average exists. 
If  $\exists^{\infty}i, \Sum{A[0,i-1]} \geq \Sum{B[0,i-1]}$ then $\LA{A} \geq \LA{B}$.

\end{lemma}
\begin{proof}
	Let the limit average of sequence $A$, $B$ be $a$, $b$ respectively. Since the limit average of $A$ and $B$ exists, for every $\epsilon >0$, there exists $N_{\epsilon}$ s.t. for all $n> N_{\epsilon}$ , $|\Av{A[0,n-1]} - a| < \epsilon$ and $|\Av{B[0,n-1]} - b| < \epsilon$. 
	
	Let $a - b = k > 0$.

	Take $\epsilon = \frac{k}{4}$. Then for all $n>N_{\frac{k}{4}}$, since $|\Av{A[0,n-1]} - a| < \epsilon$, $|\Av{B[0,n-1]} - b| < \epsilon$ and that $a-b=k>0$, $\Av{A[0,n-1]} - \Av{B[0,n-1]} > \frac{k}{2} \implies \frac{\Sum{A[0,n-1]}}{n} - \frac{\Sum{B[0,n-1]}}{n} > \frac{k}{2} \implies  \Sum{A[0,n-1]} - \Sum{B[0,n-1]} > 0$.
	
	Specifically, $\ExistInf{i}{A}{B}$. Furthermore, since there is no index greater than $N_{\frac{k}{4}}$ where $\Sum{A[0,n-1]} \leq \Sum{B[0,n-1]}$, $\ExistFin{i}{B}{A}$.
	\qed
\end{proof}
\begin{lemma}
\label{Lemma:LAExistsProp}
Let $A$, $B$ be sequences s.t their limit-average exists. 
If $\LA{A} >  \LA{B}$ then $\ExistFin{i}{B}{A}$ and $\ExistInf{i}{A}{B}$.
\end{lemma}
\begin{proof}
	Let the limit-average of sequence $A$, $B$ be $L_a$, $L_b$ respectively. Since, the limit average of both $A$ and $B$ exists, for every $\epsilon >0$, there exists $N_{\epsilon}$ s.t. for all $n> N_{\epsilon}$ , $|\Av{A[1,n]} - L_a| < \epsilon$ and $|\Av{B[1,n]} - L_b| < \epsilon$. 
	
	Suppose it were possible that $\LA{A}<\LA{B}$. Suppose $L_b-L_a = k>0$. Let $\epsilon = \frac{k}{4}$. By arguing as in Lemma~\ref{Lemma:LAExistFull}, it must be the case that for all $n > N_{\frac{k}{4}}$, ${\Sum{B[1, n]}} - {\Sum{A[1, n]}} > 0$. But this is not possible, since we are given that $\ExistInf{i}{A}{B}$ (or $\exists^{\infty}i, \Sum{A[0,i-1] \geq \Sum{B[0,i-1]}}$).
	Hence $\LA{A} \geq \LA{B}$.
	\qed
\end{proof}
\subsection{Prefix-average comparison and comparator}
\label{Sec:LAClassificationAll}

The previous section relates limit-average comparison with the sums of equal length prefixes of the sequences (Lemma~\ref{Lemma:LAExistFull}-\ref{Lemma:LAExistsProp}). The comparison criteria is based on the number of times sum of prefix of one sequence is greater than the other, which does not rely on the existence of limit-average. 
Unfortunately, this criteria cannot be used for limit-average comparison since it is incomplete (Lemma~\ref{Lemma:LAExistsProp}). 
Specifically,  for sequences $A$ and $B$ with equal limit-average it is possible that $\exists^{\infty} i, \Sum{A[0,n-1]} > \Sum{B[0,n-1]}$ and $\exists^{\infty} i, \Sum{B[0,n-1]} > \Sum{A[0,n-1]}$. 
Instead, we use this criteria to define {\em prefix-average comparison}. In this section, we define prefix-average  comparison and explain how it relaxes limit-average comparison. Lastly, we construct the prefix-average comparator, and prove that it is not $\omega$-regular but is $\omega$-context-free.

\begin{Def}[Prefix-average comparison for relation $\geq$]
\label{Def:PrefixAverage}
Let $A$ and $B$ be number sequences. We say $\PLA{A} \geq \PLA{B}$ if 
$\ExistFin{i}{B}{A}$ and  
$\ExistInf{i}{A}{B}$.
\end{Def}
Note that, by definition prefix average comparison is defined on inequality relation $\geq$ or $\leq$, and not for the other inequality or equality relations. 
Intuitively, prefix-average comparison states that $\PLA{A}\geq \PLA{B}$ if eventually the sum of prefixes of $A$ are always greater than those of $B$. We use $\geq$ since the average of prefixes may be equal when the difference between the sum is small.
It coincides with limit-average comparison when the limit-average exists for both sequences.
 Definition\ref{Def:PrefixAverage} and  Lemma~\ref{Lemma:LAExistFull}-\ref{Lemma:LAExistsProp} relate  limit-average comparison  and prefix-average comparison:

\begin{corollary}
When limit-average of $A$ and $B$ exists, then
\label{Coro:PLAtoLA}

\begin{itemize}
	\item $\PLA{A}\geq\PLA{B} \implies \LA{A}\geq \LA{B}$.
	\item $\LA{A}>\LA{B} \implies \PLA{A}\geq \PLA{B}$.
\end{itemize}
\end{corollary}
\begin{proof}
	The first item falls directly from definitions. 
	
	For the second,
	let $\LAInf{A}$ and $\LASup{B}$ be $a$ and $b$ respectively. 	
	For all $\epsilon>0$ there exists an $N_{\epsilon}$ s.t for all $n>N_{\epsilon}$, $\frac{\Sum{A[1,n]}}{n} > a-\epsilon$, and $\frac{\Sum{B[1,n]}}{n} < b + \epsilon$.  
	Let $a-b = k>0$. Take $\epsilon=\frac{k}{4}$. Replicate the argument from Lemma~\ref{Lemma:LAExistFull} to show that there can exist only finitely many indexes $i$ where ${\Sum{B[0,i-1]}} \geq {\Sum{A[0,i-1]}}$.
	Similarly, show there exists infinitely many prefixes where $\Sum{A[0,i-1]} > \Sum{B[0, i-1]}$
\qed
\end{proof}
Therefore, limit-average comparison and prefix-average comparison return the same result on sequences for which limit-average exists. In addition, prefix-average returns intuitive results when even when limit-average may not exist. For example, 
suppose limit-average of $A$ and $B$ do not exist, but $\LAInf{A} > \LASup{B}$, then $\PLA{A} \geq \PLA{B}$.
Therefore, prefix-average comparison relaxes limit-average comparison. 

The rest of this section describes {\em prefix-average comparator for relation $\geq$}, denoted by $\A_{\PASuc}^\geq$, an automaton that accepts the pair $(A,B)$  of sequences iff $\PLA{A}\geq \PLA{B}$.
\begin{lemma}
	\label{Lemma:PumpingLemmaRegular}
	\textbf{(Pumping Lemma for $\omega$-regular language~\cite{alur2009omega})}
	Let $L$ be an $\omega$-regular language. There exists $p \in \mathbb{N}$ 
	such that, for each $w = u_1w_1u_2w_2 \dots \in L$ such that $|w_i| \geq p$ for all $i$, there are
	sequences of finite words $(x_i)_{i\in \N}$, $(y_i)_{i\in \N}$, $(z_i)_{i\in \N}$ s.t., for all $i$, $w_i = x_iy_iz_i$, $|x_iy_i| \leq p$ and $|y_i| > 0$ and for every sequence of pumping factors $(j_i)_{i\in \N} \in \N$, the pumped word $u_1 x_1 y_1^{j_1}z_1 u_2 x_2 y_2^{j_2}z_2 \dots \in L$. 
\end{lemma}

\begin{theorem}
\label{Lemma:LANotRegular}
The prefix-average comparator for $\geq$ is not $\omega$-regular.
\end{theorem}
\begin{proof}
We use Lemma~\ref{Lemma:PumpingLemmaRegular} to prove that $\A_{\PASuc}^\geq$ is not $\omega$-regular. 
Suppose $\A_{\PASuc}^\geq$ were $\omega$-regular. 
For $p >0 \in \N$, let $w = (A,B) = ((0,1)^p(1,0)^{2p})^{\omega}$. 
The segment $(0,1)^*$ can be pumped s.t the resulting word is no longer in $\A_{\PASuc}^\geq$.

Concretely, $A = (0^p 1^{2p})^{\omega}$, $B = (1^p0^{2p})^{\omega}$, $\LA{A} = \frac{2}{3}$, $\LA{B} = \frac{1}{3}$.
So, $w = (A,B) \in \A_{\PASuc}^\geq$.
Select as factor $w_i$ (from Lemma~\ref{Lemma:PumpingLemmaRegular}) the sequence $(0,1)^p$.
Pump each $y_i$ enough times so that the resulting word is $\hat{w} = (\hat{A}, \hat{B}) = ((0,1)^{m_i}(1,0)^{2p})^{\omega}$  where $m_i>4p$.
It is easy to show that $\hat{w} = (\hat{A}, \hat{B})\notin \A_{\PASuc}^\geq$. 
\qed
\end{proof}
We discuss key ideas and sketch the construction of the prefix average comparator. 
The term {\em prefix-sum difference at $i$} indicates $\Sum{A[0,i-1]} - \Sum{B[0,i-1]}$, i.e. the difference between sum of $i$-length prefix of $A$ and $B$.

\subsubsection*{Key ideas}
For sequences $A$ and $B$ to satisfy $\PLA{A}\geq\PLA{ B}$, $\ExistFin{i}{B}{A}$ and  $\ExistInf{i}{A}{B}$.   
This occurs iff there exists an index $N$ s.t. for all indices $i>N$, $\Sum{A[0,i-1]} - \Sum{B[0,i-1]}>0$.
While reading a word, the prefix-sum difference is maintained by states and the stack of $\omega$-PDA: states maintain whether it is negative or positive,
while number of tokens in the stack equals its absolute value.
The automaton non-deterministically guesses the aforementioned index $N$, 
beyond which the automaton ensure that prefix-sum difference remains positive.

\subsubsection*{Construction sketch}

The push-down comparator $\A_{\PASuc}^\geq$ consists of three states: (i) State $s_P$ and (ii) State $s_N$ that indicate that the prefix-sum difference is greater than zero and  or not respectively, (iii)  accepting state $s_F$. 
An execution of $(A,B)$ begins in state $s_N$ with an empty stack. On reading letter $(a,b)$, the stack pops or pushes $|(a-b)|$ tokens from the stack depending on the current state of the execution. From state $s_P$, the stack pushes tokens if $(a-b)>0,$ and pops otherwise. The opposite occurs in state $s_N$. 
State transition between $s_N$ and $s_P$ occurs only if the stack action is to pop but the stack consists of $k < |a-b|$ tokens. In this case, stack is emptied, state transition is performed and $|a-b|-k$ tokens are pushed into the stack. 
For an execution of $(A,B)$ to be an accepting run, the automaton non-deterministically transitions into state $s_F$. 
State $s_F$ acts similar to state $s_P$ except that execution is terminated if there aren't enough tokens to pop out of the stack. 
$\A_{\PASuc}^\geq$ accepts by accepting state.

To see why the construction is correct, it is sufficient to prove that at each index $i$, the number of tokens in the stack is equal to $|\Sum{A[0,i-1]} - \Sum{B[0,i-1]}|$. Furthermore, in state $s_N$, $\Sum{A[0,i-1]} - \Sum{B[0,i-1]}\leq 0$, and in state $s_P$ and $s_F$, $\Sum{A[0,i-1]} - \Sum{B[0,i-1]}>0$. 
Next, the index at which the automaton transitions to the accepting state $s_F$ coincides with index $N$. The execution is accepted if it has an infinite execution in state $s_F$, which allows transitions only if $\Sum{A[0,i-1]} - \Sum{B[0,i-1]} > 0$.

\subsubsection*{Construction}
We provide a sketch of the construction of the B\"uchi push-down autoamaton $\A_{\PASuc}^\geq$, and then prove that it corresponds to the prefix average comparator. 

Let $\mu$ be the bound on sequences. Then  $\Sigma = \{0,1,\dots, n\}$ is the alphabet of sequences. 
Let $ \A_{\PASuc}^\geq = (\Statess, \Sigma\times\Sigma, \Gamma, \delta, s_0, Z_0)$ where:

\begin{itemize}
	\item $\Statess = \{s_N, s_P, s_F\}$ is the set of states of the automaton.
	
	\item $\Sigma\times\Sigma$ is the alphabet  of the language. 
	
	\item $\Gamma = \{Z_0, \alpha\}$ is the push down alphabet.  
	
	\item $s_0 = s_N$ is the start state of the push down automata. 
	
	\item $Z_0$ is the start symbol of the stack. 
	
	\item $s_F$ is the accepting state of the automaton. Automaton $\A_{\PASuc}^\geq$ accepts words by final state. 
	
	\item Here we give  a sketch of the behavior of the transition function $\delta$. 
	
	\begin{itemize}
		\item When $\A_{\PASuc}^\geq$ is in configuration $(s_P, \tau)$ for $\tau \in \Gamma$, push $a$ number of $\alpha$-s into the stack. 
		
		Next, pop $b$ number of $\alpha$-s. If after popping $k$ $\alpha$-s where $k<b$, the PDA's configuration becomes $(s_P, Z_0)$, then first move to state $(s_N, Z_0)$ and then resume with pushing $ b-k$ $\alpha$-s into the stack.
		
		\item When $\A_{\PASuc}^\geq$ is in configuration $(s_N, \tau)$ for $\tau \in \Gamma$, push $b$ number of $\alpha$-s into the stack
		
		Next, pop $a$ number of $\alpha$-s. If after popping $k$ $\alpha$-s where $k<a$, the PDA's configuration becomes $(s_N, Z_0)$, then first move to state $(s_P, Z_0)$ and then resume with pushing $ a-k$ $\alpha$-s into the stack.
		
		\item When $\A_{\PASuc}^\geq$ is in configuration $(s_P, \tau)$ for $\tau \neq Z_0$, first move to configuration $(s_F, \tau)$ and then push $a$ number of $\alpha$-s and pop $b$ number of $\alpha$-s. Note that there are no provisions for popping $\alpha$ if the stack hits $Z_0$ along this transition. 
		
		\item When $\A_{\PASuc}^\geq$ is in configuration $(s_F, \tau)$ for $\tau \neq Z_0$, push $a$ $\alpha$-s then pop $b$ $\alpha$-s.
		
		Note that there are no provisions for popping $\alpha$ if the stack hits $Z_0$ along this transition. 
		
	\end{itemize}
\end{itemize}

\begin{lemma}
	\label{Lemma:PAComparator}
	Push down automaton $\A_{\PASuc}^\geq$ accepts a pair of sequences $(A,B)$ iff $\PLA{A} \geq \PLA{B}$.
\end{lemma}
\begin{proof}
	To prove this statement, it is sufficient to demonstrate that $\A_{\PASuc}^\geq$ accepts a pair of sequences $(A,B)$ iff there are only finitely many indexes where $\Sum{B[1,i]} > \Sum{A[1,i]}$. This is  true by definition of $\mathsf{PrefixAvg}$ itself. 
	
	On $\A_{\PASuc}^\geq$ this corresponds to the condition that there being only finitely many times when the PDA is in state $N$ during the run of $(A,B)$. This is ensured by the push down automaton since the word can be accepted only in state $F$ and there is no outgoing edge from $F$. Therefore, every word that is accepted by $\A_{\PASuc}^\geq$ satisfies the condition $\ExistFin{i}{B}{A}$. 
	
	Conversely, for every word $(A,B)$ that satisfies $\ExistFin{i}{B}{A}$ there is a point, call it index $k$, such that for all indexes $m>k$, $\Sum{B[1,m]} \ngeq \Sum{A[1,m]}$. If a run of $(A,B)$ switches to $F$ at this $m$, then it will be accepted by the push down automaton. Since $\A_{\PASuc}^\geq$ allows for non-deterministic move to $(F, \tau)$ from $(P, \tau)$, the run of $(A,B)$ will always be able to move to $F$ after index $m$. Hence, every $(A,B)$ satisfying $\ExistFin{i}{B}{A}$ will be accepted by $\A_{\PASuc}^\geq$.	
	\qed
\end{proof}

\begin{theorem}
	\label{Thrm:PAComparator}
	The prefix-average comparator for relation $\geq$ is an $\omega$-CFL.
\end{theorem}
While $\omega$-CFL can be easily expressed, they do not possess closure properties, and problems on $\omega$-CFL are easily undecidable. Hence, the application of $\omega$-context-free comparator will require further investigation.

\section{$\omega$-Regular aggregate functions}
\label{Sec:RegularAggFunctions}

So far, we have investigated aggregate functions individually in order to determine if their comparators are $\omega$-regular. In this section, we investigate a broader class of aggregate functions, called $\omega$-regular aggregate functions~\cite{chaudhuri2013regular}. Intuitively, $\omega$-regular functions are those aggregate functions for which arithmetic can be conducted on an automaton. Examples of $\omega$-regular functions include discounted-sum with integer discount factors, limsup and liminf are examples of aggregate functions that are $\omega$-regular. 

The question we ask is whether a necessary and sufficient condition for an aggregate function to have an $\omega$-regular compartor is that the function should be $\omega$-regular? We prove one side of the argument. We show that if a function is $\omega$-regular, then its comparator will be $\omega$-regular. A proof/disproof of the other direction is open for investigation.

The argument showing that every $\omega$-regular function will have $\omega$-regular comparators is given below:

\begin{theorem}
	\label{thrm:functionthencomparator}
	Let $\mu>0$ be the upper-bound on weight sequences, and $\beta\geq 2$ be the integer base. Let $f: \{0,1,\dots,\mu\}^\omega\rightarrow \Re$ be an aggregate function. If aggregate function $f$ is $\omega$-regular under base $\beta$, then its comparator for all inequality and equality relations is also $\omega$-regular. 
\end{theorem}
\begin{proof}
	We show that if an aggregate function is $\omega$-regular under base $\beta$, then its comparator for relation $>$ is $\omega$-regular. By closure properties of $\omega$-regular comparators, this implies that comparators of the aggregate function are $\omega$-regular for all inequality and equality relations. 
	
	But first we prove that for a given integer base $\beta\geq 2$ there exists an automaton $\A_\beta$ such that for all $a,b\in \Re$, $\A_\beta$ accepts $(\mathsf{rep}(a,\beta), \mathsf{rep}(b, \beta))$ iff $a>b$.
	Let $a,b\in\Re$, and $\beta>2$ be an integer base. Let $\mathsf{rep}(a,\beta) = \mathsf{sign}_a\cdot(\mathsf{Int}(a, \beta), \mathsf{Frac}(a, \beta))$ and $ \mathsf{rep}(b,\beta)  = \mathsf{sign}_b\cdot(\mathsf{Int}(b, \beta), \mathsf{Frac}(b, \beta))$. Then, the following statements can be proven using simple evaluation from definitions:
	\begin{itemize}
		\item When $\mathsf{sign}_a =+$ and $\mathsf{sign}_b = -$. Then $a > b$.
		\item When $\mathsf{sign}_a = \mathsf{sign}_b = +$
		\begin{itemize}
			\item  If $\mathsf{Int}(a, \beta) \neq \mathsf{Int}(b, \beta)$: Since $\mathsf{Int}(a, \beta)$ and $\mathsf{Int}(b, \beta)$ eventually only see digit $0$ i.e. they are necessarily identical eventually, there exists an index $i$ such that it is the last position where $\mathsf{Int}(a, \beta)$ and $\mathsf{Int}(b, \beta)$ differ. 
			If $\mathsf{Int}(a, \beta)[i]>\mathsf{Int}(b, \beta)[i]$, then $a > b$. 
			If $\mathsf{Int}(a, \beta)[i]<\mathsf{Int}(b, \beta)[i]$, then $a < b$.
			
			\item If $\mathsf{Int}(a, \beta) = \mathsf{Int}(b, \beta)$ but  $\mathsf{Frac}(a, \beta) \neq \mathsf{Frac}(b, \beta)$: 
			Let $i$ be the first index where $\mathsf{Frac}(a, \beta)$ and $\mathsf{Frac}(b, \beta)$ differ. 
			If $\mathsf{Frac}(a, \beta)[i] > \mathsf{Frac}(b, \beta)[i]$ then $a > b$.
			If $\mathsf{Frac}(a, \beta)[i] < \mathsf{Frac}(b, \beta)[i]$ then $a < b$.	
			
			\item Finally, if $\mathsf{Int}(a, \beta) = \mathsf{Int}(b, \beta)$ and $\mathsf{Frac}(a, \beta) = \mathsf{Frac}(b, \beta)$: Then $a = b$.
		\end{itemize}
		\item When $\mathsf{sign}_a = \mathsf{sign}_b = -$
		\begin{itemize}
			\item  If $\mathsf{Int}(a, \beta) \neq \mathsf{Int}(b, \beta)$: Since $\mathsf{Int}(a, \beta)$ and $\mathsf{Int}(b, \beta)$ eventually only see digit $0$ i.e. they are necessarily identical eventually. Therefore, there exists an index $i$ such that it is the last position where $\mathsf{Int}(a, \beta)$ and $\mathsf{Int}(b, \beta)$ differ. 
			If $\mathsf{Int}(a, \beta)[i]>\mathsf{Int}(b, \beta)[i]$, then $a < b$. 
			If $\mathsf{Int}(a, \beta)[i]<\mathsf{Int}(b, \beta)[i]$, then $a > b$.
			
			\item If $\mathsf{Int}(a, \beta) = \mathsf{Int}(b, \beta)$ but  $\mathsf{Frac}(a, \beta) \neq \mathsf{Frac}(b, \beta)$: 
			Let $i$ be the first index where $\mathsf{Frac}(a, \beta)$ and $\mathsf{Frac}(b, \beta)$ differ. 
			If $\mathsf{Frac}(a, \beta)[i] > \mathsf{Frac}(b, \beta)[i]$ then $a < b$.
			If $\mathsf{Frac}(a, \beta)[i] < \mathsf{Frac}(b, \beta)[i]$ then $a > b$.	
			
			\item Finally, if $\mathsf{Int}(a, \beta) = \mathsf{Int}(b, \beta)$ and $\mathsf{Frac}(a, \beta) = \mathsf{Frac}(b, \beta)$: Then $a = b$.
		\end{itemize}
		\item When $\mathsf{sign}_a =-$ and $\mathsf{sign}_b = +$. Then $a < b$.
	\end{itemize}
	Since the conditions given above are exhaustive and mutually exclusive, we conclude that for all $a,b\in\Re$ and integer base $\beta\geq 2$, let $\mathsf{rep}(a,\beta) = \mathsf{sign}_a\cdot(\mathsf{Int}(a, \beta), \mathsf{Frac}(a, \beta))$ and $ \mathsf{rep}(b,\beta)  = \mathsf{sign}_b\cdot(\mathsf{Int}(b, \beta), \mathsf{Frac}(b, \beta))$. Then $a>b$ iff one of the following conditions occurs:
	\begin{enumerate}
		\item $\mathsf{sign}_a =+$ and $\mathsf{sign}_b = -$. 
		\item $\mathsf{sign}_a = \mathsf{sign}_b = +$, $\mathsf{Int}(a, \beta) \neq \mathsf{Int}(b, \beta)$, and $\mathsf{Int}(a, \beta)[i]>\mathsf{Int}(b, \beta)[i]$ when $i$ is the last index where $\mathsf{Int}(a, \beta)$ and $\mathsf{Int}(b, \beta)$ differ. 
		\item $\mathsf{sign}_a = \mathsf{sign}_b = +$, $\mathsf{Int}(a, \beta) = \mathsf{Int}(b, \beta)$, $\mathsf{Frac}(a, \beta) \neq \mathsf{Frac}(b, \beta)$, and $\mathsf{Int}(a, \beta)[i]>\mathsf{Int}(b, \beta)[i]$ when $i$ is the first index where $\mathsf{Frac}(a, \beta)$ and $\mathsf{Frac}(b, \beta)$ differ. 
		\item $\mathsf{sign}_a = \mathsf{sign}_b = +$, $\mathsf{Int}(a, \beta) \neq \mathsf{Int}(b, \beta)$, and $\mathsf{Int}(a, \beta)[i]<\mathsf{Int}(b, \beta)[i]$ when $i$ is the last index where $\mathsf{Int}(a, \beta)$ and $\mathsf{Int}(b, \beta)$ differ. 
		\item $\mathsf{sign}_a = \mathsf{sign}_b = +$, $\mathsf{Int}(a, \beta) = \mathsf{Int}(b, \beta)$, $\mathsf{Frac}(a, \beta) \neq \mathsf{Frac}(b, \beta)$, and $\mathsf{Int}(a, \beta)[i]<\mathsf{Int}(b, \beta)[i]$ when $i$ is the first index where $\mathsf{Frac}(a, \beta)$ and $\mathsf{Frac}(b, \beta)$ differ. 
	\end{enumerate}
	
	Note that each of these five condition can be easily expressed by a B\"uchi automaton over alphabet $\mathsf{AlphaRep(\beta)}$ for an integer $\beta\geq 2$. For an integer $\beta\geq 2$, the union of all these B\"uchi automata will result in a B\"uchi automaton $\A_\beta$ such that for all $a,b\in \Re$ and $A = \mathsf{rep}(a, \beta)$ and $B = \mathsf{rep}(b,\beta)$, $a>b$ iff interleaved word $(A,B) \in \L(\A_\beta)$. 
	
	Now we come to the main part of the proof. Let $f:\Sigma^\omega\rightarrow \Re$ be an $\omega$-regular aggregate function with aggregate function automata $\A_f$. We will construct an $\omega$-regular comparator for $f$ with relation $>$. Note that $(X,Y)$ is present in the comparator iff $(X, M), (Y,N)\in \A_f$ for $M,N \in \mathsf{AlphaRep}(\beta)^\omega$ and $(M,N) \in \A_\beta$, for $\A_\beta$ as described above. Since $\A_f$ and $\A_\beta$ are both B\"uchi automata, the comparator for function $f$ with relation $>$ is also a B\"uchi auotmaton. Therefore, the comparator for aggregate function $f$ with relation $> $ is $\omega$-regular. 
	\qed
\end{proof}

The converse direction is still open 
For all aggregate functions considered in this paper for which the comparator is $\omega$-regular, the function has also been $\omega$-regular. But that is not a full proof of the converse direction. For now, due to the lack of any counterexample, we present the converse direction as a conjecture:

\begin{conjecture}
	\label{Conjecture:comparatortofunction}
	Let $\mu>0$ be the upper-bound on weight sequences, and $\beta\geq 2$ be the integer base. Let $f: \{0,1,\dots,\mu\}^\omega\rightarrow \Re$ be an aggregate function. If the comparator for an aggregate function $f$ is $\omega$-regular for all inequality and equality relations, then its aggregate function is also $\omega$-regular under base $\beta$.
\end{conjecture}

\section{Chapter summary}
This chapter studied comparator automata for well known aggregate functions, namely discounted-sum and limit average. Among these, only discounted-sum with integer discount factors  can be represented by $\omega$-regular comparators. In later chapters, we will observe that once can design $\omega$-regular comparators for approximations to discounted-sum with non-integer discount factor (Chapter~\ref{ch:anytime}). To obtain a a broader classification of aggregate functions that permit $\omega$-regular comparators, we show that $\omega$-regular aggregate functions exhibit $\omega$-regular comparators. However, we do not whether $\omega$-regular functions is the sufficient condition for the existence of $\omega$-regular comparators. We conjecture that may be the case, but leave that as an open question.
\part{Quantitative inclusion with discounted-sum}
\label{part:DSinclusion}

Having laid out the theoretical framework of comparator automata in Part~\ref{part:framework}, Part~\ref{part:DSinclusion} and Part~\ref{part:games} demonstrate the efficacy of the integrated approach proposed by comparators.

This Chapter~\ref{ch:Comparator} studies Discounted-sum inclusion or DS inclusion in detail. Recall, in Theorem~\ref{Thrm:DSInSEq} we use comparator automata  to establish that DS inclusion is \cct{PSPACE}-complete when the discount factor is an integer; The decidability of DS inclusion is still unknown when the discount factor is not an integer. This part will delve into solving DS inclusion in practice. 

\begin{itemize}[label=Chapter]
	\item\ref{ch:empirical} conducts the first comparative study of the empirical performance of existing algorithms for DS inclusion with integer discount factors. These are a separation-of-techniques algorithm from prior work, and our $\omega$-regular comparator algorithm for DS inclusion.
	Our analysis shows how the two approaches complement each other. This is a nuanced picture that is much richer than the
	one obtained from the complexity-theoretic study alone.
	\item \ref{ch:safetycosafety} picks up from where Chapter~\ref{ch:empirical} ends. We prove that DS comparison languages are either safety or co-safety langauges.  We show that when this property is utilized to solve DS inclusion using comparators, then comparator-based approach outperforms the separation-of-techniques algorithm on all accounts. 
	\item \ref{ch:anytime} shifts focus to solving DS inclusion with non-integer discount factors. Currently even its decidability is unknown. Therefore, to solve DS inclusion in practice, we design an {\em anytime algorithm}, which will either terminate with a crisp $\mathsf{True}$ or $\mathsf{False}$ answer after a finite amount of time, or continuously generate a tighter approximation. This algorithm makes use of comparator automata for approximations of DS with non-integer  discount factor, which are shown to exhibit regularity.
\end{itemize}

\subsubsection{Terminology and notation}

We  refer to a weighted $\omega$-automata with the discounted-sum aggregate function by {\em discounted-sum automata}.
In detail, a {\em discounted-sum automaton} with discount factor $d>1$, {\em DS automaton} in short,  is a tuple  $\A= (\mathcal{M},  \gamma)$, where $\mathcal{M} = (\Statess, \Sigma, \delta, \StartState, \Statess) $ is a B\"uchi automaton, and     $\gamma : \delta \rightarrow \N$ is the {\em weight function} that assigns a weight to each transition of automaton $\mathcal{M}$. {\em Words} and {\em runs} in weighted $\omega$-automata are defined as they are in B\"uchi automata. Note that all states are accepting states in this definition.  
The {\em weight sequence} of run $\rho = s_0 s_1 \dots$ of word $w = w_0w_1\dots$ is given by $wt_{\rho} = n_0 n_1 n_2\dots$ where $n_i = \gamma(s_i, w_i, s_{i+1})$ for all $i$. 
The {\em weight of a run}  $\rho$ is given by $\DSum{wt_{\rho}}{d}$. For simplicity, we denote this by $\DSum{\rho}{d}$. 
The {\em weight of a word} in DS automata is defined as $wt_{\A}(w) = sup \{\DSum{\rho}{d} | \rho$ is a run of $w$ in $\A \}$.
By convention, if a word $w\not\in \L(\mathcal{\A})$, then
$wt_{\A}(w)=0$ \cite{chatterjee2010quantitative}.  
A DS automata is said to be {\em complete} if from every state there is at least one transition on every alphabet. Formally, for all $p\in \Statess$ and for all $a \in \Sigma$, there exists $q \in \Statess$ s.t $(p, a, q) \in \delta$. 
We abuse notation, and use $w \in \A$ to mean $w \in \L(\A)$ for B\"uchi automaton or DS-automaton $\A$.         
Given DS automata $P$ and $Q$ and discount-factor $d>1$, the {\em discounted-sum inclusion problem}, denoted by  $P \subseteq_d Q$, determines whether  for all words $w \in \Sigma^{\omega}$, $wt_{P}(w) \leq wt_{Q}(w)$. 

\chapter{Analysis of DS inclusion with integer discount factor}
\label{ch:empirical}

Discounted-sum inclusion, or DS inclusion, is when quantitative inclusion is performed with the discounted-sum aggregation function. Prior work has demonstrated the applicability of DS inclusion in computation of rational solutions in multi-agent systems with rational agents~\cite{MastersThesis}. 
Yet, the focus on DS inclusion has mostly been from a complexity-theoretic perspective, and not on algorithmic performance.
Even in this thesis so far, comparator automata have been used to establish that DS inclusion with integer discount factor is \cct{PSPACE}-complete. 
But whether these comparator-based algorithms are scalable and efficient has not been evaluated yet. 
 
To this end, this chapter undertakes a thorough theoretical and empirical analysis of two contrasting approaches for DS inclusion with integer discount factors: our comparator-based integrated algorithm, and the prior known separation-of-techniques algorithm.
We present the first implementations of these algorithms,
and perform extensive experimentation to compare between the two approaches. Our analysis shows how the two approaches complement each other. This is a nuanced picture that is much richer than the
one obtained from the complexity-theoretic study alone.

\section{Saga of theoretical vs empirical analysis }
\label{Sec:Intro}

The hardness of  quantitative inclusion for nondeterministic DS automata, or DS inclusion, is evident from \cct{PSPACE}-hardness of language-inclusion (\LI) problem for nondeterministic B\"uchi automata \cite{thomas2002automata}. Decision procedures for DS inclusion were first investigated in~\cite{chatterjee2010quantitative}, and subsequently through target discounted-sum~\cite{boker2015target}, \DS-determinization~\cite{boker2014exact}. The comparator-based argument~\cite{BCVFoSSaCS18}, presented in Chapter~\ref{ch:applicationcomparator}, finally established its \cct{PSPACE}-completeness. However, these theoretical advances in DS inclusion have not been accompanied with the development of efficient and scalable tools and algorithms. This is the focus of this chapter; our goal is to develop practical algorithms and tools for DS inclusion.

Theoretical advances have lead to two algorithmic approaches for DS inclusion. The first approach, referred to as \cct{DetLP}, combines automata-theoretic reasoning with linear-programming (\LP). This method first determinizes the DS automata~\cite{boker2014exact}, and reduces the problem of DS inclusion for deterministic DS automata to \LP~\cite{andersen2013fast,andersson2006improved}.
Since determinization of DS automata causes an exponential blow-up, \DetLPt yields an exponential time algorithm. An essential feature of this approach is the separation of automata-theoretic reasoning-- determinization--and numerical reasoning, performed by an \LP-solver. Because of this separation, it does not seem easy to apply on-the-fly techniques to this approach and perform it using polynomial space, so this approach uses exponential time and space.

In contrast, the second algorithm for DS inclusion, referred to as \BCVt (after name of authors) is purely automata-theoretic~\cite{BCVFoSSaCS18}, was presented in Chapter~\ref{ch:applicationcomparator}-Section~\ref{Sec:DSApplication}. The component of numerical reasoning between costs of executions is handled by a special B\"uchi automaton, called the {\em comparator}, that enables an on-line comparison of the discounted-sum of a pair of weight-sequences. Aided by the comparator, \cct{BCV} reduces DS inclusion to language-equivalence between B\"uchi automata. Since language-equivalence is in \cct{PSPACE}, \BCVt is a polynomial-space algorithm.

While the complexity-theoretic argument may seem to suggest a clear advantage for the pure automata-theoretic approach of \cct{BCV}, the perspective from an implementation point of view is more nuanced. \BCVt relies on \LI-solvers as its key algorithmic component. The polynomial-space approach for \cct{LI} relies on Savitch's Theorem, which proves the equivalence between deterministic and non-deterministic space complexity~\cite{savitch1970relationships}. This theorem, however, does not yield a practical algorithm. Existing efficient \LI-solvers~\cite{GOAL,RABIT-Reduce} are based on Ramsey-based inclusion testing~\cite{abdulla2011advanced} or rank-based approaches~\cite{kupferman2001weak}. These tools actually use exponential time and space. In fact, the exponential blow-up of Ramsey-based approach seems to be worse than that of \DS-determinization. Thus, the theoretical advantage \BCVt seems to evaporate upon close examination.
Thus, it is far from clear which algorithmic approach is superior. To resolve this issue, we provide in this paper  the first implementations for both algorithms and perform exhaustive empirical analysis to compare their performance.

Our first tool, also called \cct{DetLP}, implements its namesake algorithm as it is. 
We rely on existing \LP-solver \glpsol to perform  numerical reasoning.
Our second tool, called \QuIP, starts from \cct{BCV}, but improves on it. 
The key improvement arises from the construction of an improved comparator with fewer states. 
We revisit the reduction to language inclusion in \cite{BCVFoSSaCS18} accordingly. The new reduction reduces the transition-density of the inputs to the \LI-solver (Transition density is the ratio of transitions to states), improving the overall performance of \cct{QuIP} since 
\LI-solvers are known to scale better at lower  transition-density inputs~\cite{mayr2013advanced}

Our empirical analysis reveals that theoretical complexity does not provide a full picture. Despite its poorer complexity, \cct{QuIP} scales significantly better than \cct{DetLP}, although \DetLPt solves more benchmarks. Based on these observations, we propose a method for DS inclusion that leverages the complementary strengths of these tools to offer a scalable tool for DS inclusion. Our evaluation also exposes the limitations of both approaches, and opens up avenues for improvement in tools for DS inclusion.

\subsubsection{Motivating example}
\label{Sec:Example}

\begin{figure}[t]
	\centering
	\begin{subfigure}{0.48\textwidth}
		\centering
		\begin{tikzpicture}[shorten >=1pt,node distance=4cm,on grid,auto] 
		
		\node[state,initial, accepting] (q_0)   {  \textsf{OFF}}; 
		\node[state, accepting] (q_2) [above left =of q_0, xshift=0.5cm, yshift=0cm] {\textsf{slow}}; 
		\node[state, accepting] (q_1) [right=of q_2] { \textsf{ON}}; 
		
		\path[->] 
		(q_0)   edge [loop right] node {  \textsf{off, 0}} ()
		edge node [above]{ \textsf{on, 10}}  (q_1)
		edge [bend left = 20] node {\textsf{slow, 5}} (q_2)
		(q_1) edge  [loop right] node  [above] { \textsf{on,2}} ()
		edge [bend left = 20]node {\textsf{off, 10}} (q_0)
		edge [bend right = 20] node [above] {\textsf{slow, 5}} (q_2)
		(q_2) edge  node [above]{ \textsf{off, 5}} (q_0)
		edge node [below] {\textsf{on, 5}} (q_1)
		edge [loop left ] node [above]{\textsf{slow, 1}} ();
		\end{tikzpicture} 
		\caption{System $S$}
		\label{Fig:SysS}
	\end{subfigure}
	\hfill
	\centering
	\begin{subfigure}{0.48\textwidth}
		\centering
		\begin{tikzpicture}[shorten >=1pt,node distance=3.5cm,on grid,auto] 
		\node[state,initial, accepting] (q_0)   { \textsf{OFF}}; 
		\node[state, accepting] (q_1) [right=of q_0] {\textsf{ON}}; 
		\path[->] 
		(q_0)   edge [loop above ] node {  \textsf{off, 0}} ()
		edge [bend left = 20]node { \textsf{on, slow, 10}}  (q_1)
		(q_1) edge  [loop above] node  { \textsf{on,2}} ()
		edge [bend left = 20]node {\textsf{off, slow, 10}} (q_0);
		
		\end{tikzpicture}
		\caption{Specification $ P $}
		\label{Fig:SpecP}
	\end{subfigure}
\caption{DS inclusion: Motivating example}
\label{Fig:DSInclusionExample}
\end{figure}
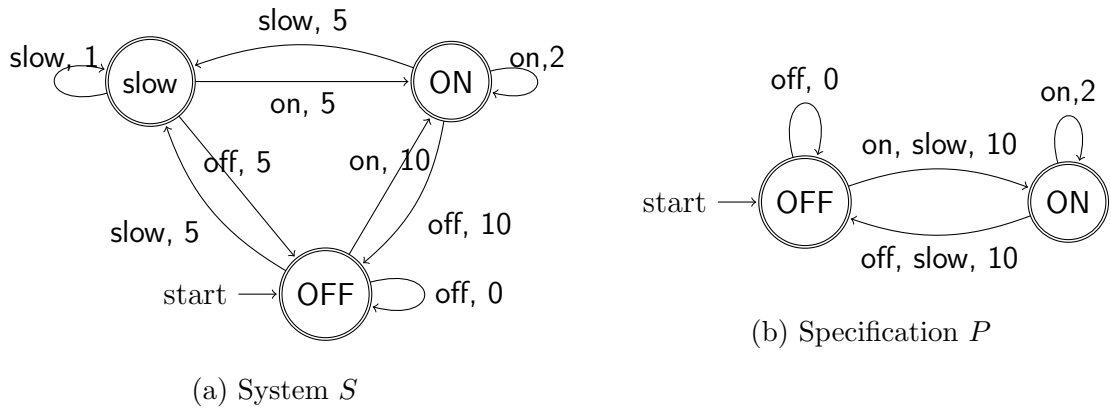

As an example of such a problem formulation, consider the system  and specification in Figure~\ref{Fig:SysS} and Figure~\ref{Fig:SpecP}, respectively~\cite{chatterjee2010quantitative}. Here, the specification $P$ depicts the worst-case energy-consumption model for a motor, and the system $S$ is a candidate implementation of the motor.  Transitions in $S$ and $P$ are labeled by transition-action and transition-cost. The cost of an execution (a sequence of actions) is given by an {\em aggregate} of the costs of transitions along its run (a sequence of automaton states). In non-deterministic automata, where each execution may have multiple runs, cost of the execution is the cost of the run with maximum cost. A critical question here is to check whether  implementation $S$ is more energy-efficient than specification $P$. This problem can be framed as a problem of quantitative inclusion between $S$ and $P$.

\section{Theoretical analysis of existing algorithms}
\label{Sec:RelatedWork}

A purely complexity-theoretic analysis will indicate that \DetLPt will fare poorer than \BCVt in practice since the former is exponential in time and space whereas the later is only polynomial in space. However, an empirical evaluation of these algorithms reveals that opposite: \DetLPt outperforms \BCVt. 

This section undertakes a closer theoretical examination of both algorithms in order to shed light on the seemingly anomalous behavior.  Not only does our analysis explain the behavior but also uncovers avenues for improvements to \BCVt.


\subsection{\cct{DetLP}: DS determinization and Linear programming}

\DetLPt follows a separation-of-techniques approach wherein the first step consists of determinization of the DS automata and the second step reduces DS inclusion to linear programming. As one may notice, the first step is automata-based while the second is based on numerical methods. 

B\"oker and Henzinger studied complexity and decision-procedures for determinization of DS automata in detail~\cite{boker2014exact}. They proved that a DS automata can be determinized if it is complete, all its states are accepting states and the discount-factor is an integer. Under all other circumstances, DS determinization may not be guaranteed. DS determinization extends subset-construction for automata over finite words. Every state of the determinized DS automata is represented by an $|S|$-tuple of numbers, where $S = \{ q_1,\dots q_{|S|}\}$ denotes the set of states of the original \DS-automaton. The value stored in the $i$-th place in the $|S|$-tuple represents the ``gap" or extra-cost of reaching state $q_i$ over a finite-word $w$ compared to its best value so far. The crux of the argument lies in proving that when the DS automata is complete and the discount-factor is an integer, the ``gap" can take only finitely-many values, yielding  finiteness of the determinized DS automata, albeit exponentially larger than the original. 

\begin{theorem}{\rm \cite{boker2014exact} [DS determinization analysis]}
	\label{thrm:DetComplexity}
	Let $A $ be a complete DS automata with maximum weight $\mu$ over transitions and $s$ number of states. DS determinization of $A$ generates a \DS-automaton with at most $\mu^{s}$ states.	
\end{theorem}

Chatterjee et al. reduced $P \subseteq_d Q$ between  non-deterministic DS automata $P$ and deterministic DS automata $Q$ to  linear-programming~\cite{andersen2013fast,andersson2006improved,chatterjee2010quantitative}. 
First, the product DS automata $P\times Q$ is constructed so that 
$(s_P, s_Q)\xrightarrow{a}(t_P,t_Q)$ is a transition with weight $ w_P-w_Q$ if transition $s_M\xrightarrow{a}t_M$ with weight $w_M$ is present in $M$, for $M \in \{P, Q\}$. $P\subseteq_q Q$ is \cct{False} iff the weight of any word in $P\times Q$ is greater than 0. 
Since $ Q$ is deterministic, it is sufficient to check if the maximum weight of all infinite paths from the initial state in $P\times Q$ is greater than 0. 
For discounted-sum, the maximum weight of paths from a given state can be determined by a linear-program: Each variable (one for each state) corresponds to the weight of  paths originating in this state, and transitions decide the constraints which relate the values of variables (or states) on them. The objective is to maximize weight of variable corresponding to the initial state.

Therefore, the \DetLPt method for $P \subseteq_d Q$ is as follows: Determinize $Q$ to $Q_D$ via DS determinization method from ~\cite{boker2014exact}, and reduce $P \subseteq_d Q_D$ to linear programming following~\cite{chatterjee2010quantitative}.
Note that since determinization is possible only if the DS automaton is complete, \DetLPt can be applied only if $Q$ is complete.

\begin{lemma}
	\label{lemma:SizeOfLP}
	Let $P$ and $Q$ be non-deterministic DS automata with $s_P$ and $s_Q$ number of states respectively, $\tau_P$ states in $P$.  Let the alphabet be $\Sigma$ and maximum weight on transitions be $\mu$.
 Then    $P\subseteq_d Q$ is reduced to  linear programming with $\O(s_P \cdot \mu^{s_Q})$ variables and $\O(\tau_P \cdot \mu^{s_Q} \cdot |\Sigma|)$ constraints. 
\end{lemma}

Therefore, the final complexity of \DetLPt is as follows:
\begin{theorem}{\rm ~\cite{andersen2013fast,chatterjee2010quantitative} [Complexity of \cct{DetLP}]}
\label{thrm:DetLPComplexity}
Let $P$ and $Q$ be DS automata with $s_P$ and $s_Q$ number of states respectively, $\tau_P$ states in $P$.  Let the alphabet be $\Sigma$ and maximum weight on transitions be $\mu$.
Complexity of \DetLPt is $\O(s_P^2\cdot \tau_P \cdot \mu^{s_Q} \cdot |\Sigma|)$.
\end{theorem}
\begin{proof}
	Anderson and Conitzer~\cite{andersen2013fast} proved that this system of linear equations can be solved in $\O(m\cdot n^2)$ for $m $ constraints and $n$ variables. 
	\qed
\end{proof}
\subsection{\cct{BCV}: Comparator-based approach}

$\BCVt$ is based on an integrated approach for DS inclusion wherein the entire algorithm uses automata-based reasoning only. 
Strictly speaking, \cct{BCV} is based on a generic algorithm for inclusion under a general class of aggregate functions which permit $\omega$-regular comparators, presented in Chapter~\ref{ch:applicationcomparator}-Section~\ref{Sec:DSApplication}. \cct{BCV} (Algorithm~\ref{Alg:BCV}) refers to its adaptation to \DS. It is described in complete detail, for sake of clarity. 

Recall, a run $\rho \in P$ of word $w \in \L(P)$ is a said to be {\em dominated w.r.t $Q$} if there exists a run $\sigma\in Q$ over the same word $w$ s.t. $\DSum{\rho}{d} < \DSum{\sigma}{d}$. 
The key idea behind $\BCV$ is that $P\subseteq_d Q$ holds iff every run of $P$ is a dominated run w.r.t $Q$. As a result, $ \BCV $ constructs an intermediate B\"uchi automaton $ \firstI $ that consists of all dominated runs of $P$ w.r.t $Q$. It then checks whether 
$\firstI$ consists of all runs of $P$, by determining language-equivalence between $\firstI$ and an automaton $\hat{P}$ that consists of all runs of $P$.
The comparator $\A^{\mu, d}_{\leq}$ is utilized in the construction of $\firstI$ to compare weight of runs in $P$ and $Q$. 

\begin{algorithm}[t]
\label{Alg:BCV}
\caption{ $\BCV(P,Q, d)$, Is $P \subseteq_d Q$?}\label{Alg:BCV}
\begin{algorithmic}[1]
\STATE \textbf{Input: } Weighted automata $P$, $Q$, and discount-factor $d$
\STATE \textbf{Output: } $\mathsf{True}$ if $P \subseteq_d Q$, $\mathsf{False}$ otherwise		
		
\STATE $\hat{P} \leftarrow  \AugmentWtAndLabel(P)$ \label{alg-line:AugmentP}
\STATE $\hat{Q} \leftarrow  \AugmentWtAndLabel(Q)$  \label{alg-line:AugmentQ}
\STATE $\hat{P}\times\hat{Q} \leftarrow \MakeProductI(\hat{P}, \hat{Q})$ \label{alg-line:Prod}		
\STATE $\mu \leftarrow \cc{MaxWeight}(P, Q)$ 	
\STATE $\A^{\mu, d}_{\leq}\leftarrow \mathsf{MakeComparator}(\mu,d)$
\STATE $\lessI \leftarrow  \cc{Intersect}(\hat{P}\times\hat{Q}, \A^{\mu, d}_{\leq} )$  \label{alg-line:Intersect}		
\STATE $\firstI \leftarrow \Project(\lessI)$ \label{alg-line:Project}
\RETURN {$\hat{P} \equiv \firstI$} \label{alg-line:ensure}			
\end{algorithmic}
\end{algorithm}

Procedure $\AugmentWtAndLabel$ separates between runs of the same word in DS automata by assigning a unique transition-identity to each transition. It also appends the transition weight, to enable weight comparison afterwards. Specifically, it transforms \DS-automaton $\A$ into B\"uchi automaton $\hat{\A}$, with all states as accepting, by converting transition $\tau = s\xrightarrow{a}t$ with weight $wt$ and unique transition-identity $l$ to transition $\hat{\tau} =  s\xrightarrow{(a, wt, l)}t $ in $\hat{\A}$. 
Procedure $\MakeProductI(\hat{P}, \hat{Q})$ takes the product of $\hat{P}$ and $\hat{Q}$ over the same word 
i.e., transitions $s_\A\xrightarrow{(a, n_\A, l_\A)}t_\A$ in $\A$, for $\A \in \{\hat{P}, \hat{Q}\}$,
generates transition $(s_P, s_Q)\xrightarrow{(a, n_P, l_P, n_Q, l_Q)}(t_P, t_Q)$ in $ \hat{P}\times\hat{Q}$.
The comparator $\A^{\mu,d}_{\leq}$ is constructed with upper-bound $\mu$ that equals the maximum weight of transitions in $P$ and $Q$, and discount-factor $d$. 
$\cct{Intersect}$ matches the alphabet of $\hat{P}\times\hat{Q}$ with $\A^{\mu,d}_{\leq}$, and intersects them. The resulting  automaton $\lessI$ accepts word $(w, wt_P, id_P, wt_Q, id_Q)$ iff $\DSum{wt_P}{d}\leq \DSum{wt_Q}{d}$. The projection of $\lessI$ on the first three components of $\hat{P}$ returns $\firstI$ which contains the word $(w,wt_P,id_P)$ iff it is a dominated run in $P$. Finally, language-equivalence between $\firstI$ and $\hat{P}$ returns the answer.  

\subsubsection{Analysis of \BCVt}
\label{Sec:BCVDrawbacks}

The proof for \cct{PSPACE}-complexity of \BCVt relies on \cct{LI} to be \cct{PSPACE}. In practice, though, implementations of \cct{LI} apply Ramsey-based inclusion testing~\cite{abdulla2011advanced}, rank-based methods~\cite{kupferman2001weak} etc. All of these algorithms are exponential in time and space in the worst case. Any implementation of \BCVt will have to rely on an \LI-solver.
Therefore, in practice \BCVt is also exponential in time and space. In fact, we show that its worst-case complexity (in practice) is poorer than \cct{DetLP}.

Another reason that prevents \BCVt from practical implementations is that it does not optimize the size of intermediate automata. Specifically, we show that the size and transition-density of $\firstI$, which is one of the inputs to \LI-solver,  is very high (Transition density is the ratio of transitions to states). Both of these parameters are known to be deterrents to the performance of  existing \LI-solvers~\cite{abdulla2010simulation}, subsequently to \BCVt as well: 
\begin{lemma}
	\label{Lemma:SizeBCV}
	Let $s_P$, $s_Q$, $s_d$ and $\tau_P$, $\tau_Q$, $\tau_d$ denote the number of states and transitions in $P$, $Q$, and $\A^{\mu,d}_\leq$, respectively. 
	Number of states and transitions in $ \firstI $ are $\O(s_P  s_Q  s_d)$ and
	$\O(\tau_P^2  \tau_Q^2 \tau_d |\Sigma|)$, respectively.
\end{lemma}

\begin{proof}
	It is easy to see that the number of states and transitions of $\hat{P}$ $\hat{Q}$ are the same as those of $P$ and $Q$, respectively. 
	Therefore, the number of states and transitions in $\hat{P}\times\hat{Q}$ are $\O(s_P s_Q)$ and $\O(\tau_P \tau_Q)$, respectively. 
	The alphabet of $\hat{P}\times\hat{Q}$ is of the form $(a, wt_1,id_1,wt_2,id_2)$ for $a \in \Sigma$, $wt_1, wt_2$ are non-negative weights bounded by $\mu$ and $id_i$ are unique transition-ids in $P$ and $Q$ respectively. The alphabet of comparator $\A^{\mu,d}_\leq$ is of the form $(wt_1,wt_2)$. To perform intersection of these two, the alphabet of comparator needs to be matched to that of the product, causing a blow-up in number of transitions in the comparator by a factor of $|\Sigma|\cdot\tau_P\cdot\tau_Q$. Therefore, the number of states and transitions in $\lessI$ and $\firstI$ is given by $\O(s_Ps_Qs_d)$ and $\O(\tau_P^2 \tau_Q^2 \tau_d |\Sigma|)$.
	\qed
\end{proof}

The comparator is a non-deterministic B\"uchi automata with $\O(\mu^2)$ states over an alphabet of size $\mu^2$~\cite{BCVFoSSaCS18}. Since transition-density $\delta = |S|\cdot |\Sigma|$ for non-deterministic B\"uchi automata, the transition-density of the comparator is $\O (\mu^4)$. Therefore,

\begin{corollary}
	\label{coro:TransDensityBCV}
	Let $s_P$, $s_Q$, $s_d$ denote the number of states in $P$, $Q$, $\A^{\mu,d}_\leq$, respectively, and $\delta_P$, $\delta_Q$ and $\delta_d$ be their transition-densities. 
	Number of states and transition-density of $\firstI$ are $ \O(s_Ps_Q\mu^2)$ and $\O(\delta_P \delta_Q \tau_P  \tau_Q   \cdot \mu^4 \cdot  |\Sigma|)$, respectively.
\end{corollary}

The corollary illustrates that the transition-density of $\firstI$ is very high even for small inputs. The blow-up in number of transitions of $\lessI$ (hence $\firstI$) occurs during alphabet-matching for B\"uchi automata intersection (Algorithm~\ref{Alg:BCV}, Line~\ref{alg-line:Intersect}). However, the blow-up can be avoided by performing intersection over a substring of the alphabet of $\hat{P} \times\hat{Q}$. Specifically, if $ s_1\xrightarrow{(a,n_P, id_P, n_Q,id_Q)} s_2$  and $ t_1 \xrightarrow{(wt_1, wt_2)} t_2$ are transitions in $\hat{P}\times\hat{Q}$ and comparator $\A^{\mu,d}_\leq$ respectively, then $(s_1,t_1,i) \xrightarrow{(a,n_P, id_P, n_Q,id_Q)} (s_2,t_2,j)$ is  a transition in the intersection iff $n_P = wt_1$ and $n_Q = wt_2$, where $j = (i+1)\mod 2$ if either $s_1$ or $t_1$ is an accepting state, and $j=i$ otherwise. We call intersection over substring of alphabet $\Intersect$. The following is easy to prove:

\begin{lemma}
	\label{lemma:IntersectSubstring}
	Let $\A_1 = \mathsf{Intersect}(\hat{P}\times\hat{Q}, \A^{\mu,d}_{\leq})$, and $\A_2 = \Intersect(\hat{P}\times\hat{Q}, \A^{\mu,d}_{\leq})$. $\mathsf{Intersect}$ extends alphabet of $\A^{\mu,d}_{\leq}$ to match the alphabet of $\hat{P}\times\hat{Q}$ and $\Intersect$ selects a substring of the alphabet of $\hat{P}\times\hat{Q}$ as defined above. 
	Then,
	$\L(\A_1) \equiv \L(\A_2)$.
\end{lemma}

$\Intersect$ prevents the blow-up by $|\Sigma|\cdot \tau_P\cdot \tau_Q$, resulting in only $\O(\tau_P\tau_Q\tau_d)$ transitions in $\firstI$
Therefore,

\begin{lemma}{\rm [Trans. Den. in \BCVt]}
	\label{lemma:TransDensityBCVIntersect}
	Let $\delta_P$, $\delta_Q$ denote transition-densities of $P$ and $Q$, resp., and $\mu$ be the upper bound for comparator $\A^{\mu,d}_\leq$.
	Number of states and transition-density of $\firstI$ are $ \O(s_Ps_Q\mu^2)$ and $\O(\delta_P \delta_Q  \cdot \mu^4 )$, respectively.
\end{lemma}

Language equivalence is performed by tools for language inclusion. 
The most effective tool for language-inclusion \cct{RABIT}~\cite{RABIT-Reduce} is based on Ramsay-based inclusion testing~\cite{abdulla2011advanced}. The worst-case complexity for $A \subseteq B$ via Ramsay-based inclusion testing is known to be $2^{\O(n^2)}$, when $B$ has $n$ states. Therefore, 

\begin{theorem}{\rm [Practical complexity of \BCVt]}
	\label{thrm:ComplexityBCVRamsay}
	Let $P$ and $Q$ be DS automata with $s_P$, $s_Q$ number of states respectively, and maximum weight on transitions be $\mu$.
	Worst-case complexity for $\BCV$ for integer discount-factor $d>1$ when language-equivalence is performed via Ramsay-based inclusion testing is $2^{\O(s_P^2 \cdot s_Q^2 \cdot \mu^4)}$.
\end{theorem}

Recall that language-inclusion queries are $\hat{P}\subseteq \firstI$ and $\firstI\subseteq \hat{P}$. Since $\firstI$ has many more states than $\hat{P}$, the complexity of $\hat{P}\subseteq \firstI$ dominates.

Theorem~\ref{thrm:DetLPComplexity}  and Theorem~\ref{thrm:ComplexityBCVRamsay} demonstrate that the complexity of \BCVt (in practice) is worse than \cct{DetLP}. This explains the inferior performance of \BCVt in practice.

\section{$ \QuantIP $: Optimized $\BCV$-based solver for DS inclusion}
\label{Sec:QuIP}
The ealrier investigate explains why \BCVt does not lend itself to a practical implementation for DS inclusion (\textsection~\ref{Sec:BCVDrawbacks}), and also identifies its drawbacks. This section proposes an improved comparator-based algorithm \cct{QuIP} for DS inclusion, as is described in \textsection~\ref{Sec:ImproveBCV}. \cct{QuIP} improves upon \BCVt by means of a new optimized comparator that we describe in \textsection \ref{Sec:OptimziedDSComp}.

\subsection{An optimized DS comparator}

\label{Sec:OptimziedDSComp}

\renewcommand{\max}{\mu}
\renewcommand{\MaxC}{\mu_C}
\renewcommand{\MaxX}{\mu_X}

The $2^{\O(s^2)}$ dependence of \BCVt on the number of states $s$ of the DS comparator motivates us to construct a more compact comparator. Currently a DS comparator consists of $\O(\mu^2)$ number of states for upper bound $\mu$~\cite{BCVFoSSaCS18}. 
In this section, we re-define DS comparison languages and DS comparators so that they consist of only $\O(\mu)$-many states and have a transition density of $\O(\mu^2)$.

\begin{Def}[DS comparison language]
	\label{def:newDSlanguage}
	For an integer upper bound $\mu>0$, discount factor $d >1$, and equality or inequality relation $\mathsf{R} \in \{<,>,\leq, \geq, =, \neq\}$,
	the {\em DS comparison language with upper bound $\mu$, relation $\mathsf{R}$, and discount factor $d$} is a language of infinite words over the alphabet $\Sigma = \{-\mu, \dots, \mu\}$ that accepts  $A \in \Sigma^{\omega}$ iff $\DSum{A}{d}$ $\mathsf{R}$ $0$ holds. 
\end{Def}
\begin{Def}[DS comparator automata]
	\label{def:newDScomparator}
	For an integer upper bound $\mu>0$, discount factor $d >1$, and equality or inequality relation $\mathsf{R} \in \{<,>,\leq, \geq, =, \neq\}$,
	the {\em DS comparator automata with upper bound $\mu$, relation $\mathsf{R}$, and discount factor $d$} is an automaton that accepts the DS comparison language with upper bound $\mu$, relation $\mathsf{R}$, and discount factor $d$.
\end{Def}


Semantically, Definition~\ref{def:newDSlanguage} and Definition~\ref{def:newDScomparator} with upper bound $\mu$, discount-factor $d$ and inequality relation $\mathsf{R}$ is the language  and automaton, respectively, of all integer sequences bounded by $\mu$ for which their discounted-sum is related to 0 by the relation $\mathsf{R}$.
They differ from the definitions presented in Chapter~\ref{ch:applicationcomparator}-Section~\ref{Sec:IntegerDF} since those definitions relate a pair of integer sequences with each other, while the current definitions related one sequence with the constant value 0. However, these definitions are equivalent since 
$\DSum{A}{d} \leq \DSum{B}{d} \equiv \DSum{A-B}{d} \leq 0$, where the sequence $(A-B)$ refers to the sequence generated by taking a point-wise difference of elements in $A$ and $B$. 
Now onwards, we will always use the new definitions for DS comparison languages and DS comparators. 

Note that this  equivalence does not hold for all aggregation functions such as limsup and liminf. Hence, this modified definition for DS comparison languages and DS comparators  specifically apply to the discounted-sum aggregation function. 
Another repercussion of thi equivalence is that the results on $\omega$-regularity of DS comparison languages and comparators on the older definitions from Chapter~\ref{ch:applicationcomparator}-Section~\ref{Sec:RationalDF}-~\ref{Sec:IntegerDF} apply to these new definitions as well. Therefore, we obtain that DS comparison languages are $\omega$-regular iff the discount factor is an integer.  
In fact, the automaton for DS comparator with upper bound $\mu$, integer discount factor $d>1$ and relation $\mathsf{R}$, denoted by $\U^{\mu,d}_{\mathsf{R}}$ under Definition~\ref{def:newDScomparator} can be derived from the DS comparator with the same parameters from the old definition, denoted $\A^{\mu,d}_{\mathsf{R}}$, by transforming the alphabet from $(a,b)$ to $(a-b)$ along every transition. 
The first benefit of the modified alphabet is that its size is reduced from $\mu^2$ to $2\cdot \mu -  1$.  
In addition, it coalesces all transitions between any two states over alphabet $(a,a+v)$, for all $a$, into one single transition over $v$, thereby also reducing  transitions. 
However, this direct transformation results in a comparator with  $\O(\mu^2)$ states.
This section presents  a new construction of the comparator with  $\O(\mu)$ states only.


In principle, the current construction adapts the construction from Chapter~\ref{ch:applicationcomparator}-Section~\ref{Sec:IntegerDF} to the modified alphabett.
The key idea behind the construction of the ew DS comparator, similar to the earlier construction, is that the discounted-sum of sequence $V$ can be treated as a number in base $d$ i.e. $\DSum{V}{d}  =  \Sigma_{i=0}^{\infty} \frac{V[i]}{d^i} = (V[0].V[1]V[2]\dots)_d$. 
So, there exists a non-negative value $C$ in base $d$ s.t. $V + C = 0$ for arithmetic operations in base $d$. This value $C$ can be represented by a non-negative sequence $C$ s.t. $\DSum{C}{d} + \DSum{V}{d}= 0$.
Arithmetic in base $d$ over sequences $C$ and $V$ result in a sequence of carry-on $X$ such that:
\begin{lemma}
 \label{Lemma:discount-invariant1}
Let $V, C, X$ be the number sequences, $ d > 1 $ be a positive integer such that following equations holds true:
 \begin{enumerate}
 \item  \label{eq:initial} When $ i = 0 $, $ V[0] + C[0] + X[0] = 0$
 \item \label{eq:invariant} When $ i\geq 1 $, $V[i] + C[i] + X[i] = d \cdot X[i-1]$
 \end{enumerate}
 Then $ \DSum{V}{d} + \DSum{C}{d} =0$.
 \end{lemma}
 \begin{proof}
 	This follows proof by expansion of all terms. Specifically, expand out $\DSum{V}{d} + \DSum{C}{d}$ to $\Sigma_{i=0}^{\infty} \frac{(V[i] + C[i])}{d^i}$ and replace each $(V[i] + C[i])$ with substitutions given by the equations. Most terms will cancel each other, and by re-arrangement  we will get $\DSum{V}{d} + \DSum{C}{d}  = 0$. 
	\qed
 \end{proof}
 
In the construction of the comparator in Chapter~\ref{ch:applicationcomparator}-Section~\ref{Sec:IntegerDF}, it has been proven that when $A$ and $B$ are bounded non-negative integer sequences s.t. $\DSum{A}{d} \leq \DSum{B}{d}$, the corresponding sequences $C$ and $X$ are also bounded integer-sequences~\cite{BCVFoSSaCS18}. The same argument transcends here: When $V$ is a bounded integer sequence s.t. $\DSum{V}{d}\leq 0$, there exists a corresponding pair of bounded integer sequence $C$ and $X$. In fact, the bounds used for the comparator carry over to this case as well. Sequence $C$ is non-negative and is bounded by $\MaxC = \mu\cdot\frac{d}{d-1}$ since $-\MaxC$ is the minimum value of discounted-sum of $V$, and integer-sequence $X$ is bounded by $\MaxX  = 1+ \frac{\mu}{d-1}$. 
%
On combining Lemma~\ref{Lemma:discount-invariant1} with the bounds on $X $ and $C$ we get: 
\begin{lemma}
	\label{lemma:RemoveC}
	Let $V$ and be an integer-sequence bounded by $\mu$ s.t. $\DSum{V}{d} \leq 0$, and $X$ be an integer sequence bounded by $(1 + \frac{\mu}{d-1})$, then there exists an $X$ s.t. 
	\begin{enumerate}
		\item \label{eq:initialRemoveC} When $i=0$, $0\leq -(X[0] + V[0]) \leq \mu\cdot\frac{d}{d-1}$
		\item  \label{eq:invariantRemoveC} When $i\geq 1$, $ 0\leq (d \cdot X[i-1] - V[i] - X[i] )\leq  \mu\cdot\frac{d}{d-1}$
	\end{enumerate}
	
\end{lemma}
\begin{proof}
Equations~\ref{eq:initialRemoveC}-\ref{eq:invariantRemoveC} from Lemma~\ref{lemma:RemoveC} have been obtained by expressing $C[i]$ in terms of $X[i]$, $X[i-1]$, $V[i]$ and $d$, and imposing the non-negative bound of $\MaxC = \mu\cdot\frac{d}{d-1}$ on the resulting expression. 
Therefore, Lemma~\ref{lemma:RemoveC} implicitly captures the conditions on $C$ by expressing it only in terms of $V$, $X$ and $d$ for $\DSum{V}{d} \leq 0$ to hold.
\qed
\end{proof}

In construction of the new DS comparator, the values of $V[i]$ is part of the alphabet, upper bound $\mu$ and discount-factor $d$ are the input parameters. The only unknowns are the value of $X[i]$. However, we know that it can take only finitely many values i.e.  integer values $|X[i]|\leq \MaxX$. So, we store all possible values of $X[i]$ in the states. Hence, the state-space $S$ comprises of $\{(x) | |x|\leq \MaxX\}$ and a start state $s$. Transitions between these states are possible iff the corresponding $x$-values and alphabet $v$ satisfy the conditions of Equations~\ref{eq:initialRemoveC}-\ref{eq:invariantRemoveC} from Lemma~\ref{lemma:RemoveC}. 
There is a transition from start state $s$ to state $(x)$ on alphabet $v$ if $0\leq -(x+v) \leq \mu\cdot\frac{d}{d-1}$, and from state $(x)$ to state $(x')$ on alphabet $v$ if $0\leq (d\cdot x - v -x') \leq \mu\cdot\frac{d}{d-1}$. All $(x)$-states are accepting. This completes the construction for DS comparator $\U^{\mu,d}_{\leq}$. Clearly $\U^{\mu,d}_{\leq}$ has only $\O(\mu)$ states. The formal construction is given below:

\subsubsection{Construction}
Let $ \MaxC = \max \cdot \frac{d}{d-1}   \leq 2\cdot \max$ and $ \MaxX = 1 + \frac{\max}{d-1}$.
$ \U^{\mu,d}_< = (\Statess, \Sigma, \delta_d, \StartState, \Final) $ 
\begin{itemize}
	
	\item $ \Statess = \StartState \cup \Final \cup S_{\bot} $ where \\ 
	$ \StartState = \{s\} $, 
	$\Final = \{x ||x| \leq \MaxX\} $, and \\
	$S_{\bot} = \{(x, \bot) | | x| \leq \MaxX\}$ where $\bot$ is a special character, and $x \in \mathbb{Z}$. 
	
	
	\item $ \Sigma = \{v : |v| \leq \max \} $ where $v$ is an integer.
	
	\item $\delta_d \subset \Statess \times \Sigma \times \Statess$ is defined as follows:
	\begin{enumerate}
		\item Transitions from start state $ s $:
		\begin{enumerate}[label = \roman*]
			\item $ (s ,v, x) $ for all $ x \in \Final $ s.t. $0<-(x+v) \leq \MaxC$
			
			\item $ (s ,v, (x, \bot)) $ for all $ (x, \bot) \in S_{\bot} $ s.t. $x+v = 0$

		\end{enumerate}
		
		\item Transitions within $ S_{\bot} $: $ ((x, \bot) ,v, (x', \bot) )$ for all $(x, \bot)$, $(x', \bot) \in S_{\bot} $, if  $ d\cdot x  = v + x'$
		
		
		\item Transitions within $ \Final $: $(x ,v, x' )$ for all $ x, x' \in \Final $ if $0\leq d\cdot x -v-x'<d  $
		
		
		\item Transition between $ S_{\bot} $ and $ \Final $: $ ((x, \bot),v, x') $ for $ (x,\bot) \in S_{\bot} $, $ x' \in \Final $ if  $0< d\cdot x -v-x'<d$
		
		%
	\end{enumerate}
	
	
	
\end{itemize}

\begin{theorem}
	The B\"uchi automaton $\U^{\mu,d}_{<}$ constructed above is  
	DS comparator automata with with upper bound $\mu$, integer discount factor $d>1$ and relation $<$. DS comparator $\U^{\mu,d}_{<}$  consists of $\O(\mu)$-states, and a transition-density of $\O(\mu^2)$.
\end{theorem}

Since B\"uchi automata are closed under set-theoretic operations, a counterpart of closure of $\omega$-regular comparison languages under all relations (Theorem~\ref{lem:closure}) applies to the new definition as well. Furthermore, simple modifications to the automaton constructed above will result in the automaton for all other relations. As a result, The DS comparator automata wth upper bound $\mu>0$, integer discount factor $d>1$ and relation $\mathsf{R} \in \{<,>,\leq, \geq, =, \neq\}$ will have $\O(\mu)$ states, alphabet size of $2\cdot\mu-1$, and transition-density of $\O(\mu^2)$.

\subsection{\QuIP: Algorithm description}
\label{Sec:ImproveBCV}
The construction of the DS comparator automata in Section~\ref{Sec:OptimziedDSComp} leads to an implementation-friendly \cct{QuIP} from \cct{BCV}. The core focus of \cct{QuIP} is to ensure that the size of intermediate automata is small and they have fewer transitions to assist the LI solvers. 
Technically, \cct{QuIP} differs from \cct{BCV} by incorporating the new comparator automata and an appropriate $\Intersect$ function, rendering  $\cct{QuIP}$ theoretical improvement over \cct{BCV}.
\begin{algorithm}[t]
	\caption{ $\QuIP(P,Q, d)$, Is $P \subseteq_d Q$?}
	\label{Alg:Quip}
	\begin{algorithmic}[1]
		\STATE \textbf{Input: } DS automata $P$ and $Q$ with integer discount factor $d$
		\STATE \textbf{Output: } $\mathsf{True}$ if $P \subseteq_d Q$, $\mathsf{False}$ otherwise
		
		\STATE $\hat{P} \leftarrow  \AugmentWtAndLabel(P)$ \label{alg-line:AugmentPQuip}
		\STATE $\hat{Q} \leftarrow  \AugmentWt(Q)$  \label{alg-line:AugmentQQuip}
		
		\STATE $\hat{P}\times\hat{Q} \leftarrow \MakeProductI(\hat{P}, \hat{Q})$ \label{alg-line:ProdQuip}
	
		\STATE $\A\leftarrow \mathsf{MakeBaseline}(\mu,d,\leq)$
		\STATE $\lessI \leftarrow  \mathsf{IntersectSelectAlpha}(\hat{P}\times\hat{Q}, \A)$  \label{alg-line:IntersectQuip}

		\STATE $\firstI \leftarrow \ProjectOutWt(\lessI)$ \label{alg-line:ProjectProd}
		\STATE $\hat{P}_{-wt} \leftarrow \ProjectOutWt(\hat{P})$ \label{alg-line:ProjectP}
		
		\RETURN {$\hat{P}_{-wt} \subseteq \firstI$} \label{alg-line:ensureQuip}

	\end{algorithmic}
\end{algorithm}
Like \cct{BCV}, \cct{QuIP} also determines all diminished runs of $P$. So, it disambiguates $P$ by appending weight and a unique label to each of its transitions. Since, the identity of runs of $Q$ is not important, we do not disambiguate between runs of $Q$, we only append the weight to each transition (Algorithm~\ref{Alg:Quip}, Line~\ref{alg-line:AugmentQQuip}). The DS comparator is constructed for discount factor $d$, maximum weight $\mu$ along transitions in $P$ and $Q$, and the inequality $\leq$. Since the alphabet of the DS comparator are integers between $-\mu$ to $\mu$, the alphabet of the product $\hat{P}\times\hat{\Q}$ is adjusted accordingly. Specifically, the weight recorded along transitions in the product is taken to be the difference of weight in $\hat{P}$ to that in $\hat{Q}$ i.e. if $\tau_P : s_1 \xrightarrow{a_1,wt_1, l} s_2$ and $\tau_Q : t_1 \xrightarrow{a_2, wt_2} t_2$ are transitions in $\hat{P}$ and $\hat{Q}$ respectively, then $\tau = (s_1,t_1) \xrightarrow{a_1, wt_1-wt_2, l} (s_2,t_2)$ is a transition in $\hat{P}\times\hat{Q}$ iff $a_1 = a_2$ (Algorithm~\ref{Alg:Quip}, Line~\ref{alg-line:ProdQuip}). In this case, $\Intersect$ intersects the DS comparator $\A$ and product  $\hat{P}\times\hat{Q}$ only on the weight-component of alphabet in $\hat{P}\times\hat{Q}$. 
Specifically, if $ s_1\xrightarrow{(a,wt_1,l)} s_2$  and $ t_1 \xrightarrow{wt_2} t_2$ are transitions in $\hat{P}\times\hat{Q}$ and comparator $\A^{\mu,d}_\leq$ respectively, then $(s_1,t_1,i) \xrightarrow{a,wt_1,l} (s_2,t_2,j)$ is  a transition in the intersection iff $wt_1=wt_2$, where $j = (i+1)\mod 2$ if either $s_1$ or $t_1$ is an accepting state, and $j=i$ otherwise. 
Automaton $\firstI$ and $\hat{P}_{-wt}$ are obtained by project out the weight-component from the alphabet of $\hat{P}\times\hat{Q}$ and $\hat{P}$ respectively. The alphabet of $\hat{P}\times\hat{Q}$ and $\hat{P}$ are converted from $(a, wt, l)$ to only $(a, l)$. It is necessary to project out the weight component since in $\hat{P}\times\hat{Q}$ they represent the difference of weights and and in $\hat{P}$ they represent the absolute value of weight. 

Finally, the language of $\firstI$ is equated with that of $\hat{P}_{-wt}$ which is the automaton generated from $\hat{P}$ after discarding weights from transitions. 
However, it is easy to prove that $\firstI \subseteq 
\hat{P}_{-wt}$. Therefore, instead of language-equivalence between $\firstI$ and $\hat{P}_{-wt}$ and, it is sufficient to check whether $\hat{P}_{-wt} \subseteq \firstI$. As a result, $\QuantIP$ utilizes LI solvers as a black-box to perform this final step.  

\begin{lemma}{\rm [Trans. Den. in \QuIP]}
	\label{lemma:TransDensityQuip}
	Let $\delta_P$, $\delta_Q$ denote transition-densities of $P$ and $Q$, resp., and $\mu$ be the upper bound for DS comparator $\U^{\mu,d}_\leq$.
	Number of states and transition-density of $\firstI$ are $ \O(s_Ps_Q\mu)$ and $\O(\delta_P \delta_Q  \cdot \mu^2 )$, respectively.
\end{lemma}

\begin{theorem}{\rm [Practical complexity of \QuIP]}
	\label{thrm:ComplexityQuipRamsay}
	Let $P$ and $Q$ be DS automata with $s_P$, $s_Q$ number of states, respectively, and maximum weight on transitions be $\mu$.
	Worst-case complexity for \cct{QuIP} for integer discount-factor $d>1$ when language-equivalence is performed via Ramsay-based inclusion testing is $2^{\O(s_P^2 \cdot s_Q^2 \cdot \mu^2)}$.
\end{theorem}

Theorem~\ref{thrm:ComplexityQuipRamsay} demonstrates that while complexity of \cct{QuIP} (in practice) improves upon \cct{BCV} (in practice), it is  still worse than \cct{DetLP}.

\section{Empirical analysis of DS inclusion algorithms}
\label{Sec:ExpEvalI}

We provide implementations of our tools \cct{QuIP} and \cct{DetLP} and conduct  experiments on a large number of synthetically-generated benchmarks to compare their performance.
We seek to find answers to the following questions: 
\begin{enumerate}
	\item  Which tool has better performance, as measured by runtime, and number of benchmarks solved?
	\item How does change in transition-density affect performance of the tools? 
	\item How dependent are our tools on their underlying solvers?  
\end{enumerate}

\subsubsection{Implementation details}
\label{Sec:ImplementDetail}
We implement our tools \cct{QuIP} and \cct{DetLP} in \cplus, with compiler optimization \texttt{o3} enabled. 
We implement our own library for all B\"uchi-automata and \DS-automata operations, except for language-inclusion for which we use the state-of-the-art \LI-solver \RABITt~\cite{RABIT-Reduce} as a black-box.
We enable the \texttt{-fast} flag in  \cct{RABIT}, and tune  its $ \Java $-threads with \cct{Xss}, \cct{Xms}, \cct{Xmx} set to 1GB, 1GB and 8GB respectively. 
We use the large-scale \LP-solver \glpsol provided by \cct{GLPK} (\cct{GNU} Linear Programming Kit)~\cite{glpk} inside \cct{DetLP}. We did not tune \glpsol since it consumes a very small percentage of total time in \cct{DetLP}, as we see later in Fig~\ref{Fig:ToolTimeRatio}.

We also employ some implementation-level optimizations.
Various steps of \cct{QuIP} and \DetLPt such as product, \DS-determinization, \univ construction, involve the creation of new automaton states and transitions.
We reduce their size by adding a  new state only if it is reachable from the initial state, and a new transition only if it originates from such a state.    

The universal automata is constructed on the restricted alphabet of only those weights that appear in the product $\hat{P}\times\hat{Q}$ to include only necessary transitions. We also reduce its size with B\"uchi minimization tool \cct{Reduce}~\cite{RABIT-Reduce}. 


Since all states of $\hat{P}\times\hat{Q}$ are accepting, we conduct the intersection so that it avoids doubling the  number of product states. This can be done, since it is sufficient to keep track of whether words visit accepting states in the universal.

\subsubsection{Benchmarks}
To the best of our knowledge, there are no standardized  benchmarks for \DS-automata. We attempted to experimented with examples that appear in research papers. However, these examples are too few and too small, and do not render an informative view of  performance of the tools. 
Following a standard approach to performance evaluation of automata-theoretic tools~\cite{abdulla2010simulation,mayr2013advanced,tabakov2005experimental},  we experiment with our tools on {\em randomly generated} benchmarks. 

\paragraph{Random weighted-automata generation}
The parameters for our random  weighted-automata generation procedure are the number of states $N$, transition-density $\delta$ and upper-bound $\mu$ for weight on transitions. The states are represented by the set $\{0,1,\dots, N-1\}$. All states of the weighted-automata are accepting, and they have a unique initial state 0. The alphabet for all weighted-automata is fixed to $\Sigma = \{a,b\}$. Weight on transitions ranges from $0$ to $\mu-1$.
For our experiments we only generate complete weighted-automata. 
These weighted automata are generated only if the number of transitions $\lfloor N\cdot \delta \rfloor$ is greater than $N\cdot |\Sigma|$, since there must be at least one transition on each alphabet from every state. We first complete the weighted-automata by creating a transition from each state on every alphabet. In this case the destination state and weight are chosen randomly. The remaining $(N\cdot |\Sigma| - \lfloor N\cdot \delta \rfloor)$-many transitions are generated by selecting all parameters randomly i.e. the source and destination states from $\{0,\dots N-1\}$, the alphabet from $\Sigma$, and weight on transition from $\{0,\mu-1\}$.
\subsubsection{Design and setup for experimental evaluation}
\label{Sec:setup}

Our experiments were designed with the objective to compare  \DetLPt and \cct{QuIP}. Due to the lack of standardized benchmarks, we conduct our experiments on randomly-generated benchmarks. Therefore, the parameters for $P\subseteq_d Q$ are the number of states $s_P$ and $s_Q$, transition density $\delta$, and maximum weight $wt$.
We seek to find answers to the questions described at the beginning of \textsection~\ref{Sec:ExpEvalI}.

Each instantiation of the parameter-tuple $(s_P, s_Q, \delta, wt)$ and a choice of tool between \cct{QuIP} and \DetLPt corresponds to one experiment. In each experiment, the weighted-automata $P$ and $Q$ are randomly-generated with the parameters $(s_P, \delta, wt)$ and $(s_Q, \delta, wt)$, respectively, and language-inclusion is performed by the chosen tool. 
Since all inputs are randomly-generated, each experiment is repeated for 50 times to obtain statistically significant data. 
Each experiment is run for a total of $1000\sec$ on for a single node of a high-performance cluster. Each node of the cluster consists of two quad-core Intel-Xeon processor running at 2.83GHz, with 8GB of memory per node. 
The runtime of experiments that do not terminate within the given time limit is assigned a runtime of $\infty$.
We report the median of the runtime-data collected from all  iterations of the experiment. 

These experiments are scaled-up by increasing the size of inputs. 
The worst-case analysis of $\QuIP$ demonstrates that it is symmetric in $s_P$ and $s_Q$, making the algorithm impartial to which of the two inputs is scaled (Theorem~\ref{thrm:ComplexityQuipRamsay}). On the other hand, complexity of \DetLPt is dominated by $s_Q$ (Theorem~\ref{thrm:DetLPComplexity}). Therefore, we scale-up our experiments by increasing $s_Q$ only. 

Since \DetLPt is restricted to complete automata, these experiments are conducted on complete weighted automata only. We collect data on total runtime of each tool, the time consumed by the underlying solver, and the number of times each experiment terminates with the given resources. We experiment with $s_P = 10$, $\delta$  ranges between 2.5-4 in increments of 0.5 (we take lower-bound of 2.5 since $|\Sigma| = 2$), $wt \in \{4,5\}$, and $s_Q$ ranges from 0-1500 in increments of 25, $d=3$. 
These sets of experiments also suffice for testing scalability of both tools. 

\subsubsection{Observations}
\label{Sec:observation}

\begin{figure*}[t]
\centering
\begin{minipage}{0.8\textwidth}
\includegraphics[width=\textwidth]{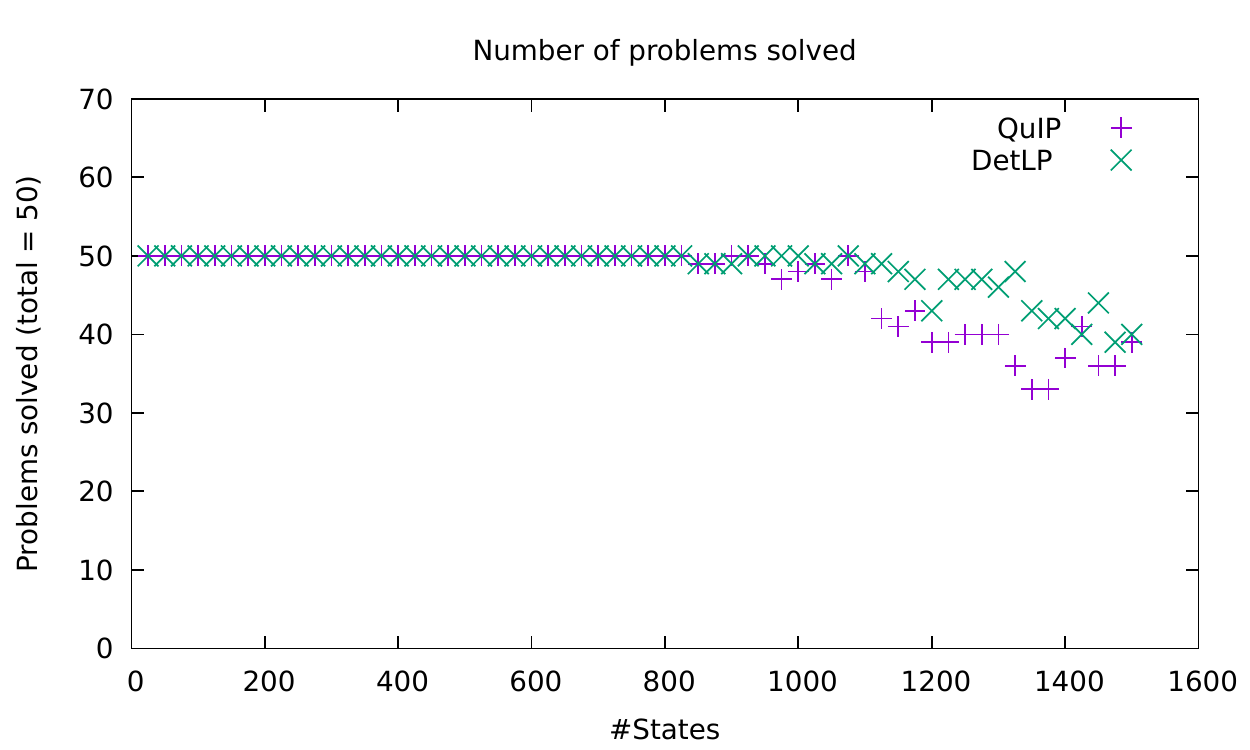}
\subcaption{}
\label{Fig:NumberR25}
\end{minipage}
\hfill
\begin{minipage}{0.8\textwidth}

\includegraphics[width=\textwidth]{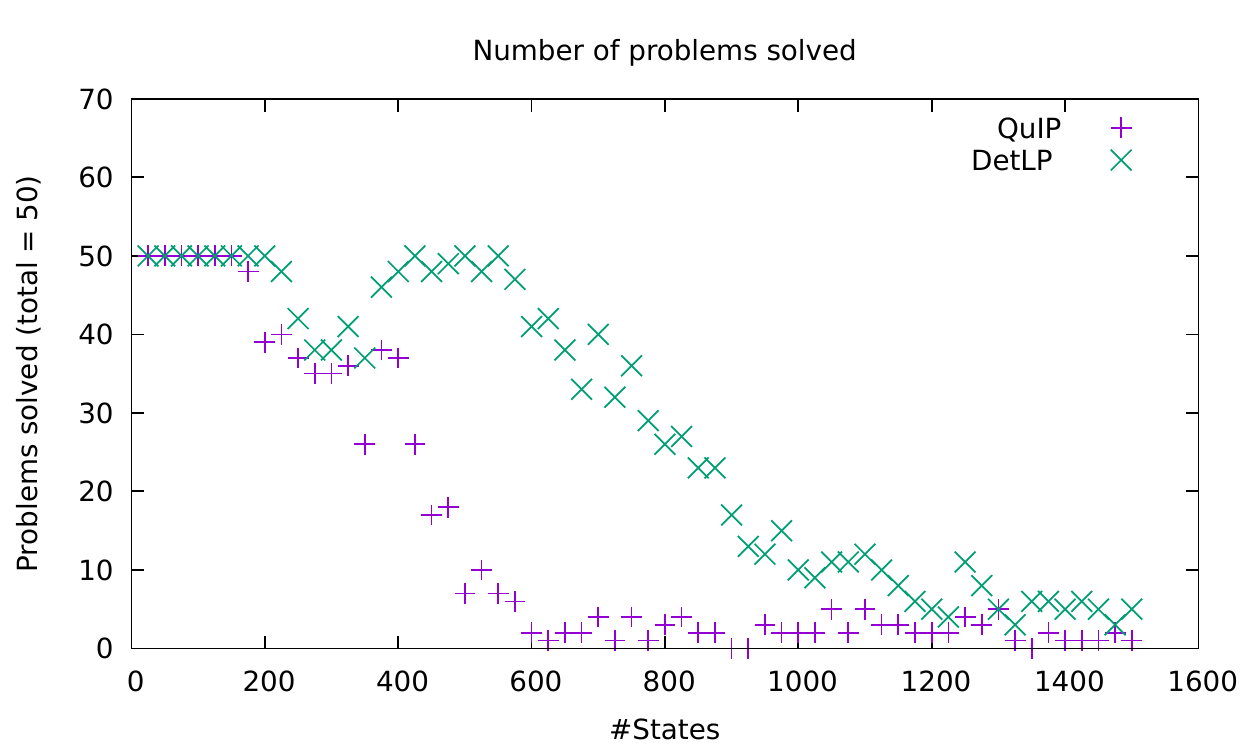}
\subcaption{}
\label{Fig:NumberR40}
\end{minipage}
\caption[Number of benchmarks solved by \cct{QuIP} and \cct{DetLP}]{Number of benchmarks solved by \cct{QuIP} and \cct{DetLP} out of 50 as $s_Q$  increases with $ s_P = 10$, $ \mu=4 $. $\delta = 2.5$ and $\delta = 4$ in Fig~\ref{Fig:NumberR25} and Fig~\ref{Fig:NumberR40}, respectively.}
\end{figure*}

We first compare the tools based on the number of benchmarks each can solve. We also attempt to unravel the main cause of failure of each tool.
Out of the 50 experiments for each parameter-value,  \DetLPt consistently solves more benchmarks than \cct{QuIP} for the same parameter-values (Fig.~\ref{Fig:NumberR25}-\ref{Fig:NumberR40})\footnote{Figures are best viewed online and in color}. 
The figures also reveal that both tools solve more benchmarks at lower transition-density. 
The most common, in fact almost always, reason for \cct{QuIP} to fail before its timeout was reported to be memory-overflow inside \cct{RABIT} during language-inclusion between $ \hat{P}_{-wt}$ and $\first$. 
On the other hand, the main cause of failure of \DetLPt was reported to be memory overflow during \DS-determinization and preprocessing of the determinized \DS-automata before \glpsol is invoked. This occurs due to the sheer size of the determinized \DS-automata, which can very quickly become very large. 
These empirical observations indicate that the bottleneck in \cct{QuIP} and \DetLPt may be language-inclusion and explicit \DS-determinization, respectively. 

\begin{figure*}[t]
\centering
\begin{minipage}{0.8\textwidth}
\includegraphics[width=\textwidth]{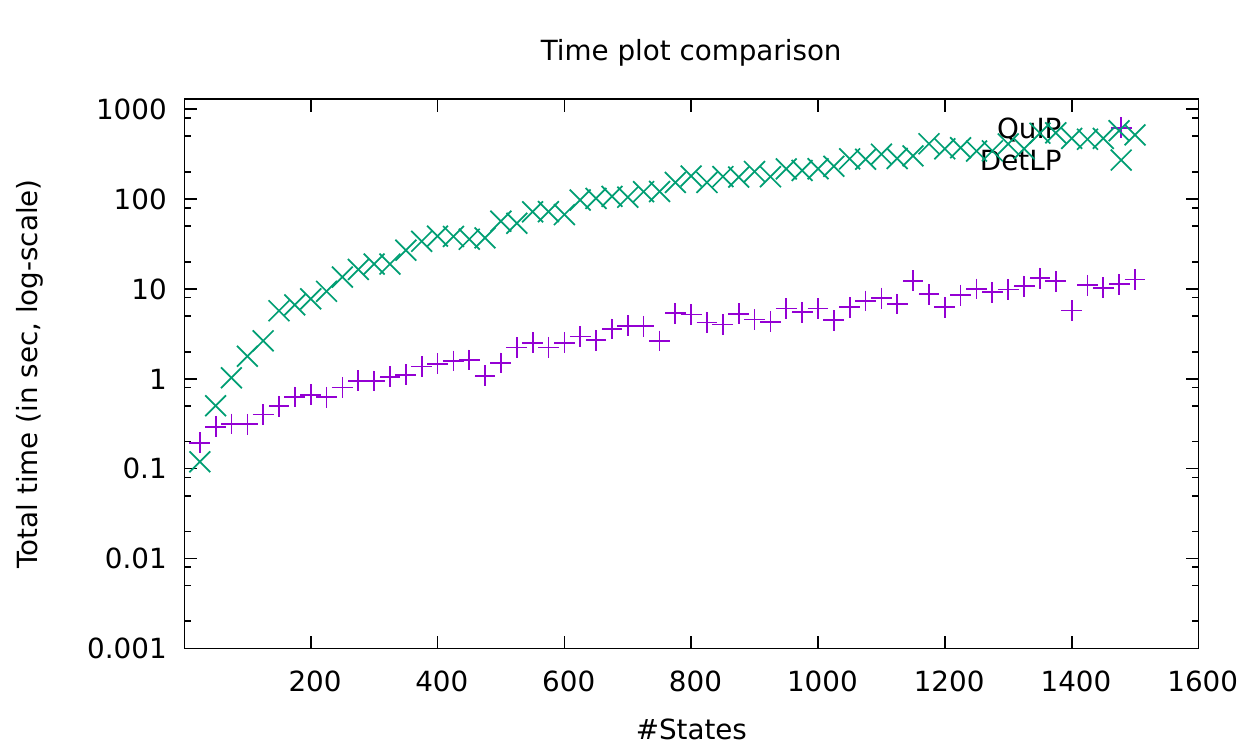}
\subcaption{}
\label{Fig:TimeR25}
\end{minipage}
\hfill
\begin{minipage}{0.8\textwidth}
\includegraphics[width=\textwidth]{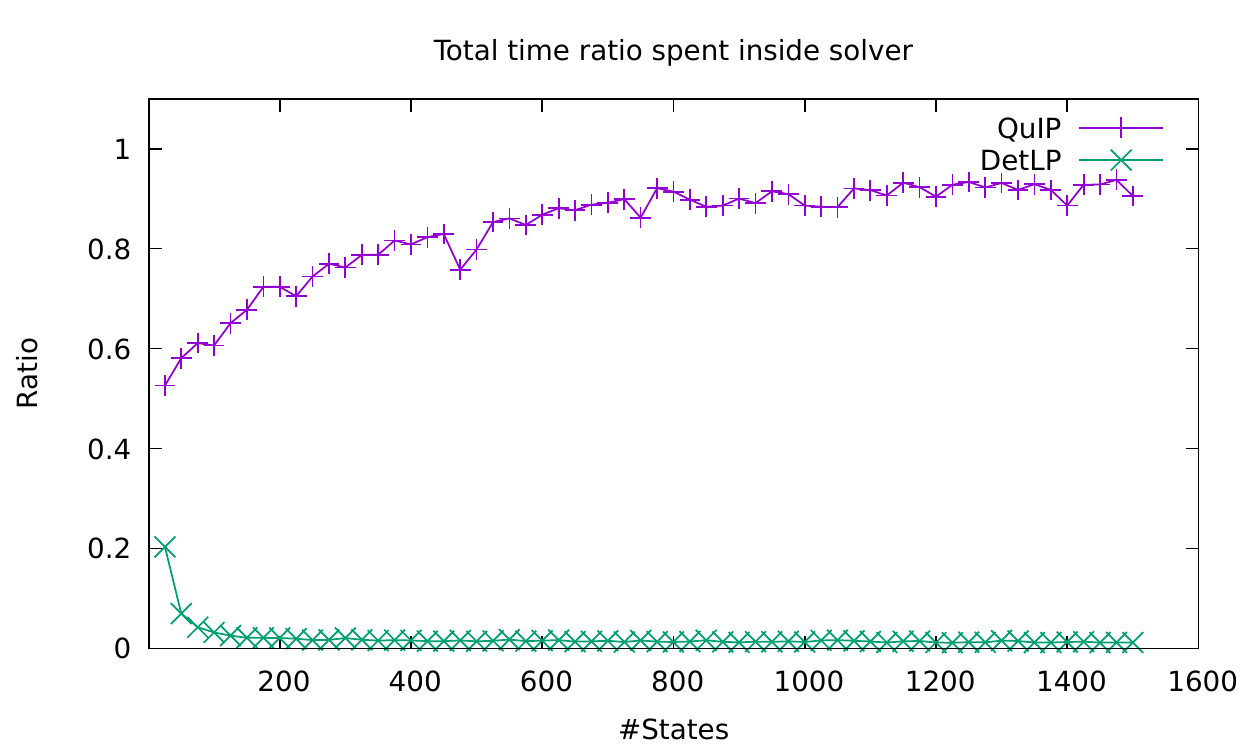}
\subcaption{}
\label{Fig:ToolTimeRatio}
\end{minipage}
\caption[Runtime performance trends of \cct{QuIP} and \cct{DetLP}]{Runtime performance trends of \cct{QuIP} and \cct{DetLP}: Fig~\ref{Fig:TimeR25} plots total runtime as $s_Q$ increases $ s_P = 10$,$ \mu=4 $, $\delta = 2.5$. 
Figure shows median-time for each parameter-value. 
Fig~\ref{Fig:ToolTimeRatio} plots the ratio of time spent by tool inside its solver at the same parameter values.}
\end{figure*}

We investigate the above intuition by analyzing the runtime trends for both tools. 
 Fig.~\ref{Fig:TimeR25} plots the runtime for both tools.  
The plot shows that \cct{QuIP} fares significantly better than \DetLPt in runtime at $\delta = 2.5$. 
The plots for both the tools on logscale seem curved (Fig.~\ref{Fig:TimeR25}), suggesting a sub-exponential runtime complexity.
These were observed at higher $\delta$ as well. 
However, at higher $\delta$ we observe very few outliers on the runtime-trend graphs of \cct{QuIP} at larger inputs when just a few more than 50\% of the runs are successful. This is expected since effectively, the median reports the runtime of the slower runs in these cases. 
Fig~\ref{Fig:ToolTimeRatio} records the ratio of total time spent inside \cct{RABIT} and \glpsol. The plot reveals that \cct{QuIP} spends most of its time inside \cct{RABIT}. We also observe that most memory consumptions in \cct{QuIP} occurs inside \cct{RABIT}. In contrast, \glpsol consumes a negligible amount of time and memory in \cct{DetLP}.
Clearly, performance of \cct{QuIP} and \cct{DetLP} is dominated by \cct{RABIT} and explicit \DS-determinization, respectively. 

We also determined how runtime performance of tools changes with increasing discount-factor $d$. Both tools consume lesser time as $d$ increases. 

\begin{figure*}[t]
\centering
\begin{minipage}{0.8\textwidth}
\includegraphics[width=\textwidth]{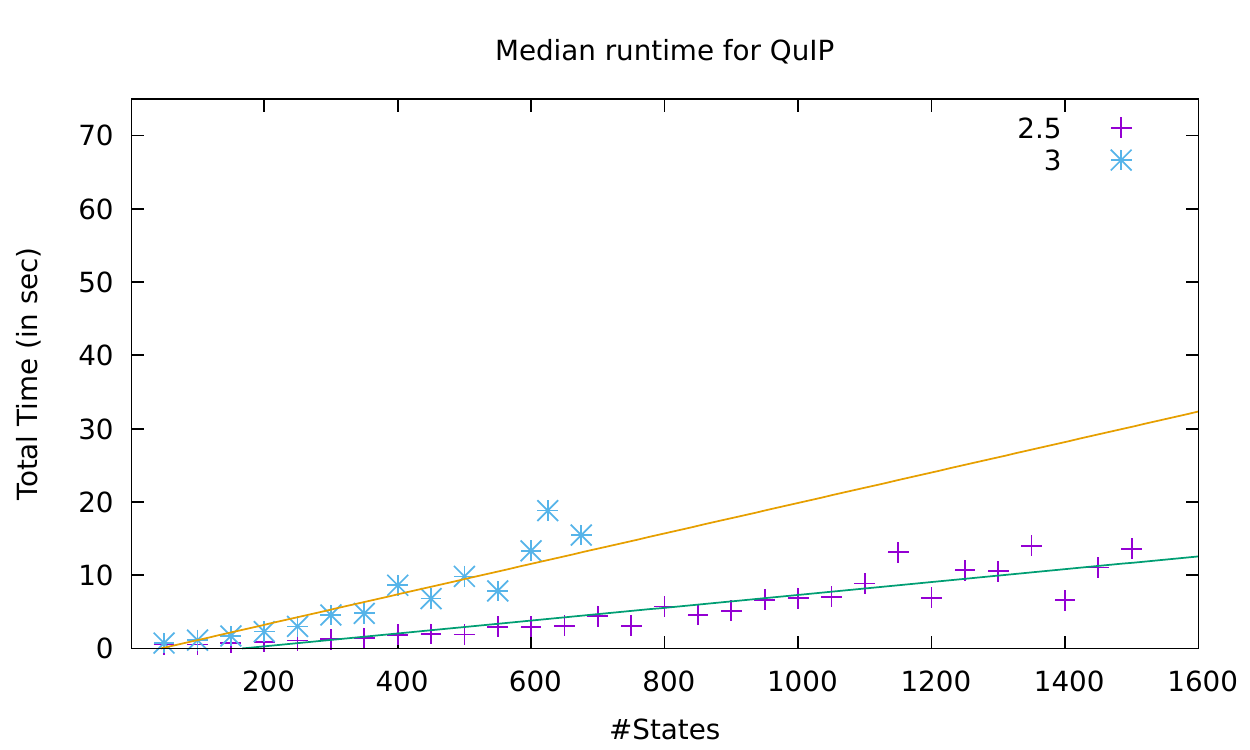}
\subcaption{}
\label{Fig:quip}
\end{minipage}
\hfill
\begin{minipage}{0.8\textwidth}
\includegraphics[width=\textwidth]{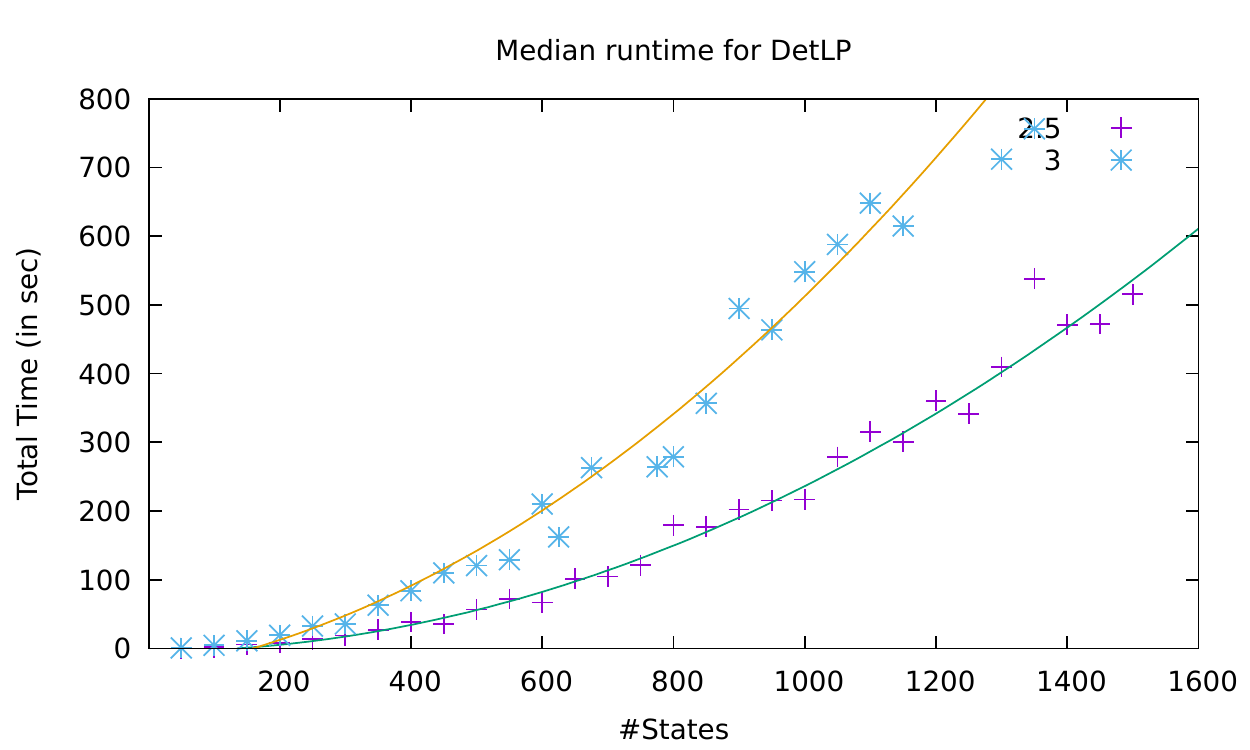}
\subcaption{}
\label{Fig:detlp}
\end{minipage}
\caption[Scalability of \cct{QuIP} and \cct{DetLP}]{Scalability of \cct{QuIP} (Fig~\ref{Fig:quip}) and \cct{DetLP} (Fig~\ref{Fig:detlp}) at $\delta = 2.5,3$. Figures show median-time for each parameter-value.}
\end{figure*}

Finally, we test for scalability of both tools. 
In Fig.~\ref{Fig:quip}, we plot the median of total runtime as $s_Q$ increases at $\delta = 2.5, 3$ $(s_P = 10, \mu=4)$ for \cct{QuIP}. 
We attempt to best-fit the data-points for each $\delta $ with functions that are linear, quadratic and cubic in $s_Q$ using squares of residuals method. 
Fig~\ref{Fig:detlp} does the same for \cct{DetLP}.
We observe that \cct{QuIP} and \DetLPt are best fit by functions that are linear and quadratic in $s_Q$, respectively. 

\subsubsection{Inferences and discussion}

Our empirical analysis arrives at conclusions that a purely theoretical exploration would not have. First of all, we observe that despite having a the worse theoretical complexity, the median-time complexity of \cct{QuIP} is better than \DetLPt by an order of $n$. In theory, \cct{QuIP} scales exponentially in $s_Q$, but only linearly in $s_Q$ in runtime. Similarly, runtime  of \cct{DetLP} scales quadratically in $s_Q$. The huge margin of complexity difference emphasizes why solely theoretical analysis of algorithms is not sufficient. 

Earlier empirical analysis of \LI-solvers had made us aware of their dependence on transition-density $\delta$. As a result, we were able to design \cct{QuIP} cognizant of parameter $\delta$. Therefore, its runtime dependence on $\delta$ is not surprising. 
However, our empirical analysis reveals runtime dependence of \DetLPt on $\delta$. 
This is unexpected since $\delta$ does not appear in any complexity-theoretic analysis of \DetLPt (Theorem~\ref{thrm:DetComplexity}). 
We suspect this behavior occurs because the creation of each  transition, say on alphabet $a$, during \DS-determinization requires the procedure to analyze every transition on alphabet $a$ in the original \DS-automata. 
Higher the transition-density, more the transitions in the original \DS-automata, hence more expensive is the creation of transitions during \DS-determinization.

We have already noted that the performance of \cct{QuIP} is dominated by \cct{RABIT} in space and time. Currently, \cct{RABIT} is implemented in \cct{Java}. Although \cct{RABIT} surpasses all other \LI-solvers in overall performance, we believe it can be improved significantly via a more space-efficient implementation in a  more performance-oriented language like \cplus. This would, in-turn, enhance \cct{QuIP}.

The current implementation of \DetLPt utilizes the vanilla algorithm for \DS-determinization. 
Since \DS-determinization dominates \cct{DetLP}, there is certainly merit in designing efficient algorithms for \DS-determinization. 
However, we suspect this will be of limited advantage to \DetLPt since it will persist to incur the complete cost of explicit \DS-determinization due to the separation of automata-theoretic and numeric reasoning.  

Based on our observations, we propose to extract the complementary strengths of both tools: First, apply \cct{QuIP} with a small timeout; Since \DetLPt solves more benchmarks, apply \DetLPt only if \cct{QuIP} fails.

\section{Chapter summary}
\label{Sec:Result}

This chapter presents the first empirical evaluation of algorithms and tools for DS inclusion. We present two tools \cct{DetLP} and \cct{QuIP}. Our first tool \DetLPt is based on explicit \DS-determinization and linear programming, and renders an exponential time and space algorithm. Our second tool \cct{QuIP} improves upon a previously known comparator-based automata-theoretic algorithm \cct{BCV} by means of an optimized comparator construction. Despite its \cct{PSPACE}-complete theoretical complexity, we note that all practical implementations of \cct{QuIP} are also exponential in time and space.

The focus of this work is to investigate these tools in practice. In theory, the exponential complexity of \cct{QuIP} is worse than \cct{DetLP}. Our empirical evaluation reveals the opposite: The median-time complexity of \cct{QuIP} is better than \DetLPt by an order of $n$. Specifically, \cct{QuIP} scales linearly while \cct{DetLP} scales quadratically in the size of inputs. This re-asserts the gap between theory and practice, and aserts the need of better metrics for practical algorithms. 
Further emprirical analysis by by scaling the right-hand side automaton will be beneficial. 

Nevertheless, \DetLPt consistently solves more benchmarks than \cct{QuIP}. 
Most of \cct{QuIP}'s experiments fail due to memory-overflow within the \LI-solver, indicating that more space-efficient implementations of \LI-solvers would boost \cct{QuIP}'s performance. We are less optimistic about \DetLPt though. Our evaluation highlights the impediment of explicit \DS-determinization, a cost that is unavoidable in \cct{DetLP}'s separation-of-techniques approach. This motivates future research that integrates automata-theoretic and numerical reasoning by perhaps combining implicit \DS-determinzation with baseline automata-like reasoning to design an on-the-fly algorithm for \DS-inclusion. 

Last but not the least, our empirical evaluations lead to discovering dependence of runtime of algorithms on parameters that had not featured in their worst-case theoretical analysis, such as the dependence of \DetLPt on transition-density. 
Such evaluations build deeper understanding of algorithms, and will hopefully serve a guiding light for theoretical and empirical investigation in-tandem of algorithms for quantitative analysis

\chapter{Safety/co-safety comparators for DS inclusion}
\label{ch:safetycosafety}

Despite being an integrated method, the comparator-based algorithm for DS inclusion with integer discount factors does not outperform separation-of-techniques on all accounts. In particular, despite exhibiting improved runtime they did not manage to solve as many benchmarks as  separation-of-techniques. This chapter aims at improving the comparator-based method for DS inclusion by diving deeper into language-theoretic properties of DS comparison languages and DS comparators. To this end, we explore safety and co-safety properties of comparison languages, and demonstrate comparator-based solutions are able to mitigate their weaknesses by leveraging the benefits of safety/co-safety languages/automata.

\section{The predicament of B\"uchi complementation}


In theory, DS inclusion for integer discount factors is \cct{PSPACE}-complete~\cite{BCVFoSSaCS18}. 
Recent algorithmic approaches have tapped into language-theoretic properties of discounted-sum aggregate function~\cite{BCVFoSSaCS18,chaudhuri2013regular} to design practical algorithms for DS inclusion~\cite{BCVCAV18,BCVFoSSaCS18}.
These algorithms use {\em DS comparator automata} ({\em DS comparator}, in short) as their main technique, and are {\em purely} automata-theoretic. While these algorithms outperform other existing approaches for DS inclusion in runtime~\cite{boker2014exact,chatterjee2010quantitative},  even these do not scale well on weighted-automata with more than few hundreds of states~\cite{BCVCAV18}.

An in-depth examination of the DS comparator based algorithm exposes their scalability bottleneck. 
DS comparator is a B\"uchi  automaton that relates the discounted-sum aggregate of two (bounded) weight-sequences $A$ and $B$ by determining the membership of the interleaved pair of sequences $(A,B)$ in the language of the comparator. As a result, DS comparators reduce DS inclusion to language inclusion between (non-deterministic) B\"uchi automaton. In spite of the fact that many techniques have been proposed to solve B\"uchi language inclusion efficiently in practice~\cite{abdulla2011advanced,doyen2010antichain}, none of them can avoid at least an exponential  blow-up of $2^{\O(n\log n)}$, for an $n$-sized input, caused by a direct or indirect involvement of 
B\"uchi complementation~\cite{safra1988complexity,vardi2007buchi}.

This work meets the scalability challenge of DS inclusion by eradicating the dependence on B\"uchi language inclusion and B\"uchi complementation. 
We discover that DS comparators  can be expressed by specialized B\"uchi automata called {\em safety automata} or {\em co-safety automata}~\cite{kupferman1999model} (\textsection~\ref{Sec:SafeCoSafe}).
Safety and co-safety automata have the property that their complementation is performed by simpler and lower $2^{\O(n)}$-complexity subset-construction methods~\cite{kupferman2001weak}. 
As a result, they facilitate a procedure for DS inclusion that uses subset-construction based intermediate steps instead of  B\"uchi complementation, yielding an improvement in theoretical complexity from $2^{\O(n\cdot \log n)}$ to $2^{\O(n)}$. Our subset-construction based procedure has yet another advantage over B\"uchi complementation as they support efficient on-the-fly implementations, yielding practical scalability as well (\textsection~\ref{Sec:ExpEval}).  

An empirical evaluation of our prototype tool $\cct{QuIPFly}$ for the proposed procedure against the prior DS-comparator algorithm and other existing approaches for DS inclusion shows that \cct{QuIPFly} outperforms them by orders of magnitude both in runtime and the number of benchmarks solved  (\textsection~\ref{Sec:ExpEval}).

Therefore, this chapter meets the scalability challenge of DS inclusion by delving deeper into language-theoretic properties of discounted-sum aggregate functions. By doing so, we obtain algorithms for DS inclusion that render both tighter theoretical complexity and improved scalability in practice. Hence, we can conclude that comparator-based methods are effective in practice as well. 

\noindent
\paragraph{Organization}
This chapter goes as follows. Section~\ref{Sec:PrelimsSC} reviews classes of B\"uchi automata for which complementation can be performed without using Safra's complementation method~\cite{safra1988complexity}. Section~\ref{Sec:SafetyCosafetyLanguages} introduces safety/co-safety comparison languages and correspondingly their safety/co-safety comparator automata. We prove that quantitative inclusion performed with safety/co-safety comparators is more efficient than quantitative inclusion with $\omega$-regular comparators. Section~\ref{Sec:DSInclusionFull} studies DS aggregation in depth, and shows that DS comparion languages are either safety/co-safety languages. We use this newly discovered property to design much compact and deterministic constructions for DS comparators, and finally utilize these comparators to re-design DS inclusion. Finally, Section~\ref{Sec:ExpEval} conducts the empirical evaluation, and affirms the superiority of comparator-based methods for DS inclusion with integer discount factors.

\section{Safraless automata}
\label{Sec:PrelimsSC}
We coin the terms {\em Safraless automata} to refer to B\"uchi automata for which complementation can be performed without Safra's B\"uchi complementation~\cite{safra1988complexity}.
We review two classes of Safraless automata~\cite{kupferman1999model}: (a). Safety and co-safety automata, and (b). Weak B\"uchi automata. 
For both of these classes of Safraless automata, complementation can be performed using simpler methods that use subset-construction. As a result, these automata incur a $2^{\O(n)}$ blow-up as opposed to a $2^{\O(n\log n)}$ blow-up, where n is the number of states in the original automaton. In addition, subset-construction based methods are more amenable to efficient practical implementations since they can be performed on-the-fly.

\subsection{Safety and co-safety automata}

Let $ \L \subseteq \Sigma^\omega$ be a language  over alphabet $\Sigma$. 
A finite word $w \in \Sigma^*$ is a {\em bad prefix} for $\L$  if for all infinite words $y \in \Sigma^\omega$, $x\cdot y \notin \L$.
A language $\L$ is a {\em safety language} if every word $ w \notin \L$ has a bad prefix for $\L$. 
A language $\L$ is a {\em co-safety language} if its complement language  is a safety language~\cite{alpern1987recognizing}.
When a safety or co-safety language is an $\omega$-regular language, the B\"uchi automaton representing it is called a safety or co-safety automaton, respectively~\cite{kupferman1999model}. 
Wlog, safety and co-safety automaton contain a {\em sink state} from which every outgoing transitions loops back to the sink state and there is a transition on every alphabet symbol. 
All states except the sink state are accepting in a safety automaton, while only the sink state is accepting in a co-safety automaton.
The complementation of safety automaton is a co-safety automaton, and vice-versa.  
Safety automata are closed under intersection, and co-safety automata are closed under union.  

\subsection{Weak B\"uchi automaton}

A B\"uchi automaton $\A=(\State, \Sigma, \delta$, $\Start, \Final)$ is {\em weak}~\cite{kupferman2001weak} if its states $\State$ can be partitioned into disjoint sets $\State_i$, such that (a). For each set $\State_i$, either $\State_i \subseteq \Final$ or $\State_i \cap \Final = \emptyset$ i.e.  partitions are either accepting or rejecting, and (b). There exists a partial order $\leq$ on the collection of the $\State_i$ such that for every $s \in \State_i$ and $t \in \State_j$ for which $t \in \delta(s, a)$ for some $a \in \Sigma$, $\State_j \leq \State_i$ i.e. The transition function is restricted so that in each transition, the automaton either stays at the same set or moves to a set smaller in the partial order. 

It is worth noting that safety and co-safety automata are also weak B\"uchi automata. In both of these cases, the states of the regular safety/co-safety automata are partitioned into two disjoint sets: The first set contains all but the sink state, and the second set contains the sink state only. Note how in both cases, the sets have either all accepting states or all non-accepting states, and the transitions are unidirectional order w.r.t to the partitions. 

We make the observation that the intersection of a safety and co-safety automaton is indeed a weak B\"uchi automata. We prove it as follows:
\begin{theorem}
	\label{Lem:IntersectSafetyAndCo}
	The intersection of a safety automata and a co-safety automata is a weak B\"uchi automata. 
\end{theorem}
\begin{proof}
	Wlog, 
	let the two partitions in the co-safety automaton be $\State_C$ and  $\{\sink_C\}$, and the partitions in the safety automaton be  $\State_S$ and $\{\sink_S\}$. Their accepting partitions are $\{\sink_C\}$ and $\State_S$, respectively. 
	
	We construct the intersection by taking simple product of both automata without the cyclic counter.
	The product automaton is a weak B\"uchi automaton, with four partitions obtained from all combinations of partitions of the  co-safety and safety automata.
	Transitions from partition $(\State_C$,$\State_S)$ could go to all four partitions. This partition also contains the initial state. 
	Transitions from partitions ($\State_C$,$\{\sink_S\})$ or 
	$ (\{\sink_C\}$,$\State_S)$ either go to themselves or to the sink partition $(\{\sink_C\},\{\sink_S\})$.
	There are no transitions between ($\State_C$,$\{\sink_S\})$ and
	$ (\{\sink_C\}$,$\State_S)$.
	
	Based on the transition relation, we assign the order of partitions as follows: 
	\begin{itemize}
		\item Partition $(\State_C$,$\State_S)$ has highest order
		
		\item Parition $(\{\sink_C\},\{\sink_S\})$ has lowest order
		
		\item Other two partitions take any order in between, since there are no transitions in between these paritions. 
	\end{itemize}
	Clearly, the partition ($\State_C$,$\{\sink_S\})$ is the only accepting partition.	
	\qed
\end{proof}

\section{Safety and co-safety comparison languages}
\label{Sec:SafetyCosafetyLanguages}
We introduce comparator automata that are safety and co-safety automata Section~\ref{Sec:DefinitionSafetyCoSafety}.   
Section~\ref{Sec:Inclusion} designs an algorithm for quantitative inclusion with regular safety/co-safety comparators. We observe that the complexity of quantitative inclusion with safety/co-safety comparator automata is complexity is lower than that with $\omega$-regular comparators. 

\subsection{Safety and co-safety comparison languages and comparators}
\label{Sec:DefinitionSafetyCoSafety}

We begin with formal definitions of safety/co-safety comparison languages and safety/co-safety comparators:

\begin{Def}[Safety and co-safety comparison languages]
	Let $\Sigma$ be a finite set of integers, $f:\Z^{\omega}\rightarrow \mathbb{R}$ be an aggregate function, and $\mathsf{R}  \in \{\leq, <, \geq, >, =, \neq\}$ be a relation. 
	A comparison language $L$ over $\Sigma\times\Sigma$ for aggregate function $f$ and relation $\mathsf{R} $ is said to be a safety comparison language (or a co-safety comparison language) if $L$ is a safety language (or a co-safety language).
	\end{Def}
	
	\begin{Def}[Safety and co-safety comparators]
		Let $\Sigma$ be a {finite set} of integers,  $f:\Z^{\omega}\rightarrow \mathbb{R}$ be an aggregate function, and $\mathsf{R}  \in \{\leq, <, \geq, >, =, \neq\}$ be a relation. 
		A comparator for aggregate function $f$ and relation $\mathsf{R}$ is a {\em safety comparator} (or {\em co-safety comparator}) is the comparison language for $f$ and $\mathsf{R}$ is a safety language (or co-safety language).
		
	\end{Def}
	
	A safety comparator is {\em regular} if its language is  $\omega$-regular (equivalently, if its automaton is a safety automaton). 
	Likewise, a co-safety comparator is {\em regular} if its language is  $\omega$-regular (equivalently, automaton is a co-safety automaton). 
		
	Just like the closure property of $\omega$-regular comparison languages, safety and co-safety comparison languages exhibit closure under safety or co-safety languages as well. 	
	By complementation duality of safety and co-safety languages,  comparison language for an aggregate function $f$ for non-strict inequality $\leq$ is safety iff the comparison language for $f$ for strict inequality $<$ is co-safety. 
	Since safety languages and safety automata are closed under intersection,
	safety comparison languages and regular safety comparator for non-strict inequality renders the same for equality.
	Similarly, since co-safety languages and co-safety automata are closed under union,  co-safety comparison languages and regular co-safety comparators for non-strict inequality render the same for the inequality relation. 
	As a result, if the comparison language for any one relation is either a safety or co-safety language, the comparison languages for all relations will also be safety or co-safety languages. Therefore, it suffices to examine the comparison language for one relation only. 

\subsection{Quantitative inclusion with regular safety/co-safety comparators}
\label{Sec:Inclusion}

This section covers the first technical contributions of this work. It studies safety and co-safety properties of comparison languages and comparators, and utilizes them to obtain tighter theoretical upper-bound for quantitative inclusion with these comparators that exhibit safety/co-safety and $\omega$-regular properties as opposed to comparators that are only $\omega$-regular.


\subsubsection{Key Ideas}
A run of word $w$ in a weighted-automaton is {\em maximal} if its weight is the supremum weight of all runs of $w$ in the weighted-automaton. 
 A run $\rho_P$ of $w$ in $P$ is a {\em counterexample} for $P\subseteq Q$ (or $ P \subset Q$) iff 
 there exists a maximal run $\mathit{sup}_Q$ of $w$ in $Q$ such that $\wt(\rho_P) > wt(\mathit{sup}_Q) $ (or $\wt(\rho_P) \geq wt(\mathit{sup}_Q) $). 
 Consequently, $P\subseteq Q$ (or  $P\subset Q$) iff there are no counterexample runs in $P$. 
 Therefore, the roadmap to solve quantitative inclusion for regular safety/co-safety comparators is as follows:
 \begin{enumerate}
     \item Use regular safety/co-safety comparators to construct the {\em maximal automaton} of $Q$ i.e. an automaton that accepts all maximal runs of $Q$ (Corollary~\ref{Coro:SizeMaximal}).
     \item Use the regular safety/co-safety comparator and the maximal automaton  to construct a {\em counterexample automaton} that accepts all counterexample runs of the inclusion problem $P \subseteq Q$ (or $P\subset Q$) (Lemma~\ref{Lemma:counterexample}). 
     \item Solve quantitative inclusion for safety/co-safety comparator by checking for emptiness of the counterexample (Theorem~\ref{thrm:quantI}). 
    
 \end{enumerate}
 
Let $W$ be a weighted automaton. Then the {\em annotated automaton} of $W$, denoted by $\hat{W}$, is the B\"uchi automaton obtained by transforming transition $s\xrightarrow{a}t $ with weight $v$ in $W$ to transition $s\xrightarrow{a, v} t$ in $\hat{W}$. Observe that $\hat{W}$ is a safety automaton since all its states are accepting. 
 A run on word $w$ with weight sequence $wt$ in $ W$ corresponds to an {\em annotated word} $(w, wt)$ in $\hat{W}$, and vice-versa.
 
 \subsubsection{Maximal automaton}
  \label{Sec:Max}
  
  This section covers the construction of the {\em maximal automaton} from a weighted automaton.
  Let $W$ and $\hat{W}$ be a weighted automaton and its annotated automaton, respectively. 
  We call an annotated word $(w, wt_1)$ in $\hat{W}$  {\em maximal} if for all other words of the form $(w, wt_2)$ in $\hat{W}$, $\wt(wt_1)\geq \wt(wt_2)$. Clearly, $(w,wt_1)$ is a maximal word in $\hat{W}$ iff word $w$ has a run with weight sequence $wt_1$ in $W$ that is maximal. 
 We define  {\em maximal automaton} of weighted automaton $W$, denoted $\maximal{W}$, to be the automaton that accepts all maximal words of its annotated automata $\hat{W}$. 
  
 
  We show that when the comparator is regular safety/co-safety, the construction of the maximal automata incurs a $2^{\O(n)}$ blow-up. 
  This section exposes the construction for maximal automaton when comparator for non-strict inequality is regular safety.  The other case when the comparator for strict inequality is regular co-safety is similar, hence skipped.

 \begin{lemma}
 \label{Lem:MaximalSafety}
  Let $W$ be a weighted automaton with regular safety comparator for non-strict inequality. 
Then the language of $\maximal{W}$ is a safety language. 
 \end{lemma}
\begin{proof}

Intuitively, an annotated word $(w, wt_1)$ is not maximal in $\hat{W}$ for one of the following two reasons: Either $(w, wt_1)$ is not a word in $\hat{W}$, or there exists another word $(w, wt_2)$ in $\hat{W}$ s.t. $\wt(wt_1) < \wt(wt_2)$ (equivalently $(wt_1, wt_2)$ is not in the comparator  non-strict inequality). Both $\hat{W}$ and comparator for non-strict inequality are safety languages, so the language of maximal words must also be a safety language. 

In detail, 
we show that every annotated word that is not a maximal word in $\hat{W}$ must have a bad-prefix. 
An annotated word $(w, wt_1)$ is not maximal in $\hat{W}$ for one of two reasons: Either $(w, wt_1)$ is not a word in $\hat{W}$, or there exists another word $(w, wt_2)$ in $\hat{W}$ s.t. $\wt(wt_1) < \wt(wt_2)$.

If $(w, wt_1)$ is not a word in $\hat{W}$, since $\hat{W}$ is a safety automata as well, $(w, wt_1)$ must have a bad-prefix w.r.t $\hat{W}$. The same bad-prefix serves as a bad-prefix w.r.t. the maximal language since none of its extensions are in $\hat{W}$ either. 

Otherwise since the comparator for $<$ is $\omega$-regular co-safety, there exists an $n$-length prefix of $(wt_1, wt_2)$ s.t. all extensions of $(wt_1, wt_2)[n]$ are present in the $\omega$-regular co-safety automaton.
We claim the  prefix of $(w,wt_1)$ of same length $n$ is a bad-prefix for the  maximal language. 
Consider the $n$-length prefix $(w,wt_1)[n]$ of $(w,wt_1)$. 
Let $(w_\mathit{ext}, wt_\mathit{ext})$ be an infinite extension of $(w,wt)[n]$. 
If $(w_\mathit{ext}, wt_\mathit{ext})$ is not a word in $\hat{W}$, it is also not a maximal word in $\hat{W}$.

If $(w_\mathit{ext}, wt_\mathit{ext})$ is a word in $\hat{W}$,  
Let $(w_{ext}, wt_{ext}')$ be and extensions of $(w, wt_2)[n]$. Note that the first component of extension of $(w,wt_1)[n]$ and $(w, wt_2)[n]$ are the same. This is possible since the underlying B\"uchi automata of $W$ is a complete automaton. Since $(wt_{ext}, wt_{ext}')$ is an extension of $(wt_1, wt_2)[n]$, $\wt(wt_{ext}) < \wt(wt_{ext}')$. Therefore, $(w_\mathit{ext}, wt_\mathit{ext})$ is not a maximal word.
\qed
 \end{proof}
 
 We now proceed to construct the safety automata for $\maximal{W}$
 
 \paragraph{Intuition}
 The intuition behind the construction of maximal automaton follows directly from the definition of maximal words. 
 Let $\hat{W}$ be the annotated automaton for weighted automaton $W$. Let $\hat{\Sigma}$ denote the alphabet of $\hat{W}$. Then an annotated word $(w,wt_1) \in \hat{\Sigma}^{\omega}$ is a word in $\maximal{W}$ if (a) $(w,wt_1) \in \hat{W}$, and (b) For all words $(w,wt_2) \in \hat{W}$, $\wt(wt_1) \geq \wt(wt_2)$.
 
 The challenge here is to construct an automaton for condition (b). Intuitively, this automaton simulates the following action: As the automaton reads word $(w, wt_1)$, it must spawn all words of the form $(w, wt_2)$ in $\hat{W}$, while also ensuring that $\wt(wt_1) \geq \wt(wt_2)$ holds for every word $(w,wt_2)$ in $\hat{W}$. Since $\hat{W}$ is a safety automaton, for a word $(w,wt_1)\in \hat{\Sigma}^{\omega}$, all words of the form $(w,wt_2)\in\hat{W}$ can be traced by subset-construction.  Similarly since the comparator $C$ for non-strict inequality ($\geq$) is a safety automaton, all words of the form $(wt_1, wt_2) \in C$ can be traced by subset-construction as well. The construction needs to carefully align the word $(w, wt_1)$ with the all possible $ (w,wt_2) \in \hat{W}$ and   their respective weight sequences $(wt_1, wt_2)\in C$.

 \paragraph{Construction and detailed proof}
Let $W$ be a weighted automaton, with annotated automaton $\hat{W}$ and $C$ denote its regular safety comparator for non-strict inequality. 
Let $S_W$ denote the set of states of $W$ (and $\hat{W}$) and $S_C$ denote the set of states of $C$.
We define $ \maximal{{W}}  = (\State, \Start, \hat{\Sigma}, \delta, \Final)$ as follows: 
 \begin{itemize}
 \item Set of states $\State$ consists of tuples of the form $(s,X)$, where  $s \in S_W$, and $X = \{(t,c) | t\in S_W, c\in S_C\}$
 \item $\hat{\Sigma}$ is the alphabet of $\hat{W}$
 \item Initial state $\Start  = (s_w, \{(s_w, s_c)\})$, where $s_w$ and $s_c$ are initial states in $\hat{W}$ and $C$, respectively.
 \item Let states $(s,X), (s,X') \in S$ such that $X = \{(t_1,c_1),\dots, (t_n, c_n)\}$ and  $X' = \{(t'_1,c'_1),\dots, (t'_m, c'_m)\}$ . Then $(s,X) \xrightarrow{(a,v)} (s',X') \in \delta$ iff
 
 	\begin{enumerate}
 	\item $s \xrightarrow{(a,v)} s'$ is a transition in $\hat{W}$, and
 	\item $(t'_j,c'_j) \in X'$ if there exists $(t_i, c_i)\in X$, and a weight $v'$  such that $t_i \xrightarrow{a,v'} t'_j$ and $c_i \xrightarrow{v,v'} c'_j$ are transitions in $\hat{W}$ and $C$, respectively. 
 	\end{enumerate}
 \item $(s, \{(t_1,c_1),\dots, (t_n, c_n)\}) \in \Final$ iff $s$ and all $t_i$ are accepting in $\hat{W}$, and all $c_i$ is accepting in $C$.
 \end{itemize}

 \begin{lemma}
 	\label{lem:Maximal}
 	Let $W$ be a weighted automaton with regular safety comparator $C$ for non-strict inequality. 
 	Then the size of $	\maximal{{W}}$ is $|W|\cdot 2^{\O(|W|\cdot|C|)}$.

\end{lemma}
\begin{proof}
First, we prove that the automaton constructed above is a safety automaton. To do so, we prove that all outgoing transitions from an accepting states go into an accepting state. 
A state $(s, \{(t_1,c_1),\dots, (t_n, c_n)\})$ is non-accepting in the automata if one of $s$,$t_i$ or $c_j$ is non-accepting in underlying  automata $\hat{W}$ and the comparator. Since $\hat{W}$ and the comparator automata are safety, all outgoing transitions from a non-accepting state go to non-accepting state in the underlying automata. Therefore, all outgoing transitions from a non-accepting state in $\maximal{W}$ go to non-accepting state in $\maximal{W}$. Therefore, $\maximal{W}$ is a safety automaton.

An analysis on the number of possible states of the form $(s,X)$ is enough to see why the number of states is $|W|\cdot 2^{\OE(|W|\cdot |C|)}$. This is because there are $|W|$ number of possibilities for $s$ and $2^{\O(|W|\cdot |C|)}$ number of possibilities for $X$.
\qed
 \end{proof}
 
 A similar construction proves that the maximal automata of weighted automata $W$ with regular safety comparator $C$ for strict inequality contains $|W|\cdot 2^{\O(|W|\cdot|C|)}$ states. The difference is that in this case the maximal automaton may not be a safety automaton. But it will still be a weak B\"uchi automaton. Therefore, Lemma~\ref{lem:Maximal} generalizes to:
 
 
\begin{corollary}
	\label{Coro:SizeMaximal}
	Let $W$ be a weighted automaton with regular safety/co-safety comparator $C$.
	Then $\maximal{W}$  is either a safety automaton or a weak B\"uchi automaton of size $|W|\cdot 2^{\O(|W|\cdot|C|)} $.
\end{corollary}

\subsubsection{Counterexample automaton}
\label{Sec:QI}
This section covers the construction of the counterexample automaton. 
Given weighted automata $P$ and $Q$, 
an annotated word $(w, wt_P)$ in annotated automata $\hat{P}$ is a {\em counterexample word} of $P\subseteq Q$ (or $ P\subset Q$)  if there exists $(w,wt_Q)$ in $\maximal{Q}$ s.t. $\wt(wt_P) > \wt(wt_Q)$ (or $\wt(wt_P) \geq \wt(wt_Q)$). 
Clearly, annotated word $(w, wt_P)$ is a counterexample word iff there exists a counterexample run of $w$ with weight-sequence $wt_P$ in $P$.

For this section, we abbreviate strict and non-strict to $\strict$ and  $\nstrict$, respectively. 
For $\R \in \{\strict, \nstrict\}$, the {\em counterexample automaton} for $\R$-quantitative inclusion, denoted by $\counter{\R}$, is the automaton that contains all counterexample words of the problem instance. 
We construct the counterexample automaton as follows:
\begin{lemma}
\label{Lemma:counterexample}
Let $P$, $Q$ be weighted-automata with regular safety/co-safety comparators. For $\R \in \{\strict, \nstrict \}$, $\counter{\R}$  is a B\"uchi automaton.
\end{lemma}
\begin{proof}
We construct B\"uchi automaton $\counter{\R}$ for $\R \in \{\strict, \nstrict\}$ that contains the counterexample words of $\R$-quantitative inclusion. Since the comparator are regular safety/co-safety, $\maximal{Q}$ is a B\"uchi automaton (Corollary~\ref{Coro:SizeMaximal}). Construct the product $\hat{P}\times\maximal{Q}$ such that 
transition $(p_1,q_1)\xrightarrow{a,v_1,v_2}(p_1,q_2)$ is in the product iff 
 $p_1 \xrightarrow{a, v_1} p_1$ and $q_1 \xrightarrow{a, v_2} q_2$ are transitions in  $\hat{P}$ and $\maximal{Q}$, respectively. 
 A state $(p, q)$ is accepting if both $p$ and $q$ are accepting in $\hat{P}$ and $\maximal{Q}$. 
 One can show that the product accepts $(w, wt_P, wt_Q)$ iff $(w,wt_P)$ and $(w, wt_Q)$ are words in $\hat{P}$ and $\maximal{Q}$, respectively.
 
If $\R = \strict$, intersect $\hat{P}\times\maximal{Q}$ with comparator for $\geq$. If $\R = \nstrict$, intersect $\hat{P}\times\maximal{Q}$ with comparator for $>$. Since the comparator is a safety or co-safety automaton, the intersection is taken without the cyclic counter. Therefore, $(s_1,t_1)\xrightarrow{a,v_1,v_2}(s_2,t_2)$ is a transition in the intersection iff $s_1\xrightarrow{a,v_1,v_2} s_2$ and $t_1\xrightarrow{v_1,v_2} t_2$ are transitions in the product and the appropriate comparator, respectively. State $(s,t)$ is accepting if both $s$ and $t $ are accepting. 
The intersection will accept $(w, wt_P, wt_Q)$ iff $(w, wt_P)$ is a counterexample of $\R$-quantitative inclusion. 
$\counter{\R}$ is obtained by projecting out the intersection as follows: Transition $ m\xrightarrow{a,v_1,v_2} n $ is transformed to $m\xrightarrow{a,v_1} n $.
\qed
\end{proof}

\subsubsection{Quantitative inclusion }
In this section, we give the final algorithm for quantitative inclusion with regular safety/co-safety comparators, and contrast the worst-case complexity of quantitative inclusion with $\omega$-regular and regular safety/co-safety comparators.

\begin{theorem}
	\label{thrm:quantI}
	Let $P$, $Q$ be weighted-automata with regular safety/co-safety comparators.  Let $C_\leq$ and $C_<$ be the comparators for $\leq$ and $<$, respectively. Then
	\begin{itemize}
		\item Strict quantitative  inclusion $P\subset Q$ is reduced to emptiness checking of a B\"uchi automaton of size $|P||C_\leq||Q|\cdot 2^{\O(|Q|\cdot|C_<|)}$.
		\item Non-strict quantitative  inclusion $P\subseteq Q$ is reduced to emptiness checking  of a B\"uchi automaton of size $|P||C_<||Q|\cdot 2^{\O(|Q|\cdot|C_<|)}$. 
	\end{itemize}	
\end{theorem}

\begin{proof}
Strict and non-strict are abbreviated to $\strict$ and  $\nstrict$, respectively. 
For $\R \in \{\strict, \nstrict\}$,
	$\R$-quantitative inclusion holds iff $\counter{\R}$ is empty.
 Size of $\counter{\R} $ is the product of size of $P$, $\maximal{Q}$ (Corollary~\ref{Coro:SizeMaximal}), and the appropriate comparator as described in Lemma~\ref{Lemma:counterexample}. 
\qed
\end{proof}

In contrast, quantitative inclusion with $\omega$-regular comparators reduces to emptiness of a B\"uchi automaton with $|P|\cdot 2^{\O(|P||Q||C|\cdot \log(|P||Q||C|))}$ states~\cite{BCVFoSSaCS18}. The $2^{\O(n\log n)}$ blow-up is unavoidable due to B\"uchi complementation. Clearly, quantitative inclusion with regular safety/co-safety  has lower worst-case complexity.

\section{DS inclusion with safety/co-safety comparators}
\label{Sec:DSInclusionFull}
This section covers the second technical contributions of this chapter. It uncovers that DS comparison languages are safety/co-safety, and utilizes this information to obtain tighter theoretical upper-bound for DS inclusion when the discount factor is an integer. Unless mentioned otherwise, the discount-factor is an integer. 

In \textsection~\ref{Sec:SafeCoSafe} we prove that  DS comparison languages are either safety or co-safety for all rational discount-factors. Since DS comparison languages are $\omega$-regular for integer discount-factors~\cite{BCVFoSSaCS18}, we obtain that DS  comparators for integer discount-factors form safety or co-safety automata. Next, \textsection~\ref{Sec:DS} makes use of newly obtained safety/co-safety properties of DS comparator to present the first deterministic constructions for DS comparators. These deterministic construction are compact in the sense that they are the present in their minimal  automata form.  
Finally, since DS comparators are regular safety/co-safety, our analysis shows that the complexity of DS inclusion is improved as a consequence of the complexity observed for quantitative inclusion with regular safety/co-safety comparators.

Recall the definitions of DS comparison languages and DS comparators from Definition~\ref{def:newDSlanguage} and Definition~\ref{def:newDScomparator}, respectively.
Here {\em DS comparison language} with upper bound $\mu$, discount-factor $d> 1$, and relation $\mathsf{R}$ is the language that accepts an infinite-length and bounded weight sequence $A$ over the alphabet $\{-\mu, \dots, \mu\}$ iff $\DSum{A}{d}$ $\mathsf{R}$ $0$ holds. Similarly, DS comparator with the same parameters $\mu$, $d> 1$, accepts the DS comparison language with parameters $\mu$, $d$ and $\mathsf{R}$. We adopt these definitions in the Section~\ref{Sec:DSInclusionFull}.
 
A note on notation: Throughout this section, the concatenation of finite sequence $x$ with finite or infinite sequence $y$ is denoted by $x\cdot y$ in the following. 

\subsection{DS comparison languages and their safety/co-safety properties}
\label{Sec:SafeCoSafe}
The central result of this section is that  DS comparison languages are safety or co-safety languages for all (integer and non-integer) discount-factors (Theorem~\ref{Thrm:DSFullS}). 
In particular, since DS comparison languages are $\omega$-regular for integer discount-factors~\cite{BCVFoSSaCS18}, this implies that DS comparators for integer discount-factors form safety or co-safety automata (Corollary~\ref{Coro:DSFull}). 
The argument for safety/co-safety of DS comparison languages depends on the property that the discounted-sum aggregate of all bounded weight-sequences exists for all discount-factors $d>1$~\cite{rudin1964principles}.
\begin{theorem}
\label{Thrm:DSFullS}
Let $\mu>1$ be the upper bound.  
For rational discount-factor $d>1$
\begin{enumerate}
\item {\label{Item:Safety}}  DS-comparison languages are safety languages for relations $\mathsf{R} \in \{\leq, \geq, =\}$
\item {\label{Item:CoSafety}} DS-comparison language are co-safety languages for relations $\mathsf{R} \in \{<, >, \neq\}$.
\end{enumerate}
\end{theorem}
\begin{proof}
	Due to duality of safety/co-safety languages, it is sufficient to show that DS-comparison language with $ \leq$ is a safety language.

	Let us assume that DS-comparison language with $ \leq$  is not a safety language. Let $W$ be a weight-sequence in the complement of DS-comparison language with $ \leq$  such that it does not have a bad prefix. 
	
	Since $W $ is in the complement of DS-comparison language with $ \leq$, $\DSum{W}{d} > 0$. By assumption, every $i$-length prefix $W[i]$ of $W$ can be extended to a bounded weight-sequence $W[i]\cdot Y^i$ such that $\DSum{W[i]\cdot Y^i}{d} \leq 0$. 
	
	Note that  $\DSum{W}{d} = \DSum{\pre{W}{i}}{d} + \frac{1}{d^i} \cdot \DSum{\suf{W}{i}}{d}$, and $\DSum{\pre{W}{i}\cdot Y^i}{d} = \DSum{\pre{W}{i}}{d} + \frac{1}{d^i} \cdot \DSum{Y^i}{d}$. The contribution of tail sequences $\suf{W}{i}$ and $Y^i$ to the discounted-sum of $W$ and $\pre{W}{i}\cdot Y^i$, respectively diminishes exponentially as the value of $i$ increases. In addition, since and $W$ and  $\pre{W}{i}\cdot Y^i$ share a common $i$-length prefix $\pre{W}{i}$, their discounted-sum values must converge to each other.
	The discounted sum of $W$ is fixed and greater than 0, due to convergence there must be a $k\geq  0$ such that $\DSum{\pre{W}{k}\cdot Y^k}{d} > 0$. Contradiction. 
	Therefore, DS-comparison language with $ \leq$  is a safety language. 
	
	The above intuition is formalized as follows:
	Since $\DSum{W}{d} > 0$ and $\DSum{\pre{W}{i}\cdot Y^i}{d} \leq 0$, the difference $\DSum{W}{d} - \DSum{\pre{W}{i}\cdot Y^i}{d} > 0$. 
	
	By expansion of each term, we get $\DSum{W}{d}- \DSum{\pre{W}{i}\cdot Y^i}{d} = \frac{1}{d^i}(\DSum{\suf{W}{i}}{d} - \DSum{Y^i}{d}) \leq  \frac{1}{d^i}\cdot(\mod{\DSum{\suf{W}{i}}{d}}  + \mod{\DSum{Y^i}{d}} )$. Since the maximum value of discounted-sum of sequences bounded by $\mu$ is $\frac{\mu\cdot d}{d-1}$, we also get that $\DSum{W}{d}- \DSum{\pre{W}{i}\cdot Y^i}{d} \leq 2\cdot \frac{1}{d^{i}} \mod{ \frac{\mu\cdot d}{d-1}} $. 
	
	Putting it all together, for all $i\geq 0$ we get
	\[
	0< \DSum{W}{d} - \DSum{W[i]\cdot Y^i}{d} \leq 2\cdot \frac{1}{d^{i}} \mod{\frac{\mu\cdot d}{d-1}} 
	\]
	As $i \rightarrow \infty$, $2\cdot \mod{\frac{1}{d^{i-1}}\cdot \frac{\mu}{d-1}} \rightarrow 0$. So, $\lim_{i \rightarrow \infty} (\DSum{W}{d} - \DSum{W[i]\cdot Y^i}{d}) = 0$. Since $\DSum{W}{d}$ is fixed, $\lim_{i \rightarrow \infty}\DSum{W[i]\cdot Y^i}{d} = \DSum{W}{d}$.
	
	By definition of convergence, there exists an index $k\geq 0$ such that $\DSum{W[k]\cdot Y^k}{d}$ falls within the  $\frac{\mod{\DSum{W}{d}}}{2}$ neighborhood of $\DSum{W}{d}$. Finally since $\DSum{W}{d}> 0$, $\DSum{W[k]\cdot Y^k}{d} > 0$ as well. But this contradicts our assumption that for all $i \geq 0$, $\DSum{W[i]\cdot Y^i}{d} \leq 0$.
	
	Therefore, DS-comparator with $ \leq$  is a safety comparator. 
	\qed 
\end{proof}

Semantically this result implies that for a bounded-weight sequence $C$ and rational discount-factor $d>1$, if $\DSum{C}{d} > 0$ then $C$ must have a finite prefix $C_{\mathsf{pre}}$ such that the discounted-sum of the finite prefix is so large that no infinite extension by bounded weight-sequence $Y$  can reduce the discounted-sum of $C_{\mathsf{pre}}\cdot Y$ with the same discount-factor $d$ to zero or below.

Prior work shows that DS-comparison languages are expressed by B\"uchi automata iff the discount-factor is an integer~\cite{BCVlmcs2019}. Therefore:
\begin{corollary}
\label{Coro:DSFull}
Let $\mu>1$ be the upper bound.  
For {\em integer} discount-factor $d>1$
\begin{enumerate}
\item {\label{Item:Safety}}  DS comparators are regular safety for relations $\mathsf{R} \in \{\leq, \geq, =\}$
\item {\label{Item:CoSafety}} DS comparators are regular  co-safety  for relations $\mathsf{R} \in \{<, >, \neq\}$.
\end{enumerate}
\end{corollary}
\begin{proof}
	Immediate from Theorem~\ref{Thrm:DSFullS}, Theorem~\ref{Thrm:DSNotRegular} and Corollary~\ref{thm:discountedSumComparator}.
	\qed
\end{proof}

Lastly, it is worth mentioning that for the same reason~\cite{BCVlmcs2019} DS comparators for non-integer rational discount-factors do not form safety or co-safety automata.

\subsection{Deterministic DS comparator for integer discount-factor}
\label{Sec:DS}

This section issues deterministic safety/co-safety constructions for DS comparators with integer discount-factors. This is  different from prior works since they supply non-deterministic B\"uchi constructions  only~\cite{BCVCAV18,BCVFoSSaCS18}. An outcome of DS comparators being regular safety/co-safety (Corollary~\ref{Coro:DSFull}) is a proof that DS comparators permit deterministic B\"uchi constructions, since non-deterministic and deterministic safety automata (and co-safety automata) have equal expressiveness~\cite{kupferman1999model}.
Therefore, one way to obtain deterministic B\"uchi construction for DS comparators is to determinize the non-deterministic constructions using standard procedures~\cite{kupferman1999model,safra1988complexity}. However, this will result in exponentially larger deterministic constructions. 
To this end, this section offers  direct deterministic safety/co-safety automata constructions for DS comparator that 
not only avoid an exponential blow-up but also match their non-deterministic counterparts in number of states (Theorem~\ref{Thrm:DBA}).

\subsubsection{Key ideas}
Due to duality and closure properties of safety/co-safety automata, we only present the construction of 
deterministic safety automata for DS comparator with upper bound $\mu$, integer discount-factor $d>1$ and relation $\leq$, denoted by $\A^{\mu,d}_\leq$.
We proceed by obtaining a {\em deterministic finite automaton}, (DFA), denoted by $\baut{\mu}{d}{\leq}$, for the language of  bad-prefixes of $\A^{\mu,d}_\leq$ (Theorem~\ref{Thrm:DFABad}). Trivial modifications to $\baut{\mu}{d}{\leq}$ will furnish the coveted deterministic safety automata for $\A^{\mu,d}_\leq$ (Theorem~\ref{Thrm:DBA}).

\subsubsection{Detailed construction}
We begin with some definitions. Let $W$ be a {\em finite} weight-sequence. 
By abuse of notation, the discounted-sum of finite-sequence $W$ with discount-factor $d$ is defined as $\DSum{W}{d} = \DSum{W\cdot 0^{\omega}}{d}$.
The {\em recoverable-gap} of a finite weight-sequences $W$ with discount factor $d$, denoted $\gap{W}$, is its normalized discounted-sum: If $W = \varepsilon$ (the empty sequence), $\gap{\varepsilon} = 0$, and $\gap{W} = d^{|W|-1}\cdot \DSum{W}{d}$ otherwise~\cite{boker2014exact}. 
Observe that the recoverable-gap has an inductive definition i.e. $\gap{\varepsilon} = 0$, where $\varepsilon$ is the empty weight-sequence, and $\gap{W\cdot v} = d\cdot \gap{W} + v$, where  $v \in \{-\mu,\dots,\mu\}$. 

This observation influences a sketch for $\baut{\mu}{d}{\leq}$.
Suppose all possible values for recoverable-gap of weight sequences forms the set of states. 
Then, the transition relation of the DFA can mimic the inductive definition of recoverable gap i.e. there is a transition from state $s$ to $t$ on alphabet $v \in \{-\mu, \dots, \mu\}$ iff $t = d\cdot s + v$, where $s$ and $v$ are recoverable-gap values of weight-sequences. 
There is one caveat here: There are infinitely many possibilities for the values of recoverable gap. 
We need to limit the recoverable gap values to finitely many values of interest. 
The core aspect of this construction is to identify these values. 

First, we obtain a lower bound on recoverable gap for bad-prefixes of $\A^{\mu,d}_\leq$:
\begin{lemma}
	\label{Lem:GapThresholdS}
	Let $\mu$ and $d>1 $ be the bound and discount-factor, resp.
	Let $\thresh = \frac{\mu}{d-1}$ be the threshold value. 
	Let $W$ be a non-empty, bounded, finite weight-sequence.   
	Weight sequence $W$ is a bad-prefix of $\A^{\mu,d}_\leq$ iff $\gap{W} > \thresh$.


\end{lemma}
\begin{proof}
Let a finite weight-sequence $W$ be a bad-prefix of $\A^{\mu,d}_\leq$.
Then, $\DSum{W\cdot Y}{d} > 0$ for all infinite and bounded weight-sequences $Y$.
Since $\DSum{W\cdot Y}{d}  = \DSum{W}{d} + \frac{1}{d^{|W|}}\cdot \DSum{Y}{d}$, we get  $\inf (\DSum{W}{d} + \frac{1}{d^{|W|}}\cdot \DSum{Y}{d}) > 0 \implies \DSum{W}{d} +  + \frac{1}{d^{|W|}}\cdot \inf(\DSum{Y}{d})>0$ as $W$ is a fixed sequence. Hence $\DSum{W}{d} + \frac{-\thresh }{d^{|W|-1}}> 0\implies \gap{W} - T>0$.
Conversely, for all infinite, bounded, weight-sequence $Y$, $\DSum{W\cdot Y}{d} \cdot d^{|W|-1} = \gap{W} + \frac{1}{d}\cdot\DSum{Y}{d}$. Since $\gap{W}> T$,  $\inf( \DSum{Y}{d}) = -\thresh \cdot d$, we get $\DSum{W\cdot Y}{d}>0$.
\qed
\end{proof}

Since all finite and bounded extensions of bad-prefixes are also bad-prefixes, Lemma~\ref{Lem:GapThresholdS} implies that if the recoverable-gap of a finite sequence is strinctly lower that threshold $\thresh $, then recoverable gap of all of its extensions also exceed $\thresh$. Since recoverable gap exceeding threshold $\thresh$ is the precise condition for bad-prefixes, all states with recoverable gap exceeding $\thresh$ can be merged into a single state. Note, this state forms an accepting sink in $\baut{\mu}{d}{\leq}$.

Next, we attempt to merge very low recoverable gap value into a single state. 
For this purpose, we define {\em very-good prefixes} for $\A^{\mu,d}_\leq$: A finite and bounded weight-sequence $W$ is a {\em very good} prefix for language of $\A^{\mu,d}_\leq$ if
for all infinite, bounded extensions of $W$ by $Y$,  $\DSum{W\cdot Y}{d}\leq 0$. 
A proof similar to Lemma~\ref{Lem:GapThresholdS} proves an upper bound for the recoverable gap of very-good prefixes of $\A^{\mu,d}_\leq$: 
\begin{lemma}
	\label{Lem:GapThresholdVeryGoodS}
		Let $\mu$ and $d>1 $ be the bound and discount-factor, resp.
		Let $\thresh = \frac{\mu}{d-1}$ be the threshold value. 
		Let $W$ be a non-empty, bounded, finite weight-sequence.   
		Weight-sequence $W$ is a very-good prefix of $\A^{\mu,d}_\leq$ iff $\gap{W} \leq -\thresh$.
		
\end{lemma}
\begin{proof}
	Proof is similar to that in Lemma~\ref{Lem:GapThresholdS}.
	\qed
\end{proof}

Clearly, finite extensions of very-good prefixes are also very-good prefixes. 
Further, $\baut{\mu}{d}{\leq}$ must not accept  very-good prefixes.
Thus, by reasoning as earlier we get that all recoverable gap values that are less than or equal to $-\thresh$ can be merged into one non-accepting sink state in $\baut{\mu}{d}{\leq}$. 

Finally, for an integer discount-factor the recoverable gap is an integer. 
Let $\floor{x}$ denote the floor of $x\in \mathbb{R}$ e.g. $\floor{2.3} = 2$, $\floor{-2} = -2$, $\floor{-2.3} = -3$.
Then,

\begin{corollary}
\label{Coro:BoundInteger}
Let $\mu$ be the bound and $ d>1$ an integer discount-factor.
		Let $\thresh = \frac{\mu}{d-1}$ be the threshold. 
		Let $W$ be a non-empty, bounded, finite weight-sequence.  
		\begin{itemize}
		\item $W$ is a bad prefix of $\A^{\mu,d}_\leq$ iff $\gap{W} > \floor{\thresh}$
		\item $W$ is a very-good prefix of $\A^{\mu,d}_\leq$ iff $\gap{W} \leq \floor{-\thresh}$
		\end{itemize} 
\end{corollary}

\begin{proof}
	Simply combining the statements in Lemma~\ref{Lem:GapThresholdS} and Lemma~\ref{Lem:GapThresholdVeryGoodS}, and using the observation that the gap value is always an integer when the discount factor is an integer. 
	\qed
\end{proof}

So, the recoverable gap value is either one of $\{\floor{-\thresh}+1, \dots , \floor{\thresh}\}$, or less than or equal to $\floor{-\thresh}$, or greater than $\floor{\thresh}$.
This curbs the state-space to $\O(\mu)$-many values of interest, as $\thresh = \frac{\mu}{d-1} < \frac{\mu\cdot d}{d-1}$ and $1<\frac{d}{d-1}\leq 2$. 
Lastly, since $\gap{\varepsilon} = 0$, state 0 must be the initial state. 

\paragraph{Construction of $\baut{\mu}{d}{\leq}$}
Let $\mu$ be the upper bound, and $d>1$ be the integer discount-factor. Let $\thresh = \frac{\mu}{d-1}$ be the threshold value. The finite-state automata $\baut{\mu}{d}{\leq} = (\State, \Start, \Sigma, \delta, \Final)$ is defined as follows:
\begin{itemize}
\item States $\State = \{\floor{-\thresh}+1, \dots , \floor{\thresh}\} \cup \{\mathsf{bad}, \mathsf{veryGood}\}$
\item Initial state $\Start = 0$
\item Accepting states $\Final = \{\mathsf{bad}\}$
\item Alphabet $\Sigma = \{-\mu, -\mu+1,\dots, \mu-1, \mu\}$
\item Transition function $\delta\subseteq \State \times \Sigma \rightarrow \State$ where $(s,a,t) \in \delta$ then:
	\begin{enumerate}
	\item \label{Trans:SelfLoop} If $s \in \{\mathsf{bad}, \mathsf{veryGood}\}$, then $t = s$ for all $a \in \Sigma$
	\item If $s\in \{\floor{-\thresh}+1, \dots , \floor{\thresh}\}$, and $a \in \Sigma$
		\begin{enumerate}
		\item \label{Trans:IntState} If $\floor{-\thresh} <  d\cdot s+a \leq \floor{\thresh}$, then $t = d\cdot s+a$
		\item \label{Trans:leq} If $d\cdot s+a > \floor{\thresh}$, then $t = \mathsf{bad}$
		\item \label{Trans:geq} If $d\cdot s+a \leq \floor{-\thresh}$, then $t = \mathsf{veryGood}$
		\end{enumerate}
	
	\end{enumerate}
\end{itemize}
\begin{theorem}
\label{Thrm:DFABad}
Let $\mu$ be the upper bound, $d>1$ be the integer discount-factor. 
  $\baut{\mu}{d}{\leq}$ accepts finite, bounded, weight-sequence iff it is a bad-prefix of $\A^{\mu, d}_\leq$. 
\end{theorem}
\begin{proof}
	First note that the transition relation is deterministic and complete. 
	Therefore, every word has a unique run in $\baut{\mu}{d}{\leq}$.
	Let $\mathsf{last}$ be the last state in the run of finite, bounded weight-sequence $W$ in the DFA. We use induction on the length of $W$ to prove the following:
	\begin{itemize}
		\item $\mathsf{last} \in \{\floor{-\thresh}+1, \dots , \floor{\thresh}\}$ iff $\gap{W} = \mathsf{last}$
		\item $\mathsf{last} = \mathsf{bad}$ iff $\gap{W} > \floor{\thresh} $
		\item $\mathsf{last} = \mathsf{veryGood}$ iff $\gap{W} \leq \floor{-\thresh} $
	\end{itemize}
	Let $W = \varepsilon$ be a word of length 0. By definition of recoverable gap, $\gap{w} = 0$. Also, its run in the DFA $\baut{\mu}{d}{\leq}$ is the single-state sequence 0, where 0 is the initial state of the DFA.Therefore, the hypothesis is true for the base case. 
	
	Let the hypothesis be true for all words of length $n$. We prove that the hypothesis can  be extended to words of length $n+1$. 
	
	Let $W=V\cdot a$ be a word of length $n+1$, where $V$ is a word of length $n$ and $\mathsf{last}_V$ be its last state . By induction hypothesis, 
	\begin{itemize}
		\item $\mathsf{last}_V \in \{\floor{-\thresh}+1, \dots , \floor{\thresh}\}$ iff $\gap{V} = \mathsf{last}_V$
		\item $\mathsf{last}_V = \mathsf{bad}$ iff $\gap{V} > \floor{\thresh} $
		\item $\mathsf{last}_V = \mathsf{veryGood}$ iff $\gap{V} \leq \floor{-\thresh} $
	\end{itemize}
	Since, the DFA is deterministic, the last state of $W$ is $\mathsf{last} = \delta(\mathsf{last}_V, a)$. 
	
	\newcommand{\last}{\mathsf{last}}
	\renewcommand{\bad}{\mathsf{bad}}
	
	If $\gap{W} > \floor{\thresh}$, we show that $\last = \bad$. 
	\begin{enumerate}
		\item If $\last_V = \bad$, then by construction $\bad$ is a sink state, in which case $\last = \bad$.
		
		\item If $\last_V = s \in \{\floor{-\thresh}+1, \dots , \floor{\thresh}\}$.  
		By I.H, we know that $s = \gap{V}$, therefore by definition of recoverable gap, $\gap{W} = d\cdot s + a > \floor{\thresh}$. Therefore, by the transition function rules, $\last = \bad$.

		\item If $\last_V = \mathsf{veryGood}$. Since $\mathsf{veryGood}$ is a sink state, none of its transitions will go to state $\mathsf{bad}$. 
	\end{enumerate}
	Conversely, let $\last = \bad$. We show that $\gap{W} > \floor{\thresh}$.
	\begin{enumerate}
		\item If $\last_V = \bad$ , then by I.H. $\gap{V} > \floor{\thresh}$. Since the discount-factor is an integer, $V$ is  a bad prefix (Corollary~\ref{Coro:BoundInteger}).
		Since extensions of bad-prefixes are also bad-prefixes, we get that W is also a bad-prefix. By the same corollary, we get $\gap{W}>\floor{\thresh}$. 
		
		\item If $\last_V = s\in \{\floor{-\thresh}+1, \dots , \floor{\thresh}\}$.
		A transition from s goes to $\bad$ only if $d \cdot s  + a > \floor{\thresh}$.
		By I.H $s = \gap{V}$, therefore $d\cdot s + a = \gap{W} > \floor{\thresh}$.
		
		\item If $ \last_v = \mathsf{veryGood}$. Not relevant since there are no transition from $\mathsf{veryGood}$ to $\bad$.
	\end{enumerate}

	The proof for $\last = \mathsf{veryGood}$ iff $\gap{W} \leq \floor{-\thresh}
	$ is similar. It uses the fact that extensions of a very good prefix are also very good prefixes. 
	
	The final case is when $\last \in  \{\floor{-\thresh}+1, \dots , \floor{\thresh}\}$. This is the simplest case since it simply follows the transition relation and inductive definition of recoverable gap. 
	
	Sink $\bad$ is the only accepting state, a words visits $\bad$ iff its recoverable gap is strictly above $ \floor{\thresh}$, the DFA accepts the language of bad-prefixes.  
	\qed
\end{proof}

Finally, the DS comparators are constructed as follows:

\begin{theorem}
\label{Thrm:DBA}
Let $\mu$ be the upper bound, and $d>1$ be the integer discount-factor. 
DS comparator for all inequalities and equality are either deterministic safety or deterministic co-safety automata with $\O(\mu)$ states. 
\end{theorem}
\begin{proof}
	
	This proof should also be treated as an exercise that illustrates that DFA $\baut{\mu}{d}{\leq}$ is the basis of all our constructions. In fact, the analysis of bad-prefixes and very-good prefixes for the one case of the $\leq$ relation is sufficient. 
	
	For $>$:	
	Syntactically, DFA $\baut{\mu}{d}{\leq}$ is also a deterministic co-safety automaton. We claim that this  deterministic co-safety automaton accepts the DS comparison language for the relation $>$. 
	This is true due to the complementation duality of the definitions of safety and co-safety languages. A sequence $A$ is in a co-safety language $L$ iff it has a finite length prefix $B$ that is a bad prefix for the complement safety language $\overline{L}$. The missing link here is that DS comparison languages for $>$ and DS comparison language for $\leq$ are complements of each other.

	For $\leq$:
	Flipping accepting and non-accepting states of the deterministic co-safety comparator for $>$ will do the trick to generate the deterministic safety comparator for $\leq$ with the same number of states. Another way to think of the construction of the coveted safety comparator is from DFA $\baut{\mu}{d}{\leq}$  itself. First of all, observe that syntactically the complement of DFA $\baut{\mu}{d}{\leq}$, denoted by $\overline{\baut{\mu}{d}{\leq}}$ is also a deterministic safety automata with a single non-accepting sink state. Second of all, semantically DFA $\overline{\baut{\mu}{d}{\leq}}$ accepts all finite sequences that are not bad-prefixes of the DS comparison language with relation $\leq$. Hence, DFA $\overline{\baut{\mu}{d}{\leq}}$ is also the deterministic safety automata for DS comparison language for relation $\leq$.
	
	For $<$ and $\geq$: The co-safety and safety automata for $<$ and $\geq$ is obtained by negating the alphabet on all transitions of the DS comparator automata for $>$ and $\leq$, respectively. Alternately, once may construct these comparators from first principles by reasoning over DFA $\baut{\mu}{d}{\leq}$, as done in the previous two cases. 
	
	For $=$ and $\neq$: The default for construction of DS comparators $=$ is to take the intersection of those for $\leq$ and $\geq$. However, that would result in a safety comparator with $\O(\mu^2)$. Similarly, default for $\neq$ would also have $\O(\mu)$ states. Instead, we construct the safety and co-safety comparator for $=$ and $\neq$ in $\O(\mu)$ states as well. 
	
	Observe that the bad-prefixes of the DS comparison language for $=$ with either be bad-prefixes of the same for $\leq$ or the bad-prefixes of $\geq$. Recall, that a bounded, finite-length weight sequence $W$ is  a bad-prefix of $\leq$ iff $\gap{W} > {\thresh} $ (Lemma~\ref{Lem:GapThresholdS}). Analogously, one can prove that $W$ is a bad-prefix of $\geq$ iff $\gap{W} < - \thresh$. Therefore, if an infinite sequence $A$ has a finite prefix $B$ such that either $\gap{B} > \thresh$ or $\gap{B} < -\thresh$, then the $\neq (\DSum{A}{d} = 0) \equiv \DSum{A}{d} \neq 0$, i.e., is $A$ is present in the DS comparison language for $\neq$. Therefore, in DFA $\baut{\mu}{d}{\leq}$, if both $\mathsf{bad}$ and $\mathsf{veryGood}$ are accepting, the resulting deterministic B\"uchi automaton accepts the DS comparison language for $\neq$. Now, this can be converted into a deterministic co-safety automaton, by fusing both the accepting states into one.
	
	Finally, the safety automata for DS comparison language with $=$ simply flips the accepting and non-accepting states of the co-safety automata for $\neq$.
	\qed
\end{proof}

As a matter of fact, the most compact non-deterministic DS comparator constructions with parameters $\mu$, $d$ and $R$ also contain $\O(\mu)$ states~\cite{BCVCAV18}.
In fact, we can go a step further and prove that the deterministic B\"uchi automaton constructed in Thereom~\ref{Thrm:DBA} are the {\em minimal} forms, i.e., none of the languages represented by these deterministic automata can be represented by other deterministic automata with fewer number of states. It is worth noting that not all (non-deterministic) B\"uchi automata have minimal forms. However, deterministic B\"uchi automata are known to have minimal forms
~\cite{ehlers2010minimising,loding2001efficient}.

\begin{theorem}
	\label{Thrm:DBAMinimal}
The deterministic B\"uchi automata constructed in Theorem~\ref{Thrm:DBA} for each DS comparator are their respective minimal deterministic B\"uchi automata.
\end{theorem}
\begin{proof}
	Each of the DS comparator constructed in Theorem~\ref{Thrm:DBA} is either a deterministic safety or co-safety automata. Therefore, if these DBAs were not in minimal form, then the underlying deterministic DFA corresponding to the appropriate language of bad-prefixes or their complement would not be in minimal form either. 
	
	For instance, if the DBA for DS comparator for  $\leq$ or $>$ were not minimal, then the DFA $\baut{\mu}{d}{\leq}$ would not be a minimal DFA either. 	
	We claim that DFA $\baut{\mu}{d}{\leq}$ is a minimal DFA, and hence it must be the case that the safety/co-safety automata for DS comparison language for $\leq $ or $>$ are minimal DBA as well. 
	
	In order to prove that DFA $\baut{\mu}{d}{\leq}$ is minimal, we prove that no two states are equivalent as per Myhill-Nerode equivalence relation~\cite{nerode1958linear}, and proving minimal form~\cite{hopcroft1969formal}. As per this relation, two states $s$ and $t$ are said to be equivalent, denoted $s\sim t$, if (a). Either both $s$ and $t$ are accepting states, or both are non-accepting states, and (b). For all alphabet $a \in \Sigma$, if $s \xrightarrow{a} s'$ and $t \xrightarrow{a} t'$ are transitions in the DFA, then $s'\sim t'$ must hold. The core idea behind proving that no two states of $\baut{\mu}{d}{\leq}$  are equivalent is to realize that the states of $\baut{\mu}{d}{\leq}$ represent values of the gap value, and transitions are governed by arithmetic on them. So, one can easily find an alphabet $a\in\Sigma$ such that outgoing transitions do not go to equivalent states. 
	
	The formal proof can be completed by induction on the minimum distance of states from the single accepting sink state. The inductive hypothesis is that no two states that are at a minimum distance of $k\geq 0$ from the accepting sink state are equivalent. In the base case, k = 0. There is only one such state - the accepting sink state itself- therefore the I.H. holds vacuously. For larger values of $k$, the intuition described above will demonstrate the veracity of the I.H.
	
	Similar arguments will prove that that the DBAs for all other relations are also minimal.
	\qed
\end{proof}

Hence, in this section, not only did we construct safety/co-safety forms for DS comparators, our constructions are also the minimal deterministic representations. These are the most efficient deterministic constructions possible for DS comparators. 
\subsection{\cct{QuIPFly}: DS inclusion with integer discount factor}

Finally, we solve DS inclusion with integer discount factors with using the safety and co-safety comparators presented in the previous Section~\ref{Sec:DS}.
Here we will apply the generic algorithm for quantitative inclusion with regular safety/co-safety comparators (Section~\ref{Sec:Inclusion}) to the case of discounted-sum with integer discount factors. 
As seen before in Theorem~\ref{thrm:quantI}, the expectation is that the resulting algorithm for DS inclusion with integer discount factor will have an improved complexity. 
Lastly, we make use of the subset construction based methods in order to present an improved on-the-fly algorithm for DS inclusion, called \cct{QuIPFly}.

The following statement instantiates Theorem~\ref{thrm:quantI} for discounted-sum aggregation. Note that the difference in the formal statement is that here we are able to completely characterize the type of the B\"uchi automaton as well:
\begin{theorem}
\label{Thrm:DSInclusionStict}

Let $P$ and $Q$ be weighted-automata, and $C$ be a regular safety/co-safety comparator for an inequality with integer discount factor $d>1$.

\begin{itemize}
	
	\item Strict DS-inclusion $P\subset Q$ is reduced to emptiness checking of a {\em safety automaton} of size $|P||C||Q|\cdot 2^{\O(|Q|\cdot|C|)}$.
	
	\item Non-strict DS-inclusion $P\subseteq Q$  is reduced to emptiness checking of a {\em weak-B\"uchi automaton~\cite{kupferman2001weak}} of size $|P||C||Q|\cdot 2^{\O(|Q|\cdot|C|)}$. 
\end{itemize}	
\end{theorem}
\begin{proof}
	Unlike Theorem~\ref{thrm:quantI}, 
	the B\"uchi automaton for emptiness checking has a specific form: safety or weak-B\"uchi~\cite{kupferman2001weak}. We can characterize these forms since we know that the DS comparator for strict inequality and equality are co-safety and safety automata respectively (Corollary~\ref{Coro:DSFull}).
	
	Recall, that the B\"uchi automata that is being checked for emptiness is the counterexample automaton. In essence, the counterexample automaton is formed by an intersection of the maximal automaton with an appropriate comparator (Lemma~\ref{Lemma:counterexample}). 
	Furthermore, we know that when the comparator for non-strict inequality is safety, as for discounted sum, the maximal automaton will be a safety automata (Lemma~\ref{Lem:MaximalSafety}). 
	
	Consequently, in case of strict DS inclusion, the counterexample automata is constructed by the intersection of a safety maximal automaton with the regular safety DS-comparator for non-strict inequality. Due to closure of safety automata under intersection, the resulting counterexample automata will be a safety automata. 
	
	In case of non-strict DS-inclusion, the counterexample automata is constructed by the intersection of the safety maximal automaton with the regular co-safety DS-comparator for strict inequality. As a consequence of Theorem~\ref{Lem:IntersectSafetyAndCo}, the counterexample automaton is weak-B\"uchi. 
	
	Lastly, the size of the resulting safety/weak B\"uchi automaton is given by the product of the minimal automata (Corollary~\ref{Coro:SizeMaximal}) and the comparator. Recall, that the size of all comparators for discounted sum are identical (Theorem~\ref{Thrm:DBA} and Theorem~\ref{Thrm:DBAMinimal}). Hence, all are given a size of $|C|$.
	\qed
\end{proof}

Therefore, the final word is:
\begin{corollary}[\rm [DS-inclusion with safety/co-safety comparator]
\label{Coro:DSInclusionSafety}
	Let $P$, $Q$ be weighted-automata, and $C$ be a regular (co)-safety DS-comparator  with integer discount-factor $d>1$.The complexity of DS-inclusion  is  $|P||C||Q|\cdot 2^{\O(|Q|\cdot|C|)}$.	
\end{corollary}

Clearly, this is an improvement over the worst-case complexity of $\cct{QuIP}$.

\subsubsection{\cct{QuIPFly}: An on-the-fly implementation}

The algorithm for DS inclusion with integer discount factor $d>1$ proposed in Theorem~\ref{Thrm:DSInclusionStict} checks for emptiness of the counterexample automata. 
A naive algorithm will construct the counterexample automata fully, and then check if they are empty by ensuring the absence of an {\em accepting lasso}.

In \cct{QuIPFly}, we implement a more efficient algorithm. In our implementation, we make use of the fact that counterexample automata is a safety/weak B\"uchi that is constructed from the intersection of a safety maximal automata and an appropriate determinisitc safety/co-safety DS comparator. Recall that maximal automata can be constructed in an on-the-fly fashion since it involves subset construction like techniques. Since, the comparators are also deterministic, they can also be designed in an on-the-fly manner.

In all, this facilitates an {\em on-the-fly procedure} for DS inclusion, since successor states of state in the counterexample automata can be determined directly from input weighted automata and the comparator automata. 
The algorithm terminates as soon as an accepting lasso is detected. When an accepting lasso is absent, the algorithm traverses all states and edges of the counterexample automata.
Note that for both strict and non-strict DS inclusion, not every lasso in the counterexample automata  is an accepting lasso. Only those lassos are accepting for which the loop lies entirely inside one accepting partition of the counterexample automata (recall that states of weak B\"uchi automata are partitioned, and safety automata are also weak B\"uchi automata). Tracking whether the loop is in an accepting partition by simply looking at the states in the counterexample automata. 

Note that the on-the-fly mechanism does not alter the worst-case performance of DS inclusion,  but it contribute to making the algorithm (a) more efficient as it can terminate as soon as an accepting lasso is found, and (b) less memory exhaustive, since it usually doesn't require to store the entire counterexample automata in memory at any given time.

\section{Empirical evaluation}
\label{Sec:ExpEval}

The goal of the empirical analysis is to examine performance of DS-inclusion with integer discount-factor with safety/co-safety comparators against existing tools to investigate the practical merit of our algorithm. 
Our optimized on-the-fly algorithm for safety/co-safety comparator-based approach for DS inclusion with integer discount factors is implemented in our prototype called \textsf{QuIPFly}. \textsf{QuIPFly} is written in \textsf{Python 2.7.12}. 
\textsf{QuIPFly} employs basic implementation-level optimizations to avoid excessive re-computation. 

We compare against (a) Regular-comparator based tool \cct{QuIP}, and (b) DS-determinization and linear-programming tool \cct{DetLP}. 
Recall from Chapter~\ref{ch:empirical}, \textsf{QuIP} is written in \textsf{C++}, and invokes state-of-the-art B\"uchi language inclusion-solver \textsf{RABIT}~\cite{RABIT-Reduce}.
We enable the \texttt{-fast} flag in  \textsf{RABIT}, and tune  its $ \textsf{Java} $-threads with \textsf{Xss}, \textsf{Xms}, \textsf{Xmx} set to 1GB, 1GB and 8GB, respectively. 
\cct{DetLP} is  also written in \textsf{C++}, and uses linear programming solver \textsf{GLPSOL} provided by \cct{GLPK} (\cct{GNU} Linear Prog. Kit)~\cite{glpk}.
We compare these tools along two axes: runtime and number of benchmarks solved.

Figures are best viewed online and in color.

\begin{figure}[t]
\centering
\includegraphics[width=0.85\textwidth]{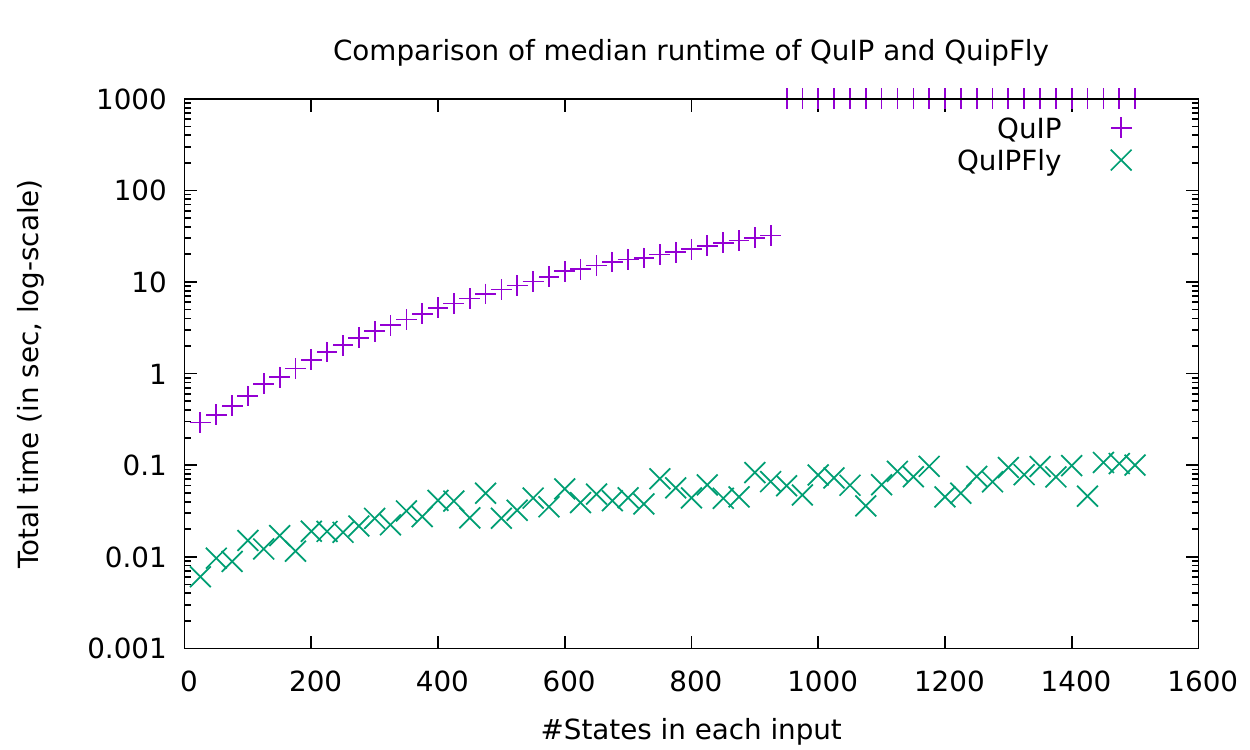}
\caption[Runtime comparion of $\cct{QuIP}$ and$ \cct{QuIPFly}$]{Runtime comparion of $\cct{QuIP}$ and$ \cct{QuIPFly}$. $s_P = s_Q$ on $x$-axis, $wt= 4$, $\delta = 3$, $d=3$, $P\subset Q$}
\label{Fig:QuIPvsQuIPFly}
\end{figure}


\subsubsection{Design and setup for experiments}

Due to lack of standardized  benchmarks for weighted automata, we
follow a standard approach to performance evaluation of automata-theoretic tools~\cite{abdulla2010simulation,mayr2013advanced,tabakov2005experimental} by experimenting with  {\em randomly generated} benchmarks, using random benchmark generation procedure described in~\cite{BCVCAV18}. 

The parameters for each experiment are number of states $s_P$ and $s_Q$ of weighted automata, transition density $\delta$, maximum weight $wt$, integer discount-factor $d$, and $\R \in \{\strict, \nstrict\}$. 
In each experiment, weighted automata $P$ and $Q$ are randomly generated, and runtime of $\R$-DS-inclusion for all three tools is reported with a timeout of 900$\sec$. 
We run the experiment for each parameter tuple 50 times. 
All experiments are run on a single node of a high-performance cluster consisting of two quad-core Intel-Xeon processor running at 2.83GHz, with 8GB of memory per node.
We experiment with $s_P = s_Q$ ranging from 0-1500 in increments of 25, $\delta\in \{3,3.5,4\}$, $d = 3$, and $wt \in \{d^1+1, d^3-1, d^4-1\}$.

\subsubsection{Observations and Inferences}
\begin{figure}[t]
\centering
\includegraphics[width=0.85\textwidth]{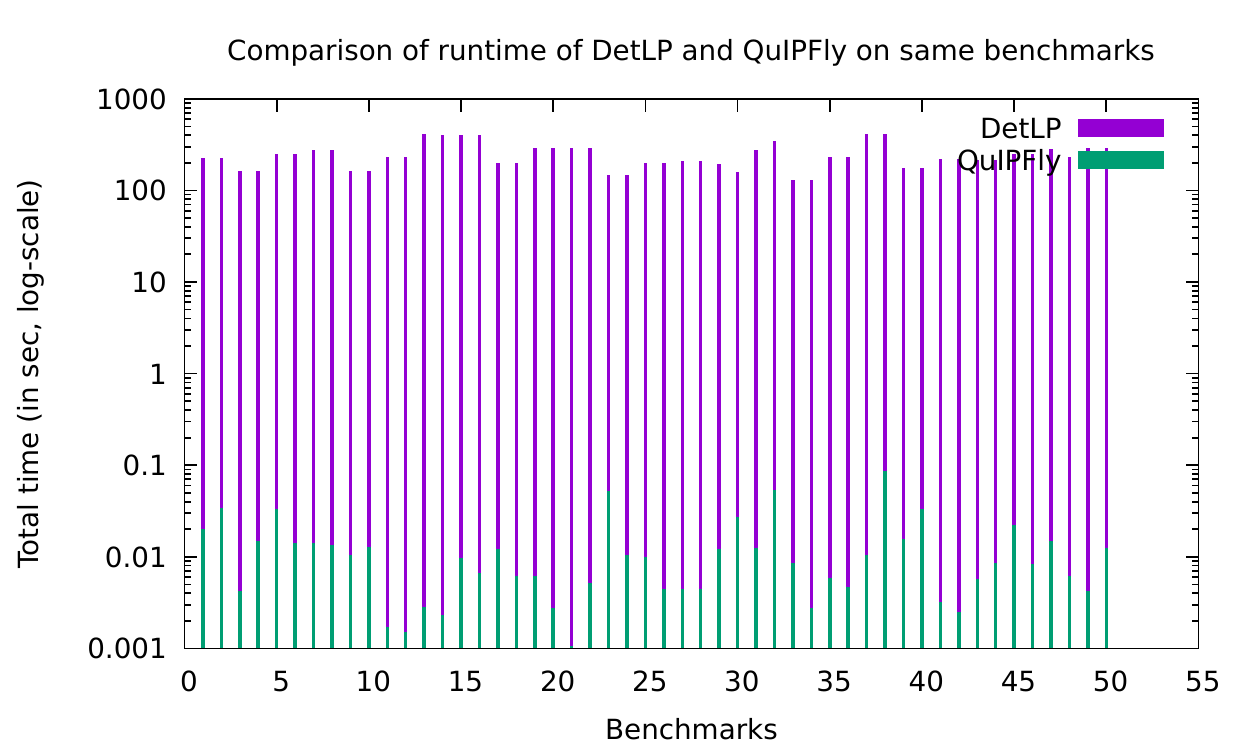}
\caption[Number of benchmarks solved between \cct{DetLP} and \cct{QuIPFly}]{Number of benchmarks solved between \cct{DetLP} and \cct{QuIPFly}. $s_P = s_Q= 75$, $wt= 4$, $\delta = 3$, $d=3$, $P\subset Q$}
\label{Fig:DetLPvsQuIPFly}
\end{figure}
For clarity of exposition, we present the observations for only one parameter-tuple. Trends and observations for other parameters were similar.

\paragraph{\cct{QuIPFly} outperforms \cct{QuIP}}
by at least an order of magnitude in runtime. Fig~\ref{Fig:QuIPvsQuIPFly} plots the median runtime of all 50 experiments for the given parameter-values for \cct{QuIP} and \cct{QuIPFly}. More importantly, \cct{QuIPFly} solves all of our benchmarks within a fraction of the timeout, whereas \cct{QuIP} struggled to solve at least 50\% of the benchmarks with larger inputs (beyond $s_P = s_Q =  1000$). Primary cause of failure is memory overflow inside \cct{RABIT}. We conclude that regular safety/co-safety comparators outperform their regular counterpart, giving credit to the simpler subset-constructions vs. B\"uchi complementation. 

\paragraph{\cct{QuIPFly} outperforms \cct{DetLP}}
comprehensively in runtime and in number of benchmarks solved. We were unable to plot \cct{DetLP} in Fig~\ref{Fig:QuIPvsQuIPFly} since it solved fewer than 50\% benchmarks even with small input instances.
Fig~\ref{Fig:DetLPvsQuIPFly} compares the runtime of both tools on the same set of 50 benchmarks for a representative parameter-tuple on which all 50 benchmarks were solved. The plot shows that \cct{QuIPFly} beats \cct{DetLP} by 2-4 orders of magnitude on all benchmarks.

\paragraph{Overall verdict}
Overall, \cct{QuIPFly} outperforms \cct{QuIP} and \cct{DetLP} by a significant margin along both axes, runtime and number of benchmarks solved.
This analysis gives unanimous evidence in favor of our safety/co-safety approach to solving DS-inclusion.

\section{Chapter summary}

The goal of this chapter was to improve comparator-based solutions for DS inclusion. 
To this end, here we further the understanding of language-theoretic properties of discounted-sum aggregate function by demonstrating that  DS-comparison languages form safety and co-safety languages. These properties are utilized to obtain a decision procedure for  DS-inclusion that offers both tighter theoretical complexity, improved scalability, and better performance than separation-of-techniques.
To the best of our knowledge, this is the first work that applies language-theoretic properties such as safety/co-safety in the context of quantitative reasoning. 
In doing so, we demonstrate that the close integration of structural analysis and numerical analysis using comparator automata can enhance algorithms in quantitative reasoning since they can benefit from the abundance of work in qualitative reasoning.

\chapter{DS inclusion with non-integer discount factors}
\label{ch:anytime}
The theoretical and practical progress in DS inclusion so far have been made for the case when the discount factor is an integer. The problem of complexity of DS inclusion with non-integer discount factors is still very much open - so much so that even its decidability is a mystery so far. In wake of the unknown decidability of DS inclusion with non-integer discount factor, this chapter takes the approach of developing pragmatic solutions to the problem. In this case, we develop an {\em anytime algorithm}~\cite{dean1988analysis} for DS inclusion for non-integer discount factors. 

\section{Introduction}

In its most generality, the decidability of DS inclusion is still unknown~\cite{chatterjee2010quantitative}. 
The goal of this work is to build a pragmatic solution for DS inclusion despite its unknown decidability.
Prior results have shown that the parameter that determines the rate at which weights loose their significance, called the {\em discount factor}, 
primarily governs the hardness of DS inclusion:
When the discount factor is an integer, DS inclusion is \cct{PSPACE}-complete~\cite{BCVFoSSaCS18}; On the contrary, when the discount factor is a non-integer,  decidability of DS inclusion is  unknown, and has been an open problem for more than a decade~\cite{chatterjee2010quantitative}.
This work strives to address DS inclusion for  non-integer discount factors, since 
most practical applications of DS aggregate function, such as reinforcement learning~\cite{sutton1998introduction}, planning under uncertainty~\cite{puterman1990markov},   and game theory~\cite{osborne1994course},  make use of non-integer discount factors.
In particular, these applications require that the discount factor $d>1$ to be very close to 1, 
i.e. $ 1<d<2$.
Therefore, this work focuses on DS inclusion when $1<d<2$.

This work addresses DS inclusion for  non-integer discount factors.
In particular, we investigate the case when the discount factor $d>1$ is very close to 1, 
i.e. $ 1<d<2$, since that is the value used in practice. Our focus is not on resolving whether DS inclusion is decidable for $1<d<2$ but to design {\em pragmatic solutions} for DS inclusion despite its  decidability being unknown. 

A natural step in this direction is to solve approximations of DS inclusion. However, we discover that approximations of DS inclusion is at least as hard as DS inclusion itself (Theorem~\ref{thrm:reducetoOver}), since DS inclusion reduces to its approximations in polynomial time. Hence, 
the decidability of approximations of DS inclusion is also currently unknown. 
As a result, we are faced with the challenge to design a solution for DS inclusion with $1<d<2$ without the ability to develop algorithms for DS inclusion or its approximations. 

To address this challenge, we turn to  {\em anytime algorithms} that have been used extensively to design pragmatic solution in AI for problems with high computational complexity or even undecidability~\cite{dean1988analysis}. 
{Anytime algorithms} are a class  of  algorithms that generate approximate answers quickly and proceed to construct progressively better approximate solutions over time~\cite{dean1988analysis}. In the process, they may either generate an exact solution and terminate, or continuously generate a better approximation. 
In the context of designing a pragmatic solution for DS inclusion with $1<d<2$ , we design an anytime algorithm for the same. In addition, our anytime algorithm is {\em co-computational enumerable}, i.e., it is guaranteed to terminate on input instances on which DS inclusion does not hold.
The algorithm gives only approximate answers otherwise.

The formalism is detailed here.
Let $P$ and $Q$ be weighted automata, 
and let  $1<d<2$ be the discount factor.  We say  $P$ {\em is DS included in } $Q$, denoted by $P \subseteq Q$, if  weight of all executions in $P$ is less than or equal to that in $Q$.  So, our anytime algorithm, called $\rationalInc$, either terminates and returns a crisp $\mathsf{True}$ or $\mathsf{False}$ answer to $P \subseteq Q$, or it continuously generates an approximation that $P$ is  {\em $d\cdot \varepsilon$-close} to $Q$, 
in a precise sense defined below.
If an execution of $\rationalInc$ is interrupted due to external factors such as manual interference or resource overflow, then the algorithm will return the most recently computed $d\cdot\varepsilon$-close approximation. 
Additionally, if $P\subseteq Q$ does not hold, then the algorithm is guaranteed to terminate after a finite amount of time.
Note that we do not prove termination when $P \subseteq Q$ holds.
But this is not surprising, because if we could then we would have proven decidability of DS inclusion for $1<d<2$.

The core of our 
algorithm is $\approxInc$  (Overview: \textsection~\ref{Sec:AnytimeFull}, Description: \textsection~\ref{Sec:anytimeInc}), a tail recursive algorithm that
recurses on the approximation factor $0< \varepsilon < 1$ by halving it in each new invocation. 
Conceptually, $\approxInc$ over- and under-approximates DS inclusion in each new invocation of the algorithm, and continues until either an exact solution is obtained or the algorithms' execution is interrupted. 
Due to the unknown decidability of both approximations, 
{\em partial solutions} for the over- and under-approximations are developed. {\em Comparator automata}~\cite{BCVFoSSaCS18} based techniques for aggregate functions that represent lower- and upper- approximations of discounted sum are used, respectively for the partial solutions (\textsection~\ref{Sec:lowerInc}  and \textsection~\ref{Sec:upperInc}, respectively).

Another advantage of our anytime algorithm is that it can be run upto a desired precision $0<\varepsilon_c<1$.
For this the recursive procedure $\rationalInc$ will be run till it is invoked with approximation factor $\varepsilon_c$ in the worst case, i.e., in the worst-case $\rationalInc$ will be invoked 
for $\O(\log \frac{1}{\varepsilon_c})$ times, each time with a lower approximation factor. 
In this case, the complexity of running $\rationalInc$ is linear in $\frac{1}{\varepsilon_c}$, exponential in size of the input DS automata and exponential in $\frac{1}{(d-1)^2}$. 
This shows that as $d$ and $\varepsilon_c$ become smaller solving upto a desired precision explodes rapidly.
These observations are expected, as they corroborate with the known result on undecidability of {\em sum-inclusion} (Loosely speaking, DS inclusion with $d =1$, and $\varepsilon=0$)~\cite{almagor2011whats,krob1992equality}. 

In conclusion, this work does not resolve the decidability debate on DS inclusion with $1<d<2$. Instead, it designs an anytime algorithm that renders approximations with theoretical guarantees, and time-bounds when the precision value is fixed. Thus, our algorithm is suitable for pragmatic purposes. 

\section{Preliminaries}

This section defines terminology and notation used in the rest of this chapter. A key difference in this chapter is that we solve DS inclusion over finite words. A similar proof can be adapted for infinite words.  

A finite sequence of weights is said to be {\em bounded} by $\mu>0$ if the absolute value of all weights in the sequence are less than or equal to $\mu$.

A {\em finite-state automaton}~\cite{thomas2002automata} is a tuple $\A = (\Statess, \Sigma, \delta, \StartState, \Final)$, where
$ \Statess $ is a finite set of \emph{states}, $\Sigma $ is a finite {\em input alphabet},  $ \delta \subseteq (\Statess \times \Sigma \times \Statess)$ is the   {\em transition relation}, $ \StartState \subseteq \Statess $ is the set of {\em initial states}, and $ \Final \subseteq \Statess $ is the set of {\em accepting states}. A finite-state automaton is {\em deterministic} if for all states $ s $ and
inputs $a$, $ |\{s'|(s, a, s') \in \delta\}|
\leq 1 $ and $|\StartState|=1$; otherwise, it is {\em nondeterministic}. Deterministic and non-determinsitic finite-state automata are denoted by DFA and NFA, respectively. An NFA is {\em complete} if for all $s \in \Statess$ and $a \in \Sigma$, there exists a transition $(s, a, t) \in \delta$ for some state $t \in \Statess$. For a word $ w = w_0w_1\dots w_m \in \Sigma^* $, a {\em run} $ \rho$ of $ w $ is a sequence of states $s_0s_1\dots s_{m+1}$,  such that $ s_0 \in \StartState$, and $ \tau_i =(s_i, w_i, s_{i+1}) \in \delta $ for all $i$. A run $\rho$ is {\em accepting} if its last state  $s_{m+1}\in\mathcal{F}$.  A word $w$ is \emph{accepted} by NFA/DFA if it has an accepting run. {\em Regular languages} are languages accepted by DFA/NFA. 

A {\em weighted automaton over finite words} (weighted automaton, in short), is a tuple $\A= (\mathcal{M},  \gamma, f)$, where $\mathcal{M} = (\Statess, \Sigma, \delta, \StartState, \Statess) $ is a 
complete NFA with all states as accepting, $\gamma: \delta \rightarrow \Z$ is a {\em weight function}, and $f: \Z^* \rightarrow \Re$ is an {\em aggregation function}. {\em Words} and {\em runs} in weighted automata are defined as they are in an NFA.
The {\em weight sequence} of a run $\rho = s_0 s_1 \dots s_{m+1}$ of word $w = w_0w_1\dots w_m$ is  $wt_{\rho} = n_0 n_1 \dots n_m$ where $n_i = \gamma(s_i, w_i, s_{i+1})$ for all $i$. The {\em weight of a run}  $\rho$, denoted by $f(\rho)$, is  $f{(wt_{\rho})}$. 
The {\em weight of word} $w \in \Sigma^{*}$ in weighted automata is defined as $\wt(w, \A) = max \{f(\rho) | \rho$ is a run of $w$ in $\A \}$. 
The problem of
{\em $f$-inclusion} compares the weight of words in two weighted automata with the same aggregate function.
Given weighted automata $P$ and $Q$ with aggregate function $f:\N^* \rightarrow \mathbb{R}$, 
$P$ is said to be {\em $f$-included} in $Q$ if for all words $w \in \Sigma^*$, $\wt(w, P) \leq \wt(w, Q)$.

This work studies the {\em discounted-sum inclusion} problem. The discounted sum (DS) of a finite sequence $A = a_0\dots a_n$ for discount factor $d>1$, denoted $\DSum{A}{d} = a_0 + \frac{a_1}{d} \dots + \frac{a_n}{d^n}$.
{\em DS automata} with discount factor $d>1$ are
weighted automata with discounted sum aggregation with discount factor $d$.
Therefore, given two DS automata $P$ and $Q$ with the {\em same} discount factor, $P$ is {\em DS-included} in $Q$ if weight of every word in $P$ is less than or equal to that in $Q$. 
We reserve the notation $P \subseteq Q$ to denote  $P$ is DS-included in $Q$.
While DS inclusion is $\cct{PSPACE}$-complete when the discount factor is an integer, its decidability is still open for non-integer discount factors~\cite{boker2015target,chatterjee2010quantitative}.
For sake of clearer exposition, this paper  assumes that the weight along transitions in a DS  automata are non-negative integers. There are no technical differences in extending the result to integer weights.

In a similar vein to recent progess on DS inlcusion for integer discount factors~\cite{BCVCAV18,BVCAV19}, this work makes use of {\em comparator automata} to design the anytime algorithm. Comparator automata and comparison languages are defined as follows: For a finite set of integers $\Sigma$, an aggregate function  $f:\Z^*\rightarrow \mathbb{R}$, and equality or inequality relation $\mathsf{R} \in \{<,>,\leq, \geq, =, \neq\}$, the {\em comparison language for $f$ and  $\mathsf{R}$} is a language over the alphabet $\Sigma$ that accepts a word $A\in\Sigma^*$ iff $f(A)$ $\mathsf{R}$ $0$ holds. A {\em comparator automaton} (comparator, in short) for $f$ and  $\mathsf{R}$ is an automaton that accepts the comparison language for $f$ and  $\mathsf{R}$~\cite{BCVFoSSaCS18}. A comparator is {\em regular} if  its automaton is an NFA.
Prior work has shown that comparator for discounted sum (DS comparator, in short) is regular for all $\mathsf{R}$ iff the discount factor is an integer~\cite{BCVlmcs2019}. Regularity of the comparator has been crucial to the progress in DS inclusion for integer discount factors. Therefore, another contribution in this work is to define an approximation for discounted-sum with non-integer discount factor so that its comparator is regular. We will use these regular comparators to design the anytime algorithm.

\section{Overview}
\label{Sec:AnytimeFull}

\sloppy

Before delving into the technical details of the anytime algorithm, we give a roadmap of the approach.  Notions of approximations are central to anytime algorithms, so we begin with defining approximations to DS inclusion:

\begin{Def}[$d\cdot\varepsilon$-close approximation]
	Given DS automata $P$ and $Q$ with discount factor $d>1$, and an approximation parameter $\varepsilon>0$, $P$ is said to be {\em $d\cdot\varepsilon$-close to $Q$}, denoted by $P \subseteq Q + d\cdot\varepsilon$, if for all words $w \in \Sigma^{\omega}$, $\wt(w, P) \leq \wt(w, Q) + d\cdot\varepsilon$.
\end{Def}

\begin{Def}[$d\cdot\varepsilon$-far approximation]
	Given DS automata $P$ and $Q$ with discount factor $d>1$, and an approximation parameter $\varepsilon>0$, $P$ is said to be {\em $d\cdot\varepsilon$-far from $Q$}, denoted by $P \subseteq Q - d\cdot\varepsilon$, if for all words $w \in \Sigma^{\omega}$, $\wt(w, P) \leq \wt(w, Q) - d\cdot\varepsilon$.
\end{Def}

Given the open decidability status of DS inclusion, the next alternative is to develop algorithms for the aforementioned approximations. That is not possible either, however, for the following reason:
\begin{theorem}
	\label{thrm:reducetoOver}
	{\rm [Unknown decidability of approximations]}
	\begin{enumerate}
		\item There exists a polynomial time reduction from 
		DS inclusion to  $d\cdot\varepsilon$-close approximation for an approximation factor $\varepsilon>0$.
		\item There exists a polynomial time reduction from 
		DS inclusion to  $d\cdot\varepsilon$-far approximation for an approximation factor $\varepsilon>0$.
	\end{enumerate}
	
\end{theorem}

\begin{proof}
	
	The intuitive argument behind showing that $P \subseteq Q$ reduces to $P' \subseteq Q' + d\cdot\varepsilon$ for an $\varepsilon>0$ is to show that one can transform $Q$ into a weighted automaton $Q'$ such that $Q' = Q - d\cdot\varepsilon$. Similarly, to show that $P \subseteq Q$ reduces to $P' \subseteq Q' - d\cdot\varepsilon$, we show that one can transform $Q$ into $ Q'$ such that $Q' = Q + d\cdot\varepsilon$.
	An direct corollary of Theorem~\ref{thrm:reducetoOver} is that the decidability of approximations of DS inclusion is currently unknown as well. 
	
	DS inclusion is decidable iff over approximation of DS is decidable: 
	First we prove that every instance of DS inclusion can be reduced to an instance of over approximation of DS inclusion. 
	Let $d>1$ be a discount factor and $P$ and $Q$ be DS  automata. Then we will show that there exists alternate DS automata $R$ and $S$, and an approximation factor $\varepsilon>0$ such that $P \subseteq Q \iff R \subseteq S + d\cdot\varepsilon$.
	Let $\#$ be an character that is not present in the alphabet of $P$ and $Q$
	First of all, we generate a new DS automaton $P_{\#}$ from $P$ such that every word in P is prefixed with the character $\#$ and its weight is multiplied by $\frac{1}{d}$.
	This can be done by a simple automata-theoretic transformation of $P$: Include all states and transitions from $P$ in $P_{\#}$. Retain all accepting states of $P$ in $P_{\#}$. Add a new state $s_{\#}$. Add a transition from $s_{\#}$ to state $s_{\init}$ of $P$, and assign it a weight of 0. Make the new state $s_{\#}$ the accepting state. This is $P_{\#}$. 
	Similarly, construct DS automaton $Q_{\#}$ from $Q$. It is easy to see that $P \subseteq Q$ iff $P_{\#} \subseteq Q_{\#}$.
	Finally, construct $Q'$ from $Q_{\#}$ by assigning the transition from $s_{\#}$ to $s_{init}$ a weight of $-1$. Then it is easy to see that 
	$P_{\#} \subseteq Q_{\#}$ iff $P_{\#} \subseteq Q' + 1$. Let $\varepsilon = \frac{1}{d}$, then it is easy to see that 
	$P_{\#} \subseteq Q_{\#}$  iff 
	$P_{\#} \subseteq Q' + d\cdot\varepsilon$.
	Therefore, $P \subseteq Q$ iff 
	$P_{\#} \subseteq Q' + d\cdot\varepsilon$.
	
	Next, we prove that every instance of over-approximation of DS inclusion can be reduced to an instance of DS inclusion. 
	Let $P$ and $Q $ be DS automata with discount factor $d>1$, and let $\varepsilon>0$ be its approximation factor. Suppose $P \subseteq Q + d\cdot \varepsilon$ holds. Let $d\cdot\varepsilon = \frac{r}{s}$. Generate $P_{\#}$ and  $Q_{\#}$ as earlier. Since $P \subseteq Q + d\cdot \varepsilon$  holds,
	we get that $P_{\#} \subseteq Q_{\#} +  \varepsilon$  holds. Modify $Q_{\#}$ to $Q'$ so that the weight of the transition from its inital state is now $\varepsilon$. Suppose $\varepsilon = \frac{m}{n}$ for natural numbers $m,n>0$. Then, 
	Multiply the weight of all edges in $P_{\#}$ and $Q'$ with $n$ to obtain new DS automata $R$ and $S$, respectively. Then, it is easy to see that $P \subseteq Q + d\cdot \varepsilon$ holds then $R \subseteq S$ holds. 
	
	DS inclusion is decidable iff under approximation of DS is decidable: 
	Similar to the previous proof except that the ``+" will be replace by ``-".	
	\qed 
\end{proof}

Given these challenges, we propose an anytime algorithm for DS inclusion. The anytime algorithm either terminates after a finite amount of time with a crisp $\mathsf{True}$ or $ \mathsf{False}$ answer to DS inclusion, or it continuously generates  $d\cdot\varepsilon$-close approximations, where the approximation factor $0<\varepsilon<1$ decreases in time. If an execution of the algorithm is interrupted at anytime before a natural termination, then it returns the most recently computed $d\cdot\varepsilon$-close approximation. 

\begin{algorithm}[t]
	\caption{ $\approxInc(P,Q,d,\varepsilon)$\\\textsf{Inputs:} DS automata $P$, $Q$, discount factor $1<d <2$, and approximation factor $0<\varepsilon<1$ }
	\label{Alg:approxInclusion}
	\begin{algorithmic}[1]
		
		\IF {$\lowerInc(P,Q,d,\varepsilon)$ returns $P \subseteq Q = \mathsf{False}$ }
	\RETURN $P\subseteq Q = \mathsf{False}$ 
	\ENDIF 
	
	
	\IF {$\upperInc(P,Q, d, \varepsilon)$ returns $P\subseteq Q = \mathsf{True}$}
	
	\RETURN $P\subseteq Q = \mathsf{True}$
	\ENDIF
	
	\STATE {\textbf{if} Interrupt \textbf{then return} $P\subseteq Q + d\cdot\varepsilon= \mathsf{True}$} 
	
	\STATE $\approxInc(P,Q,d, \frac{\varepsilon}{2})$
	
\end{algorithmic}
\end{algorithm}

Algorithm~\ref{Alg:approxInclusion} outlines our anytime procedure.
On receiving DS automata $P$  and $Q$ with discount factor $1<d<2$, the algorithm 
invokes $\approxInc$ with an initial approximation factor $0<\varepsilon_{\mathsf{init}}<1$. As is clear from Algorithm~\ref{Alg:approxInclusion}, $\approxInc$ is a tail recursive procedure in which the approximation factor is halved in each new invocation. In the invocation with approximation factor $0<\varepsilon<1$, $\approxInc$ calls two functions $\lowerInc$ and $\upperInc$. 
Ideally, these procedures would  over- and under- approximate DS inclusion using  $d\cdot\varepsilon$-close  and $d\cdot\varepsilon$-far, respectively. Unfortunately, that is not possible due to Theorem~\ref{thrm:reducetoOver}.
Therefore, a challenge here is the design of these two subprocedures. 
At this point it is sufficient to know that in response to the challenge we design $\lowerInc$ so that it combines partial solutions of DS inclusion and $d\cdot\varepsilon$-close approximation. Specifically, given approximation factor $\varepsilon>0$, its outcomes are either  $P \subseteq Q = \mathsf{False}$ or $P \subseteq Q + d\cdot\varepsilon  = \mathsf{True}$. Similarly, the algorithm $\upperInc$ is designed to  return either $P \subseteq Q = \mathsf{True}$ or $P \subseteq Q - d\cdot\varepsilon = \mathsf{False}$, hence combining DS inclusion and  $d\cdot\varepsilon$-far approximation.
The algorithm design for $\lowerInc$ and $\upperInc$ use regular comparator automata for aggregate functions that represent the lower and upper approximations of discounted sum, respectively. These subprocedures have been presented in detail in  \textsection~\ref{Sec:lowerInc} and \textsection~\ref{Sec:upperInc}, respectively. 


Equipped with descriptions of   $\lowerInc$ and $\upperInc$, we can finally describe $\approxInc$ (Algorithm~\ref{Alg:approxInclusion}). 
Without loss of generality, suppose that $\approxInc$ can be interrupted only on completion of  $\upperInc$. 
Consider the invocation with approximation factor $0<\varepsilon<1$. 
First, $\lowerInc$ will be called. If it returns $P \subseteq Q = \mathsf{False}$, then $\approxInc$ is terminated, and it returns the crisp outcome that $P \subseteq Q = \mathsf{False}$. Otherwise, $\lowerInc$ must have returned $P\subseteq Q + d\cdot\varepsilon = \mathsf{True}$, i.e. the $d\cdot\varepsilon$-close approximation holds.
Therefore, if $\approxInc$ is interrupted here onward, it can return this approximate result. 
But if $\approxInc$ is not interrupted,  it proceeds to solve $\upperInc$. If $\upperInc$ returns $P \subseteq Q = \mathsf{True}$, once again $\approxInc$ is terminated, and the crisp $P\subseteq Q = \mathsf{Ture}$ solution is returned.  
If $\approxInc$ is interrupted at this point, the algorithm returns the $d\cdot\varepsilon$-close approximation result obtained from $\lowerInc$. Finally, if the algorithm has not been interrupted yet, $\approxInc$ is invoked with lower approximation factor $\frac{\varepsilon}{2}$.
That completes the description of our anytime algorithm. 

Finally, to see why this algorithm is co-recursively enumerable, observe that if $P \subseteq Q = \mathsf{False}$, then there must be an approximation factor $0<\gamma<1$ such that for all $0<\delta < \gamma$, $P \subseteq Q + d\cdot\delta = \mathsf{False}$. Therefore, as the approximation factor is halved in each invocation of $\approxInc$, it will eventually be smaller than the aforementioned $\gamma$. When this happens, then subprocedure $\lowerInc$ will be forced to return $P \subseteq Q = \mathsf{False}$. Hence, if $P \subseteq Q = \mathsf{False}$, then the anytime algorithm will necessarily terminate.

\section{Algorithm $\lowerInc$}
\label{Sec:lowerInc}

This section describes  Algorithm $\lowerInc$.
Recall, given inputs DS automata $P$ and $Q$, discount factor $1<d<2$ and approximation factor $0<\varepsilon<1$, $\lowerInc(P,Q,d,\varepsilon)$ either returns $P \subseteq Q$ does not hold or  $P\subseteq Q + d\cdot\varepsilon$ holds. 
Note that these outcomes are not {\em mutually exclusive}, i.e., there exist input instances for which both of the outcomes may hold. In these cases, the algorithm may return either of the outcomes; the procedure will still be sound.

Intuitively,
$\lowerInc$ solves whether  $P$ is $f$-included in $Q$, where $f$ is an aggregate function that approximates the discounted-sum from below. 
Let this aggregate function, denoted $\mathsf{DSLow}$, be defined such that  given a weight-sequence $W$, discount factor $1<d<2$ and approximation factor $0<\varepsilon<1$,  $0\leq \DSum{W}{d} - \DSumL{W} < d\cdot \varepsilon$ holds. Then, we argue that if $P$ {\em is} $\mathsf{DSLow}${\em-included in} $Q$, then 
$P\subseteq Q + d\cdot\varepsilon$ holds. Otherwise, if $P$ {\em is not} $\mathsf{DSLow}${\em-included in} $Q$, then $P \subseteq Q$ will not hold. Therefore, $\lowerInc$ solves $\mathsf{DSLow}$-inclusion. We use techniques from regular comparators~\cite{BCVFoSSaCS18} to solve $\mathsf{DSLow}$-inclusion. 

{\bf Organization and notation.} First, the lower approximation of discounted-sum $\mathsf{DSLow}$ is formally defined in \textsection~\ref{Sec:lowerApprox}. Second, a regular comparator for $\mathsf{DSLow}$ is constructed in \textsection~\ref{Sec:lowerComparator}. Finally, 
we use the regular comparator to design $\lowerInc$ \textsection~\ref{Sec:lowerIncAlgo}.
Let $k,p>0$ be positive rationals
such that the discount factor $1<d<2$ and approximation factor $0<\varepsilon<1$ are expressed as $d = 1+2^{-k}$ and $\varepsilon=2^{-p}$, respectively.

\subsection{Lower approximation of discounted-sum}
\label{Sec:lowerApprox}

This section defines lower approximation of discounted sum  when $1<d<2$. A consideration while defining the aggregate function is that its comparator should be regular. 
Therefore, our definition of lower approximation of discounted sum is motivated from the  notion of {\em recoverable gap}~\cite{boker2014exact} which is known to play an important role in guaranteeing regularity of comparators~\cite{BCVlmcs2019}. 

The {\em recoverable gap} of a weight sequence $W$ w.r.t discount factor $d>1$ is  $d^{|W|-1}\cdot\DSum{W}{d}$. In other words,  it is the normalized DS of a weight sequence.
Intuitively, the recoverable gap of a weight sequence gives a measure of how far its  discounted-sum is from 0, and hence is a building block for designing the comparator automata. 
A property of recoverable gap that results in the regularity of comparator for DS with integer discount factor is that 
the minimum non-zero difference between the recoverable gap of sequences is fixed. 
Specifically, this difference is 1 when the discount factor is an integer. 
But for non-integer discount factors, this difference can become arbitrarily small~\cite{akiyama2005representation}. 
This explains why DS comparator are not regular for non-integer discount factors.

To this effect, we begin by defining an approximation of the recoverable gap such that the aforementioned difference is fixed under the new definition. 
This is guaranteed by rounding-off the recoverable gap to a fixed {\em resolution} $r = (d-1)\cdot \varepsilon = 2^{-(p+k)}$, where $d=1+2^{-k}$ is the  discount factor  and $\varepsilon = 2^{-p}$ is the approximation factor. 
Formally, let $\roundL{x}$ denote the largest integer multiple of the resolution that is less than or equal to $x$, for $x \in \Re$. 
Then, $\roundL{x} = i\cdot 2^{-(p+k)}$ for an integer $i \in \Z$  such that for all integers $j\in \Z$ for which $j\cdot 2^{-(p+k)} \leq x$, we get that $j\leq i$. Then, 

\begin{lemma}
	\label{lem:LowerRound}
	Let $k,p >0$ be rational-valued parameters. Then,
	for all real values $x\in \Re$, $0\leq x - \roundL{x} < 2^{-(p+k)}$.
\end{lemma}
\begin{proof}
	There exists a unique  integer $i \in \Z$ and $0\leq b  < 2^{-(p+k)}$ such that $x = i\cdot 2^{-(p+k)} + b$. Then, $ \roundL{x} = i\cdot 2^{-(p+k)}$. Therefore, we get that $0\leq x-\roundL{x} < 2^{-(p+k)}$.
	\qed
\end{proof}

\begin{lemma}[Monotonicity]
	\label{Lem:LowerMono}
	Let $k,p >0$ be rational-valued parameters. Then,
	if $x\leq y$, then $\roundL{x} \leq \roundL{y}$. 
\end{lemma}
\begin{proof}
	There exist unique integers $i,j,\in\mathbb{Z}$, and positive values $0\leq a,b < 2^{-(p+k)}$ such that
	$ x = i\cdot 2^{-(p+k)} + a$ and $ y = j\cdot 2^{-(p+k)} + b$. 
	By definition of $\mathsf{roundLow}$, $\roundL{x} = i\cdot 2^{-(p+k)}$ and $\roundL{y} = j\cdot 2^{-(p+k)}$. 
	Then, if $x \leq y$ then one of the two must have occurred:
	\begin{itemize}
		\item $i < j$. In this case, $\roundL{x} < \roundL{y}$.
		\item $i = j$ and $a \leq b$. In this case, $\roundL{x} = \roundL{y}$
	\end{itemize}
	Therefore, if $x \leq y$ then $\roundL{x} \leq \roundL{y}$.
	\qed
\end{proof}

Then, for all real values $x \in \Re$, $0\leq x - \roundL{x} < 2^{-(p+k)}$.
Then, {\em lower gap} is defined as follows:

\begin{Def}[Lower gap]
	\label{def:lowergap}
	Let $k,p>0$.
	Let $W$ be a finite weight sequence. The {\em lower  gap} of $W$ with discount factor $d = 1+2^{-k}$ and approximation factor $\varepsilon=2^{-p}$, denoted $\gapL{W}$, is 
	\[
	\gapL{W} = 
	\begin{cases}	
	0, \text{ for } |W| = 0 \\
	\roundL{\gapL{U}+u} \text{ for } W = U\cdot u
	\end{cases}
	\]
\end{Def}

Note that the minimum non-zero difference between the lower gap of weight sequences is the resolution $r = 2^{-(p+k)}$.
Similar to the relationship between recoverable gap and DS, lower
approximation of DS is defined as follows:

\begin{Def}[Lower approximation of discounted-sum]
	\label{def:approxDSLow}
	Let $k,p>0$.
	Let $W$ be a finite weight sequence.
	The {\em lower approximation of discounted sum}, called lower DS, for weight sequence $W$ with discount  factor $d = 1+2^{-k}$ and approximation factor $\varepsilon=2^{-p}$ is denoted by and defined as 
	$$\DSumL{W} = {\gapL{W}}/{d^{|W|-1}}$$
\end{Def}

To complete the definition, we prove that the value computed by $\mathsf{DSLow}$  in Definition~\ref{def:approxDSLow} corresponds to a value close to the discounted-sum. To prove that we first establish the following"
Given discount factor $d = 1+2^{-k}$ and approximation factor $\varepsilon = 2^{-p}$, a {\em resolution sequence of length $n>0$}, denoted $R_n$, is the $n$-length sequence in which all elements are $r= 2^{-(p+k)}$.

\begin{lemma}
	\label{lem:LowerGapDifBounded}
	Let $k,p>0$ be rational-valued parameters. Let $d = 1+2^{-k}$ be the non-integer, rational discount factor and $\varepsilon=2^{-p}$ be the approximation  factor. Let $W$ be a finite non-empty weight sequence.
	Then $0\leq \gap{W} - \gapL{W} < \gap{R_{|W|}}$.
\end{lemma}
\begin{proof}
	We prove the above by induction on the length sequence $W$.
	
	Base case: When $|W| = 1$. Let $W = w_0$.
	In this case, $\gap{W} = w_0$ and $\gapL{W}  = \roundL{w_0}$. Then, from Lemma~\ref{lem:LowerRound} we get that $0\leq \gap{W} - \gapL{W} < r$, which in turn is the same as $0\leq \gap{W} - \gapL{W} < \gap{R_1}{d}$.
	
	Inductive hypothesis: For all weight-sequences $W$ of length $n\geq1$,  it is true that $0\leq \gap{W} - \gapL{W} < \gap{R_n}$.

	Induction step: We extend this result to weight-sequences of length $n+1$. Let $W$ be an $n+1$-length weight-sequence. Then $W = W[n] \cdot w_n$, where $W[n]$ is the $n$-length prefix of $W$ and $w_n$ is $n+1$-th element.
	
	We first show that $\gap{W} - \gapL{W}\geq 0$:
	\begin{align*} 
		& \gap{W} - \gapL{W} \\
		= &  d\cdot \gap{W[n]} + w_n - \roundL{d\cdot\gapL{W[n]} + w_n }\\
		& \text{ Using monotonicity of } \mathsf{roundLow} \text{ and the inductive hypothesis, we get} \\
		\geq &     d\cdot \gap{W[n]} + w_n - \roundL{d\cdot\gap{W[n]} + w_n } \\
		& \text{ From Lemma~\ref{lem:LowerRound}, we get the desired result.} 
	\end{align*} 
	
	Next, we show that $\gap{W} - \gapL{W} < \gap{R_{n+1}}$.
	\begin{align*} 
		& \gap{W} - \gapL{W} \\
		= &  d\cdot \gap{W[n]} + w_n - \roundL{d\cdot\gapL{W[n]} + w_n }\\
		& \text{ From Lemma~\ref{lem:LowerRound}, we get} \\
		< &     d\cdot \gap{W[n]} + w_n - {(d\cdot\gapL{W[n]} + w_n)} + 2^{-(p+k)} \\
		= &   d\cdot \gap{W[n]}  - {d\cdot\gapL{W[n]}} + 2^{-(p+k)} \\
		& \text{ From the inductive hypothesis, we get} \\
		< & d\cdot\gap{R_n} + 2^{-(p+k)} \text{ where } R_n \text{ is the } n\text{-length resolution sequence} \\
		= & \gap{R_{n+1}} \text{ where } R \text{ is the } n+1\text{-length resolution sequence}
	\end{align*} 
	This completes both sides of the proof. 
	\qed
\end{proof}

\begin{theorem}
	\label{thrm:ApproxDSLower}
	Let $d = 1+2^{-k}$ be the  discount factor and $\varepsilon=2^{-p}$ be the approximation  factor, for rationals $p,k>0$. 
	Then for all weight sequences $W$, 
	$0\leq \DSum{W}{d} - \DSumL{W} < d\cdot \varepsilon$.
\end{theorem}
\begin{proof}
	The statement clearly holds when $|W| = 0$ since $\DSum{W}{d} = \DSumL{W} =0$.
	For $|W|>n$, we have proven that $0\leq\gap{W}  -\gapL{W}< \gap{R_{|W|}}$, where $R_{|W|}$  is the $|W|$-length sequence in which all elements are equal to $r = 2^{-(p+k)}$ (Lemma~\ref{lem:LowerGapDifBounded}). 
	Finally, division by $d^{|W|-1}$ completes the proof.
	
	The complete details are as follows: 
		When $|W| = 0$, $\DSum{W}{d} = \DSumL{W} = 0$, since $\gap{W} = \gapL{W} = 0$. Therefore, $0\leq \DSum{W}{d} - \DSumL{W} < d\cdot 2^{-p}$ holds when $|W| = 0$.
		
		Otherwise, from Lemma~\ref{lem:LowerGapDifBounded}, we get that $0\leq \gap{W} - \gapL{W} < \gap{R_{n}}$, where $n = |W|$.
		On division by $d^{n-1} $, we get that $0\leq \DSum{W}{d} - \DSumL{W} < \DSum{R}{d}$. Now $\DSum{R}{d} \leq \DSum{R_\infty}{d}$, where  $R_\infty$ is the $\infty$-length resolution sequence. Now, $\DSum{R_\infty}{d} = \frac{2^{-(p+k)}\cdot d}{d-1}  = \frac{(d-1)\cdot\varepsilon \cdot d}{d-1}< \varepsilon\cdot d$. Therefore, we get the desired result that
		$0\leq \DSum{W}{d} - \DSumL{W} < d\cdot\varepsilon$.
		
	\qed
\end{proof}

Therefore, the lower approximation of DS is well defined in this section. 

\subsection{Comparator for lower approximation of DS}
\label{Sec:lowerComparator}

This section covers the construction of the comparator automaton for the lower approximation of discounted-sum from Defintion~\ref{def:approxDSLow}. We show that the comparator is regular by explicitly constructing its DFA. The construction will utilize the fixed non-zero minimum property of the lower gap.

We begin with formal definitions of the comparison language and comparator automata for lower DS.

\begin{Def}[Comparison language for lower approximation of DS]
	\label{def:comparisonAutLow}
	Let $\mu>0$ be an integer bound, and $k,p$ be positive integers.
	The {\em comparison language for lower approximation of discounted sum} with discount factor $d = 1+2^{-k}$, approximation factor $\varepsilon=2^{-p}$, upper bound $\mu$ and inequality relation $\mathsf{R} \in \{\leq, \geq \}$ is a language of finite weight sequences $W$ over the alphabet $\Sigma = \{-\mu,\dots, \mu\}$ that accepts $W$ iff $\DSumL{W}$ $\mathsf{R}$ $0$ holds. 
\end{Def}

\begin{Def}[Comparator automata for lower approximation of DS]
	\label{def:comparatorAutLow}
	Let $\mu>0$ be an integer bound, and $k,p$ be positive integers.
	The {\em comparator automata for lower approximation of discounted sum} with  discount factor $d = 1+2^{-k}$, approximation factor $\varepsilon=2^{-p}$, upper bound $\mu$ and inequality relation $\mathsf{R} \in \{\leq, \geq \}$ is an automaton that accepts 
	the corresponding comparison language. 
\end{Def}

Next, we construct a DFA for the comparator for lower DS. 

The first observation towards the construction is that $\DSumL{W}$ $\mathsf{R}$ $0$ iff $\gapL{W}$ $\mathsf{R}$ $0$, for all  finite weight sequences $W$.
Therefore, it is sufficient to construct a DFA that accepts weight sequence $W$ iff $\gapL{W}$ $\mathsf{R}$ $0$.
We achieve this by 
(a). Creating one state of the DFA for every possible value of lower gap.
(b). Note that the definition of lower gap (Definition~\ref{def:lowergap}) is inductive on the length of the weight sequence. So, transitions between states is defined so that they obey the inductive definition. 
For (a), note that  lower gap are of the form $i\cdot r$ where $i \in \Z$. Therefore, for all $i \in \Z$, we introduce a state $i$ to represent the lower gap  $i \cdot r$.
For (b). we include a transition from state $i$ to state $j$ on symbol $a \in \Sigma = \{-\mu,\dots, \mu\}$ iff $j\cdot r = \gapL{i\cdot r + a}$, thereby following Definition~\ref{def:lowergap}.
Note that this equation makes the transition relation  deterministic as for all $i\in\Z$ and $a \in \Sigma$ there is a unique $j\in \Z$ that satisfies it.
Since $\gapL{W} = 0$ when $|W| = 0$, state 0 is made the initial state. Finally, a state $i$ is an accepting state   iff $i$ $\mathsf{R}$ $0$ holds.

The automaton created above has infinitely many states as every possible value of lower gap corresponds to a state. 
To obtain finitely many states, we show that it is sufficient to consider finitely many values of the lower gap:

\begin{lemma}[Bounds on lower gap-value]
	\label{lem:compartorLowerThresholds}
	Let $\mu>0$ be an integer bound. 
	Let $k,p$ be positive integers  s.t. $d = 1+2^{-k}$ is the discount factor, and $\varepsilon=2^{-p}$ is the approximation factor.
	Let $W$ be a finite and bounded weight sequence. 
	\begin{enumerate}
		\item If $\gapL{W}\leq -\mu \cdot 2^{k}$ then for all $u \in \{-\mu, \dots, \mu\}$, $\gapL{W\cdot u}\leq -\mu \cdot 2^{k}$.
		\item If $\gapL{W} \geq \mu \cdot 2^{k} + 	2^{-p}$ then for all $u \in \{-\mu, \dots, \mu\}$, $\gapL{W\cdot u} \geq \mu \cdot 2^{k}+ 	2^{-p}$.
	\end{enumerate}
\end{lemma}

\begin{proof}
	Proof for (1.). 
	Recall, $\gapL{W\cdot u} = \roundL{d\cdot\gapL{W} + u}$. From the definition of $\mathsf{roundLow}$, we get that 
	$\gapL{W\cdot u} \leq d \cdot \gapL{W}+ u $. Since $\gapL{W}\leq -\mu \cdot 2^{k}$ holds, 
	we get  $\gapL{W\cdot u} \leq d \cdot (-\mu\cdot 2^{k})+ u  = (1+ 2^{-k}) \cdot(-\mu\cdot 2^{k}) + u = -\mu \cdot 2^{k}  - \mu + u $. Since $ u$ is at most $\mu$, we get that $\gapL{W\cdot u} \leq -\mu \cdot 2^{k}$.
	
	Proof of (2.) follows similarly, and hence has been omitted. 
	\qed
\end{proof}

The outcome of Lemma~\ref{lem:compartorLowerThresholds} is that it is sufficient to track the lower gap value only when it is between $-\mu \cdot 2^{k}$ and $ \mu\cdot2^{k}+ 	2^{-p}$ in the construction. 

\subsubsection{Construction}
Let $\mu>0$,  $d = 1+2^{-k}$, $\varepsilon=2^{-p}$  be the upper bound, discount factor and approximation factor, respectively.
Let $\thresh_l$ be the largest integer such that $\thresh_l\cdot 2^{-(p+k)} \leq -\mu \cdot 2^{k}$ (Lemma~\ref{lem:compartorLowerThresholds}- Part 1). Let $\thresh_u$ be the smallest integer such that
$\thresh_u \cdot 2^{-(p+k)} \geq \mu\cdot 2^{k} + 2^{-p}$ (Lemma~\ref{lem:compartorLowerThresholds}
- Part 2).
For relation $\mathsf{R} \in \{\leq, \geq\}$,
construct DFA $\compL(\mu,k,p,\mathsf{R}) = (\State, \Start, \Sigma, \delta, \Final)$ as follows:
\begin{itemize}
	\item  $\State = \{\thresh_l, \thresh_l+1,\dots,  \thresh_u \}$,
	$\Start = \{0\}$ and
	$\Final = \{i | i\in S \text{ and $i$ $\mathsf{R}$ $0$}\} $
	
	\item Alphabet $\Sigma = \{-\mu, -\mu+1,\dots, \mu-1, \mu\}$
	\item Transition function $\delta\subseteq \State \times \Sigma \times \State$ where $(s,a,t) \in \delta$ then:
	\begin{enumerate}
		\item \label{Trans:SelfLoop} If $s = \thresh_l$ or $s= \thresh_u$, then $t = s$ for all $a \in \Sigma$
		\item Else, let $\roundL{d\cdot s \cdot 2^{-(p+k)} + a} = i \cdot 2^{-(p+k)}$ for $i\in\Z$
		\begin{enumerate}
			\item \label{Trans:IntState} If $\thresh_l \leq i \leq \thresh_u$, then $t = i$
			
			\item \label{Trans:leq} If $i> {\thresh_u}$, then $t = \thresh_u$
			\item \label{Trans:geq} If $i< {\thresh_l}$, then $t = \thresh_l$
		\end{enumerate}
		
	\end{enumerate}
\end{itemize}
\begin{theorem}
	\label{thrm:Comparatorlower}
	Let $\mu>0$ be and integer upper bound. Let $k,p>0$ be rational parameters s.t. $d = 1+2^{-k}$ is the  discount factor and $\varepsilon=2^{-p}$ is the approximation parameter. 
	DFA $\compL(\mu,k,p,\mathsf{R}) $ accepts a finite weight sequence $W \in \Sigma^*$ iff $\DSumL{W}$ $\mathsf{R} $ $ 0$.
	DFA $\compL(\mu,k,p,\mathsf{R}) $ has $\O(\mu\cdot 2^{2k+p})$ states.
\end{theorem}
\begin{proof}
	The proof shows that the final state of a the run of a word represents its lower gap value. 
	For this we show  three things: Let $s_f$ be the final state of the run. (a). if $\thresh_l < s_f < \thresh_u$ then its lower gap value is $ s_f \cdot r$, (b). if $\thresh_l \geq s_f $ then the lower gap value is less  than or equal to $\thresh_l\cdot r$, and (c).  if $\thresh_u \leq s_f $ then the lower gap value is greater than or equal to $\thresh_u\cdot r$.
	
	The proof is very similar to Theorem~\ref{Thrm:DFABad}, and hence its details have been left for the reader to fill in. 
	\qed
\end{proof}

\subsection{Algorithm details for $\lowerInc$}
\label{Sec:lowerIncAlgo}

This section describes $\lowerInc$. Recall, given DS automata $P$ and $Q$, discount factor $d=1+2^{-k}$ and approximation factor $2^{-p}$, $\lowerInc$ returns either $P \subseteq Q = \mathsf{False}$ or $P \subseteq Q + d\cdot \varepsilon = \mathsf{True}$. In cases where both outcomes are possible, $\lowerInc$ may return either of the outcomes. In our design of $\lowerInc$, it performs $\mathsf{DSLow}$-inclusion between the DS automaton, as justified in Lemma~\ref{lem:algLowerApproxDS}-~\ref{lem:algLowerApprox}. Subsequently, the algorithm uses the regular comparator for $\mathsf{DSLow}$ to design its inclusion procedure (Algorithm~\ref{Alg:LowerInclusion}). Lastly, we illustrate a property of $\lowerInc$ that is necessary in our anytime procedure for DS inclusion to become co-recursively enumerable (Theorem~\ref{thrm:lowerIncNegCases}).  

Terminology: A run $\rho_P$ of word $w$ in $P$  is said to be  {\em dominated} by $Q$ if there exists a run $\rho_Q$ in $Q$ on the same word such that $\DSumL{\rho_P - \rho_Q} \leq 0$. 
\begin{lemma}
	\label{lem:algLowerApproxDS}
	Given DS automata $P$ and $Q$, discount factor $d = 1+2^{-k}$  and approximation factor $\varepsilon = 2^{-p}$ for rational values $k,p>0$. 
	\begin{enumerate}
		\item If all runs in $P$ are dominated by $Q$, then $P \subseteq Q + d\cdot \varepsilon$ holds.
		
		\item If there exists a run in $P$ that is not dominated by $Q$, then $P \subseteq Q$ does not hold.
	\end{enumerate}
\end{lemma}
\begin{proof}
	Proof of (1.): Let for all words $w \in \Sigma^{*}$, for all runs of $w$ $\rho_P \in P$, there exists a run of $w$ $\rho_Q \in Q$ such that $\mathsf{DSLow}(\rho_P - \rho_Q, k,p) \leq 0$ be true. Then
	$\DSumL{\rho_P - \rho_Q} \leq 0$ implies that  $\DSum{\rho_P - \rho_Q}{d} \leq d \cdot \varepsilon \equiv \DSum{\rho_P}{d} \leq \DSum{\rho_Q}{d} + d\cdot\varepsilon$. Since weight of a word is given by the maximum weight of  its all runs, we get that for all word $w \in\Sigma^*$, $\wt_P(w) < \wt_Q(w)+d\cdot\varepsilon$. Therefore, 
	$P \subseteq Q + d\cdot\varepsilon$ holds.

	Proof of (2.): Let $w\in\Sigma^*$ be the word for which there exists a run of $w$ $   \rho_P\in P$ such that for all runs of $w \rho_Q \in Q$,  $\mathsf{DSLow}(\rho_P - \rho_Q, k,p) > 0$ holds. 
	$\mathsf{DSLow}(\rho_P - \rho_Q, k,p) > 0$  implies $\DSum{\rho_P-\rho_Q}{d} >0 \equiv \DSum{\rho_P}{d} > \DSum{\rho_Q}{d}$. 
	Since weight of a word is given by the maximum weight of  its all runs, we get that there exists word $w \in\Sigma^*$, $\wt_P(w) > \wt_Q(w)$. So, $P \subseteq Q$ does not hold. \qed
\end{proof}

Therefore,  $\lowerInc(P,Q,d,\varepsilon)$ is designed such that it returns $\mathsf{True}$ iff all runs in $P$ are dominated by $Q$. To attain this, we construct NFA $\mathsf{dominated}$ in Algorithm~\ref{Alg:LowerInclusion} so that it  contains all runs in $P$ that are dominated by $Q$. This construction utilizes the comparator automata for lower DS. Lastly, language inclusion between $\mathsf{dominated}$ and NFA $\hat{P}_{-wt}$, that consists of all runs of $P$, determines if all runs of $P$ are dominated or not. 
Recall, currently we assume all weights in the DS automata are non-negative integers. As a result, the upper bound for comparator construction is set to $\mu$, where $\mu>0$ is the maximum weight along all transitions in both DS automata. In case the weights along transitions are non-negative, then the comparator will be constructed with upper bound $2\cdot\mu>0$, where $\mu$ is the maximum of absolute values of weight along all transitions in both DS automata. The rest will be identical.

\subsubsection{Algorithm Details}
For DS automaton $P$, procedure $\AugmentWtAndLabel(P)$ generates an NFA $\hat{P}$ by converting transition $s\xrightarrow{a}t$ with weight $wt$ and unique transitions label $l$ in $P$ to a transition $s\xrightarrow{a, wt,l}t$ in $\hat{P}$ (Line~\ref{alg-line:AugmentP}).  
Procedure $\MakeDifProduct(\hat{P},Q)$ generates the product of NDA $\hat{P}$ with DS automaton $Q$ in an NDA $\hat{P}-Q$ such that it also records the difference of weight of transitions. Therefore, if $s\xrightarrow{a,wt_1, l}t$ and $p\xrightarrow{a} q$ with weight $wt_2$ are transitions in $\hat{P}$ and $Q$, respectively, then $(s,p)\xrightarrow{a, wt_1 - wt_2, l}(t,q)$ is a transition in $\hat{P}-Q$.
All states in NFA $\hat{P}-Q$ are accepting states. 
From \textsection~\ref{Sec:lowerApprox} we know that $\mathsf{approxDSComp}$ is the comparator for lower approximation of DS with upper bound $\mu$, discount factor  $d$, approximation factor $\varepsilon$ and relation $\leq$.
NDA $\dominatedWitness$ forms the intersection of $\hat{P}-Q$ with the comparator by matching the weight-component in $\hat{P}-Q$ with the weight-alphabet in the comparator. Therefore, if $s\xrightarrow{a, wt, l} t$ and $p\xrightarrow{wt} q$ are transitions in $\hat{P}-Q$ and the comparator, respectively, then $(s,p)\xrightarrow{a, wt, l}(t,q)$ is a transition in the intersection. Furthermore, a state $(r,s)$ is accepting iff both $r$ and $s$ are accepting states in their respective NFA.
Finally, $\dominated$ and $\hat{P}_{-wt}$ are obtained by projecting out the weight component from $\dominatedWitness$ and $\hat{P}$, respectively. Specifically, if $s\xrightarrow{a,wt,l}t$ is a transition in the original automaton, then $s\xrightarrow{a,l}t$ is a transition in the final automata. Here, $\hat{P}_{wt}\subseteq \dominated$ refers to language inclusion between the two NFAs.

\begin{algorithm}[t]
	\caption{ $\lowerInc(P,Q, d,\varepsilon)$\\\textsf{Inputs:} DS automata $P$, $Q$, discount factor $1<d<2$, approximation factor $0<\varepsilon<1$} \label{Alg:LowerInclusion}
	\begin{algorithmic}[1]
		\STATE $\hat{P} \leftarrow  \AugmentWtAndLabel(P)$ \label{alg-line:AugmentP}		
		\STATE $\hat{P}-{Q} \leftarrow \MakeDifProduct(\hat{P}, {Q})$ \label{alg-line:ProdDiff}		
		\STATE $\dominatedWitness \leftarrow  \cct{Intersect}(\hat{P}-{Q}, \compL(\mu,\log\frac{1}{d-1}, \log\frac{1}{\varepsilon}, \leq))$, where $\mu$ is the maximum of the absolute value of weights in  $\hat{P}-{Q}$ \label{alg-line:Intersect}		
		\STATE $\dominated \leftarrow \cct{Project}(\dominatedWitness)$ \label{alg-line:Project}
		\STATE {$\mathsf{aux} \leftarrow \hat{P}_{-wt} \subseteq_{\mathsf{LI}} \dominated$ \hfill{// $\subseteq_\mathsf{LI}$ refers to Language inclusion}}
		\label{alg-line:ensure}			
		\IF {$\mathsf{aux} $}
		\RETURN $P \subseteq Q + d\cdot\varepsilon = \mathsf{True}$
		\ELSE \RETURN $P \subseteq Q = \mathsf{False}$
		\ENDIF
	\end{algorithmic}
\end{algorithm}

\begin{lemma}
	\label{lem:algLowerApprox}
	Given DS automata $P$ and $Q$, discount factor $d = 1+2^{-k}$  and approximation factor $\varepsilon = 2^{-p}$ for rational values $k,p>0$, $\lowerInc(P,Q,d,\varepsilon)$ returns $\mathsf{True}$ iff all runs in $P$ are dominated by $Q$.
\end{lemma}

\begin{proof}
	
	From the algorithm, it is clear that there is a one-one correspondence between words $(w, L) \in \hat{P}_{-wt}$ and runs $\rho_P $ of word $w$ in $P$ for all words $w \in \Sigma^*$.

	From the algorithm, it is also clear that $(w,L) \in \mathsf{dominated}$ iff $w \in \Sigma^*$, with a run $\rho_P \in P$ which has been labelled by $L$ such that there exists a run $\rho_Q \in Q$ of $w$ such that $\DSumL{\rho_P - \rho_Q} \leq 0$.

	Then $\mathsf{aux} == \mathsf{True}$ iff  $\hat{P}_{-wt} \subseteq \mathsf{dominated}$. By definition of language inclusion, this holds iff for all $(w, L) \in \hat{P}_{-wt}$ we get that $(w, L) \in \mathsf{dominated}$.
	From the one-one correspondence between words in $\hat{P}_{-wt}$ and runs in $P$, we get that for all $w \in \Sigma^*$, for all runs $\rho_P \in P$ of $w$, let $(w, L)$ be its correspondence in $\hat{P}_{-wt}$ then $(w,L) \in \mathsf{dominated}$. 
	By the condition under which a word is a member of $\mathsf{dominated}$ we get that
	$w \in \Sigma^*$, for all runs $\rho_P \in P$ of $w$, such that there exists a run $\rho_Q \in Q$ of $w$ such that $\DSumL{\rho_P - \rho_Q} \leq 0$.
	\qed
\end{proof}

\begin{theorem}{\em [ Soundness]}
	\label{thrm:lowerIncSound}
	For all inputs DS automata $P$ and $Q$, discount factor $d = 1+2^{-k}$  and approximation factor $\varepsilon = 2^{-p}$ for rational values $k,p>0$, algorithm  $\lowerInc$ is sound.
\end{theorem}
\begin{proof}
	This follows directly from Lemma~\ref{lem:algLowerApproxDS} and Lemma~\ref{lem:algLowerApprox}. \qed
\end{proof}

\begin{theorem}{\rm [Complexity]}
	\label{thrm:complexityLower}
	Given DS automata $P$ and $Q$, discount factor $d = 1+2^{-k}$  and approximation factor $\varepsilon = 2^{-p}$ for rational values $k,p>0$.
	Let $\mu$ be the absolute value of the largest weight in $P$ and $Q$.
	Then the worst case complexity of $\lowerInc$ is $2^{\O(n)}$ where $n = |P|\cdot |Q| \cdot  \frac{\mu}{(d-1)^2 \cdot \varepsilon}$. 
\end{theorem}
\begin{proof}
	The size of $\mathsf{dominated}$ is $\mathcal{O}(|P|\cdot |Q| \cdot \mathsf{compLow(\mu, \log(\frac{1}{d-1}), \log(\frac{1}{\varepsilon}))})$, which is equal to $\mathcal{O}(|P|\cdot |Q| \cdot \frac{\mu}{(d-1)^2 \cdot \varepsilon})$. Then, the complexity of  $\hat{P}_{-wt} \subseteq \mathsf{dominated}$ is $|P| \cdot 2^{ \mathcal{O}(|P|\cdot |Q| \cdot \frac{\mu}{(d-1)^2 \cdot \varepsilon})}$. This equates to $2^{ \mathcal{O}(|P|\cdot |Q| \cdot \frac{\mu}{(d-1)^2 \cdot \varepsilon } + \log(|P|) }$. Keeping the dominating terms in the exponent, we get the worst-case complexity to be $2^{\mathcal{O}(|P|\cdot |Q| \cdot \frac{\mu}{(d-1)^2 \cdot \varepsilon })}$.
	\qed
\end{proof}

Lastly, we show that if $P \subseteq Q$ does not hold, 
there exists a sufficiently small approximation factor $\varepsilon>0$ such that when $\lowerInc$ is invoked with $\varepsilon$, it returns that $P \subseteq Q$ does not hold.
This property will be crucial in proving co-recursive enumerability of our anytime procedure for DS inlcusion.

\begin{theorem}[\rm Bias]
	\label{thrm:lowerIncNegCases}
	Given DS automata $P$, $Q$,  and discount factor $1<d<2$.
	If $P \subseteq Q = \mathsf{False}$, there exists an approximation factor $0<\varepsilon<1$ such that for all $0<\gamma < \varepsilon$, $\lowerInc(P,Q,d,\gamma)$ returns  $P \subseteq Q = \mathsf{False}$.

\end{theorem}
\begin{proof}
	The core idea is that when $P \subseteq Q$ does not hold, then there must exist a word $w \in \Sigma^*$ such that $\wt(w, P) > \wt(w, Q ) + d\cdot\delta$. Therefore, for a sufficiently low value of $\varepsilon$, $P \subseteq Q + d\cdot \varepsilon = \mathsf{False}$. Then, for these values of $\varepsilon$, $\lowerInc$ will necessarily return $P\subseteq Q = \mathsf{False}$.
	
	Since  $P \subseteq Q = \mathsf{False}$
	there exists a word  $w \in \Sigma^*$ such that $\wt(w,P) = \wt(w,Q) + d\cdot \gamma$ for a rational value $\gamma>0$.
	Since weight of words is computed as the maximum of weight of its runs, there must exist a run $\rho_P$ of $w$ in $P$ such that for all runs $\rho_Q$ of $w$ in  $Q$, we get that such that $\DSum{\rho_P}{d} - \DSum{\rho_Q}{d} > d\cdot\gamma = \DSum{\rho_P - \rho_Q}{d} > d\cdot\gamma$. 
	Let $k,p>0$ be rational values such that $d = 1 + 2^{-k}$ and $\gamma  = 2^{-p}$. From Theorem~\ref{thrm:ApproxDSLower}, we get that  for all $q \geq p+1$
	$\mathsf{DSLow}(\rho_P - \rho_Q, k, q) > d\cdot\frac{\gamma}{2}$. Therefore, from Lemma~\ref{lem:algLowerApproxDS} we get that for all $q\geq p+1$, $\lowerInc(P,Q,d, q) = \mathsf{False}$.
\qed	
\end{proof}

\section{Algorithm $\upperInc$}
\label{Sec:upperInc}

This section describes  Algorithm $\upperInc$ - the second sub-procedures in  our anytime algorithm for DS inclusion. 
Given inputs DS automata $P$ and $Q$, discount factor $1<d<2$ and approximation factor $0<\varepsilon<1$, $\upperInc$ $(P,Q,d,\varepsilon)$ either returns $P \subseteq Q$ holds or  $P\subseteq Q - d\cdot\varepsilon$ does not hold.
As earlier, these outcomes are not {\em mutually exclusive}. In these cases, the algorithm may return either of the outcomes as they are both sound. 

The design of $\upperInc$ follows that of $\lowerInc$ very closely. 
Intuitively, $\upperInc$ solves whether $P$ is $f$-included in $Q$ where aggregate function $f$ is the upper approximation of discounted-sum. Notice how similar this is to the intuition behind $\lowerInc$. As a result,  $\upperInc$ follows the same three stages as earlier:
(a). Define the upper approximation of discounted-sum, (b). Construct its regular comparator, and (c). Use the regular comparator to design $\upperInc$. Each of these individual steps are very similar to those in the previous section. As a result, the critical distinctions are highlighted first, and then the details are given. One may skip the details to avoid repetition. The details are mentioned here for sake of completeness. 

The first distinction is in the definition of the upper approximation of discounted. It is similar to that of the lower approximation for DS except that it makes use of an {\em upper gap}. In turn, the upper gap of a value is defined similar to the lower gap except that the upper gap is  rounded-off to the smallest multiple of $2^{-(p+k)}$ that is greater than or equal to the value, where $p$ and $k$ are defined as earlier. Using a similar vein of reasoning as in \textsection~\ref{Sec:lowerComparator}, the comparator for the upper approximation can also be constructed.
Second, algorithm $\upperInc$ is almost identical to $\lowerInc$ in Algorithm~\ref{Alg:LowerInclusion} except that in Line~\ref{alg-line:Intersect} algorithm $\upperInc$ constructs the regular comparator for the upper approximation of discounted sum. 



Yet, another important distinction between $\lowerInc$ and $\upperInc$ is that if $P\subseteq Q = \mathsf{True}$ holds, then $\upperInc$ cannot guarantee that for a small enough value of the approximation factor $\upperInc$ with return $P\subseteq Q = \mathsf{True}$. The core idea here is that if $P\subseteq Q = \mathsf{True}$ then the difference between words in $P$ and in $Q$ could be arbitrarily small. In particular, for every possible value of the approximation factor, there may be a word for which the difference in its weight in $P$ and $Q$ is smaller than the approximation factor. As a result, the option of $P \subseteq Q - d\cdot\varepsilon = \mathsf{False}$ may get triggered, never returning the outcome that $P \subseteq Q = \mathsf{True}$ holds.

The rest of this section gives all details of $\upperInc$.

\subsection{Upper approximation of discounted-sum}
\label{Sec:upperApprox}

In the first stage we define the upper approximation of discounted-sum so that its recoverable gap obeys the bounded non-zero minimal difference property. 

For a rational number $x \in \mathbb{Q}$, let $\roundU{x}$ denote the smallest integer multiple of resolution that is more than or equal to $x$. 
Formally, $\roundU{x} = i\cdot 2^{-(p+k)}$ for an integer $i \in \Z$  such that for all $j\in \Z$,  $j\cdot 2^{-(p+k)} \geq x$ implies  $i\leq j$. 
The upper gap value and upper approximation of discounted sum are defined as follows:

\begin{lemma}
	\label{lem:UpperRound}
	Let $k,p >0$ be rational-valued parameters. Then,
	for all real values $x\in \Re$, $0\leq  \roundU{x} -x < 2^{-(p+k)}$.
\end{lemma}
\begin{proof}
	There exists a unique  integer $i \in \Z$ and $0\leq b  < 2^{-(p+k)}$ such that $x = i\cdot 2^{-(p+k)} - b$. Then, $ \roundU{x} = i\cdot 2^{-(p+k)}$. Therefore, we get that $0\leq \roundU{x} - x < 2^{-(p+k)}$.
	\qed
\end{proof}

\begin{lemma}[Monotonicity]
	\label{lem:uppermono}
	Let $k,p >0$ be rational-valued parameters. Then,
	if $x\geq y$, then $\roundL{x} \geq \roundL{y}$. 
\end{lemma}
\begin{proof}
	The proof of this is very similar to that of Lemma~\ref{Lem:LowerMono}.
	\qed
\end{proof}

\begin{Def}[Upper gap]
	\label{def:approxDSUpper}
	Let $W$ be a finite weight sequence. The {\em upper  gap} of $W$ with discount factor $d = 1+2^{-k}$ and approximation factor $\varepsilon=2^{-p}$, denoted $\gapU{W}$, is 
	\[
	\gapU{W} = 
	\begin{cases}	
	0, \text{ for } |W| = 0 \\
	\roundU{\gapU{U}+u} \text{ for } W = U\cdot u
	\end{cases}
	\]
\end{Def}

\begin{Def}[Upper approximation of discounted-sum]
	\label{def:approxDSUpper}
	Let $W$ be a finite weight sequence.
	The {\em upper approximation of discounted sum}, called upper DS, for weight sequence $W$ with discount  factor $d = 1+2^{-k}$ and approximation factor $\varepsilon=2^{-p}$ is denoted by and defined as 
	$$\DSumU{W} = \gapU{W}/d^{|W|-1}$$
\end{Def}

Definition~\ref{def:approxDSUpper} is completed by showing the it indeed corresponds to an upper approximation of discounted sum. This requires a basic lemma statement:

\begin{lemma}
	\label{lem:UpperGapDifBounded}
	Let $k,p>0$ be rational-valued parameters. Let $d = 1+2^{-k}$ be the non-integer, rational discount factor and $\varepsilon=2^{-p}$ be the approximation  factor. Let $W$ be a finite non-empty weight sequence.
	Then $0\leq \gapU{W} - \gap{W} < \gap{R_{|W|}}$.
\end{lemma}
\begin{proof}
	The proof argument follows by induction on length of weight sequence $W$. It makes use of Lemma~\ref{lem:UpperRound} and Lemma~\ref{lem:uppermono}, and closely follows the proof presented in 
	Lemma~\ref{lem:LowerGapDifBounded}.
	\qed
\end{proof}

\begin{theorem}
	\label{thrm:ApproxDSUpper}
	Let $d = 1+2^{-k}$ be the  discount factor and $\varepsilon=2^{-p}$ be the approximation  factor, for rationals $p,k>0$. 
	Then for all weight sequences $W$, 
	$0\leq \DSumU{W} - \DSum{W}{d} < d\cdot \varepsilon$.
\end{theorem} 

\begin{proof}
	The proof argument makes use of Lemma~\ref{lem:UpperGapDifBounded} and closely follows that of Theorem~\ref{thrm:ApproxDSLower}.
	\qed
\end{proof}

\subsection{Comparator automata for upper approximation of DS}
\label{Sec:UpperComparator}

This section constructs a regular comparator for the upper DS defined above. The construction here differs from that of the comparator for lower DS in only one aspect - the values of the thresholds within which it is sufficient to track the value of upper gap in. In this section, we define the comparison language and its comparator automata, prove the necessary thresholds and give the complete construction of the comparator.

\begin{Def}[Comparison language for upper approximation of DS]
	\label{def:comparisonAutUpper}
	Let $\mu>0$ be an integer bound, and $k,p$ be positive rationals.
	The {\em comparison language for upper approximation of discounted sum} with discount factor $d = 1+2^{-k}$, approximation factor $\varepsilon=2^{-p}$, upper bound $\mu$ and inequality relation $\mathsf{R} \in \{\leq, \geq \}$ is a language 
	that accepts 
	bounded and finite weight sequence $W  \in \Sigma^*$ iff $\DSumU{W}$ $\mathsf{R}$ $0$ holds. 
\end{Def}

\begin{Def}[Comparator automata for upper approximation of DS]
	\label{def:comparatorAutUpper}
	Let $\mu>0$ be an integer bound, and $k,p$ be positive rationals.
	The {\em comparator automata for upper approximation of discounted sum} with  discount factor $d = 1+2^{-k}$, approximation factor $\varepsilon=2^{-p}$, upper bound $\mu$ and inequality relation $\mathsf{R} \in \{\leq, \geq \}$ is an automaton that accepts 
	the corresponding comparison language. 
\end{Def}

We establish the range of sufficient values for the upper gap. The new bounds are as follows:

\begin{lemma}
	\label{lem:compartorUpperThresholds}
	Let $\mu>0$ be an integer bound. 
	Let $k,p$ be positive rationals  s.t. $d = 1+2^{-k}$ is the discount factor, and $\varepsilon=2^{-p}$ is the approximation factor.
	Let $W$ be a finite and bounded weight sequence. 
	
	\begin{enumerate}
		\item If $\gapU{W}\leq -\mu \cdot 2^{k} - 2^{-p}$ then for all $u \in \{-\mu, \dots, \mu\}$, $\gapU{W\cdot u}\leq -\mu \cdot 2^{k} - 2^{-p}$.
		\item If $\gapU{W} \geq \mu \cdot 2^{k}$, then for all $u \in \{-\mu, \dots, \mu\}$, $\gapU{W\cdot u} \geq \mu \cdot 2^{k}$.
	\end{enumerate}
\end{lemma}

\begin{proof}
	Part 1. 
	Let $W$ and $u$ be as defined above. 
	Then 
	$\gapU{W\cdot u} = \roundU{d\cdot \gapU{W} +  u}$. From Lemma~\ref{lem:UpperRound}, we get that
	$\gapU{W\cdot u} \leq d \cdot \gapU{W}+ u + 2^{-(p+k)}$.
	From our assumption, we further get that $\gapU{W\cdot u} \leq d \cdot -(\mu\cdot 2^{k}+2^{-p} )+ u  = (1+ 2^{-k}) \cdot-(\mu\cdot 2^{k} + 2^{-p}) + u +2^{-(p+k)}= -\mu\cdot 2^{k} - \mu - 2^{-p} - 2^{-(p+k)} + u + 2^{-(p+k)} \leq -(\mu\cdot 2^k + 2^{-p})$.
	
	Part 2. 
	Let $W$ and $u$ be as defined above. 
	Then 
	$\gapU{W\cdot u} = \roundU{d\cdot \gapU{W} + u}$. From Lemma~\ref{lem:UpperRound}, we get that $\gapU{W\cdot u} \geq  d \cdot \gapU{W}+ u $. Further, from our assumptions we get that $\gapU{W\cdot u} \geq  d \cdot \mu \cdot 2^{k}+ u
	= (1+2^{-k}) \cdot \mu \cdot 2^{k} + u  =
	\mu\cdot 2^k+  \mu +  u  \geq \mu\cdot 2^k$, since $\mu \geq u$. 
	\qed
\end{proof}

\subsubsection{Construction}

Let $\mu>0$,  $d = 1+2^{-k}$, $\varepsilon=2^{-p}$  be the upper bound, discount factor and approximation factor, respectively.
Let $\thresh_l$ be the largest integer such that $\thresh_l\cdot 2^{-(p+k)} \leq -\mu \cdot 2^{k}$. Let $\thresh_u$ be the smallest integer such that
$\thresh_u \cdot 2^{-(p+k)} \geq \mu\cdot 2^{k} + 2^{-p}$.
Note, the thresholds are from Lemma~\ref{lem:compartorUpperThresholds}.
For relation $\mathsf{R} \in \{\leq, \geq\}$,
construct DFA $\compU(\mu,k,p,\mathsf{R}) = (\State, \Start, \Sigma, \delta, \Final)$ as follows:

\begin{itemize}
	\item $\State = \{\thresh_l, \thresh_l+1,\dots, \thresh_u \}$,
	$\Start = \{0\}$ and 
	$\Final = \{i | i\in S \text{ and $i$ $R$ $0$}\}$
	
	\item Alphabet $\Sigma = \{-\mu, -\mu+1,\dots, \mu-1, \mu\}$
	\item Transition function $\delta\subseteq \State \times \Sigma \rightarrow \State$ where $(s,a,t) \in \delta$ then:
	\begin{enumerate}
		\item \label{Trans:SelfLoop} If $s = \thresh_l$ or $s= \thresh_u$, then $t = s$ for all $a \in \Sigma$
		\item Else, let $\roundU{d\cdot s \cdot 2^{-(p+k)} + a} = i \cdot 2^{-(p+k)}$ for an integer $i$
		\begin{enumerate}
			\item \label{Trans:IntState} If $\thresh_l \leq i \leq \thresh_u$, then $t = i$
			
			\item \label{Trans:leq} If $i> {\thresh_h}$, then $t = \thresh_u$
			\item \label{Trans:geq} If $i< {\thresh_l}$, then $t = \thresh_l$
		\end{enumerate}
		
	\end{enumerate}
\end{itemize}
\label{thrm:Comparatorupper}

\begin{theorem}
	\label{thrm:Comparatorupper}
	Let $\mu>0$ be and integer upper bound. Let $k,p>0$ be rational parameters s.t. $d = 1+2^{-k}$ is the  discount factor and $\varepsilon=2^{-p}$ is the approximation parameter. 
	DFA $\compU(\mu,k,p,\mathsf{R}) $ accepts a finite weight sequence $W \in \Sigma^*$ iff $\DSumU{W}$ $\mathsf{R} $ $ 0$.
	DFA $\compU(\mu,k,p,\mathsf{R}) $ has $\O(\mu\cdot 2^{2k+p})$ states.
\end{theorem}

\subsection{Algorithm $\upperInc$}
\label{Sec:Positive}

This section utilizes the comparator for upper approximation of DS to describe  $\upperInc$ (Algorithm~\ref{Alg:UpperInclusion}).
Recall, given inputs $P$, $Q$, discount factor $1<d<2$ and approximation factor $0<\varepsilon<1$, $\upperInc(P,Q,d,\varepsilon)$ returns $P \subseteq Q$ holds or  $P\subseteq Q - d\cdot\varepsilon$ does not hold.

Once again, the intuition and algorithm design fo $\upperInc$ resembles that of $\lowerInc$. Intuitively, $\upperInc$ solves whether $P$ is $f$-included in $Q$ where aggregate function $f$ is the upper approximation of discounted-sum. 
Lemma~\ref{lem:algUpperApproxDS} precisely states the intuition, and the algorithm is given in Algorithm~\ref{Alg:UpperInclusion}.

We begin with formalizing the intuition. We say, a run $\rho_P$ of word $w$ in $P$  is {\em dominated} by $Q$ if there exists a run $\rho_Q$ in $Q$ on the same word such that $\DSumU{\rho_P - \rho_Q} \leq 0$.
Then, 
\begin{lemma}
\label{lem:algUpperApproxDS}
Given DS automata $P$ and $Q$, discount factor $d = 1+2^{-k}$  and approximation factor $\varepsilon = 2^{-p}$ for rational values $k,p>0$. 
\begin{enumerate}
    \item If all runs in $P$ are dominated by $Q$, then $P \subseteq Q $ holds.
    
    \item If there exists a run in $P$ that is not dominated by $Q$, then $P \subseteq Q - d\cdot \varepsilon$ does not hold.
\end{enumerate}
\end{lemma}

\begin{proof}
Proof of (1.)
Let for all words $w \in \Sigma^*$, for all runs of $ w$ in $\rho_P \in P$, there exists a run of word $w$  $\rho_Q \in Q$ such that $\DSumU{\rho_P - \rho_Q} \leq 0$ implies that $\DSum{\rho_P - \rho_Q}{d} \leq 0$. By arguing as in Lemma~\ref{lem:algLowerApproxDS}, we get that $P \subseteq Q$.

Proof of (2.)
Let $w \in \Sigma*$ be a word for which there exists a run of $w \rho_P\in P$ such that for all runs of $w$ $\rho_Q \in Q$, $\DSumU{\rho_P - \rho_Q} > 0$ implies $\DSum{\rho_P - \rho_Q}{d} > - d\cdot\varepsilon$. Therefore, by arguing as done in Lemma~\ref{lem:algLowerApproxDS}, we get that $ P \subseteq Q -d\cdot\varepsilon$ does not hold. 
\qed
\end{proof}
\begin{algorithm}[t]
\label{Alg:UpperInclusion}
\caption{ $\upperInc(P,Q, d,\varepsilon)$\\\textsf{Inputs:} DS automata $P$, $Q$, discount factor $1<d<2$, approximation factor $0<\varepsilon<1$} \label{Alg:UpperInclusion}
\begin{algorithmic}[1]
\STATE $\hat{P} \leftarrow  \AugmentWtAndLabel(P)$ \label{alg-line:AugmentP}		
\STATE $\hat{P}-{Q} \leftarrow \MakeDifProduct(\hat{P}, {Q})$ \label{alg-line:ProdDiff}		
\STATE $\dominatedWitness \leftarrow  \cct{Intersect}(\hat{P}-{Q}, \compU(\mu,\log\frac{1}{(d-1)},\log\frac{1}{\varepsilon},\leq))$ where $\mu$ is the maximum of the absolute value of weights in  $\hat{P}-{Q}$ \label{alg-line:Intersect}		
\STATE $\dominated \leftarrow \cct{Project}(\dominatedWitness)$ \label{alg-line:Project}
\STATE {$\hat{P}_{-wt} \subseteq \dominated$} \label{alg-line:ensure}
\IF {$\mathsf{aux}$ }
\RETURN $P \subseteq Q = \mathsf{True}$
\ELSE 
\RETURN $P \subseteq Q - d\cdot\varepsilon = \mathsf{False}$
\ENDIF
\end{algorithmic}
\end{algorithm}

We design Algorithm~\ref{Alg:UpperInclusion} so that it resembles Algorithm~\ref{Alg:LowerInclusion} except that in this case the comparator refers to that of  the upper approximation of discounted-sum:

\begin{lemma}
\label{lem:algUpperApprox}
Given DS automata $P$ and $Q$, discount factor $d = 1+2^{-k}$  and approximation factor $\varepsilon = 2^{-p}$ for rational values $k,p>0$, $\upperInc(P,Q,d,\varepsilon)$ returns $\mathsf{True}$ iff all runs in $P$ are dominated by $Q$.
\end{lemma}
\begin{proof}
The proof argument is similar to that in Lemma~\ref{lem:algLowerApprox}.
\qed
\end{proof}
\begin{theorem}[\rm {Soundness}]
\label{thrm:upperIncSound}
For all inputs DS automata $P$ and $Q$, discount factor $d = 1+2^{-k}$  and approximation factor $\varepsilon = 2^{-p}$ for rational values $k,p>0$, algorithm  $\upperInc$ is sound.
\end{theorem}
\begin{proof}
The proof is similar to proof that Theorem~\ref{thrm:lowerIncSound}.
\qed
\end{proof}

\begin{theorem}[Complexity]
	\label{thrm:complexityUpper}
Given DS automata $P$ and $Q$, discount factor $d = 1+2^{-k}$  and approximation factor $\varepsilon = 2^{-p}$ for rational values $k,p>0$.
Let $\mu$ be the absolute value of the largest weight in $P$ and $Q$.
Then the worst case complexity of $\upperInc$ is $2^{\O(n)}$ where $n = |P|\cdot |Q| \cdot  \frac{\mu}{(d-1)^2 \cdot \varepsilon}$. 	
\end{theorem}
\begin{proof}
The proof is similar to that of Theorem~\ref{thrm:complexityLower}.
\qed
\end{proof}

\section{Anytime algorithm for DS inclusion}
\label{Sec:anytimeInc}

This section describes the core contribution of this work. We design an anytime algorithm for discounted-sum inclusion.  
On inputs DS automata $P$ and $Q$, and discount factor $1<d<2$, our algorithm $\rationalInc(P,Q,d)$
either terminates and returns a crisp $\mathsf{True}$ or $\mathsf{False}$ answer to $P \subseteq Q$, or  it establishes a {\em $d\cdot \varepsilon$}-close approximation, where the  approximation factor $\varepsilon>0$ decreases with time. 
In addition,  algorithm $\rationalInc$ is co-computational enumerable i.e. if $P \subseteq Q$ does not hold then $\rationalInc(P,Q,d)$ is guaranteed to terminate with that outcome after a finite amount of time.

This section proves soundness, and co-computational enumerability of $\rationalInc$. Finally, we evaluate the complexity of running $\rationalInc$  upto a desired approximation (Theorem~\ref{thrm:complexityPrecision}). The analysis reveals that as $\varepsilon$ tends to 0, the worst-case complexity grows rapidly.

\subsubsection{Algorithm details}
$\rationalInc$ invokes $\approxInc$ with an initial approximation factor $0<\varepsilon_{\textsf{init}}<1$, where $\approxInc$ is described in Algorithm~\ref{Alg:approxInclusion}. Here  $\varepsilon_{\textsf{init}}=\frac{1}{2}$.
Formally, $\rationalInc(P,Q,d)  = \approxInc(P,Q,d,0.5)$ Recall, Algorithm~\ref{Alg:approxInclusion} is a tail recursive procedure in which subprocedures $\lowerInc$ and $\upperInc$ are invoked in each round of the recursion. If the current round of recursion is invoked with approximation factor $\varepsilon>0$, then $\lowerInc$ and $\upperInc$ are invoked with $\varepsilon$. In this round, if $\lowerInc$ or $\upperInc$ returns $P \subseteq Q$ does not hold or $P \subseteq Q$ holds, respectively, $\approxInc$ terminates. Otherwise, it invokes $\approxInc$ with approximation factor $\frac{\varepsilon}{2}$.

\sloppy
In order to analyse $\rationalInc$, we begin with some useful terminology:
$\rationalInc(P,Q,d)$ is said to be in the {\em $\varepsilon$-th round} when the current invocation of $\approxInc$ occurs with approximation factor $\varepsilon$. 
$\rationalInc(P,Q,d)$ is said to {\em terminate in the $\varepsilon$-th round } if $\approxInc(P,Q,d,\varepsilon)$ returns a crisp solution to DS inclusion in the $\varepsilon$-th round. 
Finally, $\rationalInc(P,Q,d)$ is said to {\em terminate} if there exists a $\varepsilon>0$ such that it terminates in the $\varepsilon$-th round. 

\begin{theorem}{[Soundness]}
\label{thrm:rationalDSsound}
Let $P$, $Q$ be DS automata, and $1<d<2$ be the  discount factor. 
\begin{enumerate}
    \item 
    \label{thrm:rationalDSsoundTerminateFalse} If $P$ is not DS-included in $Q$ and  $\rationalInc(P,Q,d)$ terminates, then
    $\rationalInc(P,Q,d)$ returns  $P\subseteq Q = \mathsf{False}$
    
    \item \label{thrm:rationalDSsoundTerminateTrue} If $P$ is DS-included in $Q$ holds and  $\rationalInc(P,Q,d)$ terminates, then $\rationalInc(P,Q,d)$ returns  $P\subseteq Q = \mathsf{True}$
    
    \item \label{thrm:rationalDSsoundNotTerminate} If $\rationalInc(P,Q,d)$ does not terminate in the $\varepsilon$-th round in which the  approximation factor is $\varepsilon>0$, then $P \subseteq Q  + d\cdot \varepsilon$ holds.
\end{enumerate}
\end{theorem}
\begin{proof}
We reason about $\approxInc$ as that is sufficient.

Proof of (1.) and (2.):
$\approxInc$ terminates only if either $\lowerInc$ returns $P \subseteq Q = \mathsf{False}$ or $\upperInc$ returns $P \subseteq Q  = \mathsf{True}$. 
Since each of these subprocedures is individually sound (Theorem~\ref{thrm:lowerIncSound} and Theorem~\ref{thrm:upperIncSound}, respectively), statements (1.) and (2.) hold.

Proof of (3.):
$\approxInc$ does not terminate in the $\varepsilon$-th round only if $\lowerInc$ and $\upperInc$ return $P \subseteq Q  + d\cdot \varepsilon = \mathsf{True} $ and $P \subseteq Q - d\cdot\varepsilon = \mathsf{False}$, respectively. Therefore, by soundness of $\lowerInc$ (Theorem~\ref{thrm:lowerIncSound}), it holds that
$P$ is $d\cdot\varepsilon$-close to $Q$.
\qed
\end{proof}

Next, we prove that $\rationalInc$ is co-computational enumerable.

\begin{theorem}[Co-ce]
\label{thrm:rationalDScomplete}
Let $P$, $Q$ be DS automata, and $1<d<2$ be the  discount factor. 
If $P \subseteq Q = \mathsf{False}$ then $\rationalInc(P,Q,d)$ terminates after a finite number of recursions. Upon termination, $\rationalInc(P,Q,d)$ returns $P \subseteq Q = \mathsf{False}$. 
\end{theorem}

\begin{proof}
By soundness of $\upperInc$, we know that $\upperInc$ will return $P \subseteq Q - d\cdot\varepsilon = \mathsf{False}$ in every invocation. Hence, $\rationalInc$ cannot terminate due to $\upperInc$. Therefore, if $\rationalInc$ terminates, it must be because $\lowerInc$ returns $P \subseteq Q= \mathsf{False}$ for some $\varepsilon>0$.
It remains to show that such an approximation factor exists. But that is exactly the result in Theorem~\ref{thrm:lowerIncNegCases}. Therefore, if $P \subseteq Q =\mathsf{False}$, then $\rationalInc$ is guaranteed to terminate with the outcome $P \subseteq Q = \mathsf{False}$. 
\qed
\end{proof}

Note that we cannot determine, apriori, the number of recursive invocations that will be conducted for a given input instance. If we could, then we could have proved the decidability of discounted sum inclusion for discount factor $1<d<2$.
As a result, one cannot determine the worst-case complexity of $\rationalInc$ in general. However, the worst-case complexity can be computed upto a certain precision. More precisely, if a user decides that it will run $\rationalInc$  until either it terminates with a crisp solution or $d\cdot \varepsilon_c$-close approximation is established, where approximation factor $\varepsilon_c$ is pre-determined.
Note that it is not sufficient to recursively only invoke $\lowerInc$ till the approximation factor is $\varepsilon_c$ since then we would never be able to generate the outcome that DS inclusion holds. 
The worst-case complexity upto $\varepsilon_c$ is computed as follows:

\begin{theorem}[\rm Complexity given precision]
\label{thrm:complexityPrecision}
Given DS automata $P$ and $Q$,  discount factor $1<d <2$. Let $\mu$ be the largest weight in $P$ and $Q$.
Let $0<\varepsilon_{\mathsf{c}}<1$ be the desired precision.
Then, 
the worst-case complexity of solving DS inclusion upto a precision of  $\varepsilon_{\mathsf{c}}$ is $2^{\O(n)}$ where $n = |P|\cdot |Q| \cdot  \frac{\mu}{(d-1)^2} \cdot \log (\frac{1}{\varepsilon_c})$
\end{theorem}
\begin{proof}
Without loss of generality, let $\varepsilon_c = 2^{-m}$ for$ m \in \N$. Then in the worst case, $\rationalInc$ will terminate after invoking $\approxInc$ with $\varepsilon_c$.
The worst case complexity of $\rationalInc$ with precision $\varepsilon_c$ is calculate by taking the sum of the complexity of each invocation of $\approxInc$ till that point. The calculation evaluates to the expression in the statement. 
\qed
\end{proof}

So, as $\varepsilon_\mathsf{c}$ converges to 0, i.e, DS inclusion is solved exactly, the worst-case complexity explodes rapidly. This renders a quantitative measure of the difficulty of solving DS inclusion. Although this isn't concrete evidence for  undecidability of DS inclusion, it certainly points in that direction. 
Finally, if $d =1$, DS inclusion would be the same as sum. Then the above evaluation corroborates the known undecidability of quantitative inclusion with sum~\cite{almagor2011whats}.

\section{Chapter summary}
This chapter investigates  DS inclusion when the discount factor $1<d<2$ is not an integer. The decidability of this problem has been open for more that a decade now. So, this chapter focuses on designing solutions for DS inclusion that could be used in practice despite its decidability being unknown. To this effect, we design an anytime algorithm for DS inclusion. The algorithm may not always solve the problem exactly. In these cases, it will generate an approximate result, which is a meaningful outcome in practice. To the best of our knowledge, this is the first attempt to solve DS inclusion with non-integer discount factors for practical purposes. 
While this chapter looks into DS inclusion over finite words, we believe the same ideas can be extended to DS inclusion over infinite words.
Our algorithm design is motivated by designing regular comparator automata for approximations of DS with non-integer discount factors. Thus, not only are comparators able to design scalable solutions when the discount factor is an integer, as shown in Chapter~\ref{ch:safetycosafety}, they also makes algorithmic advances for non-integer discount factors. 

\part{Quantitative games with discounted-sum}
\label{part:games}

\chapter{On the analysis of quantitative games}
\label{ch:integergames}


This part continues the investigation of automata-based quantitative reasoning by studying their impact on quantitative games. We show that even here, comparator automata delivers the benefits of scalability, efficiency, and broader applicability. 

\section{Introduction}
\label{Sec:Intro}

{\em Quantitative properties} of systems are increasingly being explored in automated reasoning across diverse application areas, including software verification~\cite{baierprobabilistic,Kwi07,kwiatkowska2010advances}, security~\cite{clark2007static,finkbeiner2018model}, computer-aided education~\cite{d2016qlose}, and even  verification of deep neural networks~\cite{baluta2019quantitative,seshia2018formal}.
In decision making domains such as planning and reactive synthesis, quantitative properties have been deployed to obtain high-quality solutions~\cite{bloem2009better},  describe hard and soft constraints~\cite{lahijanian2015time}, constraints on cost and resource~\cite{he2017reactive}, and the like.  
In most cases, planning and synthesis with quantitative properties is formulated into an {\em analysis problem} over a two player, finite-state game arena with a cost model that encodes the quantitative property~\cite{chakrabarti2005verifying}.

{\em Optimization} over such games is a popular choice of analysis in literature. 
Typically, optimization over these games has polynomial time algorithms. 
However, the degree of polynomial may be too high to scale in practice; resulting in limited applicability of the synthesis task at hand~\cite{bloem2018graph}. 
Furthermore, often times these synthesis tasks are  accompanied with temporal goals. Under this extension, the games are required to generate an optimal solution while also satisfying the temporal objective. However, prior works have proven that optimal strategies may not exist under extension with temporal goals, rendering analysis by optimization incompatible with temporal goals~\cite{chatterjee2017quantitative}.

To this end, we propose an alternate form of analysis in which the objective is to search for a solution that adheres to {\em a given threshold constraint} as opposed to generating an optimal solution. 
We call this analysis the {\em satisficing problem}, a well-established notion that we borrow from economics~\cite{satisficing}. Our argument is that in several cases optimal solutions may be replaced with satisficing ones. 
For instance, a solution with {\em minimal} battery consumption can be substituted by one that operates {\em within} the battery life. 
This work contrasts between  the optimization problem and the satisficng problem w.r.t. theoretical complexity, empirical performance, and temporal extensibility. 

More specifically, this work studies the aforementioned contrast on two player, finite-state games with the discounted-sum aggregate function as the cost model, which is a staple cost-model in decision making domains~\cite{osborne1994course,puterman1990markov,sutton1998introduction}. 
In these games, players take turns to pass a token along the {\em transition relation} between the states. As the token is pushed around, the play accumulates weights along the transitions using the  discounted-sum cost model. 
The players are assumed to have opposing objectives: one player maximizes the cost while the other minimizes it. 
The optimization problem is to find the optimal cost of all possible plays in the game~\cite{shapley1953stochastic}.
We define the satisficing problem as follows:  Given a threshold value $v \in \Q$, does there exist a strategy for the minimizing player such that the cost of all resulting plays is less than (or $\leq$) to the threshold $v$?

From prior work, one can infer that the optimization problem is pseudo-polynomial~\cite{hansen2013strategy,zwick1996complexity}.
Zwick and Patterson have shown that  a value-iteration (VI) algorithm  converges to the optimal cost at infinitum~\cite{zwick1996complexity}.
Interestingly, even though the VI algorithm finds extensive usage~\cite{bloem2009better,bloem2018graph}, a thorough worst-case analysis of VI has  hitherto been absent. 
This work, first of all, amends the oversight. In \textsection~\ref{Sec:Optimization}, we present the  analysis for VI for all discount factors $d>1$.
Towards this, we observe that VI  performs many arithmetic operations. Therefore, it is crucial to account for the cost of arithmetic operations as well. 
Its significance is  emphasised as we show that there are orders of magnitude  of difference between the complexity of VI  under unit-cost and bit-cost models of arithmetic. For instance, when the discount factor is an integer, we show that VI  is $\O(|V|^2)\cdot |E|)$ and $\O(|V|^4\cdot |E|)$ under unit- and bit-cost model, respectively, where $V$ and $E$ are the set of states and transitions.  
In addition, we observe that  VI can take as many as $\Theta(|V|^2)$ iterations to compute the optimal value. Note that this bound is tight, indicating that the scalability of VI will be limited only to games with a small number of 
states and transitions.
We confirm this though an empirical analysis. We show that despite heuristics, VI-based algorithms cannot escape their worst-case behavior, adversely impacting its empirical performance (\textsection~\ref{Sec:VIForSatisfice}).   
Finally, it is known that these games may not have optimal solutions when extended with temporal goals~\cite{chatterjee2017quantitative}.

In contrast, our examination of the satisficing problem illustrates its advantages over optimization.
We solve satisficing via an automata-based approach as opposed to an arithmetic approach (\textsection~\ref{Sec:Satisfice}). 
Our approach is motivated by recent advances in automata-based reasoning of quantitative properties using {\em comparator automata}~\cite{BCVFoSSaCS18,BVCAV19}.
We show that when the discount factor is an integer, satisficing problem can be solved in $\O(|V|+|E|)$ via an efficient reduction to safety/reachability games~\cite{thomas2002automata}.
Observe that there is a fundamental separation between the complexity of satisficing and the number of iterations required in VI, indicating a computational gain of our  comparator-based solution for satisficing. As before, an empirical evaluation confirms this as well. 
Last but not the least, we show that  unlike optimization, satisficing naturally integrates with temporal goals. The reason is that since both, saisficing and temporal goals, adopt automata-based solutions, the two can be seamlessly combined with one another (\textsection~\ref{Sec:Temporalgoals}). 
Currently, our method works for integer discount factors only. We believe, these results can be extended  to the non-integer case with some approximation guarantee.

\section{Optimization problem}
\label{Sec:Optimization}

The optimization problem can be solved by a VI algorithm (\textsection~\ref{Sec:VIDescription}~\cite{zwick1996complexity}).
While a reduction from mean-payoff games proves that the VI algorithm is pseudo-polynomial,  
a thorough worst-case analysis of VI has been missing. 
This section undertakes a thorough investigation of its worst-case complexity.

Our analysis also exposes the dependence of VI on the discount factor $d>1$ and the cost-model for arithmetic operations i.e. unit-cost or bit-cost model.   We observe that these parameters bring about drastic changes to the algorithm's theoretical evaluation.
Our analysis works for all discount factors $d>1$. 

We begin by describing the VI algorithm in \textsection~\ref{Sec:VIDescription}. First, we prove that it is sufficient for VI to perform a finite-number of iterations to compute the optimal value in \textsection~\ref{Sec:NumIteration}. Finally, this result is used to compute the worst-case complexity of  VI under the unit- and bit-cost models of arithmetic in \textsection~\ref{Sec:worstcasecomplexityVI}. 

\subsection{Value-iteration  algorithm}
\label{Sec:VIDescription}

The VI algorithm plays a  min-max game between the two players~\cite{zwick1996complexity}. 
Let $\wt_k(v)$ denote the optimal cost of a $k$-length game that begins in state $v \in V$.
Then $\wt_k(v)$ can be computed using the following equations:
The optimal cost of a 1-length game beginning in  state $v \in V$ is as follows:
\begin{align*}
	\wt_1(v) =
	\begin{cases}
		\textit{max}\{ \gamma(v, w) | (v,w)\in E\} &\text{ if } v \in V_0 \\
		\textit{min}\{ \gamma(v, w) | (v,w)\in E\} &\text{ if } v \in V_1
	\end{cases}
\end{align*}

Given the optimal-cost of a $k$-length game, the optimal cost of a $(k+1)$-length game is computed as follows:
\begin{align*}
	\wt_{k+1}(v) = 
	\begin{cases}
		\textit{max}\{ \gamma(v, w) + \frac{1}{d}\cdot \wt_k(w) | (v,w) \in E\} \text{ if } v \in V_0 \\
		\textit{min}\{ \gamma(v, w) + \frac{1}{d}\cdot \wt_k(w) | (v,w) \in E\} \text{ if } v \in V_1 
	\end{cases}
\end{align*}

Then, it has been shown that the optimal cost of a $k$-length game beginning at state $v \in V$ converges to the optimal cost of an infinite-length game beginning in state $v \in V$~\cite{shapley1953stochastic,zwick1996complexity} as $k\rightarrow \infty$. In particular, let $\opt$ be the optimal value (beginning in state $\init$), then
$\opt = \lim_{k\rightarrow \infty}\wt_k(\init)$.

\subsection{Number of iterations}
\label{Sec:NumIteration}

The VI algorithm, as defined above, waits for convergence to terminate. 
Clearly, that is not sufficient for a worst-case analysis. 
To this end, in this section we establish a crisp bound on the number of iterations required to compute the optimal value.
Furthermore, we show that the bound we calculate is tight. 


\subsubsection{Upper bound on number of iterations.}
We prove the upper bound in several steps, as described below:
 
\begin{enumerate}[label=Step \arabic*]

    \item~\label{step:interval} We show that the optimal value $W$ must fall between an interval such that as the number of iterations $k$ of the VI algorithm increases, the interval length converges to 0.

    \item~\label{step:boundvalue} We show that the optimal value $W$ is a rational number with denominator at most $\mathsf{bound_\opt}$, where $\mathsf{bound_\opt}$ is parameterized by numerator and denominator of the discount factor and the number of states in the graph game.

    \item~\label{step:bounddiff} Next, we show that the denominator of the minimum non-zero difference between possible values of the optimal cost is at most $\mathsf{bound_{diff}}$, where $\mathsf{bound_{diff}}$ is also parameterized by numerator and denominator of the discount factor and the number of states in the graph game.

    \item\label{step:final} Finally, we use the previous three Steps to prove the pseudo-polynomial bound. 
    Since the interval in~\ref{step:interval} converges to 0, we can choose a $k$  such that the interval is less than than $1/\mathsf{bound_{diff}}$. 
    In our computations below, we will see that $\frac{1}{\mathsf{bound_{diff}}} < \frac{1}{\mathsf{bound_{v}}}$. 
    Therefore, 
    there can be only one rational number with denominator $\mathsf{bound_\opt} $ or less in the interval identified by the chosen $k$.
    Since this interval must also contain the optimal value $\opt$, the unique rational number with denominator less than or equal to $\mathsf{bound_\opt} $ must be the optimal value $\opt$. 
    
    Our task is to compute the value of $k$. 
\end{enumerate}

We prove all of these steps one-by-one. We begin with proof of~\ref{step:interval}.
We show that there is a finite-horizon approximation of the optimal value, i.e., for all $k\in \N$ the optimal value can be bounded using $\wt_k(\init)$ as follows:
\begin{lemma}
\label{lem:interval}
Let $\opt$ be the optimal value of a graph game $G$. Let $\mu>0$ be the maximum of absolute value of cost on all transitions in $G$. Then, for all $k \in \N$, 
\[
    \wt_k(v_{\mathsf{init}}) - \frac{1}{d^{k-1}}\cdot \frac{\mu}{d-1} \leq
    \opt \leq
    \wt_k(v_{\mathsf{init}}) + \frac{1}{d^{k-1}}\cdot \frac{\mu}{d-1}
    \]
\end{lemma}
\begin{proof}
    This holds because since $\opt$ is the limit of $\wt_k(v_{\mathsf{init}})$ as $k\rightarrow \infty$, its value must lie in between the
    minimum and maximum cost possible if the $k$-length game is extended to an infinite-length game. 
    The minimum possible extension would be when the $k$-length game is extended by iterations in which the cost incurred in each round is $-\mu$. Therefore, the resulting lowest value is $\wt_k(v_{\mathsf{init}}) - \frac{1}{d^{k-1}}\cdot \frac{\mu}{d-1} $.
    Similarly, the maximum value is $\wt_k(v_{\mathsf{init}}) + \frac{1}{d^{k-1}}\cdot \frac{\mu}{d-1} $.
\qed
\end{proof}

Clearly, as $k\rightarrow \infty$, the interval around the optimal value $\opt$ converges to 0. This way we  can find an arbitrarily small interval around the optimal value $\opt$.

To prove~\ref{step:boundvalue}, we know that there exist memoryless optimal strategies for both players~\cite{shapley1953stochastic}. Therefore, there must exists an optimal play that is in the form of a {\em simple lasso}. 
A {\em lasso} is a play represented as $v_0v_1\dots v_n(s_0 s_2\dots s_m)^\omega$.
The initial segment $v_0v_1\dots v_n$ is called the {\em head} of the lasso, and the cycle segment $s_0s_1\dots s_m$ the {\em loop} of the lasso.
A lasso is {\em simple} if
each state in $\{v_0\dots v_n, s_0,\dots s_m\}$ is distinct. 
Hence, in order to compute the optimal value, it is sufficient to examine simple lassos in the game only. 
Therefore, we evaluate the DS of cost sequences derived from simple lassos only.
Let $l = a_0\dots a_n(b_0\dots b_m)^{\omega}$ be the cost sequence of a lasso. Let $l_1 = a_0\dots a_n$ and  $l_2 = b_0\dots b_m$ correspond to the cost sequences of the head and loop of lasso, respectively. Then, 

\begin{lemma}
\label{lem:optimalplay}
Let $l = l_1\cdot (l_2)^{\omega}$ represent  an integer cost sequence of a lasso, where $l_1$ and $l_2$ are the cost sequences of the head and loop of the lasso. Let $d = \frac{p}{q}$ be the discount factor. 
Then, $\DSum{l}{d}$ is a rational number with denominator at most ($p^{|l_2|} - q^{|l_2|})\cdot (p^{|l_1|})$.
\end{lemma}
\begin{proof}
The result can proven by unrolling the expression of DS of a cost-sequence on a lasso path. 
The discounted sum of $l$ is given as follows:
\begin{align*}
\DSum{l}{d} 
= & \DSum{l_1}{d} + \frac{1}{d^{|l_1|}} \cdot(\DSum{(l_2)^\omega}{d})   \\
= & \DSum{l_1}{d} + \frac{1}{d^{|l_1|}} \cdot \Big(\DSum{l_2}{d} + \frac{1}{d^{|l_2|}}\cdot \DSum{l_2}{d} + \frac{1}{d^{2\cdot|l_2|}}\cdot \DSum{l_2}{d}+ \dots\Big) \\
& \text{Taking closed form expression of the term in the parenthesis, we get} \\
= & \DSum{l_1}{d} + \frac{1}{d^{|l_1|}} \cdot \Big(\frac{d^{|l_2|}}{d^{|l_2|} -1 }\Big)\cdot \DSum{l_2}{d}  \\
& \text{Let } l_2 = b_0b_1\dots b_{|l_2|-1} \text{ where } b_i \in \Z \\
=  &\DSum{l_1}{d} + \frac{1}{d^{|l_1|}} \cdot \Big(\frac{d^{|l_2|}}{d^{|l_2|} -1 }\Big)\cdot \Big(b_0 + \frac{b_1}{d} + \dots + \frac{b_{|l_2|-1}}{d^{|l_2|-1}}\Big) \\
=  &\DSum{l_1}{d} + \frac{1}{d^{|l_1|}} \cdot \Big(\frac{1}{d^{|l_2|} -1 }\Big)\cdot \Big(b_0\cdot d^{|l_2|} + {b_1}\cdot d^{|l_2|-1} + \dots + {b_{|l_2|-1}}{d}\Big) \\
& \text{Expressing } d =\frac{p}{q} \text{, we get} \\
= & \DSum{l_1}{d} + \frac{1}{d^{|l_1|}} \cdot \frac{q^{|l_2|}}{p^{|l_2|} - q^{|l_2|}}\cdot \Big(b_0 (\frac{p}{q})^{|l_2|} + \dots b_{|l_2|-1}\cdot \frac{p}{q}\Big) \\
& \DSum{l_1}{d} + \frac{1}{d^{|l_1|}} \cdot \frac{1}{p^{|l_2|} - q^{|l_2|}}\cdot M, \text{ where } M \in \Z \\
& \text{Expressing } d =\frac{p}{q} \text{ again, we get} \\
= & \frac{1}{p^{|l_1|}} \cdot \frac{1}{p^{|l_2|} - q^{|l_2|}} \cdot N,  \text{ where } N \in \Z
\end{align*}
\qed 
\end{proof}

The essence of Lemma~\ref{lem:optimalplay} is that the DS of the cost-sequence of a lasso, simple or non-simple, is a rational number. From here, we can immediately  derive \ref{step:boundvalue}:

\begin{corollary}
\label{coro:optimalplay}
Let $G=(V, \init, E, \gamma)$ be a graph game. Let $d = \frac{p}{q}$ be the discount factor. Then the optimal value of the game is a rational number with denominator at most  $(p^{|V|} - q^{|V|})\cdot (p^{|V|})$ 
\end{corollary}
\begin{proof}
The optimal value is obtained on a simple lasso as both players have memoryless strategies that result in the optimal value. The length of the simple lasso is at most $|V|$. Therefore, the length of the head and loop are at most $|V|$  each. Hence, the expression from Lemma~\ref{lem:optimalplay} simplifies to   $(p^{|V|} - q^{|V|})\cdot (p^{|V|})$.
\qed
\end{proof}

Similarly, \ref{step:bounddiff} is resolved as follows:

\begin{corollary}
\label{coro:optimaldiff}
Let $G=(V, \init, E, \gamma)$ be a graph game. Let $d = \frac{p}{q}$ be the discount factor.
Then the  minimal non-zero difference between the cost on simple lassos  is a rational number with denominator at most $(p^{(|V|^2)} - q^{(|V|^2)})\cdot (p^{(|V|^2)})$.
\end{corollary}
\begin{proof}
The difference of two lassos $l_1$ and $l_2$ can be represented  by another lasso $l = l_1\times l_2$ constructed from taking their product, and assigning the difference of their costs on each transition. If the maximum length of the lassos is $|V|$, then the maximum length of the difference lasso will be $|V|^2$.
Then, from Lemma~\ref{lem:optimalplay} we immediately obtain that the upper bound of the denominator of the minimum non-zero difference of optimal plays is $(p^{(|V|^2)} - q^{(|V|^2)})\cdot (p^{(|V|^2)})$.
\qed
\end{proof}

Therefore, $\mathsf{bound}_{{\opt}} = (p^{|V|} - q^{|V|})\cdot (p^{|V|})$ and $\mathsf{bound}_{\mathsf{diff}} = (p^{(|V|^2)} - q^{(|V|^2)})\cdot (p^{(|V|^2)})$.
Recall, in~\cite{zwick1996complexity} Zwick-Paterson {\em informally claim} that a pseudo-polynomial algorithm can be devised to compute the optimal value of the game. We formalize their statement in the final step as follows:

\begin{theorem}
\label{thrm:vsquareiterations}
Let $G = (V, \init,E,\gamma)$ be a graph game. The number of iterations required by the value-iteration algorithm or the length of the finite-length game
to compute the optimal value $\opt$ is
\begin{enumerate}
    \item $\O(|V|^2)$ when discount factor $d\geq 2$, 
    \item $\O\Big(\frac{\log(\mu)}{d-1} + |V|^2\Big)$ when discount factor $1<d<2$.
\end{enumerate}
\end{theorem}
\begin{proof}

Recall, the task is to find a $k$ such that the interval identified by~\ref{step:interval} is less than $\frac{1}{\mathsf{bound}_{\mathsf{diff}}}$.
Note that $\mathsf{bound}_{{\opt}} < \mathsf{bound}_{\mathsf{diff}}$. Therefore, $\frac{1}{\mathsf{bound_{diff}}} < \frac{1}{\mathsf{bound_{v}}}$. Hence, there can be only one rational value with denominator $\mathsf{bound}_{\mathsf{\opt}}$ or less in the small interval identified by the chosen $k$. Since the optimal value must also lie in this interval, the unique rational number with denominator $\mathsf{bound}_{\mathsf{\opt}}$ or less must be the optimal value. 
Let $k$ be such that the interval from~\ref{step:interval} is less than $\frac{1}{\mathsf{bound}_{\mathsf{diff}}}$.
Then, 
\begin{align*}
2\cdot\frac{\mu}{d-1\cdot d^{k-1}} \leq &  c\cdot \frac{1}{(p^{(|V|^2)} - q^{(|V|^2)})\cdot (p^{(|V|^2)})} \text{ for some  } c>0  \\
2\cdot\frac{\mu}{d-1\cdot d^{k-1}} \leq & c\cdot \frac{q^{2\cdot |V|^2}}{(p^{(|V|^2)} - q^{(|V|^2)})\cdot (p^{(|V|^2)})} \text{ for some  } c>0  \\
2\cdot\frac{\mu}{d-1\cdot d^{k-1}} 
\leq & 
c\cdot \frac{1}{(d^{(|V|^2)} - 1)\cdot (d^{(|V|^2)})} \text{ for  } c>0  \\
d-1\cdot d^{k-1} \geq &
c' \cdot \mu \cdot (d^{(|V|^2)} - 1)\cdot (d^{(|V|^2)}) \text{ for   } c'>0 \\
2\cdot\log(d-1) + k\cdot \log(d) \geq &
c'' + \log(\mu) + \log(d^{(|V|^2)} - 1) +  {|V|^2}\cdot\log(d) \text{ for   } c''>0 
\end{align*}

The following cases occur depending how large or small the values are:

\begin{enumerate} [label= Case \arabic*.]
	\item When $d\geq 2$:  
	In this case, both $d$ and $d^{|V|^2}$ are large. Then, 
	\begin{align*}
	2\cdot\log(d-1) + k\cdot \log(d) & \geq 
	c'' + \log(\mu) + \log(d^{(|V|^2)} - 1) +  {|V|^2}\cdot\log(d) \text{ for  } c''>0 \\
	2\cdot\log(d) + k\cdot \log(d) & \geq 
	c'' + \log(\mu) + {(|V|^2)}\log(d) +  {|V|^2}\cdot\log(d) \text{ for  } c''>0 \\
	k & = \O(|V|^2)
	\end{align*}

	\item When $d$ is small but $d^{|V|^2}$ is not: In this case, $\log(d) \approx (d-1)$, and $\log(d-1) \approx 2-d$. Then,
	\begin{align*}
	2\cdot\log(d-1) + k\cdot \log(d) & \geq 
	c'' + \log(\mu) + \log(d^{(|V|^2)} - 1) +  {|V|^2}\cdot\log(d) \text{ for  } c''>0 \\
	2\cdot\log(d-1) + k\cdot \log(d) & \geq 
	c'' + \log(\mu) + {(|V|^2)}\log(d) +  {|V|^2}\cdot\log(d) \text{ for  } c''>0 \\
	2\cdot (2-d) + k\cdot (d-1) & \geq 
	c'' + \log(\mu) + {(|V|^2)}(d-1) +  {|V|^2}\cdot (d-1) \text{ for  } c''>0 \\
	k = \O\Big(\frac{\log(\mu)}{d-1} + |V|^2\Big)
	\end{align*}

	\item When both $d$ and $d^{|V|^2}$ are small. Then, in addition to the approximations from the earlier case, $ \log(d^{|V|}-1) \approx (2-d^{|V|})$. So, 
	\begin{align*}
	2\cdot\log(d-1) + k\cdot \log(d) & \geq 
	c'' + \log(\mu) + \log(d^{(|V|^2)} - 1) +  {|V|^2}\cdot\log(d) \text{ for  } c''>0 \\
	2\cdot\log(d-1) + k\cdot \log(d) & \geq 
	c'' + \log(\mu) + 2-d^{(|V|^2)} +  {|V|^2}\cdot\log(d) \text{ for  } c''>0 \\
	2\cdot (2-d) + k\cdot (d-1) & \geq 
	c'' + \log(\mu) + 2-d^{(|V|^2)} +  {|V|^2}\cdot (d-1) \text{ for  } c''>0 \\
	k = \O\Big(\frac{\log(\mu)}{d-1} + |V|^2\Big)
	\end{align*}
\qed	
\end{enumerate}

\end{proof}

\subsubsection{Lower bound on number of iterations}

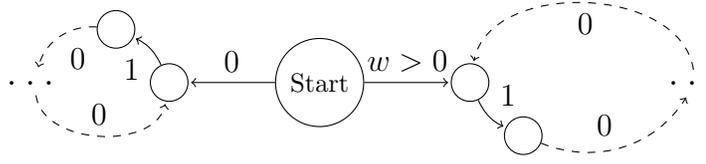
\begin{figure}[t]
    \centering
    
    \begin{tikzpicture}[shorten >=1pt,on grid,auto] 
   
    \node[state] (q1) {\footnotesize{Start}};
    \node[state] (q2)  [right of = q1, node distance=2cm, minimum size=0.5cm] {};
    \node[state] (q3)  [left of = q1, node distance=2cm, minimum size=0.5cm] {};
    \node[] (q4)  [right of = q1, node distance=5cm] {\Large{\textellipsis}};
    \node[] (q5) [left of=q3, node distance=1.8cm]
    {\Large\textellipsis};
    \node[state] (q6) [below right of = q2, minimum size=0.5cm, node distance=1cm] {};
    \node[state] (q7) [above left of = q3, minimum size=0.5cm, node distance=1cm] {};
  
    \path[->] 
    (q1) edge  node {$w>0$} (q2)
          edge  node [above] {$0$} (q3)
         
    (q2) edge [bend right = 20] node {1} (q6) 
    (q6) edge [bend right = 40, dashed] node   {0} (q4) 
    (q3) edge [bend right = 20] node {1} (q7) 
    (q7) edge [bend right = 40, dashed] node   {0} (q5) 
    (q4) edge [bend right = 80, dashed] node  {0} (q2)
    (q5) edge [bend right = 80, dashed] node {0} (q3);
    \end{tikzpicture}
    \caption{Sketch of game graph which requires $\Omega(|V|^2)$ iterations}
\label{fig:lowerboundexample}
\end{figure}
This bound is tight. 
We show this by constructing a quantitative graph game for which it is  necessary that the VI algorithm takes $\Omega(|V|^2)$  iterations. 
We give a sketch of this input instance in Fig~\ref{fig:lowerboundexample}.
Let all states in Fig~\ref{fig:lowerboundexample} belong to the maximizing player. Hence, the optimization problem reduces to searching for a {\em path} with optimal cost. 
The idea is to show that the cost of finite-length game with lesser than $\Omega(|V|^2)$-length will result in an incorrect optimal cost. 
Therefore, Fig~\ref{fig:lowerboundexample} is designed in a way so that the path for optimal cost of a $k$-length game is along the right hand side (RHS) loop when $k$ is small, but along the  left hand side (LHS) loop when $k$ is large. 
This way, the correct maximal value can be obtained only at a large value for $k$. Hence the VI algorithm runs for at least $k$ iterations. 



Fig~\ref{fig:lowerboundexample} realizes these objectives by making the loop to the RHS larger than the one on the LHS, as shown in Fig~\ref{fig:lowerboundexample}, and assigning $w>0$.
The intuition is that  when $k$ is small, the optimal path is along the RHS loop since $w>0$. However, since the RHS is larger, it accumulates cost slower than the LHS loop. As a result, as $k$ becomes larger, for an appropriate assignment to $w$, the cost on the LHS will eventually be larger than that on the PHS.
By meticulous reverse engineering of the size of both loops and the value of $w$, one can guarantee that $k = \Omega(|V|^2)$.

A concrete instance is as follows:
Let the left hand side loop have $4n$ edges, the right hand side of the loop have $2n$ edges, and $w = \frac{1}{d^{3n}} + \frac{1}{d^{7n}} + \cdots + \frac{1}{d^{m\cdot n -1}}$ such that $m\cdot n -1 = c\cdot n^2$
for a positive integer $c>0$.
One can show for a finite games of length  $(m\cdot n -1)$ or less, the optimal path arises from the loop to the right. But for games of length greater than  $(m\cdot n -1)$, the optimal path will be to due to the left hand side loop.

\subsection{Worst-case complexity analysis}
\label{Sec:worstcasecomplexityVI}
Finally, we present the complete worst-case complexity analysis. Since, VI algorithm is dominated by arithmetic operations, we take into account their cost as well. We work with the unit-cost and bit-cost models of arithmetic.
We observe that this choice has a drastic impact to the worst-case complexities.

\subsubsection{Unit-cost model}
Under the unit-cost model of arithmetic, all arithmetic operations are assumed to take constant time. 
\begin{theorem}
\label{thrm:optimizationunitcost}
Let $G = (V, \init, E, \gamma)$ be a quantitative graph game. 
The worst-case complexity of computing the optimal value under unit-cost model for arithmetic operations is  
\begin{enumerate}
     \item $\O(|V|^2\cdot |E|)$ when discount factor $d\geq 2$, 
    \item $\O\Big(\frac{\log(\mu)\cdot |E|}{d-1} + |V|^2\cdot |E|\Big)$ when discount factor $1<d<2$.
\end{enumerate}
\end{theorem}
\begin{proof}
The cost of updating the cost of all vertices from iteration $j$ to $j+1$ is linear in $|E|$ since every transition is traversed exactly once. 
Therefore, the worst-case complexity of computing the optimal values of a $k$-length game is $\O(k\cdot |E|)$. In particular, the worst-case complexity of computing the optimal value is $\O(|V|^2\cdot |E|)$ or $\O\Big(\frac{\log(\mu)\cdot |E|}{d-1} + |V|^2\cdot |E|\Big)$ depending on whether the discount factor $d \geq 2$ or $1<d<2$, respectively (Theorem~\ref{thrm:vsquareiterations}).
\qed
\end{proof}

\subsubsection{Bit-cost model}
Under the bit-cost model, the cost of arithmetic operations depends on the size of the numerical values. Iintegers are represented in their bit-wise representation. Rational numbers $\frac{r}{s}$ are represented as a tuple of the bit-wise representation of integers $r$ and $s$. For two integers of length $n$ and $m$, the cost of their addition and multiplication is $O(m+n)$ and $O(m\cdot n)$, respectively.

To compute the cost of arithmetic in each iteration of the value-iteration algorithm, we define the cost of a transition $(v,w) \in E$ in the $k$-th iteration as
\begin{align*}
    \transwt_1(v,w)  = \gamma(v,w) \text{  and  }
    \transwt_k(v,w)  =  \gamma(v,w) + \frac{1}{d}\cdot\wt_{k-1}(v)  \text{ for } k >1
\end{align*}
Then, clearly, $\wt_k(v) = \max\{\transwt_k(v,w) | w \in vE\}$ if $v \in V_0$ and $\wt_k(v) = \min\{\transwt_k(v,w) | w \in vE\}$ if $v \in  V_1$. Since, we compute the cost of every transition in each iteration, it is crucial to analyze the size and cost of computing $\transwt$.

\begin{lemma}
\label{lem:transcostform}
Let $G$ be a quantiative graph game. Let $\mu>0$ be the maximum of absolute value of all costs along transitions. Let $d = \frac{p}{q}$ be the discount factor.
Then for all $(v,w )\in E$, for all $k>0$ 
\[
\transwt_k(v,w) = \frac{q^{k-1}\cdot n_1 + q^{k-2}p\cdot n_2 + \cdots + p^{k-1}n_{k}}{p^{k-1}}
\]
where $n_i \in \Z$ such that $|n_i| \leq \mu$  for all $ i \in \{1,\dots , k\}$.
\end{lemma}
\begin{proof}
Lemma~\ref{lem:transcostform} can be proven by induction on $k$.
\qed
\end{proof}

\begin{lemma}
\label{lem:transcostcost}
Let $G$ be a quantiative graph game. Let $\mu>0$ be the maximum of absolute value of all costs along transitions. Let $d = \frac{p}{q}$ be the discount factor.
For all $(v,w) \in E$, for all $k>0$ the cost of computing $\transwt_k(v,w)$   in the $k$-th iteration is $\O(k\cdot \log p \cdot \max\{\log \mu, \log p\})$.
\end{lemma}
\begin{proof}
We compute the cost of computing $\transwt_k(v,w)$ given that optimal costs have been computed for the $(k-1)$-th iteration.
Recall,
\begin{align*}
    \transwt_k(v,w) & =  \gamma(v,w) + \frac{1}{d}\cdot\wt_{k-1}(v) 
    \text{ } =     \text{ } \gamma(v,w) + \frac{q}{p}\cdot\wt_{k-1}(v) \\
    & = \gamma(v,w) + \frac{q}{p}\cdot \frac{q^{k-2}\cdot n_1 + q^{k-3}p\cdot n_2 + \cdots + p^{k-2}n_{k-1}}{p^{k-2}}
\end{align*}
for some $n_i \in \Z$ such that $|n_i| \leq \mu$. Therefore, computation of $\transwt_k(v,w)$ involves four operations:
\begin{enumerate}
    \item Multiplication of $q$ with $(q^{k-2}\cdot n_1 + q^{k-3}p\cdot n_2 + \cdots + p^{k-2}n_{k-1})$. The later is bounded by $(k-1)\cdot \mu \cdot p^{k-1}$ since $|n_i| \leq \mu$ and $p>q$. The cost of this operation is $\O(\log ((k-1)\cdot \mu \cdot p^{k-1})\cdot \log(p)) = \O(((k-1)\cdot \log p + \log \mu + \log (k-1))\cdot(\log p))$.
    
    \item Multiplication of $p$ with $p^{k-2}$. Its cost is $\O((k-2)\cdot (\log p)^2)$.
    
    \item Multiplication of $p^{k-1}$ with $\gamma(v,w)$. Its cost is $\O((k-1)\cdot \log p \cdot \log \mu)$.
    
    \item Addition of $\gamma(v,w)\cdot p^{k-1}$ with $q\cdot (q^{k-2}\cdot n_1 + q^{k-3}p\cdot n_2 + \cdots + p^{k-2}n_{k-1})$. The cost is linear in their representations. 
\end{enumerate}
Therefore, the cost of computing $\transwt_k(v,w)$ is $\O(k\cdot \log p \cdot \max\{\log \mu, \log p\})$.
\qed
\end{proof}

Now, we can compute the cost of computing optimal costs in the $k$-th iteration from the $k-1$-th iteration. 
\begin{lemma}
\label{lem:costupdateiteation}
Let $G$ be a quantiative graph game. Let $\mu>0$ be the maximum of absolute value of all costs along transitions. Let $d = \frac{p}{q}$ be the discount factor.
The worst-case complexity of computing optimal costs in the $k$-th iteration from the $(k-1)$-th iteration is $\O(|E|\cdot k \cdot \log \mu \cdot \log p)$.
\end{lemma}
\begin{proof}
The update requires us to first compute the transition cost in the $k$-th iteration for every transition in the game. Lemma~\ref{lem:transcostcost} gives the cost of computing the transition cost of one transition. Therefore, the worst-case complexity of computing transition cost for all transitions is $\O(|E|\cdot k \cdot \log p \cdot \max\{\log \mu, \log p\})$.

To compute the optimal cost for each state, we are required to compute the maximum transition cost of all outgoing transitions from the state. Since the denominator is same, the maximum value can be computed via lexicographic comparison of the numerators on all transitions. Therefore, the cost of computing maximum for all states is $\O(|E|\cdot k \cdot \log \mu \cdot \log p)$.

Therefore, total cost of computing optimal costs in the $k$-th iteration from the $(k-1)$-th iteration is $\O(|E|\cdot k \cdot \log p \cdot \max\{\log \mu, \log p\})$.
\qed
\end{proof}

Finally, the worst-case complexity of computing the optimal value of the quantitative game under bit-cost model for arithmetic operations is as follows:
\begin{theorem}
\label{thrm:optimizationbitcost}
Let $G = (V, \init, E, \gamma)$ be a quantitative graph game. 
Let $\mu>0$ be the maximum of absolute value of all costs along transitions. Let $d = \frac{p}{q}>1$ be the discount factor.
The worst-case complexity of computing the optimal value under bit-cost model for arithmetic operations is 
\begin{enumerate}
    \item $\O(|V|^4 \cdot |E| \cdot \log p \cdot \max\{\log \mu, \log p\})$ when $d \geq 2$, 
    \item $\O(\Big(\frac{\log(\mu)}{d-1} + |V|^2\Big)^2 \cdot |E| \cdot \log p \cdot \max\{\log \mu, \log p\})$ when $1<d < 2$.
\end{enumerate}
\end{theorem}
\begin{proof}
This is the sum of computing the optimal costs for all iterations.

When $d\geq 2$, it is sufficient to perform value iteration for $O(|V|^2)$ times (Theorem~\ref{thrm:vsquareiterations}).
So, the cost is $\O((1+2+3\cdot + |V|^2) \cdot |E| \cdot \log p \cdot \max\{\log \mu, \log p\})$. This expression simplifies to $\O(|V|^4 \cdot |E| \cdot \log p \cdot \max\{\log \mu, \log p\})$.

A similar computation solves the case for $1<d<2$.
\qed
\end{proof}

\subsubsection{Final remarks (Integer discount factor)}
Our analysis from Theorem~\ref{thrm:vsquareiterations} till Theorem~\ref{thrm:optimizationbitcost} show that VI will not scale well despite being polynomial in size of the graph game. 
Even when the discount factor $d>1$ is an integer ($d\geq 2 $), the algorithm requires $\Theta(|V|^2)$ iterations as Theorem~\ref{thrm:vsquareiterations} is tight.  In this case,  VI will be $\O(|V|^2\cdot |E|)$ and $\O(|V|^4\cdot |E|)$ under the unit-cost and bit-cost models for arithmetic operation, respectively (Theorem~\ref{thrm:optimizationunitcost} and Theorem~\ref{thrm:optimizationbitcost}, respectively).

From a practical point of view, implementations of VI will use the bit-cost model as they may rely on multi-precision libraries in order to to avoid floating-point errors. Therefore, VI for optimization will be expensive in practice. One may argue that the upper bounds from Theorem~\ref{thrm:optimizationbitcost} may be tightened. But it must be noted that  since VI requires $\Omega(|V|^2)$ iterations, even a tighter analysis will not significantly improve its performance in practice.

\section{Satisficing problem}
\label{Sec:Satisfice}

This section formally defines and investigates the {\em satisficing problem}. The intuition captured by satisficing is to  determine whether a player can guarantee that the cost of all plays will never exceeds a given threshold value. Hence,  satisficing  can be perceived as a decision variant of  optimization. 
In this section, we
prove that when the discount factor is an integer, the satisficing problem can be solved in {\em linear} time in size of the game graph. Therefore, showing that satisficing is more scalable and efficient than optimization. 
A key feature of our solution for satisficing is that our solution relies on purely automata-based methods and avoids  numerical computations.

We begin by formally defining the satisficing problem as follows:
\begin{Def}[Satisficing problem]
	Given a quantitative graph game $G$ and a threshold value $v \in \Q$, the {\em satisficing problem} is to determine whether player $P_1$ has a strategy such that for all possible resulting plays of the game, the cost of all plays is less than (or $\leq$) to the threshold $v$, assuming that the objectives of $P_0$ and $P_1$ are to maximize and minimize the cost of plays, respectively. 
\end{Def}

This section is divided into two parts.  \textsection~\ref{Sec:Comparator} describes the core technique that our solution builds on. At the heart of our solution lie {\em DS comparator automata}. Prior work on DS comparator automata have been limited to representing languages that accept a bounded weight sequence $A$ iff $\DSum{A}{d}$ $\R$ $0$.
This section generalizes the definition to accept weight sequence $A$ iff $\DSum{A}{d}$ $\R$ $v$, for an {\em arbitrary but fixed threshold} $v \in \Q$.
\textsection~\ref{Sec:Reduction} presents our complete solution. Here we use the DS comparators to reduce the satisficing problem to solving a safety or reachability game, when the discount factor is an integer. In this way, we solve satisficing with automata-based methods only.

\subsection{Foundations of DS comparator automata with threshold $v \in \Q$}
\label{Sec:Comparator}

This section generalizes DS comparison languages and DS comparator automata to arbitrary  rational threshold values $v \in \Q$.
It formally defines the notions, and studies their   safety or co-safety, and $\omega$-regular properties. 

We begin with  formal definitions:

\begin{Def}[DS comparison language with threshold $v \in \Q$]
For an integer upper bound $\mu>0$, discount factor $d >1$, equality or inequality relation $\mathsf{R} \in \{<,>,\leq, \geq, =, \neq\}$, and a threshold value $v \in \Q$ 
the {\em DS comparison language with upper bound $\mu$, relation $\mathsf{R}$, discount factor $d$ and threshold value $v$} is a language of infinite words over the alphabet $\Sigma = \{-\mu, \dots, \mu\}$ that accepts  $A \in \Sigma^{\omega}$ iff $\DSum{A}{d}$ $\mathsf{R}$ $v$ holds. 
\end{Def}

\begin{Def}[DS comparator automata with threshold $v \in \Q$]
For an integer upper bound $\mu>0$, discount factor $d >1$, equality or inequality relation $\mathsf{R} \in \{<,>,\leq, \geq, =, \neq\}$, and a threshold value $v \in \Q$ 
the {\em DS comparator automata with upper bound $\mu$, relation $\mathsf{R}$, discount factor $d$ and threshold value $v$} is an automaton that accepts the DS comparison language with upper bound $\mu$, relation $\mathsf{R}$, discount factor $d$ and threshold value $v$.
\end{Def}

For sake of succinctness, we refer to the DS comparison language and DS comparator automata by {\em comparison language} and {\em comparator}, respectively.


\subsubsection{Safety and co-safety of DS comparison languages}
\label{Sec:safetycosafety}

Our finding here is all DS comparison languages are either safety langauges or co-safety langagues. More interestingly, the only parameter that decides whether a comparison langauge is safety/co-safety is the equality or inequality relation $\R$. The values of the discount factor and threshold have no implication on this property. 

These observations are formally proven next:
Recall definitions of safety/co-safety languages and bad-prefixes from Chapter~\ref{ch:safetycosafety}.

\begin{theorem}
\label{Thrm:DSFull}
Let $\mu>1$ be the integer upper bound.  
For arbitrary discount factor $d>1$ and threshold value $v$
\begin{enumerate}
\item {\label{Item:Safety}}  DS comparison languages are safety languages for relations $R \in \{\leq, \geq, =\}$.
\item {\label{Item:CoSafety}} DS comparison language are co-safety languages for relations $R \in \{<, >, \neq\}$.
\end{enumerate}
\end{theorem}
\begin{proof}
	
Due to duality of safety/co-safety languages, it is sufficient to show that DS-comparison language with $ \leq$ is a safety language.

Let us assume that DS-comparison language with $ \leq$  is not a safety language. Let $W$ be a weight-sequence in the complement of DS-comparison language with $ \leq$  such that it does not have a bad prefix. 

Since $W $ is in the complement of DS-comparison language with $ \leq$, $\DSum{W}{d} > v$. By assumption, every $i$-length prefix $W[i]$ of $W$ can be extended to a bounded weight-sequence $W[i]\cdot Y^i$ such that $\DSum{W[i]\cdot Y^i}{d} \leq v$. 

\sloppy
Note that  $\DSum{W}{d} = \DSum{\pre{W}{i}}{d} + \frac{1}{d^i} \cdot \DSum{\suf{W}{i}}{d}$, and $\DSum{\pre{W}{i}\cdot Y^i}{d} = \DSum{\pre{W}{i}}{d} + \frac{1}{d^i} \cdot \DSum{Y^i}{d}$. The contribution of tail sequences $\suf{W}{i}$ and $Y^i$ to the discounted-sum of $W$ and $\pre{W}{i}\cdot Y^i$, respectively diminishes exponentially as the value of $i$ increases. In addition, since and $W$ and  $\pre{W}{i}\cdot Y^i$ share a common $i$-length prefix $\pre{W}{i}$, their discounted-sum values must converge to each other.
The discounted sum of $W$ is fixed and greater than $v$, due to convergence there must be a $k\geq  0$ such that $\DSum{\pre{W}{k}\cdot Y^k}{d} > v$. Contradiction. 
Therefore, DS-comparison language with $ \leq$  is a safety language. 

The above intuition is formalized below:

Since $\DSum{W}{d} > v$ and $\DSum{\pre{W}{i}\cdot Y^i}{d} \leq v$, the difference $\DSum{W}{d} - \DSum{\pre{W}{i}\cdot Y^i}{d} > 0$. 

By expansion of each term, we get $\DSum{W}{d}- \DSum{\pre{W}{i}\cdot Y^i}{d} = \frac{1}{d^i}(\DSum{\suf{W}{i}}{d} - \DSum{Y^i}{d}) \leq  \frac{1}{d^i}\cdot(\mod{\DSum{\suf{W}{i}}{d}}  + \mod{\DSum{Y^i}{d}} )$. Since the maximum value of discounted-sum of sequences bounded by $\mu$ is $\frac{\mu\cdot d}{d-1}$, we also get that $\DSum{W}{d}- \DSum{\pre{W}{i}\cdot Y^i}{d} \leq 2\cdot \frac{1}{d^{i}} \mod{ \frac{\mu\cdot d}{d-1}} $. 

Putting it all together, for all $i\geq 0$ we get
\[
0< \DSum{W}{d} - \DSum{W[i]\cdot Y^i}{d} \leq 2\cdot \frac{1}{d^{i}} \mod{\frac{\mu\cdot d}{d-1}} 
\]
As $i \rightarrow \infty$, $2\cdot \mod{\frac{1}{d^{i-1}}\cdot \frac{\mu}{d-1}} \rightarrow 0$. So, $\lim_{i \rightarrow \infty} (\DSum{W}{d} - \DSum{W[i]\cdot Y^i}{d}) = 0$. Since $\DSum{W}{d}$ is fixed, $\lim_{i \rightarrow \infty}\DSum{W[i]\cdot Y^i}{d} = \DSum{W}{d}$.

By definition of convergence, there exists an index $k\geq 0$ such that $\DSum{W[k]\cdot Y^k}{d}$ falls within the  $\frac{\mod{\DSum{W}{d}}}{2}$ neighborhood of $\DSum{W}{d}$. Finally since $\DSum{W}{d}> 0$, $\DSum{W[k]\cdot Y^k}{d} > 0$ as well. But this contradicts our assumption that for all $i \geq 0$, $\DSum{W[i]\cdot Y^i}{d} \leq 0$.

Therefore, DS-comparator with $ \leq$  is a safety comparator. 

\qed
\end{proof}


\subsubsection{$\omega$-regularity of DS comparison languages}
\label{Sec:regularity}

Next, we determine $\omega$-regularity of comparison languages with threshold value $v \in \Q$. The critical parameter in this case is the discount factor $d>1$. We observe that a comparison language  is $\omega$-regular iff the discount factor is an integer. Finally, while the threshold value does not affect the $\omega$-regularity of a comparison language, it impacts the size of the $\omega$-regular comparator automata (when discount factor $d >1$ is an integer).

First, we introduce notation. Since $v \in \Q$, w.l.o.g. let us assume that the threshold value is  represented by the regular expression $v = v[0]v[1]\dots v[m](v[m+1]v[m+2]\dots v[n])^{\omega}$. By abuse of notation, we denote both the regular expression $ v[0]v[1]\dots v[m](v[m+1]v[m+2]\dots v[n])^{\omega}$ and the value $\DSum{v}{d}$ by $v$. 

Our first result is that a comparison language with $v \in \Q$ is not $\omega$-regular if the discount factor is not an integer. The result follows immediately from prior work as it is known that comparison language with $v=0$ is not $\omega$-regular if the discount factor is not an integer~\cite{BCVlmcs2019}. Therefore:

\begin{theorem}
\label{thrm:notregularWithVal}
Let $\mu>0$ be the integer upper bound, $v \in \Q$ be the threshold value, and $\R \in \{<,>,\leq, \geq, =, \neq\}$ be an equality or inequality relation. Let the discount factor $d>1$ be a non-integer.
Then, the DS comparison language with $\mu$, $d$, $\R$, and $v$ is not $\omega$-regular.  
\end{theorem}

Next, we prove that DS comparison languages are $\omega$-regular if the discount factor $d>1$ is an integer. To prove this, we construct the B\"uchi automaton corresponding to the language. Since B\"uchi automata are closed under complementation, intersection and union, it is sufficient to construct the (B\"uchi) comparator automata for one equality or inequality relation $\R$. We do so for the relation  $\leq$.
The construction technique resembles the construction of comparator automata for threshold $v = 0$ for inequality relation $\leq$ presented in~\cite{BVCAV19}. Due to space constraints, we present critical lemma statements, theorem statement and high-level intuition. Please refer to~\cite{BVCAV19} or the supplemental material for details.  

Since we know that the comparison language for $\leq$ is a safety language, we begin by characterizing the bad-prefixes of the comparison language with threshold $v \in \Q$ and relation $\leq$.
For this, we introduce notation. Let $W$ be a {\em finite} weight-sequence. 
By abuse of notation, the discounted-sum of finite-sequence $W$ with discount-factor $d$ is defined as $\DSum{W}{d} = \DSum{W\cdot 0^{\omega}}{d}$.
The {\em recoverable-gap} of a finite weight-sequences $W$ with discount factor $d$, denoted $\gap{W}$, is its normalized discounted-sum: If $W = \varepsilon$ (the empty sequence), $\gap{\varepsilon} = 0$, and $\gap{W} = d^{|W|-1}\cdot \DSum{W}{d}$ otherwise~\cite{boker2014exact}.


\begin{lemma}
	\label{Lem:GapThreshold}
	Let $\mu>0$ be the integer upper bound,  $d >1$ be an integer discount factor, and the relation $\R$ be the inequality $\leq$. 
	Let $v \in \Q$ be the threshold value such that 
	 $v = v[0]v[1]\dots v[m](v[m+1]v[m+2]\dots v[n])^{\omega}$.
	Let $W$ be a non-empty, bounded, finite weight-sequence.   
	Then, weight sequence $W$ is a bad-prefix of the DS comparison language with $\mu$, $d$, $\leq$ and $v$ iff $\gap{W-\pre{v}{|W|}} >  \frac{1}{d}\cdot \DSum{\post{v}{|W|}}{d} + \frac{\mu}{d-1}$.
	
\end{lemma}
\begin{proof}
	
	\sloppy
	Let $W$ be a bad prefix. Then for all infinite length, bounded weight sequence $Y$ we get that $\DSum{W\cdot Y}{d} > v \implies \DSum{W}{d} + \frac{1}{d^{|W|}}\cdot \DSum{Y}{d} \geq \DSum{\pre{v}{|W|}\cdot\post{v}{|W|}}{d} \implies \DSum{W}{d} - \DSum{\pre{v}{|W|}}{d} > \frac{1}{d^{|W|}}\cdot (\DSum{\post{v}{|W|}}{d} - \DSum{Y}{d})\implies \gap{W-\pre{v}{|W|}} > \frac{1}{d} (\DSum{\post{v}{|W|}}{d} + \frac{\mu\cdot d}{d-1})$.
	
	Next, we prove that if a finite weight sequence $W$ is such that, the W is  a bad prefix. Let $Y$ be an arbitrary infinite but bounded weight sequence. Then $\DSum{W\cdot Y}{d} 
	= \DSum{W}{d} + \frac{1}{d^{|W|}}\cdot \DSum{Y}{d} 
	= \frac{1}{d^{|W|-1}}\cdot \gap{W} + \frac{1}{d^{|W|}}\cdot \DSum{Y}{d} 
	= \frac{1}{d^{|W|-1}}\cdot \gap{W} + \frac{1}{d^{|W|}}\cdot \DSum{Y}{d} + \frac{1}{d^{|W|-1}}\cdot (\gap{\pre{v}{|W|}} - \gap{\pre{v}{|W|}})
	$. By re-arrangement of terms we get that 
	$\DSum{W\cdot Y}{d}
	= \frac{1}{d^{|W|-1}}\cdot \gap{W-\pre{v}{|W|}} + \frac{1}{d^{|W|}}\cdot \DSum{Y}{d} + \frac{1}{d^{|W|-1}}\cdot \gap{\pre{v}{|W|}}$. Since 
	$ \gap{W - \pre{v}{|W|}} > \frac{1}{d}\cdot (\DSum{\post{v}{|W|}}{d} + \frac{\mu\cdot d}{d-1})$ holds, we get that 
	$\DSum{W\cdot Y}{d} 
	>  \frac{1}{d^{|W|}}\cdot (\DSum{\post{v}{|W|}}{d} + \frac{\mu\cdot d}{d-1}) +  \frac{1}{d^{|W|}}\cdot \DSum{Y}{d} + \frac{1}{d^{|W|-1}}\cdot \gap{\pre{v}{|W|}}$. Since minimal value of $\DSum{Y}{d} is \frac{-\mu\cdot d}{d-1}$, the inequality simplifies to  $\DSum{W\cdot Y}{d} > \frac{1}{d^{|W|-1}}\cdot \gap{\pre{v}{|W|}} + \frac{1}{d^{|W|}}\cdot \DSum{\post{v}{|W|}}{d}  \implies \DSum{W\cdot Y}{d} > \DSum{v}{d} = v$. Therefore, $W$ is a bad prefix.
	\qed
\end{proof}

Intuitively, Lemma~\ref{Lem:GapThreshold} says that if an infinite length weight sequence $A$ has a finite prefix for which its recoverable gap is too large, then the sequence $A$ will not be present in the coveted language. This way, if we track the recoverable gap value of finite-prefixes of $A$, Lemma~\ref{Lem:GapThreshold} gives a handle for when to reject $A$ from the coveted language. 
It also says that if the recoverable gap of a finite-prefix exceeds the bounds in the lemma statement, then there is no need to track the recoverable gap of other finite-prefixes any further. 


Similarly, we define {\em very-good prefixes} for the  comparison language with $\mu$, $d$, $\leq$ and $v$ as follows: A finite and bounded weight-sequence $W$ is a {\em very good} prefix for the aformentioned language if
for all infinite, bounded extensions of $W$ by $Y$,  $\DSum{W\cdot Y}{d}\leq v$. 
A proof similar to Lemma~\ref{Lem:GapThreshold} proves an upper bound for the recoverable gap of very-good prefixes of the language:

\begin{lemma}
	\label{Lem:GapThresholdVeryGood}
	Let $\mu>0$ be the integer upper bound,  $d >1$ be an integer discount factor, and the relation $\R$ be the inequality $\leq$. 
	Let $v \in \Q$ be the threshold value such that 
	 $v = v[0]v[1]\dots v[m](v[m+1]v[m+2]\dots v[n])^{\omega}$.
	Let $W$ be a non-empty, bounded, finite weight-sequence.   
	Weight sequence $W$ is a very good-prefix of DS comparison language with $\mu$, $d$, $\leq$ and $v$  iff $\gap{W-\pre{v}{|W|}} \leq \frac{1}{d}\cdot \DSum{\post{v}{|W|}}{d} - \frac{\mu}{d-1}$.
	
\end{lemma}

\begin{proof}
	Let $W$ be a very good prefix. Then for all infinite, bounded sequences $Y$, we get that $\DSum{W\cdot Y}{d} \leq v \implies \DSum{W}{d} + \frac{1}{d^{|W|}} \cdot \DSum{Y}{d} \leq v$. By re-arrangement of terms, we get that $\gap{W-\pre{v}{|W|}} \leq \frac{1}{d}\cdot \DSum{\post{v}{|W|}}{d} - \frac{1}{d}\cdot \DSum{Y}{d}$. Since maximal value of $\DSum{Y}{d} = \frac{\mu\cdot d}{d-1}$, we get that $\gap{W-\pre{v}{|W|}} \leq \frac{1}{d}\cdot \DSum{\post{v}{|W|}}{d} - \frac{\mu}{d-1}$.
	
	Next, we prove the converse. We know 
	$\DSum{W\cdot Y}{d} 
	= \DSum{W}{d} + \frac{1}{d^{|W|}}\cdot \DSum{Y}{d}
	= \frac{1}{d^{|W|-1}}\cdot \gap{W} + \frac{1}{d^{|W|}}\cdot \DSum{Y}{d}
	= \frac{1}{d^{|W|-1}}\cdot \gap{W} + \frac{1}{d^{|W|}}\cdot \DSum{Y}{d} +   \frac{1}{d^{|W|-1}}\cdot (\gap{\pre{v}{|W|}} - \gap{\pre{v}{|W|}})
	$. 
	By re-arrangement of terms we get that 
	
	\sloppy
	$\DSum{W\cdot Y}{d}
	=\frac{1}{d^{|W|-1}}\cdot \gap{W-\pre{v}{|W|}} + \frac{1}{d^{|W|}}\cdot \DSum{Y}{d} + \frac{1}{d^{|W|-1}}\cdot\gap{\pre{v}{|W|}}$. From assumption we derive that 
	$\DSum{W\cdot Y}{d} \leq \frac{1}{d^{|W|}}\cdot \DSum{\post{v}{|W|}}{d} - \frac{\mu}{d-1} + \frac{1}{d^{|W|}}\cdot\DSum{Y}{d}  + \DSum{\pre{v}{|W|}}{d}$. Since maximal value of $\DSum{Y}{d}$ is $ \frac{\mu}{d-1}$, we get that $\DSum{W\cdot Y}{d}\leq v$. Therefore, $W$ is a very good prefix.
	\qed
\end{proof}

Intuitively, Lemma~\ref{Lem:GapThresholdVeryGood} says that if an infinite-length weight sequence $A$ has a finite-prefix for which the recoverable gap is too small, $A$ is present in the coveted language. 
This condition gives a handle on when to accept the weight sequence $A$ by tracking on the recoverable gap of its finite prefixes. It also says that if the recoverable gap of a finite-prefix falls below the bound in Lemma~\ref{Lem:GapThresholdVeryGood}, there is no need to track recoverable gaps of finite-prefixes of $A$ any further. 

The question we need to answer to obtain the targeted B\"uchi automata is how to track the recoverable gaps of finite-prefixes using a finite amount of memory (or states). 
We already know that there are upper and lower bounds on which recoverable gaps are {\em interesting}: If a recoverable gap is too large (Lemma~\ref{Lem:GapThreshold}) or too small (Lemma~\ref{Lem:GapThresholdVeryGood}), the recoverable gap value does not need to be tracked any more. Now note that since the discount factor is an integer, in our case the recoverable gap value is always an integer (from definition of recoverable gap). Therefore, between the upper and lower bounds for recoverable established by Lemma~\ref{Lem:GapThreshold} and Lemma~\ref{Lem:GapThresholdVeryGood}, there are only finitely many candidate values of recoverable gaps. These finitely many values will form the states of the automata.   

The final question is how to determine the transition function of the automata. 
For this, observe that the recoverable-gap has an inductive definition i.e. $\gap{\varepsilon} = 0$, where $\varepsilon$ is the empty weight-sequence, and $\gap{W\cdot w} = d\cdot \gap{W} + w$, where  $w \in \{-\mu,\dots,\mu\}$. Therefore, transitions between states are established by mimicking the inductive definition. 

Therefore, the formal construction of the B\"uchi automata for comparison language with threshold $v$ with relation $\leq$ is given as follows:
\begin{theorem}
\label{thrm:regularWithVal}
Let $\mu>0$ be the integer upper bound,  $d >1$ be an integer discount factor, and the relation $\R$ be the inequality $\leq$. 
	Let $v \in \Q$ be the threshold value such that 
	 $v = v[0]v[1]\dots v[m](v[m+1]v[m+2]\dots v[n])^{\omega}$.
Then the DS comparison language for with $\mu$, $d$, $\leq$ and $v$ is $\omega$-regular. 
\end{theorem}
\begin{proof}

The ideas described above are formalized to construct the desired B\"uchi automaton as follows: 


For $i \in \{0,\dots,n\}$, let $\mathsf{U}_i = \frac{1}{d}\cdot \DSum{\post{v}{i}}{d} + \frac{\mu}{d-1}$ (from Lemma~\ref{Lem:GapThreshold}).

For $i \in \{0,\dots,n\}$, let $\mathsf{L}_i = \frac{1}{d}\cdot \DSum{\post{v}{i}}{d} - \frac{\mu}{d-1}$ (from Lemma~\ref{Lem:GapThresholdVeryGood}).

The B\"uchi automata $\mathcal{A} = (\State, \Start, \Sigma, \delta, \Final)$ is defined as follows:

\begin{itemize}

\item States $\State = \bigcup_{i=0}^n S_i \cup \{\mathsf{bad}, \mathsf{veryGood}\}$ where 
$S_i = \{(s, i) | s \in \{\floor{\mathsf{L}_i}+1, \dots , \floor{\mathsf{U}_i} \}\} $

\item Initial state $\Start = (0,0)$, Accepting states $\Final = S\setminus\{\mathsf{bad}\}$
\item Alphabet $\Sigma = \{-\mu, -\mu+1,\dots, \mu-1, \mu\}$
\item Transition function $\delta\subseteq \State \times \Sigma \rightarrow \State$ where $(s,a,t) \in \delta$ then:
	\begin{enumerate}
	\item \label{Trans:SelfLoop} If $s \in \{\mathsf{bad}, \mathsf{veryGood}\}$, then $t = s$ for all $a \in \Sigma$
	\item If $s$ is of the form $(p,i)$, and $a \in \Sigma$
		\begin{enumerate}
		\item \label{Trans:leq} If $d\cdot p+a-v[i] > \floor{\mathsf{U}_i}$, then $t = \mathsf{bad}$
		\item \label{Trans:geq} If $d\cdot p+a -v[i]\leq \floor{\mathsf{L}_i}$, then $t = \mathsf{veryGood}$
		\item \label{Trans:IntState} If $\floor{\mathsf{L}_i} <  d\cdot p+a -v[i] \leq \floor{\mathsf{U}_i}$, 
		\begin{enumerate}
		    \item If $i == n$, then $t = (d\cdot p+a - v[i], m+1)$
		    \item Else, $t = (d\cdot p+a - v[i], i+1)$
		\end{enumerate}
		\end{enumerate}
	\end{enumerate}
\end{itemize}
\qed
\end{proof}

\begin{corollary}
\label{coro:regularWithVal}
Let $\mu>0$ be the integer upper bound,  $d >1$ be an integer discount factor, and $\R $ be the equality or inequality relation.
Let $v \in \Q$ be the threshold value such that 
 $v = v[0]v[1]\dots v[m](v[m+1]v[m+2]\dots v[n])^{\omega}$.
Then the DS comparator automata with with $\mu$, $d$, $\leq$ and $v$  is a safety or co-safety automata with $\O(\frac{\mu\cdot n}{d-1})$ states. 
\end{corollary}

Note that the (deterministic) B\"uchi automaton constructed in Theorem~\ref{thrm:regularWithVal} is  a safety automaton~\cite{kupferman1999model}. Safety/Co-safety automata for all other relations can be constructed by simple modifications to the one constructed above. Further, note that the number of states depends on the value $v$ (number $n$). Lastly, this construction is tight, since prior work shows that when $v=0$, the automaton constructed above is the minimal automata for its language~\cite{BVCAV19}.



\subsection{Reduction of satisficing to solving safety or reachability games}
\label{Sec:Reduction}
We arrive at the core result of this section. We show that when the discount factor is an integer, satisficing is reduced to solving a safety or reachability game. This method is linear in size of the game graph, as opposed to the higher polynomial solutions of optimization via VI. Hence, satisficing is  more efficient and scalable alternative to analyze quantitative graph games for integer discount factors. 

The key idea behind this reduction is as follows:
Recall, the satisficing  problem is to determine whether player $P_1$ has a strategy that can guarantee that all resulting plays will have cost less than (or less than or equal to) a given threshold value $v\in\Q$. When the discount factor is an integer, the criteria for satisficing to hold is the same as ensuring that every resulting play is accepted by the safety/co-safety comparator automata with the same discount factor inequality relation, and threshold value. This lets us show that when the discount factor is an integer, the satisficing problem reduces to solving a {\em new game} obtained by taking the product of the quantitative graph game with the appropriate comparator. Finally, the resulting {\em new game} will be safety or reachability depending on whether the comparator is safety or co-safety, respectively.

The formal reduction is as follows:
Let $G = (V  = V_0 \uplus V_1, \init, E, \gamma)$ be a quantitative graph game. Let the maximum weight on the graph game be the integer $\mu>0$. Let $d>1 $ be the integer discount factor. Suppose, $v = v_0v_1\cdot v_{m}(v_{m+1}\cdots v_{n})^\omega$ is the rational threshold value for the satisficing problem, and suppose the inequality in the problem is $\R \in \{<, \leq\}$.

Then, first construct the $\omega$-regular comparator automaton for $\mu$, $d$, $\R$ and $v$. Let us denote it by $\mathcal{A} = (\State, \Start, \Sigma, \delta, \Final)$.  From Corollary~\ref{coro:regularWithVal} we know that $\A$ is a deterministic safety or co-safety automaton. 
Next,  construct structure $\GA = (W = W_0 \cup W_1, s_0 \times \mathsf{init}, \delta_W, \F_W)$ from $G$ and $\A$ as follows:
\begin{itemize}
    \item $W = V \times S$. Specifically $W_0 = V_0 \times S$ and $W_1 = V_1 \times S$. 
    
    Since $V_0$ and $V_1$ are disjoint, $W_0$ and $W_1$ are disjoint too. Let sets of states $W_0$ and $W_1$ belong to players $P_0$ and $P_1$ in $\GA$.
    
    \item Let $ s_0 \times \mathsf{init}$ be the initial state of $\GA$.
    
    \item Transition relation $\delta_W = W \times W$ is defined such that transition $(w, w') \in \delta_W$ where $w = (v,s)$ and $w' = (v', s')$ if 
    \begin{itemize}
        \item transition $(v, v') \in \delta$,
        \item $n = \gamma((v, v'))$, i.e., $n$ is the cost of the transition in $G$, and
        \item $(s, n, s') \in \delta_C$ is a transition in the comparator automata.
    \end{itemize}
    \item  $\F_W = V \times \F$
\end{itemize}
Finally, let $\GA$ be a reachability game if the comparator $\A$ is a co-safety  automaton, and let $\GA$ be a safety game otherwise. Correspondingly, $\F_W$ will be referred to as accepting states or rejecting states. 

Let us call a play in the quantitative graph game $G$ to be {\em winning play for $P_1$} if its cost relates to threshold $v$ by $\R$. Recall, the definition of winning plays in reachability and safety games (Chapter~\ref{ch:background}). Then, it is easy to show the following correspondence by relating plays in $\G $ and those in $\GA$, since $\A$ is deterministic:
\begin{lemma}
\label{lem:one-one-plays}
There is a one-one correspondence between plays in $G$ and plays in $\GA$. This correspondence preserves the winning condition. 
\end{lemma}

\begin{proof}
	
	(One-one correspondence) 
	Let $\rho = v_0v_1\dots$ be a play in $G$ with cost sequence $w_0w_1\dots$. 
	Let $s_0s_1\dots$ be the run of weight cost  of $w_0w_1\dots$ in the DS comparator. Note, that since the comparator is deterministic, there is a unique run for each finite/infinite cost sequence. 
	Then, play $\rho  = v_1v_2\dots$ in $G$ and corresponds to play $(v_0,s_0)(v_1,s_1)\dots$ in $\GA$.
	Similarly, one can prove that for each play $(v_0,s_0)(v_1,s_1)\dots$ in $\GA$ uniquely corresponds to play $v_0v_1\dots$ in $G$, where the correspondence is defined by the weight sequence shared by the run $s_0s_1\dots$ in the comparator and play $v_0v_1\dots$ in $G$.
	
	(Winning-condition preservation) 
	A play $\rho = v_0v_1v_2\dots$ is a winning run for player $P_1$ in $G$ iff its cost relates to threshold value $v$ by $\R$. So, let $n_1n_2\dots$ be the cost sequence of $\rho$ in G, then $\DSum{n_1n_2\dots}{d}$ $\R$ $v$ holds. 
	Let the run of weight sequence $n_0n_1\dots$ be $s_0s_1s_2\dots$ in comparator automata $ \A$. Then, run $s_0s_1s_2\dots $ is an accepting run in $\A$. As a result, the corresponding play in $\GA$, i.e. $s_0s_1s_2\dots$ must be a winning run in $G$.
	
	\qed
\end{proof}

Next, let us call a strategy of player $P_1$ a {\em winning strategy for $P_1$} in $G$  if the strategy guarantees that all resulting plays in $\GA$ are winning for $P_1$.
Then, it is easy to see that the winning play preserving, one-one correspondence between plays in $\GA $ and  $G$ obtained in Lemma~\ref{lem:one-one-plays} can be lifted to winning strategies between both games. As a result, we get the following:

\begin{lemma}
\label{thrm:reduction}

Let $G$ be a quantitative graph game, and $d>1$ be an integer discount factor. 
Then player $P_1$ has a winning strategy for the satisficing problem with threshold $v \in Q$ and relation $\R \in \{<, \leq\}$  iff player $P_1$ has a winning strategy in the reachability or safety game $\GA$ constructed as above.
\end{lemma}
\begin{proof}
	The proof lifts Lemma~\ref{lem:one-one-plays} from plays to strategies. The same idea of winning runs-preserving one-one correspondence is lifted to complete this proof. Due to the similarities, we skip the proof here. 
	\qed
\end{proof}

Then, the final statement for our reduction is:
\begin{theorem}
Let $G = (V, \init, E, \gamma)$ be a quantitative graph game. Let $\mu >0$ be the maximum of the absolute value of costs on transitions in $G$.
Let $d>1$ be an integer discount factor $v = v[0]v[1]\cdots v[m](v[m+1]\cdots v[n])^\omega$ be the rational threshold value, and $\R \in \{\leq, <\}$ be the inequality relation.  
Then, the complexity of satisficing on $G$ with threshold $v$ and inequality $\R$ is  $\O((V + E)\cdot \mu \cdot n)$.
\end{theorem}
\begin{proof}
From  Lemma~\ref{thrm:reduction} we know satisficing $G$ is equivalent to solving the reachability/safety game $\GA$
In order to determine its complexity of solving $\GA$, we need to know its number of states and edges. Clearly, there are $\O(|V|\cdot\mu\cdot n)$ states in $\GA$, since the comparator $\A$ has $\O(\mu\cdot n)$ states. We claim that $\GA$ has $\O(|E|\cdot \mu\cdot v)$ edges. For this, we observe that a state $(v, s)$ in $\GA$ has the same number of outgoing edges as state $v$ in $G$, as comparator $\A$ is deterministic. Since there are $\O(\mu\cdot n)$ copies of each state $v$ in $\GA$, there are $\O(|E|\cdot \mu\cdot n)$ edges in $\GA$. Therefore, the solving the reachability/safety game is $\O((|V| + |E|)\cdot \mu \cdot n)$.
\qed
\end{proof}

\sloppy
Therefore, we observe that when the discount factor is an integer, our comparator-based solution for satisficing is more efficient  than optimization via VI by degrees of magnitude in size of the game graph. 

\section{Satisficing via value iteration}
\label{Sec:VIForSatisfice}

So far, we have observed that for integer discount factors, comparator-based satisficing is more efficient than VI-based optimization. However, a crucial question still remains unanswered: Does the complexity improvement between VI for optimization to comparators for satisficing arise from (a) solving the decision problem of satisficing instead of optimization, or (b) adopting the comparator approach instead of numerical methods?
To answer this, we design and analyze a VI based algorithm for satisficing: If this algorithm shows improvement over VI for optimization, then the complexity gain would have occurred from solving satisficing. Otherwise, if this algorithm does not reflect any improvement, then the gain should  have come from adhering to comparator based methods. 

The VI based algorithm for satisficing is described as follows: Perform VI as was done for optimization in \textsection 3.1. Terminate the algorithm after whichever of the following two occurs first: (a) VI has performed the number of iterations as shown in Theorem~\ref{thrm:vsquareiterations}, or (b): In the $k$-th iteration, the threshold value $v$ falls outside of the interval defined in Lemma~\ref{lem:interval}.
In either case, one can determine how the threshold value relates to the optimal value, and hence determines satisficing.

Clearly, termination by condition (a). results in the same number of iterations as optimization itself (Theorem~\ref{thrm:vsquareiterations}). We show that condition (b) does not reduce the number of iterations either.
On the contrary, it introduces {\em non-robustness} to the algorithm's performance (See \textsection~\ref{Sec:Intro}). The reason is that the number of iterations is based on the distance between the threshold value and the optimal value.
So, if this distance is large, few iterations will be required since the interval length in Lemma~\ref{lem:interval} declines exponentially. But if the distance is small (say close to 0), then it will take as many iterations as taken by optimization itself. Formally,
\begin{theorem}
\label{thrm:nonrobust}
Let $d >1$ be an integer discount factor. 
Let $G $ be a quantitative graph game with state set $V$. Let $\opt$ be the optimal cost, and  $v \in \Q$ be the threshold value. 
Then number of iterations taken by VI for satisfaction is $\min\{O(|V|^2), \log{\frac{\mu}{|W|-v}}\}$.
\end{theorem}
Hence, VI for satisficing does not improve upon VI for optimization. 
This indicates comparator-based methods may be responsible for the  improvement. 

\section{Implementation and Empirical Evaluation}

The goal of the empirical analysis is to determine whether the practical performance of these algorithms resonate with our theoretical discoveries.

\subsubsection{Implementation details}

We implement three algorithms:
(a) $\optimize$: Optimization tool based on the value-iteration,
(b). $\compSatisfice$: Satisfying tool implementing our comparator-based method, and
(c)$\visatisfice$: Satisficing tool based on value iteration.
All tools have been implemented in \textsf{C++}. 
To overcome floating-point errors in $\optimize$ and $\visatisfice$, the tools invoke open-source arbitrary precision arithmetic library \textsf{GMP} (\textsf{GNU} Multi-Precision)~\cite{gmp}. 
Arithmetic operations in $\compSatisfice$ are contained within integers only. Therefore,  $\compSatisfice$ does not call the \textsf{GMP} library.

\subsubsection{Design and setup for experiments}
\begin{figure}[t]
\centering

\begin{minipage}{0.85\textwidth}
\centering
\includegraphics[width=\textwidth]{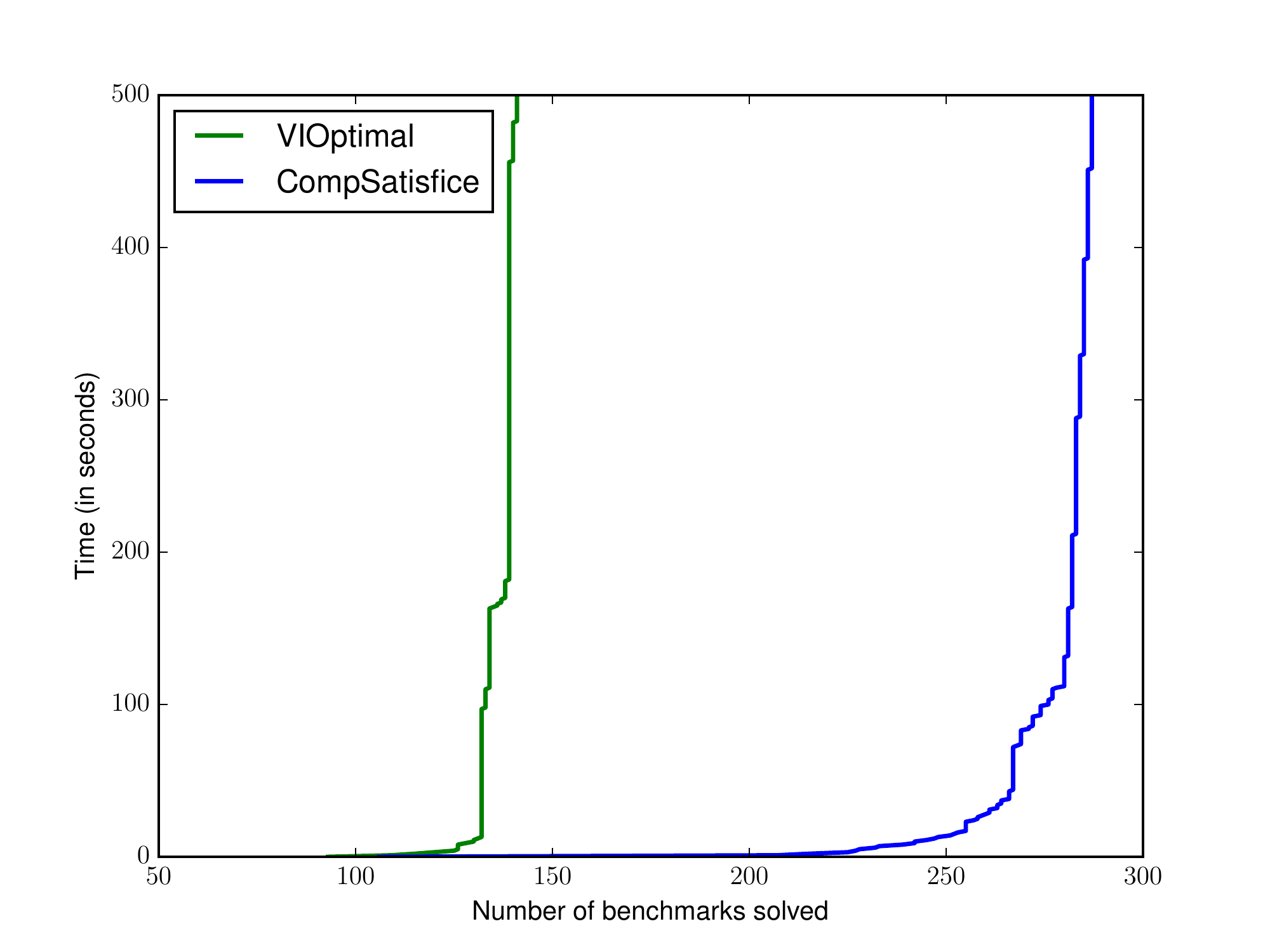}
\caption{Cactus plot.  $\mu=5, v=3$. Total benchmarks = 291}
\label{Fig:Cactus}
\end{minipage}
\hfill
\begin{minipage}{0.85\textwidth}
\centering
\includegraphics[width=\textwidth]{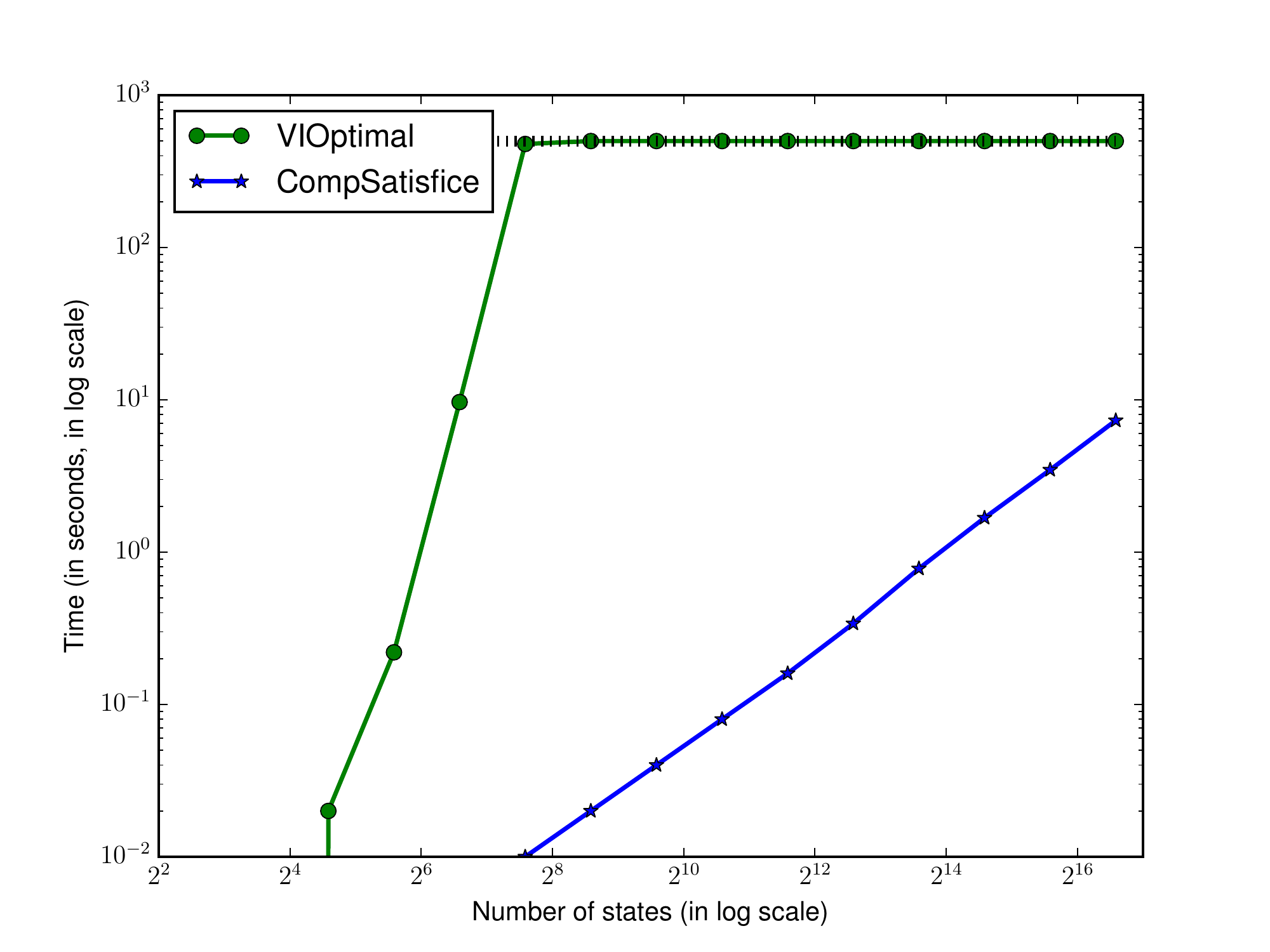}
\caption{Single counter scalable benchmark. $\mu=5, v=3$. Timeout = 500s.}
\label{Fig:Scale}
\end{minipage}

\end{figure}

Since empirical evaluations and applications of synthesis with quantitative constraints is an emerging area of research, there exist too few benchmarks of quantitative game graphs to generate a substantial amount of data to make inferences on algorithm performance.
To this end, 
our benchmarks are derived from specification used in synthesis from the temporal specifications. 
We obtain (non-quantitative) graph games by converting  temporal specifications to their automaton/game form. Then, we randomly assign an integer weight between $-\mu$ and $\mu$ (for a given $\mu>0$) over all transitions. 
In this way, the benchmarks we create retain the structural aspects of a game graph, and hence are a better fit than randomly generated graph games.

In all, we create 291 benchmarks.
We use temporal specifications 
used in  prior literature in synthesis from temporal specifications~\cite{camacho2018finite,zhu2017symbolic}. 
The state-of-the-art logic-to-automaton conversion tool $\mathsf{Lisa}$~\cite{aaai2020} has been deployed to obtain the automaton/graph game.
The number of states in our benchmark set ranges from 3 to 50000+. Discount factor $d = 2$, threshold $v$ ranges in  0-10.
All experiments were run on 8 CPU cores at 2.4GHz, 16GB RAM on a 64-bit Linux machine.

\subsubsection{Observations and Inferences}

\paragraph{$\compSatisfice$
outperforms $\optimize$}\footnote{Figures are best viewed online and in color}
 in runtime and consequently in number of benchmarks solved The cactus plot in Fig~\ref{Fig:Cactus} clearly indicates that $\compSatisfice$ is more efficient than $\optimize$.
In fact, a closer look reveals that all benchmarks solved by $\optimize$ have fewer than 200 states. In contrast,  $\compSatisfice$ solves all but two benchmarks. Thus solving benchmarks with many thousands of states.

To test scalability, we plot runtime of both tools on a set of scalable benchmarks. For integer parameter $i>0$, the $i$-th scalable benchmark has $3\cdot 2^i$ states.  
Fig~\ref{Fig:Scale} plots the observations in $\log$-$\log$ scale.
Thus, the slope of the straight line indicates the degree of polynomial (in practice). We see that in practice $\compSatisfice$ exhibits linear behavior (slope $\sim$1), whereas $\optimize$ is much more expensive (slope $>>1$) even for small values of weights and threshold. 

\paragraph{$\compSatisfice$ is more robust than $\visatisfice$.}

\begin{figure}[t]
\centering
\includegraphics[width=0.85\textwidth]{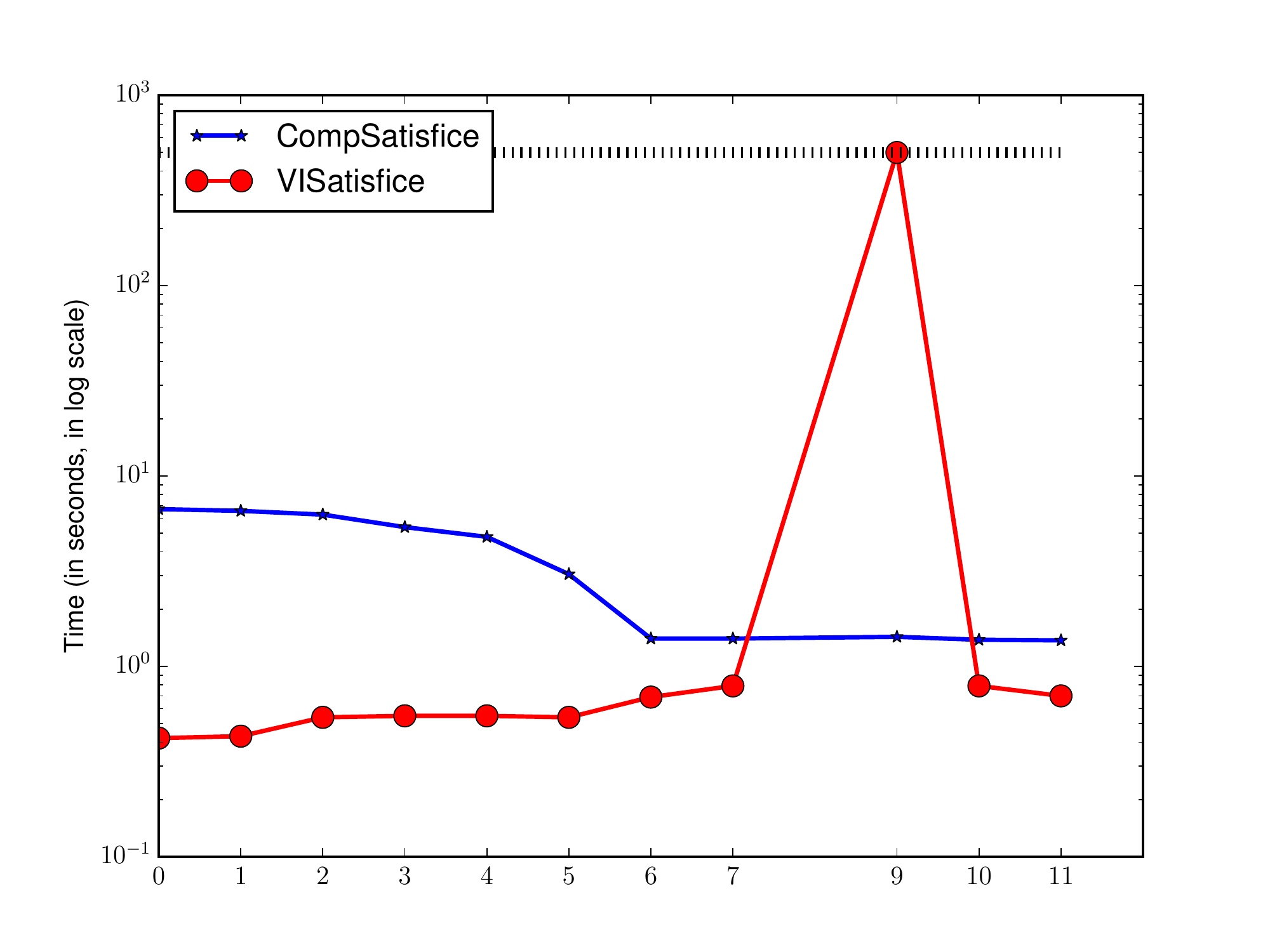}
\caption{Robustness. Fix benchmark, vary $v$.  $\mu = 5$. Timeout = 500s.}
\label{Fig:Robust}
\end{figure}

We compare $\compSatisfice$ and $\visatisfice$  as the threshold value changes. This experiment is chosen due to Theorem~\ref{thrm:nonrobust} which proves that $\visatisfice$ is non-robust. 
The goal is to see how $\visatisfice$ fluctuates in practice, while $\compSatisfice$ maintains steady performance owning to its low complexity. These are shown in Fig~\ref{Fig:Robust}.

\section{Quantitative games with temporally extended goals}
\label{Sec:Temporalgoals}

In several synthesis tasks, the objective is to solve a quantitative game while also satisfying a given temporal goal. So, the goal is to ensure that the system has a strategy that also ensures that the temporal objective is met along all possible plays resulting from the game.  
However, prior works have proven that optimal strategies may not exist under extension with temporal goals, rendering analysis by optimization incompatible with temporal goals~\cite{chatterjee2017quantitative}.
In this section, we show that our comparator-based solution for satisficing  can extend to all kinds of temporal goals.

Intuitively, a {\em quantitative game with temporal goals}  appends a quantitative game with  a {\em labeling function} that assigns states to atomic propositions. Therefore, each play of the game is associated with the sequence of atomic propositions. It is over this sequence of atomic propositions that the given temporal goal is evaluated. 
In this section, we show that our comparator-based approach to solve satisficing on quantitative games can be extended to all temporal goals expressed by {\em linear temporal logic (LTL)}~\cite{pnueli1977temporal}. More generally, our approach extends to all $\omega$-regular goals. This is in sharp contrast to existing work based on VI  which extends to a limited subclass of temporal goals only, namely safety objectives~\cite{wen2015correct}, since optimal solutions are not known to exist with $\omega$-regular objectives~\cite{chatterjee2017quantitative}.

Formally, a quantitative game with temporal goals, denoted by $\GT$, is the tuple  $(G,\ap, \L, \varphi)$ where $G$ is a quantitative game, $\ap$ is a finite set of  atomic propositions, $\L:V\rightarrow 2^{\ap}$ is a  labeling function, and $\varphi$ is an LTL formula over propositions $\ap$. Plays, cost-sequences, cost and strategies are defined as earlier for quantitative games. The difference is that plays in a quantitative game with temporal goals  are also accompanied with a sequence of propositions. If $\rho = v_0v_1v_2\dots $ is a play in game $\GT$, then $\rho_\ap = A_0A_1A_2\dots$ is the {\em proposition sequence} where $A_i = \L(v_i) \subseteq \ap$ for all $i \geq 0$. Given an LTL formula $\varphi$ over $\ap$, we say that a play {$\rho$ satisfies $\varphi$} if its corresponding proposition sequence $\rho_{\ap}$ satisfies $\varphi$.

Let us assume 
that the objectives of $P_0$ and $P_1$ are to maximize and minimize the cost of plays, respectively. 
Given threshold value $v \in \Q$ and inequality relation $\R\in\{\leq, <\}$, a strategy for $P_1$ is  said to be {\em winning} in a quantitative game with temporal goals $\GT$, threshold $v$ and relation $\R$  if the strategy is winning for $P_1$ in quantitative game $G$ with threshold $v$ and relation $\R$, and every resulting play in the game satisfies $\varphi$.

The satisficing problem with temporal goals is defined as follows:
\begin{Def}[Satisficing problem with temporal goals]
Given a quantitative game with temporal goals $\GT = (G, \ap, \L,\varphi)$, 
a threshold value $v \in \Q$, and an inequality relation $\mathsf{R} \in\{\leq, <\}$, the {\em satisficing problem with temporal goals} is to determine whether $P_1$ has a winning strategy
in $\GT$ with threshold $v$ and relation $\R$, 
assuming 
that the objectives of $P_0$ and $P_1$ are to maximize and minimize the cost of plays, respectively. 
\end{Def}

We show that solving the satisficing problem with temporal goals reduces to solving a parity game.
The key insight behind this is that when the discount factor is an integer then both the satisficing criteria and the temporal goal are represented by NBAs.  
As a result of which the criteria for the player to win a satisficing problem with temporal goals 
can be represented by a single  NBA that refers to the combination of both conditions. Thus, leading to a parity game. 

\begin{theorem}
\label{thrm:withtemporalgoals}
Let $\GT = (G, \ap, \L, \varphi) $ be a quantitative game with temporal goals. 
Let $V$ and $E$ refer to the set of states and edges in game $G$, and $\mu$ be the maximum of the absolute value of weights along transitions in $G$. Let $d >1$ be an integer discount factor, and $v = v[0]v[1]\cdots v[m](v[m+1]\cdots v[n])^\omega$ be the rational threshold value
\begin{itemize}
    \item Solving the satisficing problem with temporal goals reduces to solving a parity game. The size of the parity game is linear in $|V|$, $\mu$, $n$ and double exponential in  $|\varphi|$.
    \item If $\varphi$ can be represented by a (deterministic) safety/co-safety automata $\A$, then solving the satisficing problem with temporal goals is $\O((|V|+|E|)\cdot \mu\cdot n \cdot |\A|)$, where $|\A| = 2^{2^{\O(|\varphi|)}}$. 
\end{itemize}
\end{theorem}
\begin{proof}
This reduction can be carried out in two steps. In the first step, we reduce the quantitative graph game  $G$ to a reachability/safety game $\GA$ by means of the comparator automata for DS with integer discount factor, as done in Section~\ref{Sec:Reduction}. There is one key difference in constructing $\GA$. The labelling function of $\GT$ is extended to states in $\GA$ so that the label of state $(v,s)$ in $\GA$ is identical to the label of state $v$ in $\GT$.  
In the second step, we incorporate the temporal goal into the safety/reachability game $\GA$. 
For this, first the temporal goal $\varphi$ is converted into its equivalent {\em deterministic parity automaton (DPA)}~\cite{thomas2002automata}. Next, take the  product of the game $\GA$ with the DFA by synchronizing the labeling function of $\GA$ with transitions in the DPA. This follows standard operations for product constructions. The end result will be a parity game that is linear in the size of $\GA$ and the DPA for $\varphi$. Recall from Section~\ref{Sec:Reduction}, size of $\GA$ is linear in size of $G$, $\mu$ and $n$. Further, note that the DPA is double exponential in size of $\varphi$. Therefore, the final product game has size linear in $|V|$ and $\mu$, and double exponential in $|\varphi|$.

Finally, in the special case where the DPA for $\varphi$ is actually a deterministic safety/co-safety automaton, the resulting product game can be solved in time linear in the size of the final product game.  The reason behind this is that the final DPA will be a weak B\"uchi automata generated from the union of safety/co-safety automaton from the comparator, and a safety/co-safety autoamton from the temporal property (Theorem~\ref{Lem:IntersectSafetyAndCo}). Games with weak B\"uchi winning conditions can be solved in size linear to the underlying graph game. 

Finally, the proof of correctness is identical to that of Lemma~\ref{thrm:reduction}.
\qed
\end{proof}

\section{Chapter summary}

This work introduced the notion of satisficing for quantitative games with discounted-sum aggregation function. This is proposed as a decision variant of the optimization problem. Following a thorough analysis of both problems, we show that satisficing theoretical, empirical and practical advantages over the optimization problem. In particular, we show that when the discount factor is an integer, then the satisficing problem can be solved by automata-based solutions. Not only is our automata-based solution for satisficing more efficient than existing solutions for optimization in size of the game graph, our solutions perform more scalablely and robustly in empirical evaluations. 
Furthermore, the practical applicability of quantitative games with temporal goals can be handled naturally with our automata-based solution for satisficing as opposed to the non-existence of optimal solutions with temporal goals.
This works yet again presents the benefits of automata-based approaches in quantitative reasoning over traditional numerical approaches. 

While this chapter explored the automata-based solution for an integer discount factor only, in theory these results could be extended to non-integer discount factors as well as done in Chapter~\ref{ch:anytime}. This may come at the cost of a small approximation factor, but we can envision solving {\em approximate} satisficing using the approximate comparators for non-integer discount factors. Of course, the critical question here is to investigate the impact of the approximation in practical usage.

\chapter{Conclusion}
\label{ch:conclude}

\section{Concluding remarks}

This thesis introduces {\em comparator automata}, a novel technique based on automata for quantitative reasoning. The use of automata for quantitative reasoning is unique in its own, and infact counterintuitive as far as earlier work is concerned.

The challenges of lack of genralizability and separation-of-techniques adversely affect the practical viability of quantitative reasoning, and also prevent a clear understanding of similarities and dissimilarities across aggregate functions. The motivation behind the development  of comparator automata is to addresses these challenges. Our investigations on quantitative inclusion and quantitative games, especially an in-depth examination of the discounted-sum aggregation function, have established key theoretical results, improved upon the state-of-the-art in empirical performance, and indicated promise of approach in applications. 

An undercurrent in the progress of comparator-based approaches is that they have bridged quantitative reasoning to  qualitative reasoning. This is in stark contrast to separation-of-techniques in prior work where the two are segregated from one another. 
As demonstrated in this thesis, this digression will prove to be of immense importance in the long run, since it allows quantitative reasoning to leverage the advances in the theory, practice and applications of qualitative reasoning. 
To the best of our knowledge, this thesis is the first work to do so.

\section{Future work}

\subsection*{Theory: Building on the foundations.} 

$\omega$-regular comparators advocate for the treatment of quantitative properties as any other $\omega$-regular property. As a result, it is likely that problems that have found success with an $\omega$-regular property, such as a temporal logic~\cite{pnueli1977temporal}, could do so with $\omega$-regular comparators as well. The study of quantitative games with imperfect information gives a glimpse of the hypothesis. One avenue of interest are probabilistic systems. 
So far, probabilistic verification and synthesis involves techniques from linear programming, optimization, solving sets of equalities and inequalities, and so on.  Whether comparator-based approaches can substitute some of these operations is an open question. 

Another line of research is to identify necessary and sufficient conditions for aggregate functions to have $\omega$-regular comparators. So far, aggregate functions have been studied on a case-by-case basis. The broadest classification we have currently is that it is sufficient if an aggregate function is $\omega$-regular. A better understanding of necessary and sufficient conditions could contribute to clearer understanding of aggregate functions.

\subsection*{Practice: Scaling further and beyond.}

{\em Symbolic methods} have been gaining in popularity in qualitative reasoning due to their scalability and efficiency advantages over explicit methods\cite{burch1992symbolic}. In symbol methods, the state space is  typically represented with a logarithmic number of bits. 
Following this success, the symbolic reasoning has been adopted in quantitative reasoning. 
In most cases, however, the symbolic structures extend their qualitative counterparts. For example, Binary Decision Diagram (BDDs)~\cite{bryant1986graph} are a popular symbolic data-structure in qualitative domains. These have been extended to MTBDDs and ADDs to reason about quantitative properties. 
Since the development of quantitative symbolic data-structures and their algorithms are still in nascent stages, relative to their qualitative counterparts, their full impact is yet to be discovered. 

While investigations on quantitative symbolic methods are underway, comparators offer an alternate perspective. Since comparators are automata, existing symbolic methods such as BDD, SAT and SMT can be applied. 
Albeit promising, only a comparative analysis can reveal the strengths and weakness of
quantitative and comparator-based qualitative symbolic methods, and expose areas for improvement.

\subsection*{Application: White-boxing RL engines.}

Owing to the unprecedented rise of machine learning algorithms, reinforcement learning (RL) has come to the forefront in the design of controllers for robotics, autonomous vehicles, and several control-related domestic appliances. Due to their safety-critical nature of these applications, it has become crucial to {\em white-box} the underlying RL algorithms. In response, the formal methods community has begun looking into the quantitative and qualitative aspects of RL algorithms. 

This direction is particularly exciting for comparator automata. First of all, several RL engines make use of discounted-sum. Secondly, with comparators we can combine quantitative and qualitative properties with rigorous guarantees, as demonstrated in this thesis. Having said that, the capabilities of comparators will have to be extended further to be applicable to the domain. For example, RL algorithms have to work under the assumption that the reward function is known partially only.  To address this, comparators will have to be effective at verification and synthesis under partial information. Similarly, one could identify numerous ways in which automata-theoretic reasoning could be of use in the space of white-boxing RL engines. 

\subsubsection*{}

In all, this thesis has introduced a novel framework for the formal analysis of quantitative systems. The evidence collected throughout the thesis indicates promise of our approach. The scope in future directions of research spreads across the wide spectrum from theoretical results to current applications. In all we believe that this thesis has managed to work at the tip of automata-based reasoning of quantitative systems. The iceberg is yet to be discovered. 

\appendix

\chapter{Appendix of miscellaneous results}

\section{Discounted-sum is an $\omega$-regular function}

We use the $\omega$-regular comparator for DS-aggregate function for integer discount-factor to prove that discounted-sum with integer discount-factors is an $\omega$-regular aggregate function. 

\begin{theorem}
	\label{Cor:RationalAggregate}
	Let $d>1$ be a non-integer, rational discount-factor. The discounted-sum aggregate function with discount factor $d$ is not $\omega$-regular.
\end{theorem}
\begin{proof}
	Immediate from Lemma~\ref{lem:cutpoint} and Theorem~\ref{thrm:functionthencomparator}. 
\end{proof}

\begin{theorem}
	\label{thm:DSRegular}
	Let $d>1$ be an integer discount-factor. The discounted-sum aggregate function with discount-factor $d$ is $\omega$-regular under base $d$. 
\end{theorem}
\begin{proof}
	We define the discounted-sum aggregate function automaton (DS-function automaton, in short): For integer $\mu>0$, let $\Sigma = \{0,1,\dots \mu\}$ be the input alphabet of DS-function, and  $d>1$ be its integer base. B\"uchi automaton $\A^\mu_d$ over alphabet $\Sigma\times\mathsf{AlphaRep}(d)$ is a DS-function automaton of type $\Sigma^\omega \rightarrow \Re$ if for all $A \in \Sigma^\omega$, $ (A, \mathsf{rep}(\DSum{A}{d}), d) \in \A^\mu_d$. Here we prove that such a $\A_d^\mu$ exists. 
	
	Let $\mu>0$ be the integer upper-bound. Let $\A^=_d$ be the DS-comparator for integer discount-factor $d>1$ for relation $=$. Intersect  $\A^=_d $ with the B\"uchi automata consisting of all infinite words from alphabet $\{0,1\dots \mu\}\times \{0,\dots,d-1\}$. The resulting automaton $\mathcal{B}$ accepts $(A,B)$ for $A \in \{0,\dots,\mu\}^\omega$ and $B\in  \{0,\dots,d-1\}^\omega$ iff $\DSum{A}{d} = \DSum{B}{d}$.  
	
	Since all elements of $B$ are bounded by $d-1$, $\DSum{B}{d}$ can be represented as an $\omega$-word as follows: Let $B = B[0],B[1]\dots$, then  its  $\omega$-word representation in base $d$  is given by $ +\cdot (\mathsf{Int}(\DSum{B}{d}, d), \mathsf{Frac}(\DSum{B}{d}, d))$ where $\mathsf{Int}(\DSum{B}{d}, d) = B[0] \cdot 0^\omega$ and $ \mathsf{Frac}(\DSum{B}{d}, d) = B[1],B[2]\dots$. This transformation of integer sequence $B$ into its $\omega$-regular word form in base $d$ can be achieved with a simple transducer $\mathcal{T}$. 
	
	Therefore, application of transducer $\mathcal{T}$ to B\"uchi automaton $\mathcal{B}$ will result in a B\"uchi automaton over the alphabet $\Sigma\times\mathsf{AlphaRep}(d)$ such that for all $A \in \Sigma^\omega$ the automaton accepts $(A, \mathsf{rep}(\DSum{A}{d},d))$. This is exactly the DS-function automaton over input alphabet $\Sigma$ and integer base $d>1$. Therefore, the discounted-sum aggregate function with integer discount-factors in $\omega$-regular.
	\qed  
\end{proof}
Recall, this proof works only for the discounted-sum aggregate function with integer discount-factor. In general, there is no known procedure to derive a function automaton from an $\omega$-regular comparator (Conjecture~\ref{Conjecture:comparatortofunction}). 

\section{Connection between discounted-sum and sum}

\begin{theorem}
	\label{lem:nSquareBound}
	Let integer $\mu>0$ be the upper-bound. Let $d = 1+2^{-k}$  be the discount-factor for rational-number $k>0$.
	Let $A$ be a  non-empty, bounded, finite, weight-sequence. Then
	$\mod{\FSum{A} - \DSum{A}{d}} < 2^{-k}\cdot\mu\cdot |A|^2$
\end{theorem}
\begin{proof}
We prove the desired bound on the difference between discounted-sum and sum. Let $n>0$ be the length of $A$, and $\varepsilon = 2^{-k}$.
\begin{align*}
|{\FSum{A} - \DSum{A}{d}}| &= \mod{\Sigma_{i=0}^{n-1} A_i - \frac{A_i}{d^i}} \\
&\leq \Sigma_{i=0}^{n-1}\mod{A_i} \cdot \mod{1 - \frac{1}{d^i}} \\
& \text{  By Cauchy-Schwarz inequality, we get} \\ 
			&\leq \sqrt{(\Sigma_{i=0}^{n-1} A_i^2) \cdot (\Sigma_{i=0}^{n-1} (1 - \frac{1}{d^i})^2)}\\
			& \text{ Since } A \text{ is bounded by } \mu \\
			&\leq \sqrt{n\cdot \mu^2 \cdot (\Sigma_{i=0}^{n-1} (1 - \frac{1}{d^i})^2)} = \sqrt{n\cdot \mu^2 \cdot (\Sigma_{i=0}^{n-1} (1 - \frac{1}{(1 + \varepsilon)^i})^2)} \\
			&\leq \sqrt{n\cdot \mu^2 \cdot (\Sigma_{i=0}^{n-1} (1 - \frac{1}{(1 + \varepsilon)^{(\frac{1}{\varepsilon})\cdot \varepsilon\cdot i}})^2)} \\
			&\text{  Since } (1+x)^{\frac{1}{x}} < e^x \text{, for all } 0<x\leq 1  \\
			&< \sqrt{n\cdot \mu^2 \cdot (\Sigma_{i=0}^{n-1} (1 - \frac{1}{e^{\varepsilon\cdot i}})^2)}   \\ 
			&\leq \sqrt{n\cdot \mu^2 \cdot (\Sigma_{i=0}^{n-1} (1 - e^{- \varepsilon\cdot i})^2)} \\
			&\leq \sqrt{n\cdot \mu^2 \cdot n\cdot (1 - e^{- \varepsilon\cdot n})^2)} \text{ as } i<n \\
			& \text{ Since } 1 - x < e^{-x} \text{ holds for all } x>0  \\
			&< \sqrt{n\cdot \mu^2 \cdot n\cdot (\varepsilon\cdot n)^2)} = \varepsilon \cdot \mu \cdot n^2 
			\end{align*}
\qed
\end{proof}

The bound above is simply an upper bound. It will be good to tighten the bound to $\O(\varepsilon \cdot \mu \cdot n)$.


\bibliographystyle{plain}
\bibliography{mybib.bib}
\end{document}